\documentclass{aa}
\pdfoutput=1
\usepackage{mathtools}
\usepackage{multirow}
\usepackage{comment}
\usepackage{siunitx}
\usepackage{booktabs}
\usepackage{graphicx}
\usepackage{menukeys}
\usepackage{algorithm}
\usepackage{ulem}
\usepackage[noend]{algpseudocode}
\usepackage{ccaption}
\usepackage{caption}
\usepackage{txfonts}
\usepackage{subfigure}
\usepackage{cuted}
\usepackage{float}
\usepackage{multicol}
\usepackage{placeins}

\newcommand\blfootnote[1]{%
  \begingroup
  \renewcommand\thefootnote{}\footnote{#1}%
  \addtocounter{footnote}{-1}%
  \endgroup
}

\graphicspath{{Fig/}}
\usepackage[colorlinks=true, allcolors=blue]{hyperref}
\DeclareCaptionFormat{continued}{ #1~(continued)}
\newcommand{\iras}{IRAS\,16293--2422 }
\newcommand{\irasE}{16293E}

\newcommand{\U}{\mathrm}

\makeatletter
\renewcommand*\aa@pageof{, page \thepage{} of \pageref*{LastPage}}
\makeatother
\newcolumntype{L}[1]{>{\raggedright\arraybackslash}p{#1}}
\newcolumntype{C}[1]{>{\centering\arraybackslash}p{#1}}
\newcolumntype{R}[1]{>{\raggedleft\arraybackslash}p{#1}}
\DeclareSIUnit \parsec {pc}
\DeclareSIUnit \astrounit {au}
\begin{document} 

   \title{The molecular environment of the solar-type protostar IRAS\,16293--2422\footnotemark}
   
   \author{K. Angelique Kahle\inst{1}
          \and
          Antonio Hern\'andez-G\'omez\inst{1}
          \and
          Friedrich Wyrowski\inst{1}
          \and
          Karl M. Menten\inst{1}
          }

   \institute{\inst{1} Max-Planck-Institut f\"{u}r Radioastronomie, Auf dem H\"{u}gel 69, 53121 Bonn, Germany
   }

   \date{Received September 06, 2022; accepted December 04, 2022}
  \titlerunning{The molecular environment of the solar type protostar IRAS\,16293--2422}
   \authorrunning{K. A. Kahle et al.}
  
\abstract{Studying the physical and chemical processes leading to the formation of low-mass stars is crucial for understanding the origin of our Sun and the Solar System. In particular, analyzing the emission and absorption lines from molecules to derive their spatial distribution in the envelopes of young stellar objects is a fundamental tool to obtain information on the kinematics and chemistry at the very early stages of star formation.}%context
  {In this work we aim to examine in detail the spatial structures and molecular abundances of material surrounding the very well-known low-mass binary protostar IRAS\,16293--2422 and the prestellar core 16293E, which are embedded in the Lynds 1689\,N dark cloud. This analysis is performed to obtain information on the physical and chemical properties of these young objects and their interaction with the molecular outflows present across the region.}%aims
  {We have used the LAsMA heterodyne array installed on the Atacama Pathfinder EXperiment (APEX) 12 meter submillimeter telescope to image a region of about 0.12\,$\times$\,0.12\,pc$^2$ around IRAS\,16293--2422 and 16293E and to study their molecular environment covering $\SI{45.6}{\giga\hertz}$ in a frequency range from $\SI{277}{\giga\hertz}$ to $\SI{375}{\giga\hertz}$. We have also used the APEX FLASH+ receiver to observe and search for molecular lines in a frequency range between  $\SI{476}{\giga\hertz}$ to $\SI{493}{\giga\hertz}$}%methods
  {We have identified 144 transitions from 36 molecular species, including isotopologues. This is the first time that such a large number of species have been mapped at large scales simultaneously in this region. The maps reveal the envelope to have a complex morphology around the cloud cores and the emission peaks known as E1, E2, W1, W2, and HE2, including the outflow structure arising from IRAS\,16293--2422. Using several transitions of para-H$_2$CO, we have derived new lower limits for the kinetic temperatures toward IRAS\,16293--2422 and the surrounding emission peaks. Based on these temperatures, new column densities for all detected species were derived around the cloud cores and all emission peaks using the radiative transfer codes CLASS-Weeds, CASSIS, and RADEX. We derived H$_2$ volume densities in Lynds 1689\,N based on ortho-H$_2$CO transitions with different upper level energies, varying between $5\times 10^{6}$\,cm$^{-3}$ and $\SI{63}{\kelvin}$ at IRAS\,16293--2422 to values on the order of $1\times 10^{6}$\,cm$^{-3}$ and $\SI{35}{\kelvin}$ at the other emission peaks.} %results
  {Our new observations further confirm the scenario of an outflow arising from IRAS\,16293--2422 interacting with the prestellar core 16293E. This is inferred from the velocity and linewidth gradient shown by several deuterated species closer to the outflow-core interaction region in 16293E.
  We observe a large-scale velocity gradient across the molecular cloud which coincides with the rotation of the envelope around IRAS\,16293--2422 reported previously in the literature. A comparison with JCMT SCUBA-2 $\SI{450}{\micro\meter}$ dust continuum maps and our data suggests that emission peak W2 may be related to a colder dust source rather than a shocked region. The newly derived column densities and temperatures for different species, combined with the molecular spatial distribution in all sources, indicate clear chemical differences between the protostellar source, the prestellar core and the shocked positions as a result of the diverse physical conditions at different locations in this region.}%conclusions
 \keywords{ISM: molecules – stars: individual (IRAS\,16293–-2422).}

\maketitle
%
%________________________________________________________________
\section{Introduction}

Studying the physical and chemical processes involved in the early stages of low-mass star formation is crucial to understand the conditions that gave birth to the formation of stars such as our Sun. In particular, the analysis of the molecular inventory in low-mass protostars can shed light on the kinematics and the chemistry of these young stages.
\blfootnote{$^\star$ All integrated intensity maps shown here are available in FITS format at the CDS via anonymous ftp to cdsarc.cds.unistra.fr (130.79.128.5) or via \url{https://cdsarc.cds.unistra.fr/cgi-bin/qcat?J/A+A/}}

One of the most studied sources that has been at the center of a plethora of molecular line studies is the archetypical Class 0 so-called protostar\footnote{A protostar may be defined as ``a contracting mass of gas which represents an early stage in the formation of a star, before nucleosynthesis has begun.''; see: {\small \url{https://languages.oup.com/google-dictionary-en/}}. Strictly speaking, at least two of the compact sources within the IRAS\,16293--2422 condensation arguably do not meet this criterion and should rather be referred to as YSOs. However, following common usage, we loosely retain the protostar nomenclature for both the IRAS source and the embedded sources.} IRAS\,16293--2422, first  discussed by \citet{walker1986}.
This far-infrared source is a dust condensation located in the Lynds 1689\,N dark cloud (L1689N) which is part of the Ophiuchus dark cloud and star-forming complex. It harbors a hierarchical  system of low-mass young stellar objects (YSOs). A trigonometric parallax measured with very long baselines interferometry (VLBI) of the H$_2$O masers associated with one of the compact sources yielded a distance of $141^{+30}_{-21}$\,pc \citep{dzib2018}. 

The composite source consists of two main condensations \citep{wootten1989, Mundy1992} called IRAS\,16293--2422 A and IRAS\,16293--2422 B (sources A and B from now on) separated by $5\si{\arcsecond}$ ($\sim$705\,$\si{\astrounit}$). Source A has been found to be a double system itself \citep{wootten1989, maureira2020} which consists of the protostars A1 and A2 separated by $0\rlap{.}''3$ ($\sim$42\,$\si{\astrounit}$). On the other hand, the nature of source B still is not yet fully constrained, although several studies suggest that it is in a very early stage of star formation \citep{pineda2012, hernandez-gomez2019}.
The \iras A/B system is surrounded by an extended envelope with about a 6000--8000\,au radius \citep{Schoier2002, crimier2010, jacobsen2018} and has a mass of about 4--6\,M$_{\sun}$ \citep{jacobsen2018,ladjelate2020}. 16293E, a very well-known prestellar core also embedded in L1689N, is located $1\rlap{.}'5$ ($\sim$12700\,$\si{\astrounit}$) east (see Fig.~\ref{fig:nnh+}) of this extended envelope \citep{stark2004}.

A multilobe molecular outflow system with scales up to $\SI{0.2}{\parsec}$ can be observed toward \iras \citep{walker1988, mizuno1990}. Two bipolar outflows have been reported to arise from source A (see Fig.~\ref{fig:CO6-5}), one of them extending in the east-west (E-W) direction (P.A. $= 110^{\circ}$) and the other in the northeast-southwest (NE-SW) direction (P.A. $= 60^{\circ}$) \citep{castets2001,hirano2001,yeh2008}. There is a third outflow with an extent of $\SI{0.01}{\parsec}$ in the northwest-southeast (NW-SE) direction \mbox{(P.A. $= 145^{\circ}$)} that seems to be arising from source A and interacting with material around source B \citep{girart2014}. While some authors suggest that source B could be indeed producing one of these outflows \citep{loinard2013, Oya2018}, there is still debate on their true origin. Interestingly, one of these outflows arising from \iras A/B seems to be interacting also with \irasE, modifying the chemical abundances in this source \citep[e.g.,][]{lis2016}.

The dust emission and the molecular content of both sources, A and B, have been studied in great detail with high angular resolution (up to $\approx 0\rlap{.}''1$) with the Very Large Array (VLA) \citep[e.g.,][]{chandler2005, hernandez-gomez2019}, the SubMillimeter Array (SMA) \citep{jorgensen2011} and the Atacama Large Millimeter/submillimeter Array (ALMA) \citep{pineda2012,jorgensen2016}. In particular, the imaging spectral line surveys toward A and B \citep{jorgensen2011, jorgensen2016} with the SMA and ALMA, most recently the ALMA protostellar interferometric line survey (PILS) (at $0{\rlap .}''5$ resolution), reveal an immense number of lines from numerous species, including ``complex organic molecules'' toward both A and B \citep{jorgensen2020}. 

On larger scales, the molecules in the envelope have been studied in the course of the IRAS\,16293--2422 millimeter and submillimeter spectral survey (TIMASSS) \citep{caux2011} using pointed single-dish observations with the IRAM 30 meter telescope and the 15 meter James Clerk Maxwell Telescope (JCMT) at coarser angular resolution ($10''-30''$).
Nevertheless, our knowledge of the spatial distribution of many molecules in the envelope and its surroundings is still quite limited as published studies deal only with single pointing observations (usually some centroid position between A1, A2 or B within a few arc second variation). In contrast, the few larger scale studies on the molecular environment further from the sources mainly focus on a small, selected number of lines \citep[e.g.,][]{castets2001,lis2002}.
Indeed, a comprehensive systematic broad-band study of the more extended molecular environment around \iras A/B and E is still missing in the literature.

In this work we present a study of the molecular environment of \iras A/B and E based on single-dish APEX observations. We have imaged a $3\rlap{.}'5\times{3\rlap.}'5$ sized region ($0.12\,\times\,0.12$~pc$^2$) centered between \iras and 16293E in a large number of molecular lines covering a significant part of the $\SI{870}{\micro\meter}$ atmospheric window.

The paper is organized as follows: In Sect.~\ref{ch:observ} we describe the details of the observations and the data reduction. In Sect.~\ref{ch:results} we present the particulars of the identified molecules and show the maps with the molecular distribution in \iras A/B and 16293E. In Sect.~\ref{ch:analysis} we present a detailed analysis of the physical properties of both sources and nearby related emission peaks based on temperature and column and volume density determinations from radiative transfer models. In Sect.~\ref{ch:discussion} we discuss the consequences of our findings and compare the molecular line morphologies with dust continuum maps. Finally in Sect.~\ref{ch:summary} we summarize our results. Extensive appendices give detailed information on the observed line detections and derived quantities and present maps of the line emission.

\section{Observations and data reduction}\label{ch:observ}

The observations were carried out with the Atacama Patfinder EXperiment 12 meter diameter submillimeter telescope (APEX) located on the Llano de Chajnantor in the Chilean High Andes at an altitude of 5107\,$\si{\meter}$ \citep{gusten2006atacama}.

\subsection{Observations with LAsMA and FLASH+}
The data were taken under project M-0102.F-9519A-2018 (P.I. Karl M. Menten) during several runs between 2019 July 19--28  with the Large APEX subMillimeter Array (LAsMA) multibeam heterodyne receiver \citep{guesten2008} in the frequency range between 277 and $\SI{375}{\giga\hertz}$. The LAsMA receiver is a 7 pixel multibeam sideband separating (2SB) receiver with the lower and upper sideband (LSB and USB, respectively), each 4~GHz wide and the centers of the bands separated by 12 GHz.
Inspecting the data, we found calibration problems with the sixth LAsMA pixel. As consequence, we did not use data taken with this pixel.

The observations cover a region of $3\rlap{.}'5\times{3\rlap.}'5$ centered on a position offset ($\delta =\SI{50}{\arcsecond}$, $\alpha = \SI{-10}{\arcsecond}$) from IRAS\,16293 A/B at $\U{RA(J2000)} = 16^\U{h}32^\U{m}22.78^\U{s}$, $\U{DEC(J2000)} = -24^\circ28\si{\arcminute}38.7\si{\arcsecond}$, in order to include both cores. They were performed in the on-the-fly (OTF) mode and a total frequency coverage of \SI{45.6}{\giga\hertz} has been recorded.

In addition, single pointing observations with the First-Light APEX Submillimeter Heterodyne instrument (FLASH+) \citep{klein+2014} were made toward IRAS\,16293 A/B and were used only for line-identification purposes. These observations in total cover \SI{17}{\giga\hertz} bandwidth, of which \SI{9}{\giga\hertz} allow us to identify lines in frequency intervals between 476 and 493\,GHz, while additional \SI{2.3}{\giga\hertz} extend the frequency intervals covered by LAsMA between \SI{276.5}{\giga\hertz} and \SI{292.9}{\giga\hertz}. An overview of the observed frequency-bands and the observing dates is given in Appendix~\ref{app:obs_bands}. Both receivers were connected to fast Fourier transform spectrometers (FFTS) \citep[e.g.,][]{klein+2012}. The FFTS modules provide a total of $2^{16}$ frequency channels per 4\,GHz wide sideband, corresponding to a channel spacing (in frequency) of $61.04$\,kHz. This corresponds to $0.07$ and $\SI{0.05}{\kilo\meter\per\second}$ at our lowest and highest observed frequencies of 277 and 37\,GHz, respectively.
These numbers all are smaller than typical line widths observed (a few km\,s$^{-1}$).

During the observations, the LAsMA system temperatures varied between 150 and 250\,K at frequencies below 310\,GHz, between 200 and 400\,K from 330 to 370\,GHz, and between 400 and 600\,K above 370\,GHz. The conversion between antenna temperature $T_{\rm A}^*$ and the main-beam brightness temperature $T_{\rm MB}$ is given by \mbox{$T_{\rm MB}=T_{\rm A}^*\times \eta_\U{FW} / \eta_{\rm MB}$}, where $\eta_{\rm MB}$ is the main-beam efficiency and $\eta_\U{FW}$ the forward coupling efficiency. Based on Jupiter continuum pointings during the observations of the project, we estimate an average conversion factor of $\eta_\U{FW} / \eta_{\rm MB} = 1/0.7$ for the full array.
The APEX half power beam width (HPBW) $\theta_{\rm B}$ at frequency $\nu$ (GHz), in arcseconds is given by $\theta_{\rm B} [''] =7\rlap{.}''8 \times (800 / \nu[{\rm GHz}])$ \citep{gusten2006atacama} and thus varies at the observed frequencies between $\SI{17}{\arcsecond}$ and $\SI{23}{\arcsecond}$ (corresponding to $0.012$\,pc and $0.015$\,pc) respectively.

\subsection{Observations with SEPIA660}
Additional data were taken under project M-0102.F-9524C-2018 with the Swedish-ESO PI Instrument for APEX (SEPIA) at a tuning frequency of $\SI{691.473}{\giga\hertz}$. The SEPIA660 receiver is one of three ALMA-resembling receiver cartridges installed at APEX and operates as single pixel heterodyne 2SB receiver with two polarizations \citep{belitsky2018sepia}. Upper and lower sideband each cover $\SI{8}{\giga\hertz}$ and the centers of the two bands are separated by $\SI{16}{\giga\hertz}$.

The OTF observations with SEPIA660 cover a region of $200''\times 200''$ centered on $(50'',-10'')$ from IRAS\,16293 A/B with a total integration time of 30 minutes. In addition, a larger region of $460''\times 280''$ has been observed in the OTF mode with a total integration time of 20 minutes. System temperatures of the receiver hereby varied between 800 and 1200\,K.

The SEPIA660 receiver was connected to 8 FFTS backends that each provide $2^{16}$ channels with a channel spacing of $\SI{61}{\kilo\hertz}$ over $\SI{4}{\giga\hertz}$ of bandwidths. In this work, we present OTF maps of the CO $(J=6-5)$ transition at $\SI{691.473}{\giga\hertz}$ for which the velocity spacing is $\SI{0.03}{\kilo\meter\per\second}$, much smaller that the observed line width. The APEX HPBW at this frequency is $\SI{9}{\arcsecond}$ which results in an increased angular resolution as compared with the LAsMA observations.

\subsection{Data reduction}\label{ch:datared}
For the data reduction, we used the CLASS program that is part of the GILDAS\footnote{\url{http://www.iram.fr/IRAMFR/GILDAS}} software package developed by the Institut de Radioastronomie Millim\'etrique (IRAM). For the mapping of each spectral line, the local standard of rest (LSR) velocity-scale was first modified to correspond to the frequency of the considered transition. In order to improve the signal-to-noise ratio in the spectra, each two adjacent channels were averaged, resulting in a velocity-resolution of about $\SI{0.1}{\kilo\meter\per\second}$ per channel, which is adequate for resolving the observed spectral lines.

A linear spectral baseline was then calculated for the line surroundings, usually considering channels covering velocities below $\SI{-10}{\kilo\meter\per\second}$ and above $\SI{15}{\kilo\meter\per\second}$, such that the line emission around the systematic cloud velocity of $\SI{4}{\kilo\meter\per\second}$ is properly excluded. After the baseline-subtraction, the resulting spectrum was extracted in a velocity-range between $\SI{-25}{\kilo\meter\per\second}$ and $\SI{30}{\kilo\meter\per\second}$. The windows used for baseline-calculation and extraction were adjusted for the processing of broader spectral lines such as the CO $(J=3-2)$ transition.

\subsection{Mapping}
The maps of the \iras A/B+E area were created by extracting all the observed lines of a given transition from the APEX data and combining and gridding them using the CLASS-module \textit{xy\char`_map}. Since the edges of the resulting maps are not covered by all pixels, they show a higher noise and are therefore not considered during the analysis. In order to determine the noise of a particular map, an integrated-intensity map was created from nearby line-free channels.

Broad line wings indicate outflows, which are visualized by creating integrated intensity maps from channels that correspond to the line wing emission.
Contours of these maps were drawn in a way that regions with signal to noise ratio higher than 3 (3$\sigma$) are highlighted, for which the mean noise over the whole map was considered. The mapping of cloud cores and outflows was performed for all transitions present in the LAsMA data.

\section{Results}\label{ch:results}

To identify the detected molecular lines, we have used the Jet Propulsion Laboratory (JPL) line catalog\footnote{\url{https://spec.jpl.nasa.gov}} \citep{pickett1998} and the Cologne Database for Molecular Spectroscopy (CDMS)\footnote{\url{https://cdms.astro.uni-koeln.de}} \citep{muller2001,muller2005,endres2016}.
The line identification was carried out at the positions of IRAS\,16293 A/B and E, since occurring transitions are expected to be brightest in the immediate surroundings of the cloud cores. While we included frequency windows observed with FLASH+ for the line identification at IRAS\,16293 A/B, we note that these data were obtained as single pointings and therefore the corresponding transitions could not be mapped.

In our observations, we identify a total of 144 spectral lines from 36 different species, which subdivide into 20 distinct molecules and 16 of their isotopologues. 33 of these species show a total of 126 transitions in the frequency range covered by LAsMA, which was used to map the larger-scale emission around the YSOs. A full list with all the observed transitions can be found in Appendix~\ref{app:molecules}. Maps of the emission distribution of each detected transition were created and are shown in Appendix~\ref{app:maps}. We show only the brightest lines of species with many transitions, since the fainter transitions do not provide us with any additional information about the spatial distributions of molecules.

\subsection{Cloud cores in L1689N}
\begin{figure}[t]
    \centering
	{\includegraphics[width=0.48\textwidth]{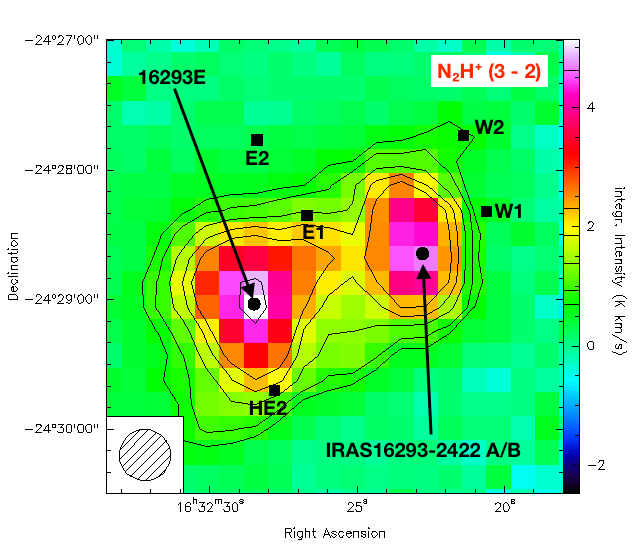}}
	\caption{N$_2$H$^+$ ($3 - 2$) integrated intensity (moment 0) map of the IRAS\,16293--2422 environment computed in a velocity range between $-1.3$ and 7.9\,km\,s$^{-1}$. The RMS of this map is $\sigma=\SI{0.23}{\kelvin\kilo\meter\per\second}$. Contours are drawn at $3 \sigma$, in steps of $2\sigma$ between $4\sigma$ and $10\sigma$ and in steps of $10\sigma$ afterwards. The right sided color bar indicates the intensity of the emission in units of $\si{\kelvin\kilo\meter\per\second}$. The position of the protostellar system IRAS\,16293--2422 A/B and the cold core 16293E are indicated. In addition, the positions of emission peaks previously identified by \citet{walker1988} and renamed by \citet{hirano2001} and \citet{castets2001} called E1, E2, W1, W2, and HE2 are also included. The FWHM beam size is shown in the lower left corner.}
	\label{fig:nnh+}
\end{figure}

In general, the molecular line maps reveal a complex morphology with larger-scale extended emission and multiple peaks, the two most prominent of them toward 16293E and the location of the protostars IRAS\,16293 A and B. Since the spatial resolution in our observations is about $\num{20}\si{\arcsecond}$, the two protostars, which are separated by $\SI{5}{\arcsecond}$, cannot be resolved individually. Instead, the observations aim to examine the large-scale molecular structure of L1689N and interactions of outflows with the cloud. Since the prestellar core 16293E is a cold source \citep[e.g.,][]{stark2004}, it is best traced by the N$_2$H$^+$ $(J=3-2)$, N$_2$D$^+$ $(J=4-3)$, and DCO$^+$ $(J=5-4)$ spectral lines, as shown in  Fig.~\ref{fig:nnh+}. In contrast, the extended envelope surrounding the A/B sources is traced by almost all molecules for which emission is detected in the data. Both cloud cores are slightly resolved in the observations and a number of molecular lines trace larger scale emission in which the cores are embedded.

\subsection{The outflow structure}\label{subsec:outflows}
In order to trace the general gas content of the cloud, CO is used as a tracer for molecular hydrogen. Its high abundance results in strong line emission with extended line wings that help us to identify molecular outflows based on the observed gas velocities relative to the cloud along the line of sight. Since the systemic cloud velocity is about $\SI{4}{\kilo\meter\per\second}$, velocities below this value trace blueshifted emission, while higher velocities were measured in case of redshifted emission from gas moving away from us.

\begin{figure}[tbp]
	\centering
	\subfigure{\includegraphics[width=0.49\textwidth]{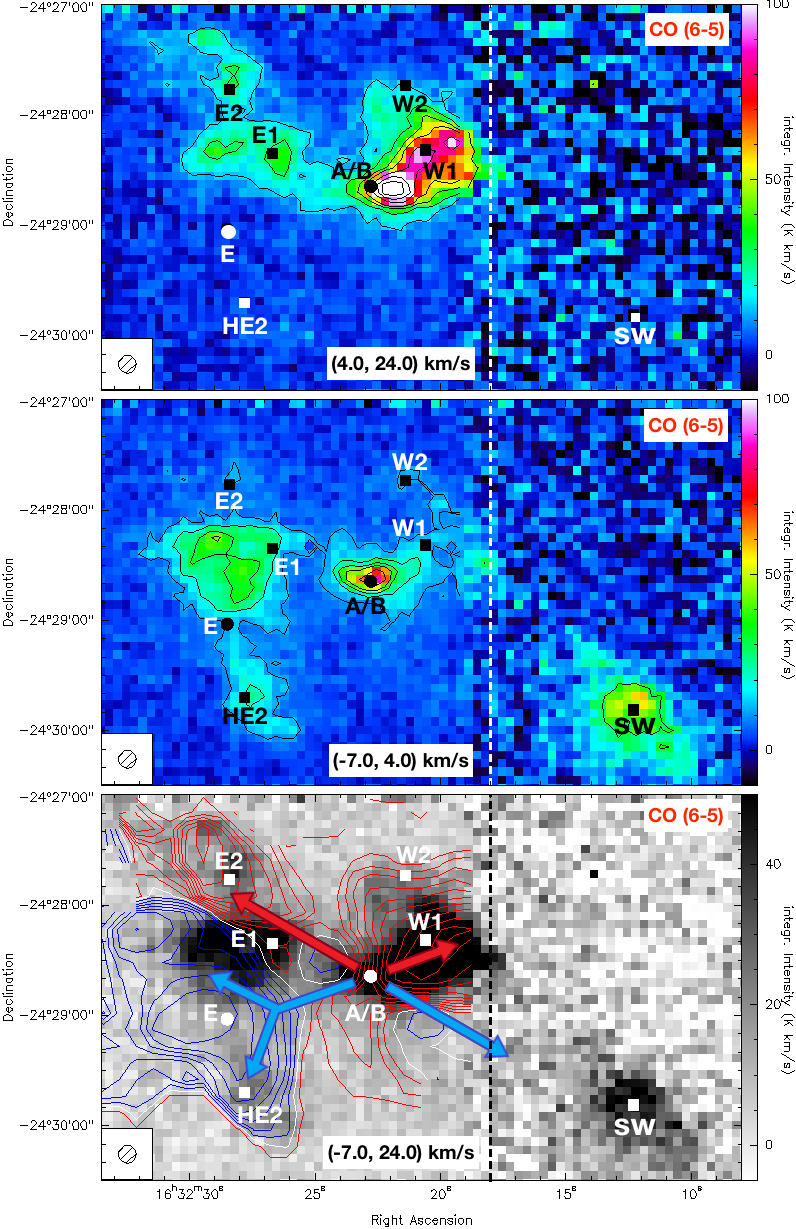}}
    \caption{CO~($6 - 5$) integrated intensity (moment 0) maps for velocities between $4$, $\SI{24}{\kilo\meter\per\second}$ (top figure), $-7$, $\SI{4}{\kilo\meter\per\second}$ (middle figure) and $-7$, $\SI{24}{\kilo\meter\per\second}$ (bottom figure). The RMS ($\sigma$) of the inner [and outer] map regions are $\SI{28.4}{\kelvin\kilo\meter\per\second}$ [$\SI{84.6}{\kelvin\kilo\meter\per\second}$] (top), $\SI{21.1}{\kelvin\kilo\meter\per\second}$ [$\SI{62.8}{\kelvin\kilo\meter\per\second}$] (middle) and $\SI{35.4}{\kelvin\kilo\meter\per\second}$ [$\SI{105.4}{\kelvin\kilo\meter\per\second}$] (bottom). Contours in the top and middle figure are drawn at 1, 2, 3, 4, 6, 8, and $10\sigma$. The bottom figure displays contours of the CO~($3-2$) moment 1 map in steps of $\SI{0.5}{\kilo\meter\per\second}$ with white contours indicating $\SI{4}{\kilo\meter\per\second}$. The beam size is indicated in the lower left corner. The lower right ascension limit of the LAsMA maps is indicated with black and white dashed lines.}
	\label{fig:CO6-5}
\end{figure}

Figure~\ref{fig:CO6-5} shows integrated intensity (moment 0) maps of the CO~($6-5$) transition, which reveal the complex large-scale structure in L1689N. In previous studies, \citet{walker1988} identified several emission peaks based on CO~($2-1$) observations of this region. Later, \citet{hirano2001} and \citet{castets2001} renamed these emission peaks as E1, E2, W1, W2, and HE2. We have included the position of these emission peaks (see Table~\ref{tab:pos_marker} in the Appendix \ref{app:positions}) in Fig.~\ref{fig:CO6-5}. In comparison to these previous works, we choose slightly different coordinates to describe the emission peaks based in our formaldehyde maps (see Sect.~\ref{sec:h2co}). In addition, the emission peak SW can be seen in the southwest of the CO~($6-5$) maps, which was not covered by our LAsMA observations. The overall CO morphology is dominated by two bipolar outflows which both originate in the vicinity of IRAS\,16293 A/B.

One of these large-scale outflows is the east-west outflow, which contributes to emission peak W1 with the western red lobe. The blue lobe of this outflow extends broadly across the region east of IRAS\,16293 A/B and appears to split around 16293E, which was already noted by~\citet{stark2004}.
As indicated by the blue arrows and velocity contours of CO~($3-2$) in the bottom panel of Fig.~\ref{fig:CO6-5}, this split results in a broader outflow region with higher velocities north of the prestellar core and a more compact region with lower relative velocities south of it. The latter corresponds to the emission peak HE2.

The second outflow extends in northeast-southwest direction and is the main contributor to emission at the peaks E1 and E2. These positions show redshifted emission with high relative velocities at the position E2. The southwestern counterpart of this outflow is not covered by our broad bandwidth LAsMA observations but instead can be seen in the further extended SEPIA maps of the CO~($6-5$) transition. Emission peaks originating from these outflows are seen in the maps of numerous species, which are analyzed in Sects.~\ref{sec:h2co}, \ref{sec:moleculdistr}, and~\ref{sec:cd}.

\subsection{Formaldehyde emission peaks}\label{sec:h2co}
\begin{figure}[tbp]
	\centering
	\includegraphics[width=0.49\textwidth]{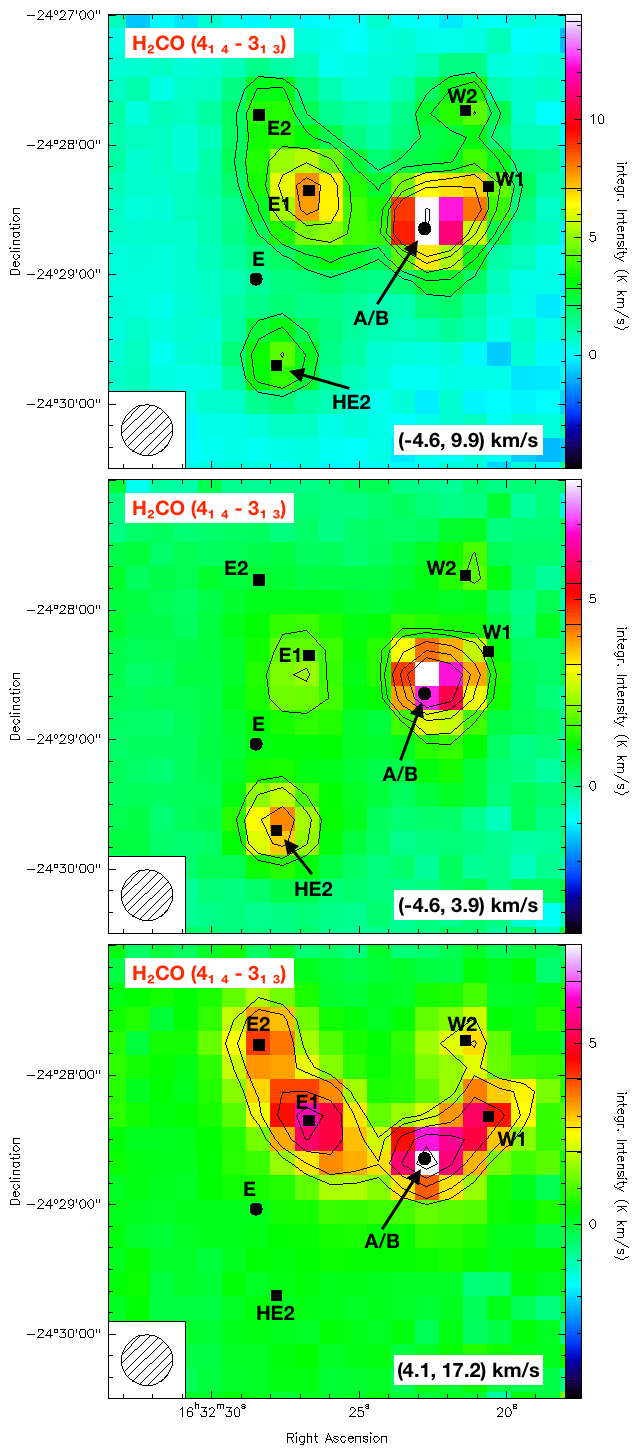}
	\caption{$\U{H}_2\U{CO}$ ($4_{1,4} - 3_{1,3}$) integrated intensity (moment 0) maps. The integrated velocity ranges are given in the lower right corner of the figures and correspond to the line core (top) and the blue (middle) and redshifted (bottom) emission. The map RMS ($\sigma$) are $\SI{0.64}{\kelvin\kilo\meter\per\second}$ (top), $\SI{0.49}{\kelvin\kilo\meter\per\second}$ (middle) and $\SI{0.61}{\kelvin\kilo\meter\per\second}$ (bottom). Contours are drawn at $3\sigma$, in steps of $2\sigma$ between $4\sigma$ and $10\sigma$ and in steps of $10\sigma$ afterwards. The right sided color bar indicates the intensity of the emission in units of K\,km\,s$^{-1}$. The black circles mark the positions of IRAS\,16293\,A/B and E, while the black squares mark positions according to Table~\ref{tab:pos_marker}. The beam size is included in the lower left corners.}
	\label{fig:h2co}
\end{figure}
\noindent
The formaldehyde (H$_2$CO) and Methanol (CH$_3$OH) molecules are known to be good tracers of molecular outflows~\citep{bachiller1997shock}. Since formaldehyde is important for the analysis in Sect.~\ref{ch:analysis}, we discuss the observational results from the H$_2$CO lines in more detail.

\begin{figure}[thbp]
	\centering
	\includegraphics[width=0.49\textwidth]{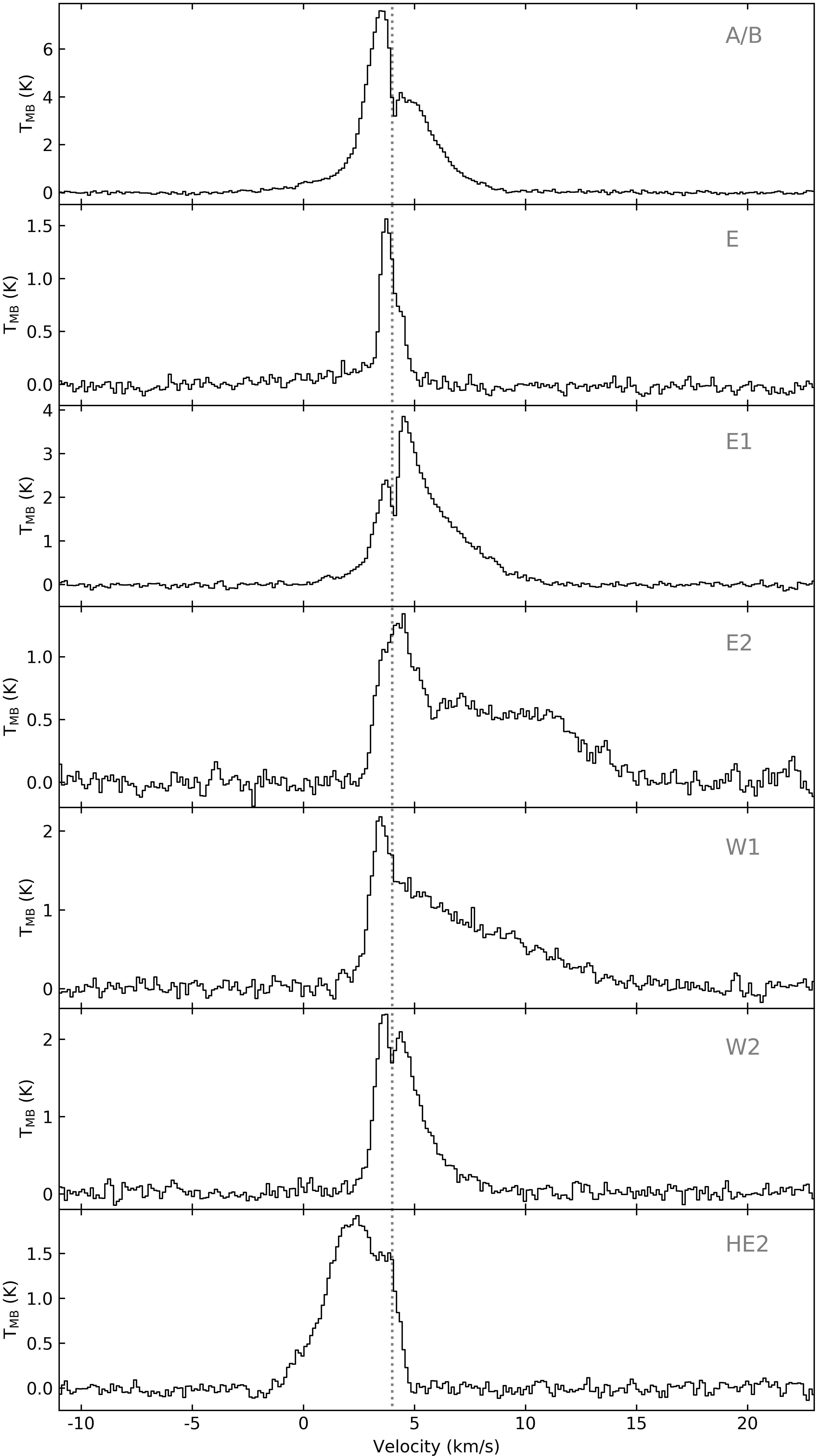}
	\caption{Spectra of the $\U{H}_2\U{CO}$ ($4_{1,4} - 3_{1,3}$) transition at the positions marked in Fig.~\ref{fig:h2co}. Dotted vertical lines mark the systematic cloud velocity of $\SI{4}{\kilo\meter\per\second}$. The extended wing profiles show the presence of molecular outflows in L1689N.}
	\label{fig:spec_E2}
\end{figure}

Figure~\ref{fig:h2co} presents the spatial distribution of the $\U{H}_2\U{CO}$ $J_{K_a,K_c}~=~4_{1,4}-3_{1,3}$ transition using different velocity ranges for integrating the emission. While the moment 0 map shown in the upper panel includes emission from the full H$_2$CO line core, the middle and bottom panels of Fig.~\ref{fig:h2co} respectively capture the blue and redshifted emission seen across the region. We note that the distribution of the emission is similar to that of CO~(6--5), concentrated around the previously defined peaks. While these maps do not show significant emission at 16293E, a narrow line can be seen in the spectra of Fig.~\ref{fig:spec_E2} at this position.

In addition, these maps clearly show a separation of the emission peak W2 north of IRAS\,16293 A/B from the remaining outflow structure. Since we do not detect a counterpart at a southern location, it is unlikely that the emission originates from a bipolar outflow. In contrast, SCUBA $\SI{450}{\micro\meter}$ maps shown in Sect.~\ref{sec:dustcontinuum} reveal emission at this position, suggesting a cold dust source to be present.

Analogous to CO, the lines of formaldehyde also show profiles with extended line wings. In order to clarify differences of the line profiles across the L1689N region, Fig.~\ref{fig:spec_E2} displays the formaldehyde spectra which occur at each of the emission peaks. While the line profile of the $\U{H}_2\U{CO}$ $J_{K_a,K_c}~=~4_{1,4}-3_{1,3}$ transition is very narrow at the position of 16293E, all other peak positions show a broader line profile. Additionally, line profiles at A/B and W2 show signs of self-absorption due to high optical depths.

Especially the redshifted emission at the emission peaks W1 and E2 extends to high velocities of about $\SI{15}{\kilo\meter\per\second}$. This is in agreement with our observations of the  CO~(3$-$2) transition, for which we detect outflow velocities of up to $\SI{20}{\kilo\meter\per\second}$. Similar velocities were also detected by~\citet{hirano2001} in SiO emission, who also found velocities up to $\SI{20}{\kilo\meter\per\second}$. These positions therefore are likely to be dominated by shocked gas as a consequence of outflow activities from IRAS\,16293 A.

\subsection{Distribution of different molecules}\label{sec:moleculdistr}
In order to obtain an overview of the molecular content in the region, we examined all emission peaks for occurring transitions. The results of this process are summarized in Table~\ref{tab:molecules_summary} and Appendix~\ref{app:molecules}.

\begin{table}[hbt]
    \caption{Summary of the molecules detected on each emission peak.}
    \label{tab:molecules_summary}
	\centering
	\begin{tabular}{c|l}
		\hline
		\hline
		Source & Species detected\\ \hline
%&& \\

A/B & All species reported in Table~\ref{tab:idlines}, except for H$_2$D$^+$ \\
& and NHD$_2$\\ \hline
%&& \\

E & H$_2$CO, CH$_3$OH, C$^{17}$O, CS, H$^{13}$CO$^+$, DCO$^+$,  H$_2$D$^+$,\\
& HNC, N$_2$H$^+$, N$_2$D$^+$, SO, CO, HCN, HCO$^+$, DNC, \\ 
& NH$_2$D, NHD$_2$, CN, NO\\ \hline

E1 & H$_2$CO, HDCO, C$_2$H, CH$_3$OH, C$^{17}$O, CS, H$^{13}$CO$^+$, \\
& DCO$^+$, HNC, N$_2$H$^+$, N$_2$D$^+$, SiO, SO, SO$_2$, \\
& CO, HCN, HCO$^+$, DNC, CN, NO \\ \hline

%&& \\
E2 & H$_2$CO, CH$_3$OH, C$^{17}$O, HNC, N$_2$H$^+$, SO, CO, HCN, \\
& HCO$^+$ \\ \hline

%&& \\
W1 & H$_2$CO, CH$_3$OH, C$^{17}$O, CS, HNC, N$_2$H$^+$, SO, CO,\\
& HCN, HCO$^+$ \\ \hline

%&& \\
W2 & H$_2$CO, CH$_3$OH, C$^{17}$O, CS, HNC, N$_2$H$^+$, SO, CO,\\
& HCN, HCO$^+$ \\ \hline
%&& \\

HE2 & H$_2$CO, CH$_3$OH, C$^{17}$O, CS, H$^{13}$CO$^+$, DCO$^+$, HNC,\\
& N$_2$H$^+$, N$_2$D$^+$, SO, CO, HCN, HCO$^+$, DNC,  \\ 
& NH$_2$D, CN, NO \\\hline

	\end{tabular}
\end{table}
In general, a majority of the considered molecules is detected around the protostars IRAS\,16293 A and B, as this is one of the positions at which the line identification was carried out. Exceptions are the transitions of doubly deuterated ammonia and the 372.4 GHz ortho-H$_2$D$^+$ line, which, at our sensitivity, can only be detected near 16293E.

Most of the sulfur bearing species are present in the surroundings of the protostars A/B and in the E1 outflow peak. The absence of strong emission in the other outflows is unusual, as those species are characteristic for warm gas and shocks, as described by~\citet{wakelam2004sulphur}. A reason for this might be that the observed SO and SO$_2$ transitions arise from high energies above the ground states and have small Einstein A coefficients, such that they are only seen toward positions with high excitation and column densities.

Previous works \citep[e.g.,][]{hirano2001,castets2001} have shown that SiO emission traces very well the outflows in the L1689N region, which is true for fast shocks in general \citep{schilke1997}. The observed frequency windows only cover the SiO ($8-7$) transition, which is detected mostly in the envelope of IRAS\,16293--2422 A/B. Weak SiO emission is also detected toward the E1 emission peak. Again, the excitation conditions at the other positions might not be sufficient for the relatively high-$J$ line of SiO to be detected.

The prestellar core 16293E shows emission of many deuterated species such as DCO$^+$, N$_2$D$^+$, and NHD$_2$. This confirms it to be a quiescent cold source, as the enhanced deuteration is a result of strong exothermic reactions that proceed in cold material \citep[see e.g.,][]{roberts2000}. In addition to these deuterated species, other molecules are detected and mapped for the first time in this source, such as CN, NO, and weak emission of CS and CH$_3$OH. 

As can be seen in Table~\ref{tab:molecules_summary} and Appendix~\ref{app:maps}, the distribution of methanol is similar to that of formaldehyde, as both molecules can be observed in all of the positions. The reason for their comparable distribution is the formation of methanol from formaldehyde through hydrogenation on CO-rich ices \citep{fuchs2009hydrogenation, fuchs2020}.

\subsection{Rotation of the envelope}\label{sec:rotation}

\begin{figure*}[thbp]
	\centering
	\includegraphics[width=0.99\textwidth]{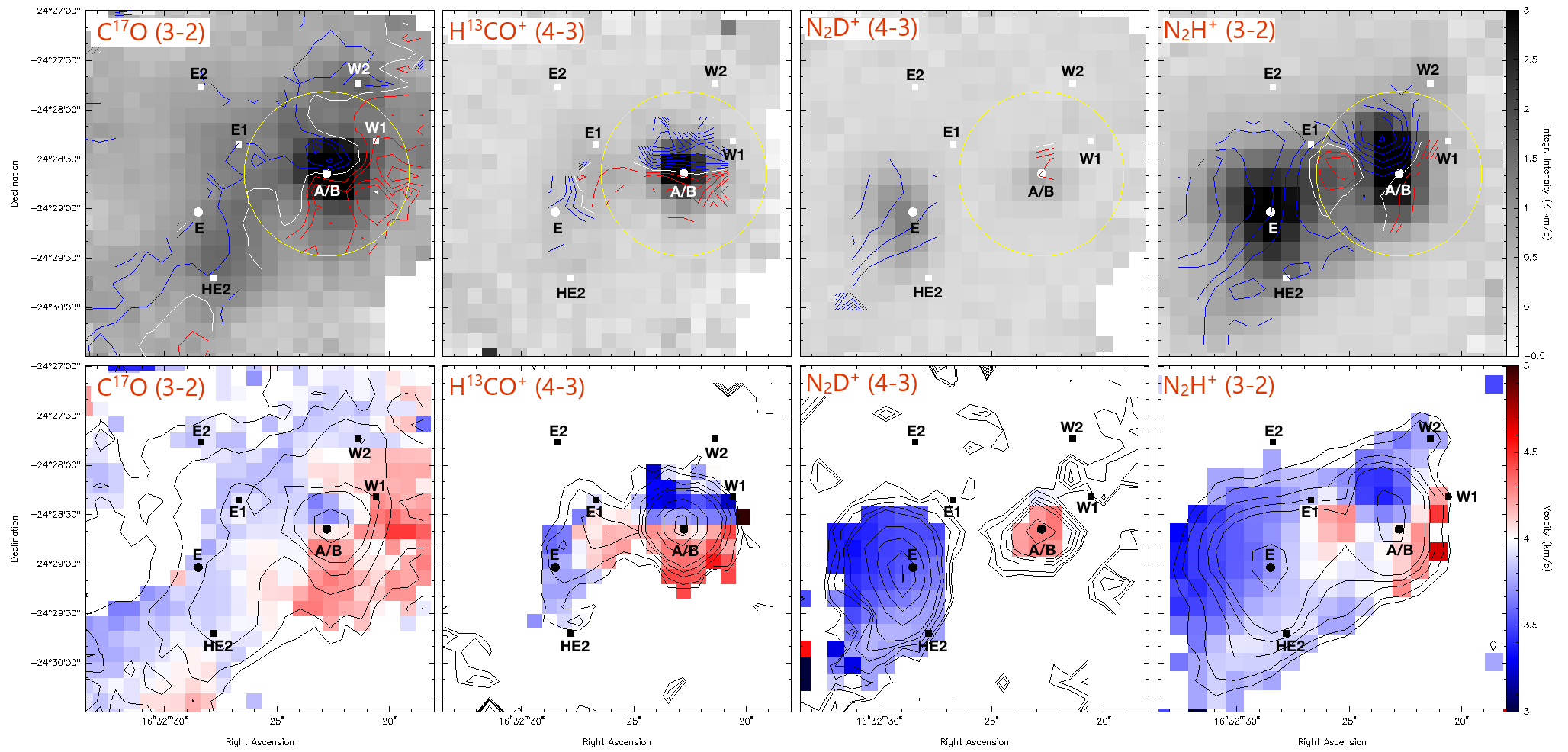}
	\caption{Velocity maps of C$^{17}$O $(3-2)$, H$^{13}$CO$^{+}$ $(4-3)$, N$_2$D$^{+}$ $(4-3)$, and N$_2$H$^{+}$ $(3-2)$. Top panels: Moment 0 maps of the respective transitions are shown in grayscale. The overlaid contours show the red-shifted and blue-shifted emission in steps of $\SI{0.1}{\kilo\meter\per\second}$ with respect to the systematic velocity of the source ($v_\U{LSR}=4\,\si{\kilo\meter\per\second}$, displayed in white). Integration ranges consider emission between $2-6\,\si{\kilo\meter\per\second}$ for C$^{17}$O~$(3-2)$ and H$^{13}$CO$^{+}$~($4-3$). Smaller velocity intervals of $2.5-5\,\si{\kilo\meter\per\second}$ and $2.3-5\,\si{\kilo\meter\per\second}$ are chosen for N$_2$D$^{+}$ $(4-3)$ and N$_2$H$^{+}$ $(3-2)$ respectively, in order to exclude nearby HDCO emission and the resolved hyperfine components of N$_2$H$^{+}$. Bottom panels: Moment 1 maps of the molecules shown in the top panels. Contours of the moment 0 maps are drawn at $3 \sigma$, in steps of $2\sigma$ between $4\sigma$ and $10\sigma$ and in steps of $10\sigma$ afterwards. The yellow circle has a radius of 50$''$ to illustrate the approximate size of the envelope around A/B sources. The plots show the morphology of the velocity field in the environment of IRAS\,16293--2422. A large-scale velocity gradient is observed in the NE--SW direction.}
	\label{fig:rotation}
\end{figure*}

The protostellar objects A/B are surrounded by a large gas and dust envelope. This structure has a mass of about 4--6\,M$_{\odot}$ \citep[][]{jacobsen2018,ladjelate2020} and radius of about 6000--$\SI{8000}{\astrounit}$ \citep{Schoier2002, crimier2010, jacobsen2018}. Several studies have focused in determining the large-scale properties of this structure based on the kinematics of a number of molecules. For example, the extended envelope was suggested to have infall motions based on the blue-skewed profile seen in CS~$(7-6)$ and $(5-4)$ molecular maps at spatial scales of about $39''$ \citep{Narayanan1998}. In addition, based on CS~$(3-2)$ and $(2-1)$ observations, \citet{Menten1987} suggested the presence of a velocity gradient that could be associated with the rotation of the envelope.

To investigate such rotational motions, we have produced velocity field maps for some species of the close environment around IRAS\,16292$-$2422 A/B. In Fig.~\ref{fig:rotation} we show the moment 0 and moment 1 maps of the C$^{17}$O~$(3-2)$, H$^{13}$CO$^{+}$~$(4-3)$, N$_2$D$^{+}$~$(4-3)$, and N$_2$H$^{+}$~$(3-2)$ transitions. These maps include emission of the line core, covering a velocity interval of $2-6\,\si{\kilo\meter\per\second}$ for C$^{17}$O~$(3-2)$, H$^{13}$CO$^{+}$~$(4-3)$. In case of N$_2$D$^{+}$~$(4-3)$, emission between $2.5-5\,\si{\kilo\meter\per\second}$ is integrated in order to exclude nearby HDCO emission. For N$_2$H$^{+}$ $(3-2)$, we only consider emission from the main hyperfine structure components, by choosing an integration range of $2.3-5\,\si{\kilo\meter\per\second}$. We note that the N$_2$D$^{+}$~$(4-3)$ transition and the main component of the N$_2$H$^{+}$~$(3-2)$ transition both consist of multiple unresolved components as a consequence of the $^{14}$N hyperfine structure. We therefore assume that possibly occurring changes in relative intensities of these components due to optical depth effects do not affect the observed velocity field.

As a reference, a yellow circle of 50$''$ radius is included to Fig.~\ref{fig:rotation} which indicates the approximate extent of the envelope. A large-scale velocity gradient is observed in the NE--SW direction in the C$^{17}$O moment 1 map. For H$^{13}$CO$^{+}$, the velocity gradient is mostly aligned in the N--S direction, while for N$_2$D$^{+}$ it is distributed in the E--W direction, although the molecular gas distribution follows the same trend as in the C$^{17}$O map. Less emission is shown due to the clipping limits used to create the map (10$\sigma$ for N$_2$D$^{+}$ and 4$\sigma$ for C$^{17}$O). For N$_2$H$^{+}$, the velocity field is more complex, but the velocity spatial distribution resembles overall that of the C$^{17}$O as well. Closer to IRAS\,16293--2422, the velocity field in the N--S direction prevails for most of the maps, which could be interpreted as rotation of the envelope itself. This would agree with the interpretation of \citet{Menten1987}. However, based on the full velocity field morphology seen in our maps, we cannot disentangle a pure rotation motion associated with the envelope and the larger cloud scale velocity field that dominates the whole studied region.

\section{Analysis}\label{ch:analysis}
\subsection{Estimation of the gas temperature in L1689N}\label{subsec:temperatures}
\begin{figure*}[tbp]
	\centering
	\includegraphics[width=0.99\textwidth]{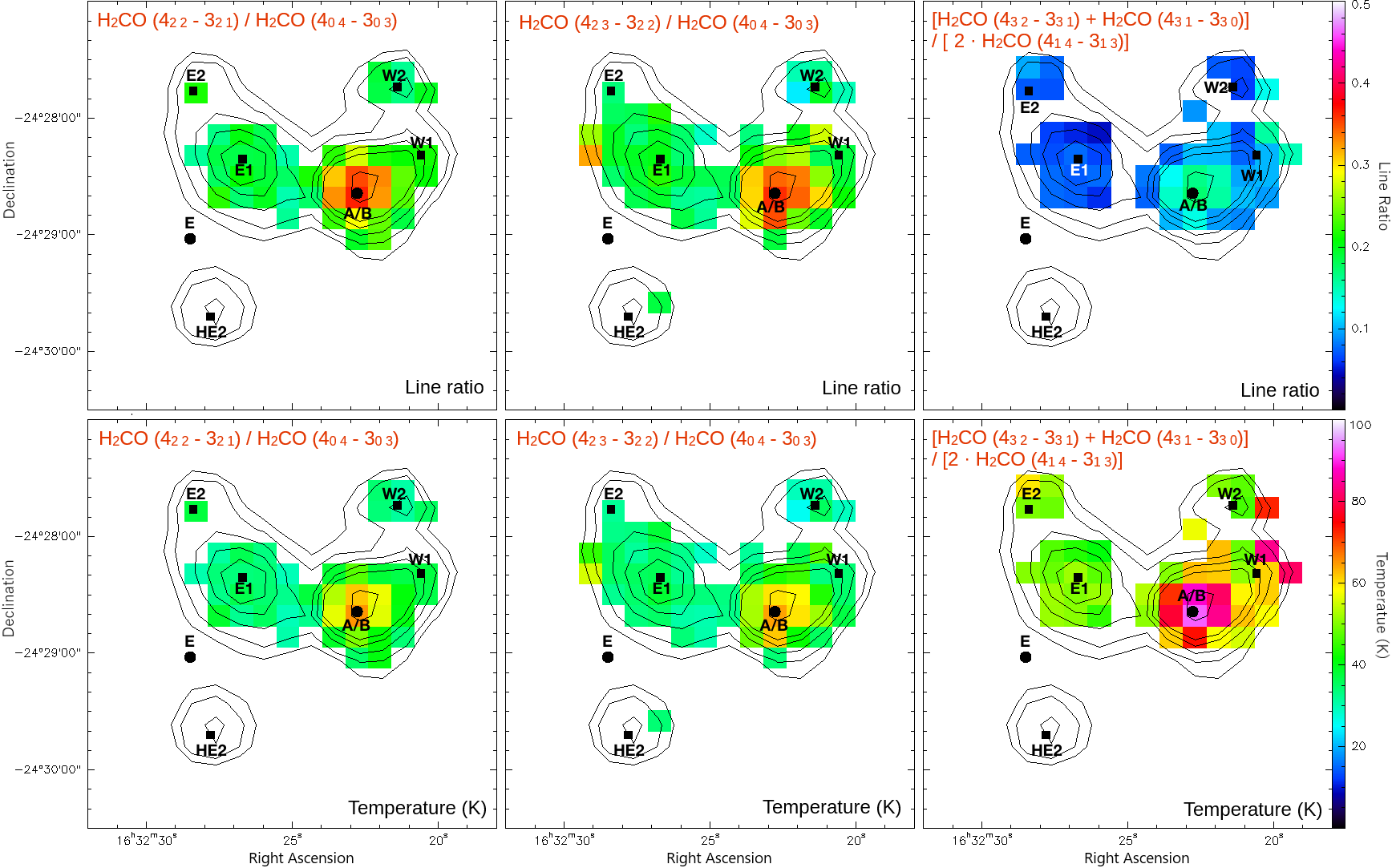}
	\caption{Temperature and line ratio maps as derived by the formaldehyde line ratios (a) ($4_{2,2} - 3_{2,1}$)/($4_{0,4} - 3_{0,3}$); (b) ($4_{2,3} - 3_{2,2}$)/($4_{0,4} - 3_{0,3}$); (c) $[(4_{3,2} - 3_{3,1})+(4_{3,1} - 3_{3,0})]/[2\times (4_{1,4}-3_{1,3})]$. The underlying moment 0 H$_2$CO ($4_{1,4} - 3_{1,4}$) contours are drawn at $3 \sigma$, in steps of  $2\sigma$ between $4\sigma$ and $10\sigma$ and in steps of $10\sigma$ afterwards. Temperatures are given in Kelvin.}
	\label{fig:temp_ratio}
\end{figure*}
The formaldehyde molecule has a long-standing history as thermometer and density probe of  molecular clouds. This is because the H$_2$CO molecule is a slightly asymmetric rotor and the transitions between levels with different $K_a$ values are dominated by collisions. Therefore, line ratios involving different $K_a$ ladders can be used to determine the kinetic temperature of the gas \citep[e.g., see][]{mangum1993formaldehyde}. To derive the excitation temperature $T_\U{ex}$, which is equal to the kinetic temperature under the assumption of a local thermodynamic equilibrium (LTE) conditions, we made use of the H$_2$CO rotational transitions detected in our observations.

In the LTE approximation, the excitation temperature between two energy levels u and l can be written as 

\begin{equation}
    T_\U{ex} = - \frac{E_\U{u}-E_\U{l}}{k}\left[\ln \left(\frac{N_\U{u}}{N_\U{l}}\frac{g_\U{l}}{g_\U{u}} \right)\right]^{-1},
\end{equation}
where $E_\U{u}$ and $E_\U{l}$ are the energies of the upper and lower levels, respectively, $g_\U{u}$ and $g_\U{l}$ are the degeneracies and $N_\U{u}$ and $N_\U{l}$ are the corresponding column densities (see Appendix \ref{app:temp_derivation} for more details).

The observed formaldehyde transitions allow the derivation of excitation temperatures for ortho- and para-H$_2$CO separately. Since only three transitions per species are available, these temperatures were obtained from the respective line ratios. For para-H$_2$CO was therefore possible to use the line ratios \mbox{($4_{2,2} - 3_{2,1}$)/($4_{0,4} - 3_{0,3}$)} and ($4_{2,3} - 3_{2,2}$)/($4_{0,4} - 3_{0,3}$). In case of ortho-H$_2$CO, the ($4_{3,2} - 3_{3,1}$) and ($4_{3,1} - 3_{3,0}$) lines are blended. For this reason, the ratio $[(4_{3,2} - 3_{3,1})+(4_{3,1} - 3_{3,0})]/[2 \times (4_{1,4}-3_{1,3})]$ was considered. \noindent In the Table~\ref{tab:h2coparams} of the Appendix \ref{app:temp_derivation}, we show the main parameters of the considered H$_2$CO transitions based on the CDMS and JPL catalogs.

The temperatures were calculated at each position, at which both of the considered transitions are detected with a $3 \sigma$ significance. In order to include most of the emission, main beam temperatures were integrated in a velocity-range between $\SI{2}{\kilo\meter\per\second}$ and $\SI{6}{\kilo\meter\per\second}$ for computing the intensities of para-H$_2$CO. To properly include the emission of both blended ortho-H$_2$CO lines, the integrations for this species were carried out on the H$_2$CO ($4_{3,2} - 3_{3,1}$) spectra in a velocity interval from $\SI{-5}{\kilo\meter\per\second}$ to $\SI{10}{\kilo\meter\per\second}$.

Based on the created line ratio and temperature maps shown in Fig.~\ref{fig:temp_ratio}, it was possible to derive the temperatures at the previously discussed positions, by including all mapped temperatures in a $10''$ radius for computing a weighted average. Statistical uncertainties were estimated based on Gaussian error propagation of the map RMS (see Table~\ref{tab:idlines}). The resulting temperatures and uncertainties are shown in Table~\ref{tab:temps}. These uncertainties do not include the calibration uncertainty, which is conservatively estimated to be about 20\%. It can be expected that this uncertainty is additionally transferred to the temperature values.

The temperatures of para-H$_2$CO were calculated by averaging the values from both para-H$_2$CO maps at the associated positions. As the transitions of formaldehyde with high upper level energies are too weak for temperature estimates in the direction of 16293E, a temperature of $\SI{12}{\kelvin}$ based on the results of \citet{stark2004} is assumed for further calculations.
\begin{table}[htb]
	\caption{Derived temperatures and statistical uncertainties at the emission peaks for ortho(o)- and para(p)-H$_2$CO.}
	\label{tab:temps}
	\centering
	\begin{tabular}{ccc}
		\hline
		\hline
		Position & T(o-H$_2$CO) & T(p-H$_2$CO) \\ 
			      & (K) & (K) \\ \hline

		IRAS\,16293\,A/B & 90.1 $\pm$ 4.4 & 62.4 $\pm$ 0.7 \\
		E1 & 50.8 $\pm$ 3.8 & 36.4 $\pm$ 1.3 \\ 
		E2 & 49.5 $\pm$ 7.8 & 34.1 $\pm$ 4.0 \\
		W1 & 52.9 $\pm$ 5.1 & 34.8 $\pm$ 1.9\\
		W2 & 47.2 $\pm$ 7.8 & 31.3 $\pm$ 2.0 \\
		HE2& - & 34.0 $\pm$ 3.8 \\ \hline
	\end{tabular}
	\tablefoot{The values of para-H$_2$CO are calculated by averaging the temperatures from both maps of para-H$_2$CO. The minus (-) symbol for HE2 indicates that no bright emission from ortho-formaldehyde lines with higher upper level energies is detected at this position. Note that the calibration uncertainty of 20\% is not considered in the stated uncertainties.}
\end{table}

It is worth noticing that the derived temperatures in the ortho-H$_2$CO map are about 20-$\SI{30}{\kelvin}$ higher than derived from the para-transitions. The reason for this deviation might be the upper level of the ortho-H$_2$CO transitions. With an upper level energy of $\SI{140.9}{\kelvin}$, these transitions trace a population which requires higher temperatures to be excited, compared to the para-H$_2$CO transitions.

Previously, \citet{Vandishoeck1995} studied the emission of H$_2$CO toward IRAS\,16293--2422 A/B based on observations obtained with the Caltech Submilimeter Observatory (CSO) and the James Clerk Maxwell Telescope (JCMT). They used the rotational diagram method and found a rotational temperature of $80\pm 10$\,K. Also, from H$_2$CO line ratios within the same ladder, they derived values for the kinetic temperature between 60 and 140\,K. Later, \citet{Ceccarelli2000} used the H$_2$CO data from \citet{Vandishoeck1995} (among others) and confirmed that the emission in IRAS\,16293--2422 A/B originates from two regions: A hot core region with temperatures above 100\,K and an outer and colder layer with lower H$_2$CO abundance. Using p-H$_2$CO, we obtained an average temperature value of 62.4 $\pm$ 0.7\,K (see Table \ref{tab:temps}), which is somewhat low compared with the values from previous studies. The o-H$_2$CO temperature of 90.1 $\pm$ 4.4\,K is in better agreement with both \citet{Vandishoeck1995} and \citet{Ceccarelli2000}. For a more detailed comparison of the temperatures for different layers in IRAS\,16293--2422, see Appendix \ref{app:twolayers}.

Furthermore, \citet{castets2001} studied the H$_2$CO and SiO emission based on observations with the IRAM 30 meter telescope, the Swedish-ESO Submillimetre Telescope (SEST), and the Infrared Space Observatory (ISO) and derived values for the other emission peaks between 80--150\,K. In this work, we derive lower p-H$_2$CO temperature values (about 35\,K on average) for the rest of the emission peaks. Since H$_2$CO is more reliable thermometer than SiO, we argue that these lower values reported in Table \ref{tab:temps} are more accurate.

In order to test the validity of the assumption of LTE, we compared the calculated temperatures and line ratios from ortho- and para-H$_2$CO with non-LTE models. These models are computed with the RADEX radiative transfer code \citep{van2007computer} considering different H$_2$ volume densities and using collision rates from \citet{wiesenfeld2013rotational}. The detailed RADEX computation can be found in Appendix \ref{app:RADEX_temp}. We find that RADEX and LTE models are in better agreement for higher volume densities, while the temperature values are underestimated by about 20\% for lower values of the density. These effects of non-LTE are partially compensated by opacity effects at the position of the protostars A/B.

\subsection{Column densities}\label{sec:cd}
To derive the column densities for all the species, we have used the LTE radiative transfer CLASS extension \textit{weeds}. This software allows us to compute a synthetic spectrum based on a set of physical parameters, namely column density, temperature, source size, velocity shift and line width~\citep{maret2011weeds}. For the computation, \textit{weeds} solves the radiative transfer equation by assuming that the emission originates from a single layer. The synthetic spectrum emulates the individual line profiles for each species and can be directly compared with the observations. 

Applying the formalism introduced in Appendix~\ref{app:temp_derivation}, the total column density $N_\U{tot}$ can be related to the column density of an upper energy level $E_\U{u}$ by
\begin{equation}\label{eq:cd_tot}
\frac{N_\U{tot}}{N_\U{u}} = \frac{Q_\U{rot}}{g_\U{u}} \exp\left(\frac{E_\U{u}}{k T_\U{ex}}\right),
\end{equation}
where \mbox{$Q_\U{rot}(T)$} is the rotational partition function which describes the sum over all rotational energy levels in a molecule for a given excitation temperature~\citep{mangum2015calculate}.

Inserting the derived temperature estimates in Eq.~\ref{eq:cd_tot} would, in combination with Eq.~\ref{eq:cd_thin}, also allow the computation of column densities based on the observed line profiles. The advantage of determining the column densities with \textit{weeds} is that the software additionally considers line blending with other molecular species and optical depths effects in the calculation of the line profiles, thus giving us a more reliable estimate.

For deriving column densities at each of the seven emission peaks listed in Table~\ref{tab:pos_marker}, we extracted and averaged the spectra in a $10''$ radius around the respective positions. This radius is motivated by the beam size of the telescope and the fact that the cores are slightly resolved in the observations. In order to obtain comparable results for all transitions, we adopted a constant source size of $\SI{20}{\arcsecond}$ in the computation of the synthetic spectra.

Since the highest upper energy level for the majority of the observed transitions is about $\SI{40}{\kelvin}$, hence much lower than the upper level energy of $\SI{140.9}{\kelvin}$ of the higher energy ortho-H$_2$CO lines, the temperatures derived from para-H$_2$CO were used in the computations of synthetic spectra.

Some of the presented lines, for example the formaldehyde transitions, show self-absorption due to nonnegligible optical thickness. In these cases, the fit is guided by the line wings instead of the self-absorbed line core which may lead to an underestimation of the derived column densities. While fitting the line wings gives reasonable results for most of the self-absorbed lines, it is not possible to compute satisfactory models for the optical thick transitions of CO~($3-2$), HCO$^+$~($4-3$), and HCN~($4-3$). Fitting these lines would require at least a two layer model, which goes beyond the scope of this work.

Instead, the column densities of these molecules were estimated based on the column densities of their less abundant isotopologues C$^{17}$O, H$^{13}$CO$^+$, and H$^{13}$CN, by converting the derived column densities according to the isotope ratios in the interstellar medium. For this we adopted a ratio of $^{12}\U{C}/^{13}\U{C} = 68$ from \citet{milam200512c} and a ratio of $^{16}\U{O}/^{17}\U{O} = 1790$ from the values suggested by \citet{wilson1994abmilam}. The observed H$^{13}$CN~(4-3) transition is blended with the SO$_2$~($13_{2,12}-12_{1,11}$) transition, due to which the derived SO$_2$ model has been subtracted from the spectrum before estimating the H$^{13}$CN column density.

In cases where we did not detect emission stronger than three times the RMS, upper limits on the column densities were constrained instead. For these, the median line width of the observed transitions at a respective position was applied to determine the column density at which the resulting computed line profile would fulfill the detection criterion.

Appendix~\ref{app:cd1} presents an overview of the derived molecular abundances and simulated line profiles. Based on a conservative estimate on the calibration uncertainty of about 20\%, it can be expected that this uncertainty is transferred to the column density values. As an example, the radiative transfer model for the H$_2$CS ($8_{1,7} - 7_{1,6}$) transition at $\SI{278.9}{\giga\hertz}$ at the position of IRAS\,16293 A/B is shown in Fig.~\ref{fig:weeds} of Appendix \ref{app:cd1}.

\begin{figure*}[tbp] 
	\centering
	\subfigure[]{\includegraphics[width=0.49\textwidth]{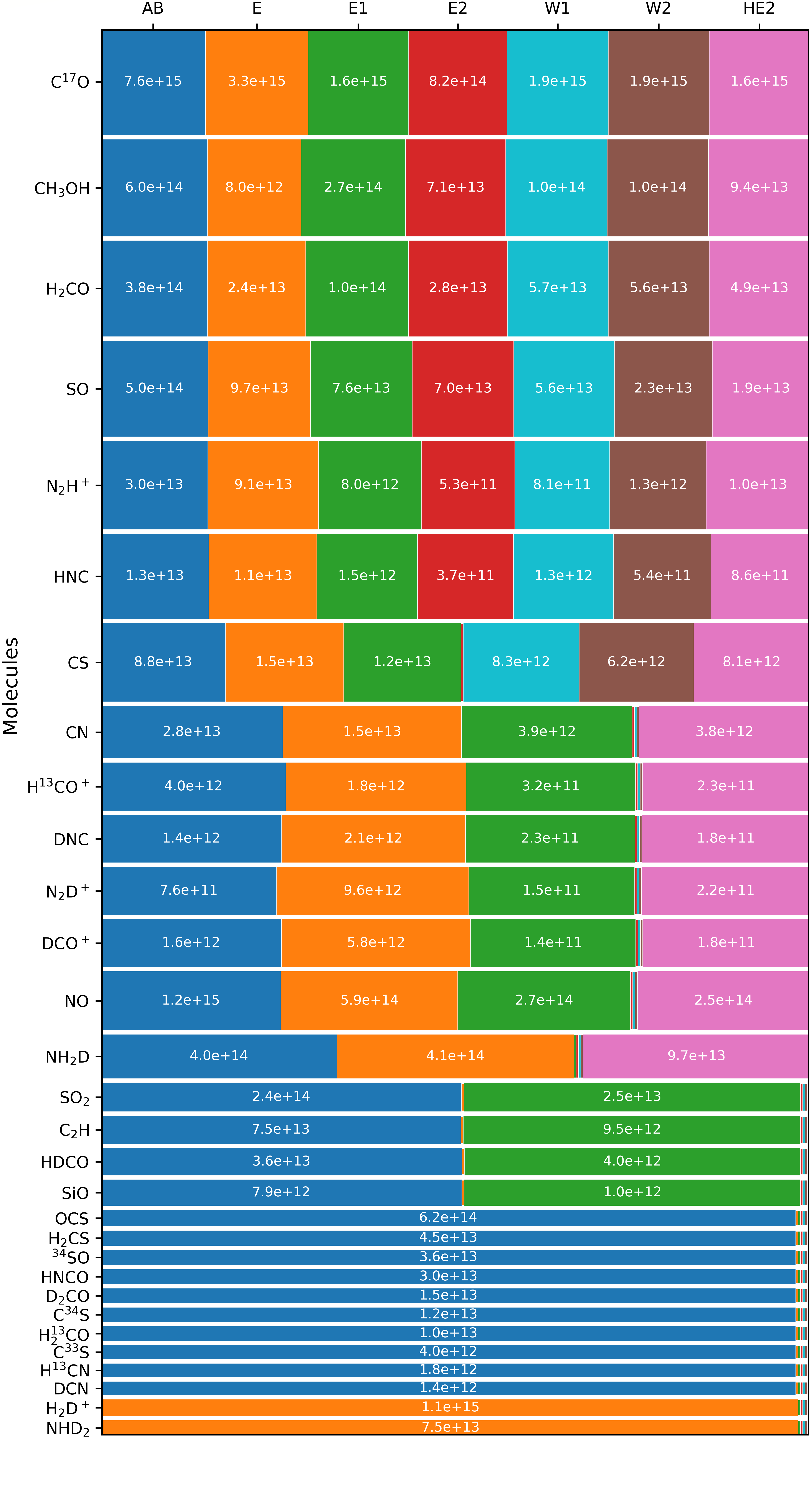}}
	\subfigure[]{\includegraphics[width=0.48\textwidth]{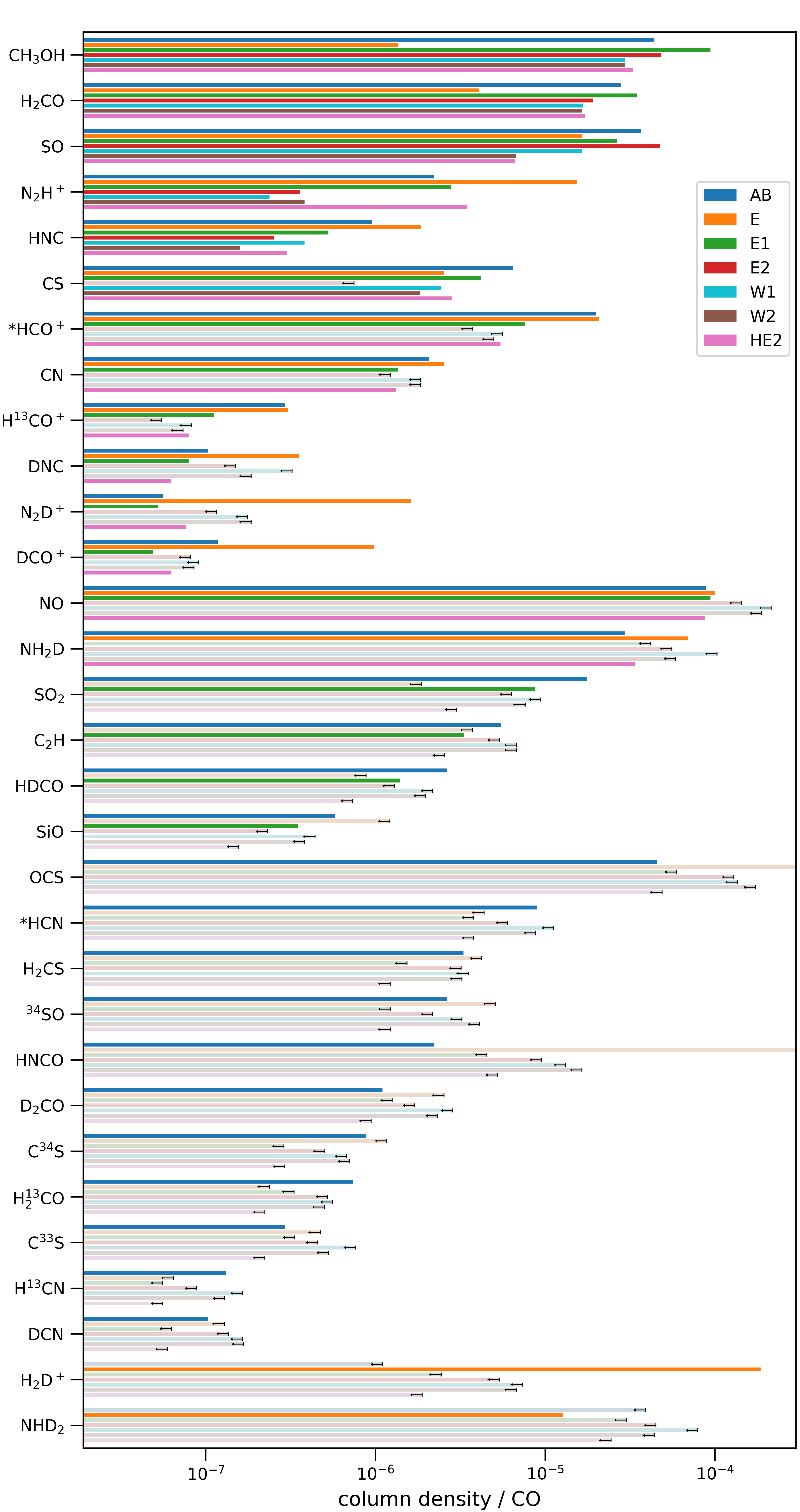}}
	\caption{Visualization of the column densities derived with the LTE CLASS-Module \textit{weeds}. a) Mosaic plot showing the column densities of species at the individual emission peaks in units of $\si{\per\centi\meter\squared}$. Tile areas represent the logarithmic scaled absolute column densities in order to also emphasize the distribution of less abundant species. CO, HCO$^+$, and HCN are not displayed, as their self-absorbed line profiles prevent a direct estimation of corresponding column densities. We note that these species show emission at all peak positions. b) Bar plot of the column densities in units of the CO column density at the respective positions, as indicated by the colors in the upper right panel. Gray bars with arrows pointing left visualize derived upper limits for the column densities, abundances of HCO$^+$ and HCN are marked with a *, as these are calculated from the less abundant isotopologues H$^{13}$CO$^+$ and H$^{13}$CN.}
	\label{fig:cd}
\end{figure*}

Figure~\ref{fig:cd} visualizes the column densities and molecular abundances relative to CO at the considered emission peaks in L1689N.
Figure~\ref{fig:cd}a displays the column densities in a mosaic plot with logarithmically scaled tiles, in order to give a quick overview of the presence and absence of species at the respective positions. While this scaling emphasizes the presence of less abundant species, the order of magnitudes differences in the molecular abundances are seen more clearly in the bar plot shown in Fig.~\ref{fig:cd}b.
These figures show that the molecular species are widely distributed across the whole cloud. The few species detected exclusively at the protostellar core IRAS\,16293 A/B are mainly less abundant isotopes of otherwise abundant species and molecules for which we only cover fainter transitions. The complexity of the distribution of the species across the cloud is discussed further in Sect.~\ref{sec:chemistry}.

Since the column densities of H$_2$CO are important for the further derivation of the H$_2$ volume densities (see Sect.~\ref{subsec:volumedens}), we also tested the plausibility of the derived values by creating multicomponent non-LTE models for the molecule using the CASSIS\footnote{\url{http://cassis.irap.omp.eu}}-RADEX software. The details of this line modeling can be found in Appendix~\ref{app:twolayers}.

We find that the self-absorbed line profiles at IRAS\,16293 A/B and W2 can be reproduced with a three and two layer model respectively. The three physical components toward A/B hereby consist of a hot corino in addition to a warm and an extended cold envelope. The derived LTE column density at this position is in between the values for the hot corino and warm envelope. A similar result is obtained by the line modeling at the W2 position, in which the column density derived with LTE assumptions is also in between the values of both considered physical components. Single layer non-LTE models at the positions E and HE2 are in agreement with the LTE models.

\begin{figure*}[htbp]
	\centering
	\subfigure[]{\includegraphics[width=0.48\textwidth]{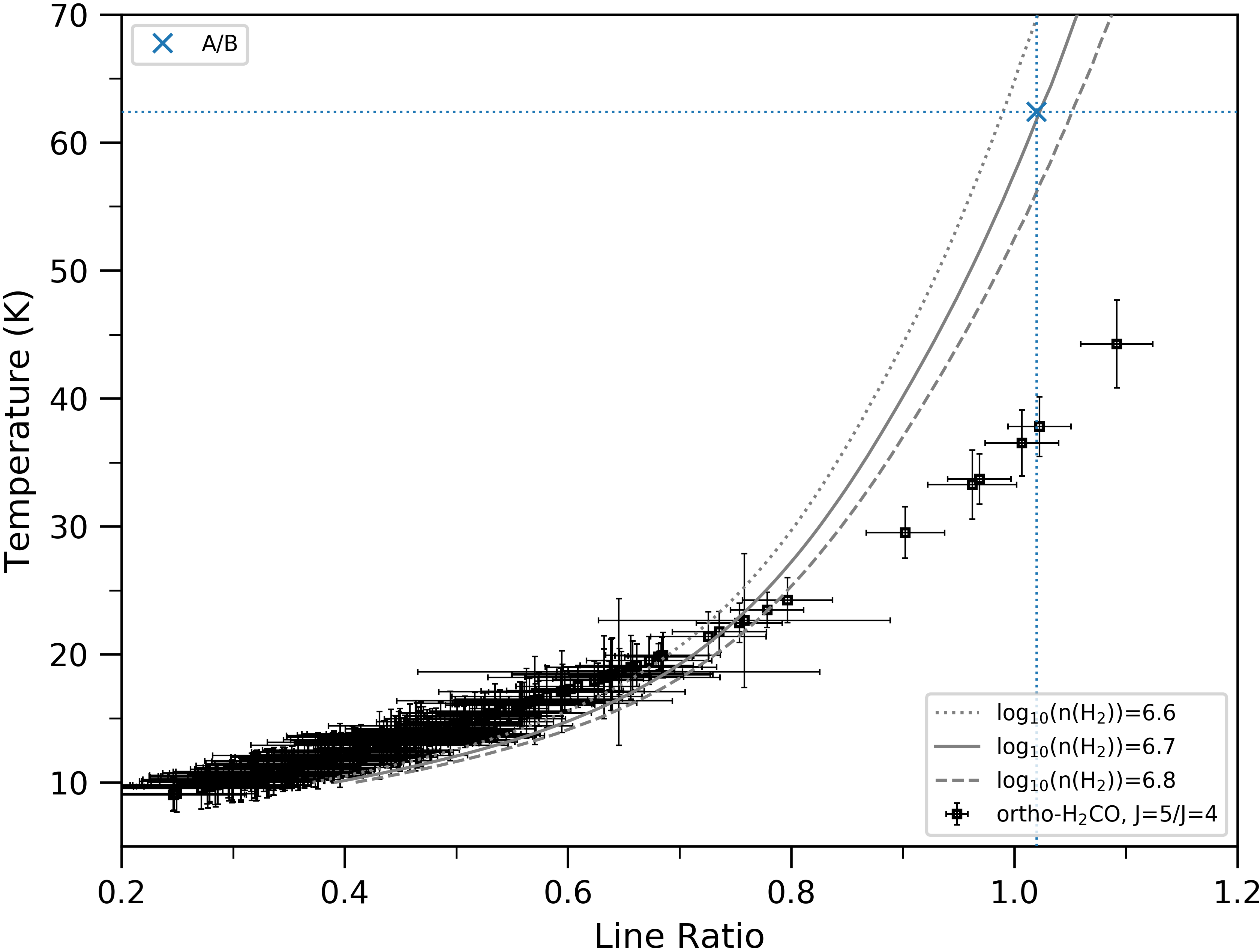}}
	\subfigure[]{\includegraphics[width=0.48\textwidth]{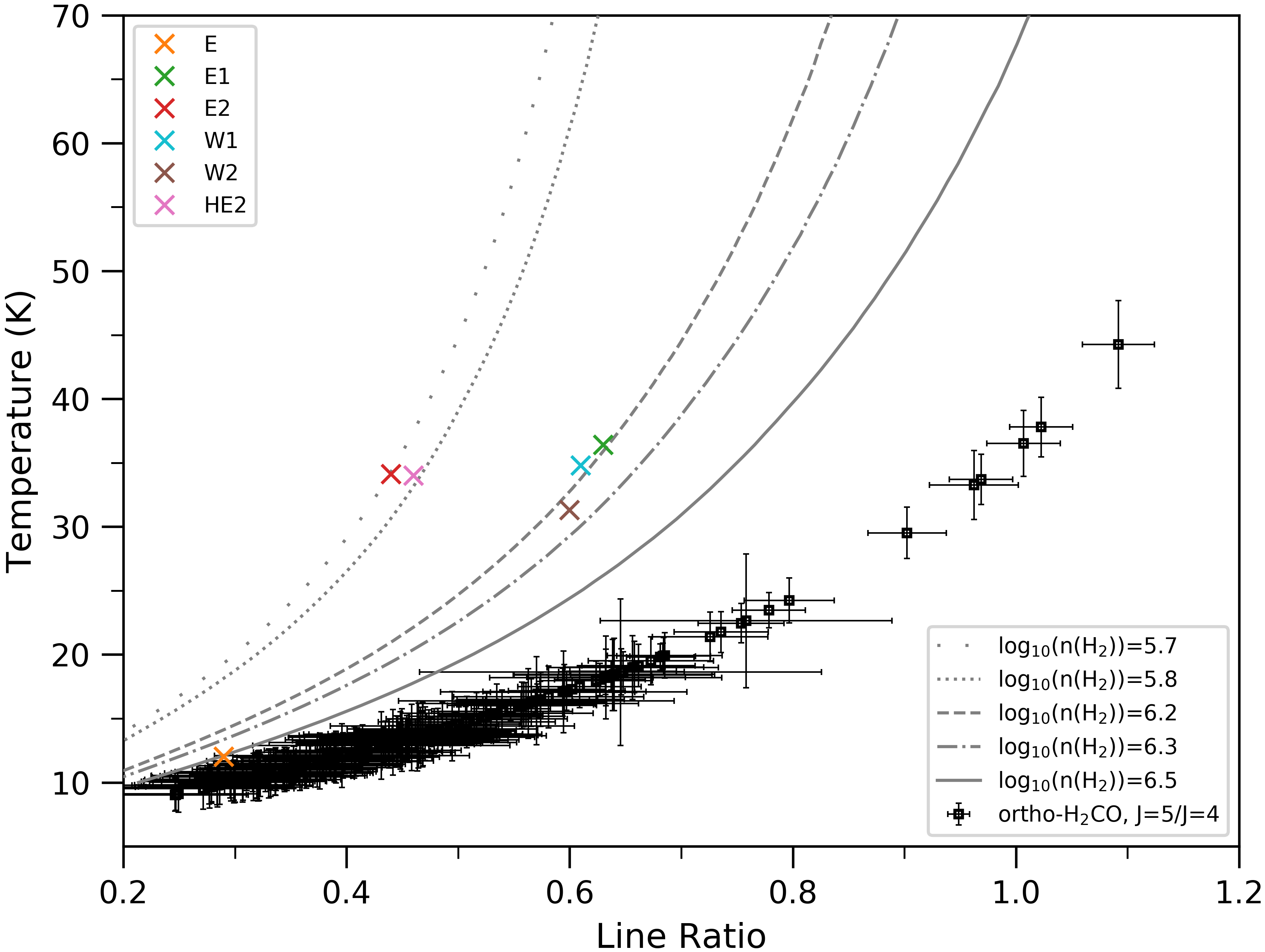}}
	\caption{Temperature (in Kelvin) as function of the H$_2$CO ($5_{1,5}-  4_{1,4}$)/($4_{1,4} - 3_{1,3}$) line ratio. The black datapoints illustrate the calculated excitation temperatures from the individual pixels of the corresponding line ratio maps. The gray lines indicate RADEX models of the kinetic temperature as function of line ratio for different values of H$_2$ volume density as indicated in the lower right panel. The models were computed using ortho-H$_2$CO column densities of (a) $\num{2.9}\times \num{10}^{14}\si{\per\centi\meter\squared}$ (b) $\num{3.9}\times \num{10}^{13}\si{\per\centi\meter\squared}$, which correspond to the values at A/B and the average on the other positions, respectively. Color markers indicate the line ratios and temperatures observed at the positions stated in the upper left corners.}
	\label{fig:radex_2}
\end{figure*}
\subsection{H\textsubscript{2} volume densities} \label{subsec:volumedens}
To determine the H$_2$ volume densities $n$ present in L1689N, a similar approach as above can be used for the line ratios and excitation temperatures derived from the H$_2$CO transitions ($5_{1,5} - 4_{1,4}$)/($4_{1,4} -  3_{1,3}$). In order to compare these transitions without the need to consider different filling factors, the maps were smoothed to the same resolution prior to the further analysis.

Since the ratios of lines with different $J$ are sensitive to density deviations~\citep{mangum1993formaldehyde}, the excitation temperature derived with these transitions is much lower than the actual kinetic temperature in the cloud.
Due to this, it is possible to get a crude estimate of the H$_2$ volume density by comparing RADEX models for the ($5_{1,5} - 4_{1,4}$)/($4_{1,4} - 3_{1,3}$) ratio with the temperatures derived in Sect.~\ref{subsec:temperatures}. In order to explain this process, the H$_2$ volume density is derived in detail for the position of IRAS\,16293 A/B.

Assuming LTE conditions, the kinetic temperature at this position is $\SI{62.4}{\kelvin}$, as derived from para-H$_2$CO transitions with $J_\U{up} = 4$ (See Sect.~\ref{subsec:temperatures}). Mapping of the H$_2$CO ($5_{1,5}-  4_{1,4}$)/($4_{1,4} - 3_{1,3}$) line ratios results in a value of $1.02$ at A/B, hence an excitation temperature of about $\SI{38}{\kelvin}$, much lower than the kinetic temperature. By comparing RADEX models with a range of H$_2$ volume densities as displayed in Fig.~\ref{fig:radex_2}a, it can be seen that a volume density estimate of $\log_{10}(n(\U{H}_2))=6.7$ predicts the kinetic temperature of $\SI{62.4}{\kelvin}$ for the observed line ratio of $\num{1.02}$. Based on the temperature uncertainties, this method eventually leads to uncertainties in the H$_2$ volume density of a factor on the order of 1.5.

RADEX models with an ortho-H$_2$CO column density of $\num{2.9}\times \num{10}^{14}\si{\per\centi\meter\squared}$ were computed for the position of IRAS\,16293 A/B, whereas a common column density of $\num{3.9}\times \num{10}^{13}\si{\per\centi\meter\squared}$ was applied for the other positions, on which we observe the transitions to be mostly optically thin. The RADEX models are shown next to the datapoints for the H$_2$CO $J_\U{up} = 5$/$J_\U{up} = 4$ ratios in Fig.~\ref{fig:radex_2}. The resulting H$_2$ volume densities can be found in Table~\ref{tab:h2}.

\begin{table}[hbt]
	\caption{Derived H$_2$ volume densities at the individual emission peaks.}
	\label{tab:h2}
	\centering
	\begin{tabular}{cccc}
		\hline
		\hline
		Position & T(p-H$_2$CO) & Line Ratio & $n$(H$_2)$ \\
		         & (K)            &             &  ($\si{\per\centi\meter\cubed}$)\\ \hline
		A/B & $62.4$ & $1.02$ & $5.0 \times 10^6$\\
		E1 & $36.4$ & $0.63$ &  $1.6 \times 10^6$\\
		E2 & $34.1$ & $0.44$ & $5.0 \times 10^5$\\
		W1 & $34.8$ & $0.61$ & $1.6 \times 10^6$\\
		W2 & $31.3$ & $0.60$ & $1.6 \times 10^6$\\
		HE2& $34.0$ & $0.46$ & $6.3 \times 10^5$\\ \hline
	\end{tabular}
	\tablefoot{The third column gives an overview about the H$_2$CO ($5_{1,5}-  4_{1,4}$)/($4_{1,4} - 3_{1,3}$) line ratios at the respective positions.}
\end{table}

While we observe high volume densities of $\num{5.0}\times\num{10}^{6}\si{\per\centi\meter\cubed}$ at the positions of the protostars IRAS\,16293 A/B, the emission peaks E2 and HE2 which relate to outer outflow positions show lower values of $\num{5.0}\times\num{10}^{5}\si{\per\centi\meter\cubed}$ and $\num{6.3}\times\num{10}^{5}\si{\per\centi\meter\cubed}$. The H$_2$ volume density estimations at the remaining emission peaks E1, W1, and W2 result in a common intermediate value of $\num{1.6}\times\num{10}^{6}\si{\per\centi\meter\cubed}$. These volume densities are in agreement with the values considered to test the assumption of LTE conditions in Appendix~\ref{app:RADEX_temp} (see Fig.~\ref{fig:radex_1}).
\section{Discussion}\label{ch:discussion}

\subsection{Dust continuum maps of L1689N}\label{sec:dustcontinuum}
\begin{figure*}[htb]
	\centering
	\begin{tabular}{c c}
	\subfigure[]{\includegraphics[width=0.47\textwidth,trim = 0 0 150 0,clip]{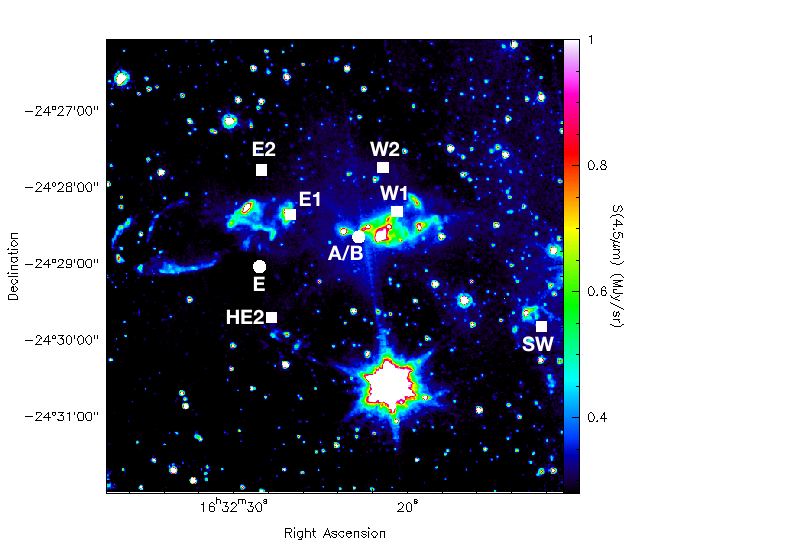}} & \subfigure[]{\includegraphics[width=0.47\textwidth,trim = 0 0 150 0,clip]{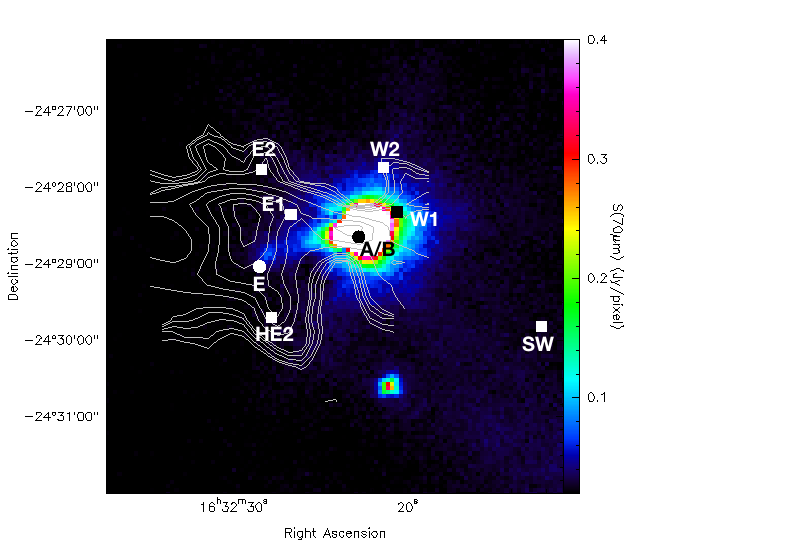}} \\
	\subfigure[]{\includegraphics[width=0.47\textwidth,trim = 0 0 150 0,clip]{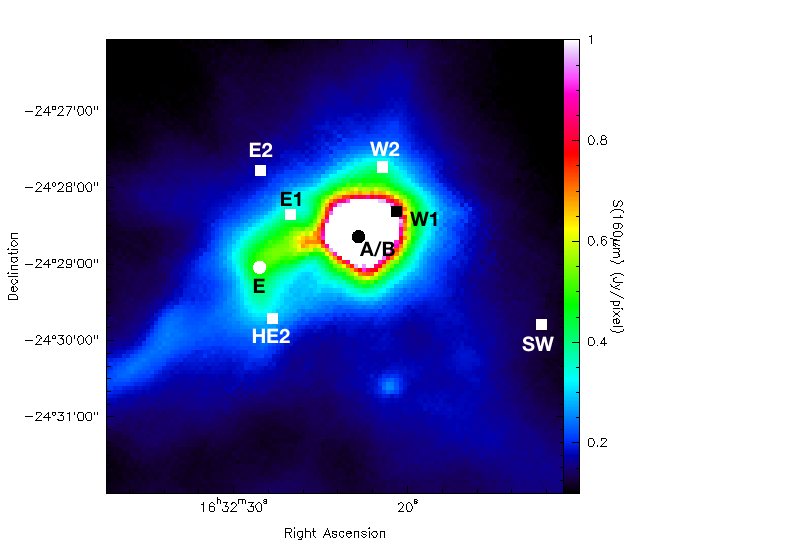}} & 	\subfigure[]{\includegraphics[width=0.47\textwidth,trim = 0 0 150 0,clip]{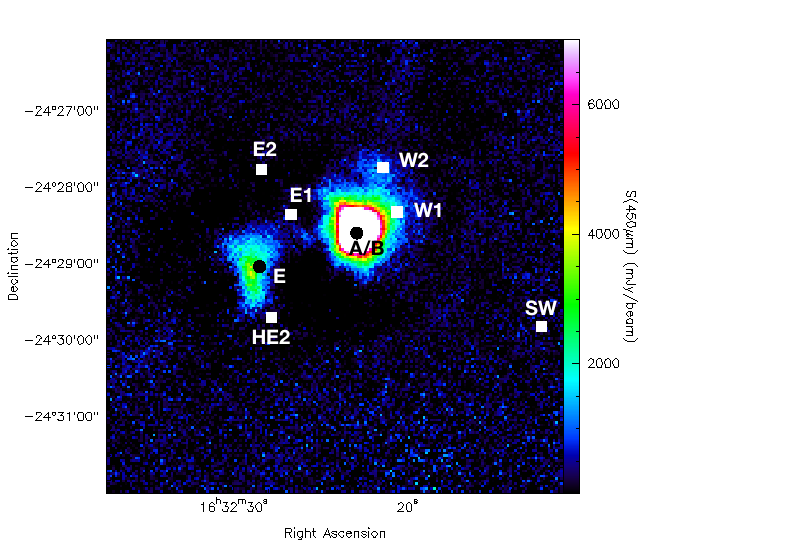}} \\
		\end{tabular}
	\caption{Continuum maps of the L1689N region at different wavelengths. a) Spitzer IRAC2 map at $\SI{4.5}{\micro\meter}$. b) Herschel PACS map at $\SI{70}{\micro\meter}$. The white contours in this panel indicate the integrated CO ($3 - 2$) intensity between $\SI{-27.0}{\kilo\meter\per\second}$ and $\SI{3.6}{\kilo\meter\per\second}$ at $\SI{1.6}{\kelvin\kilo\meter\per\second}$, $\SI{5.3}{\kelvin\kilo\meter\per\second}$, $\SI{13.2}{\kelvin\kilo\meter\per\second}$, $\SI{26.4}{\kelvin\kilo\meter\per\second}$ and $\SI{39.6}{\kelvin\kilo\meter\per\second}$ levels. c) Herschel PACS map at $\SI{160}{\micro\meter}$. d) JCMT SCUBA-2 map at $\SI{450}{\micro\meter}$. The markers show positions discussed in the previous sections.}
	\label{fig:cont_maps}
\end{figure*}
In order to compare the results of Sect.~\ref{ch:results} with the dust continuum, archival data of the region surrounding L1689N were obtained at wavelengths between $\SI{4.5}{\micro\meter}$ and $\SI{450}{\micro\meter}$, which were taken with the following instruments/telescopes: IRAC2/Spitzer\footnote{\url{https://irsa.ipac.caltech.edu/data/SPITZER/C2D}}, PACS/Herschel Space Observatory\footnote{\url{https://www.cosmos.esa.int}} and the SCUBA-2 James Clerk Maxwell Telescope\footnote{\url{http://www.cadc-ccda.hia-iha.nrc-cnrc.gc.ca/en/jcmt}} (JCMT).

A Spitzer $\SI{4.5}{\micro\meter}$ continuum map is shown in Fig.~\ref{fig:cont_maps}a. Besides continuum emission, lines from H$_2$ and vibrational excited CO are included in the $\SI{4,5}{\micro\meter}$ band, both of which can be excited by shocks~\citep{reach2006spitzer}. This band shows emission near all outflow-positions except for E2, whereas the emission does not peak at IRAS\,16293 A/B, E, and W2. In addition, the emission near 16293E indicates a splitting of the eastern blue lobe, of which the northern part is followed up by two bow-shocks in the east, slightly outside of our map coverage with LAsMA. Nevertheless, the CO~($3 - 2$) submillimeter line (see bottom panel of Fig.~\ref{fig:CO6-5}) shows strong emission in the direction of these bow-shocks, indicating that they likely originate from the E--W-outflow.

\begin{figure*}[tbp]
	\centering
	\begin{tabular}{c c}
	\subfigure[]{\includegraphics[width=0.48\textwidth]{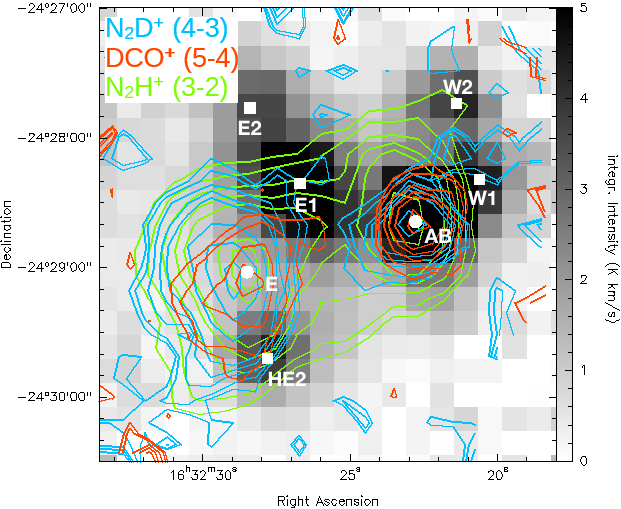}} & 	\subfigure[]{\includegraphics[width=0.48\textwidth]{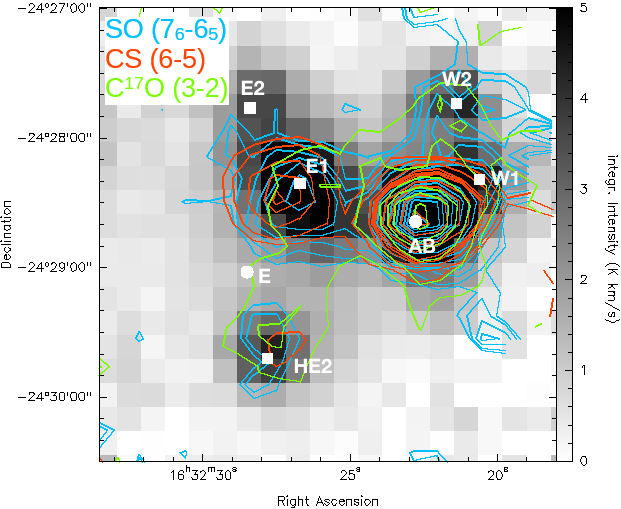}} \\
	\subfigure[]{\includegraphics[width=0.48\textwidth]{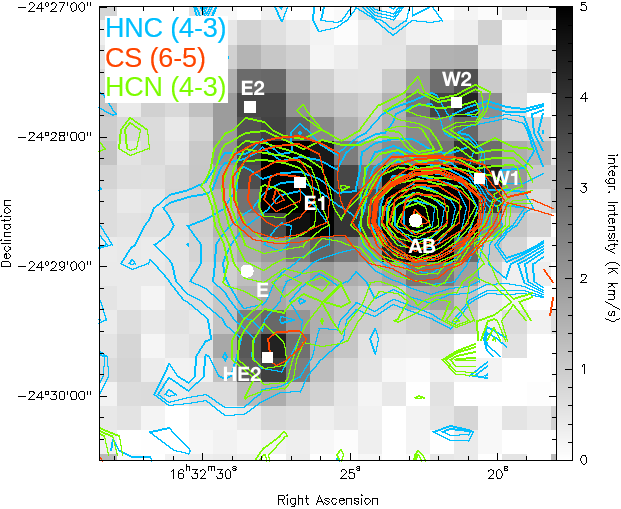}} & 	\subfigure[]{\includegraphics[width=0.48\textwidth]{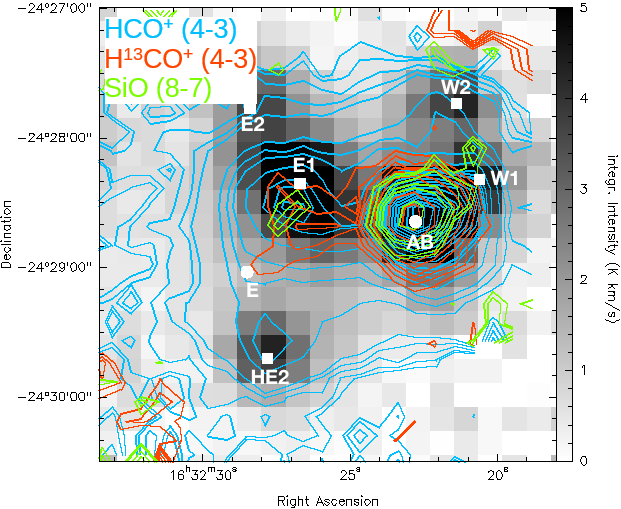}}\\
	\end{tabular}
	\caption{Gray-scale image of $\U{H}_2\U{CO}$ ($4_{1,4} - 3_{1,3}$) integrated in a velocity range between $\SI{-4.6}{\kilo\meter\per\second}$ and $\SI{9.9}{\kilo\meter\per\second}$ overlaid with contours of integrated line emission for multiple molecular species, as shown in the upper left corner. Contours are drawn at $3 \sigma$, in steps of $2\sigma$ between $4\sigma$ and $10\sigma$ and in steps of $10\sigma$ afterwards. Top panels: a) Additional contours are drawn for N$_2$H$^+$ ($18\sigma$) and DCO$^+$ ($15\sigma$). Major differences between displayed species can be seen around the region of the pre stellar core 16293E. b) Additional contours are drawn for CS ($9\sigma$), SO ($15\sigma$), and C$^{17}$O ($14\sigma$). Bottom panels: c) Additional contours are drawn for HNC ($14\sigma$), HCN ($16\sigma$), and CS ($9\sigma$). d) Additional contours are drawn for HCO$^+$ ($25, 55\sigma$) and H$^{13}$CO$^+$ ($5, 27\sigma$).  Notice the differences in peak emission around the E1 region, for which most species peak southern of formaldehyde. White markers show positions discussed in previous sections.}
	\label{fig:offsets}
\end{figure*}

Another strong continuum source exists south of IRAS\,16293 A/B, which~\citet{IRstar} identify as a young stellar object driving an outflow pointing toward the position HE2. This source can also be seen in the Herschel $\SI{70}{\micro\meter}$ map (Fig.~\ref{fig:cont_maps}b) which shows the dust continuum map together with contours of blue-shifted CO~($3 - 2$) emission. Since CO clearly shows the influence of the E--W-outflow on HE2 (also discussed in Sect.~\ref{subsec:outflows}), it is likely that the outflows of both sources contribute to the conditions in this region.

As longer wavelengths can be attributed to colder dust, the maps of $\SI{160}{\micro\meter}$ and $\SI{450}{\micro\meter}$ (Fig.~\ref{fig:cont_maps}c,d) emission show the bulk of the cloud material. As it can be seen in these maps, the cold core 16293E is more clearly detected at these wavelengths. In addition, more extended emission is observed in all directions, particularly to the southeast of 16293E at $\SI{160}{\micro\meter}$.  

Finally, the coldest dust components are probed by continuum maps with wavelengths of $\SI{450}{\micro\meter}$ and above. As shown in Fig.~\ref{fig:cont_maps}d, this long wavelength emission is mainly seen in the close vicinity of the cloud cores IRAS\,16293 A/B and E. Additional emission can be seen at the position W2 north of IRAS\,16293 A/B, which will be discussed further in Sect.~\ref{sec:W2}.

\subsection{Spatial distribution of selected species at the emission peaks}
The maps shown in Appendix \ref{app:maps} show the morphology of the molecules in the IRAS\,16293 complex. In order to visualize more easily the differences in the spatial distribution among the emission peaks, we have overlaid the emission for some species that have strong large-scale emission for a more direct comparison.

In Fig.~\ref{fig:offsets}a, we show the moment 0 maps of the H$_2$CO $(4_{1 \,4}-3_{1 \,3})$ transition (grayscale), overlaid with the N$_2$D$^+$ $(4-3)$, DCO$^+$ $(5-4)$, and N$_2$H$^+$ $(3-2)$ emission in contours. From this figure, we see that all four species have the same peak around IRAS\,16293 A/B. Indeed, the emission of all considered molecules in all panels peaks at the exact position of the embedded protostars A and B. In contrast, Fig.~\ref{fig:offsets}a shows that the emission peaks in the prestellar core 16293E differ for the presented species. The peak of DCO$^+$ emission is located 5.3$''$ toward the south of the reference position for 16293E (marked with a white point as in previous figures), while the N$_2$H$^+$ peak is located 1.1$''$ to the north and the N$_2$D$^+$ is 7.2$''$ offset toward the east of 16293E respectively. The offsets between some species such as N$_2$D$^{+}$ and ND$_3$ were noted by \citet{lis2016}, suggesting that they might be due to either intrinsic abundance gradients or optical depth effects. 

In Fig.~\ref{fig:offsets}b, we compare the spatial distribution of H$_2$CO $(4_{1 \,4}-3_{1 \,3})$ transition (grayscale), together with the SO $(7_6-6_5)$, CS $(6-5)$, and the C$^{17}$O $(3-2)$ emission peaks. We see a more pronounced difference around E1 for CS and SO, which have an offset from E1 of 12.1$''$ and 6.8$''$ respectively. The C$^{17}$O emission around E1 is weak and does not extend toward E2. It seems there is also an offset between these four species at HE2 but it is more complicated to see as the emission is not very strong at that position. 

In Fig.~\ref{fig:offsets}c we show a similar comparison between HNC $(4-3)$, CS $(6-5)$, and HCN $(4-3)$. At HE2, there is also an offset between CS and HCN. In contrast, these species are distributed quite similar around E1, while HNC which seem to peak about 30$''$ to the south of E1. In Fig.~\ref{fig:offsets}d, the HCO$^+$ $(4-3)$, H$^{13}$CO$^+$ $(4-3)$, and SiO $(8-7)$ in contours are shown. These 3 species peak at about the same position south of E1. Of these species, only H$_2$CO and HCO$^+$ are present at HE2 and seem to peak at about the same position.

\citet{hirano2001} found an anticorrelation between SiO $(2-1)$ and H$^{13}$CO$^+$ $(1-0)$ (their Fig.~4), suggesting that the SiO emission arises from the region in which the outflow from IRAS\,16293 A/B is interacting with the prestellar core 16293E. We observe higher energy transitions of these two species in our data (SiO $(8-7)$ and H$^{13}$CO$^+$ $(4-3)$) but we do not see such anticorrelation, probably because our observed transitions have larger critical densities.

\begin{figure}[tbp]
	\centering
	\includegraphics[width=0.48\textwidth]{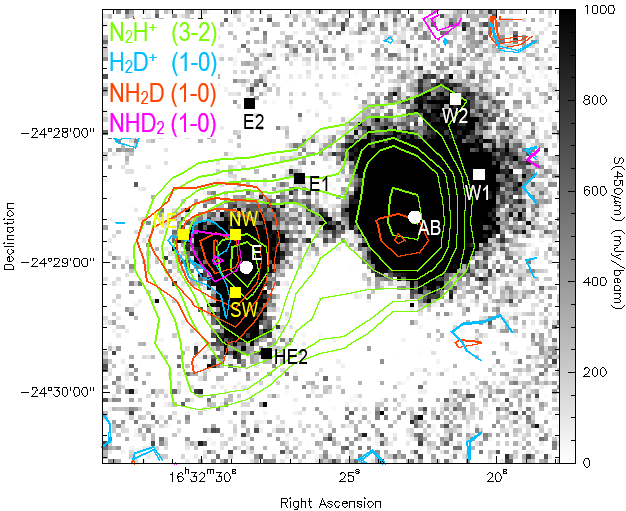}
	\caption{Continuum  JCMT SCUBA-2 image at 450 $\mu$m (grayscale), overlaid with the emission of NH$_2$D $(1-0)$ (red), NHD$_2$ (magenta), H$_2$D$^+$ $(1-0)$ (blue), and N$_2$H$^+$ $(3-2)$ (green). Contours of N$_2$H$^+$ are the same as in Fig.~\ref{fig:offsets}, while contours of NH$_2$D, NHD$_2$, and H$_2$D$^+$ are drawn at ($1\sigma$, $1.5\sigma$, $2\sigma$, $3\sigma$, $3.3\sigma$), ($1\sigma$, $1.1\sigma$), and ($1\sigma$, $1.2\sigma$)  respectively. The three yellow squares mark the positions of the extracted spectra for the analysis in Sect.~\ref{sec:kinematics}.}
	\label{fig:H2D+}
\end{figure}

It is worth noticing that the emission of H$_2$D$^+$ ($1-0$) behaves differently to other species in the prestellar core 16293E. In Fig.~\ref{fig:H2D+}, we show the continuum emission at 450\,$\mu$m obtained with the JCMT SCUBA-2 (grayscale), overlaid with emission of H$_2$D$^+$, NH$_2$D, NHD$_2$, and N$_2$H$^+$ in contours. The deuterated species show an emission peak in the northeast with respect to the reference position of 16293E as traced by N$_2$H$^+$ (see Fig.~\ref{fig:offsets}). Interestingly, the NH$_2$D emission also does not peak exactly at the position of the A/B protostars, in contrast to the majority of the molecular tracers. The H$_2$D$^+$ emission seems to be located exclusively in the eastern part of the prestellar core, without expanding further toward the N$_2$H$^+$ peak. \citet{Pagani2015} also note this behavior of H$_2$D$^+$ when comparing their data with the distribution of the continuum emission. Based on the finding that the position of the dust emission peak changes with wavelength, they suggest a temperature gradient in this region with H$_2$D$^+$ tracing the coldest and highest column density gas.

The offset between the continuum emission and the H$_2$D$^+$ peaks indeed is likely related to the difference in the physical conditions (temperature and density) between these two spots. In particular, determining the temperature from H$_2$CO line ratios at these peaks is not possible since we do not detect emission of the excited H$_2$CO transitions $(4_{2\, 3}-3_{2\, 2}, 4_{2\, 2}-3_{2\, 1}, 4_{3\, 1}-3_{3\, 0}, 4_{3\, 2}-3_{3\, 1})$ at these positions. In contrast, N$_2$H$^+$ is well detected in the vicinity of 16293E.

\begin{figure}[tbp]
	\centering
	\includegraphics[width=0.48\textwidth, trim = 0 0 130 0,clip]{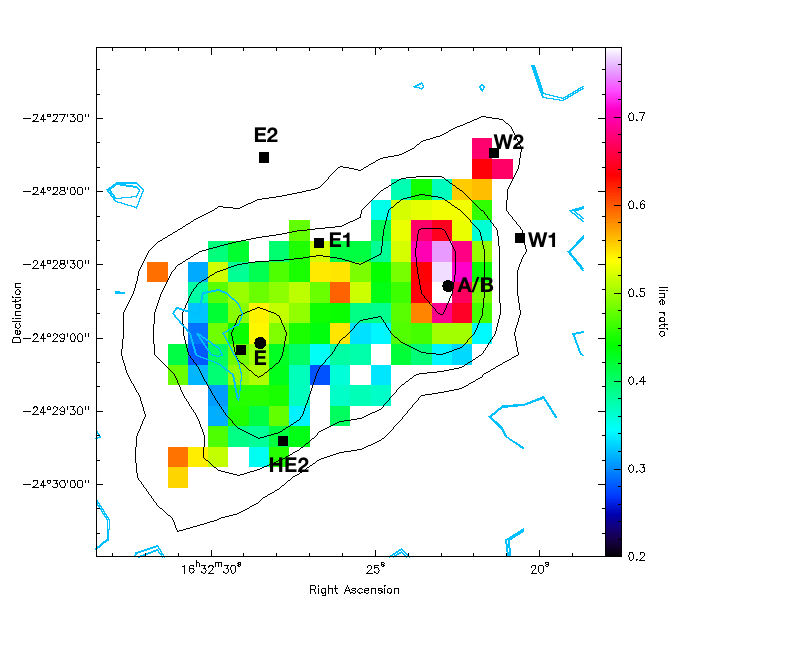}
	\caption{Line ratio between the N$_2$H$^+$ $(4-3)$ and N$_2$H$^+$ $(3-2)$ transitions (color scale). The blue contours show the emission of  H$_2$D$^+$ while the black contours indicate the extent of the N$_2$H$^+$ $(3-2)$ emission. The additional square marker east of 16293E marks the $\SI{450}{\micro\meter}$ dust continuum peak. Low line ratio values close to the H$_2$D$^+$ peak agree with the interpretation of having the coldest (and densest) part toward this spot.}
	\label{fig:ratio_N2H+}
\end{figure}

Figure~\ref{fig:ratio_N2H+} shows a map of the line-ratio between the N$_2$H$^+$ $(4-3)$ and N$_2$H$^+$ $(3-2)$ transitions, where we also display the H$_2$D$^+$ emission in blue contours as a reference. Several pixels close to the H$_2$D$^+$ peak show lower line ratios as compared to the continuum peak position. Since the N$_2$H$^+$ $(4-3)$/N$_2$H$^+$ $(3-2)$ ratio probes to first order the pressure of the gas, Fig.~\ref{fig:ratio_N2H+} indicates that H$_2$D$^+$ peaks at a position of significant change in pressure. This might also be indicative of lower temperatures in this region, which would agree with the interpretation of having the coldest part of the prestellar core close to the H$_2$D$^+$ peak. Unfortunately, it is not straightforward to assess if there are also important density differences between the two peaks. A density gradient in this region would imply variations in the observed line ratios, which prevents us from discriminating between the effects of the two physical quantities on this region.

Note that the W2 position shows large values for the N$_2$H$^+$ $(4-3)$/N$_2$H$^+$ $(3-2)$ line ratio. The spectra at the corresponding pixels show an artifact close to the N$_2$H$^+$ $(4-3)$ transition, which is likely the cause of these seemingly high line ratios. The actual N$_2$H$^+$ $(4-3)$ line emission at W2 is very faint and not detected with a 3$\sigma$ significance.

Based on the contour plots shown in Fig.~\ref{fig:offsets}, it is clear that we observe the spacial distribution of different molecular species to differ across the whole eastern cloud. This seems to be in contrast of what we observe at the position of the protostars A/B, for which we observe almost all species to peak at the exact position of the protostellar core. Also, we do not see significant differences between species in the peak positions at W1. In contrast, species do not show a clear emission peak at W2, but rather broadly distributed weak emission. 
The observed morphological differences between the studied species could be related to the interaction between the outflows from IRAS\,16293--2422 and the cold core 16293E. Therefore, we study the kinematics in 16293E with more detail in the next section.

\subsection{The interaction between 16293E and the outflows from IRAS\,16293--2422 A/B}
As previously discussed, one of the outflows arising from IRAS\,16293--2422 AB is proposed to be interacting with the prestellar core 16293E~\citep[e.g., see][]{stark2004,lis2016}. As earlier works were limited to observations of a small number of species, we aim to test this scenario with our unbiased survey of the region.

\subsubsection{Velocity offsets between the deuterated and nondeuterated species at 16293E}
Since both deuterated and nondeuterated species have been studied in the prestellar core 16293E, a difference in the velocities between them with respect to the cloud velocity (4.0\,$\si{\kilo\meter\per\second}$) has been observed. In particular, deuterium-bearing species such as ND, N$_2$D$^+$, DCO$^+$, D$_2$H$^+$, and ND$_2$H show peak velocities closer to 3.5$-$3.7\,$\si{\kilo\meter\per\second}$ \citep{lis2002,Vastel2004, Gerin2006,Vastel2012, bacmann2016}, while other tracers such as H$^{13}$CO$^+$ and C$^{17}$O show peak velocities closer to 3.8$-$4.0\,$\si{\kilo\meter\per\second}$ \citep{lis2002,stark2004}. \citet{Vastel2012} suggest that such behavior could be due to the interaction between the outflows from A/B and the prestellar core 16293E, introducing a difference between the deuterated and nondeuterated molecules. Since this effect could also be introduced by the different beam sizes corresponding to the different observations of these species, we tested if this trend can be confirmed from our observations, which have the same beam.

Based on Gaussian fits to estimate the integrated intensity of all the observed species reported in Table~\ref{tab:cdp1}, it can be concluded that we do see a similar trend in our data. Indeed, the deuterated species H$_2$D$^+$, DNC, NH$_2$D, NHD$_2$, and N$_2$D$^+$ have a peak velocity of 3.6\,$\si{\kilo\meter\per\second}$, while the DCO$^+$ and N$_2$H$^+$ have a peak velocity of 3.7\,$\si{\kilo\meter\per\second}$. On the other hand, the rest of the species have larger peak velocities between 3.8 and 4.0\,$\si{\kilo\meter\per\second}$, closer to the cloud velocity (except for CN and NO).
Interestingly, this velocity offset trend for deuterated species is not followed by the HDCO, D$_2$CO, and DCN molecules which appear to be at the cloud velocity. 

\subsubsection{Kinematics in the vicinity of 16293E}\label{sec:kinematics}
\citet{lis2016} study the emission of NH$_2$D based on HIFI Herschel Space Observatory observations at 494.454\,$\si{\giga\hertz}$. They extracted the spectrum of this transition at three different positions in 16293E and estimated the peak velocities and line widths of the individual spectra (see their Fig.~5). They conclude that the change in the peak velocity and line width of NH$_2$D in the vicinity of 16293E is an evidence of the outflowing gas interacting with this source. 

\begin{figure}[tbp]
	\centering
	\includegraphics[width=0.48\textwidth]{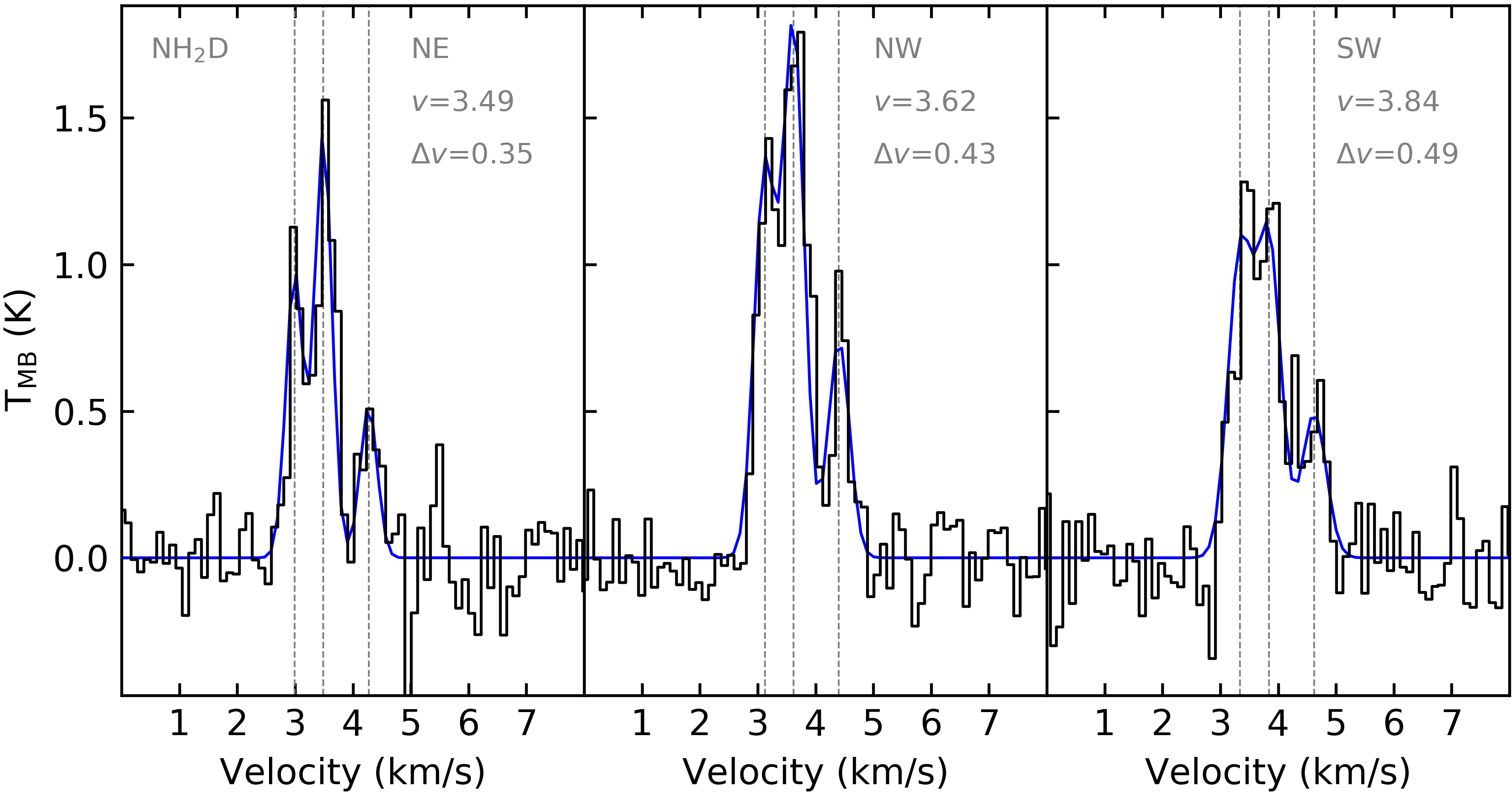}
	\caption{Line profiles of the NH$_2$D (332781.890\,$\si{\mega\hertz}$) transition in the vicinity of 16293E. The data is displayed in black while the blue line shows the computed Gaussian HFS fits to the spectra. Gray vertical lines mark the position of the fit Gaussians. A trend to larger line widths and higher velocities is visible in NE--SW direction.}
	\label{fig:gaussfits}
\end{figure}

To investigate if such behavior holds for other molecular species in this source, we made use of our unbiased survey to see if this is valid for other molecular species. As seen in Figs.~\ref{fig:offsets} and \ref{fig:H2D+}, different species are distributed differently along 16293E. Therefore, we have selected three positions where we have clear emission for all deuterated species that peak near 16293E: H$_2$D$^+$, DCO$^+$, DNC, N$_2$D$^+$, NHD$_2$, and NH$_2$D. The positions are shown in Fig.~\ref{fig:H2D+} and are located at the coordinates: NE: (16$^\U{h}$32$^\U{m}$30.6$^\U{s}$, -24$^\circ$28$'$46.7$''$), NW: (16$^\U{h}$32$^\U{m}$28.9$^\U{s}$, -24$^\circ$28$'$46.7$''$), SW: (16$^\U{h}$32$^\U{m}$28.9$^\U{s}$, -24$^\circ$29$'$13.7$''$). Although these positions are different to the ones chosen by \citet{lis2016}, they are still well separated. We have extracted the spectra at each of these positions and fit a Gaussian to the line profiles using the CLASS software to determine their velocity peak and line widths. The results from the fits are shown in Table \ref{tab:vel_offsets} in Appendix \ref{app:gauss}. It is worth mentioning that both NHD$_2$ and NH$_2$D present hyperfine structure in their spectra, resulting in three lines close in frequency for each species. We fit the visible HFS components under the assumption of same line widths for all components using the component-separations from the CDMS. The fit spectra are shown in Fig.~\ref{fig:gaussfits} and the Appendix~\ref{app:gauss}.

Most species show a similar trend as \citet{lis2016} find for NH$_2$D. More precisely, we see that the peak velocity increases from the NE to the SW position of the core, which is indicative of the interaction with the outflows powered by IRAS\,16293--2422. Although, this trend was also observed to be present on larger scales, as shown in Fig.~\ref{fig:rotation} of Sect.~\ref{sec:rotation}. Due to this larger scale velocity field, the peak velocity alone is not a clear indicator of motions from the proposed core outflow interactions. Nevertheless, a similar trend of line widths increasing from NE to SW positions is observed near 16293E, with the exception of DCO$^+$ and the weak lines of NHD$_2$. This may be a good indicator for a core-outflow interaction, as the observed velocity gradient should not influence the line widths. Since the emission of DCO$^+$ peaks further southwest compared to the other considered species, it may trace a differently effected layer of the gas.
As compared to \citet{lis2016}, we extracted the analyzed spectra further to the inner part of the cold cloud core, which may be located at the NH$_2$D emission peak. Such trends therefore might indicate that the interaction of the outflows is propagating toward the inner part of the core.

\subsection{Chemical differences at the selected positions}
\label{sec:chemistry}
In the previous sections, we analyze the distribution of molecular species across the L1689N cloud and derived column densities of these species at the visible emission peaks. Table~\ref{tab:molecules_summary} gives an overview about the detected species at all peaks, while the derived molecular abundances given in Table~\ref{tab:cd} are visualized in Fig.~\ref{fig:cd}. To emphasize further the chemical differences between pairs of sources, we created Venn diagrams that cover the occurrences of different species (see Appendix \ref{app:VennDiag}).

These diagrams show that E1 not only contains a larger number number of molecules than E2, but all the species observed in E2 are also found in E1. In a previous study, \citet{castets2001} suggested that the simultaneous presence of H$_2$CO and SiO emission in E1 might indicate that E1 is actually younger than E2, given that this source was observed to only emit SiO and this molecule is expected to last longer than H$_2$CO. From our maps, we see that E1 is indeed chemically richer than E2, while we do observe the presence of H$_2$CO emission at E2. Interestingly, both E1 and E2 show SO emission, while only E1 shows SO$_2$ emission.

It can be noticed that the emission peak W1 and W2 share the same species, although W2 shows SO, HNC, and CS less abundant, while being richer in N$_2$H$^+$. The emission peaks that show the largest number of detected species besides IRAS\,16293 A/B correspond to the prestellar core 16293E and HE2. Similar to previous studies, we mostly see probes of cold and dense gas at the E position. We find comparable molecules toward HE2 and E, with the exception of H$_2$D$^+$ and NHD$_2$ being absent in the HE2 emission peak. While 16293E is a quiescent cloud core, the line profiles at the position HE2 (see Fig.~\ref{fig:spec_E2}) show rather blueshifted shocked gas. As can be seen in Fig.~\ref{fig:cd}, all species detected at IRAS\,16293~A/B and E are also observed at HE2, which is in favor of an interaction between the E--W outflow and the prestellar core.

Based on the abundant molecules, E1 can be interpreted as prime region of core-outflow interactions, since it shows moderate line wings and both, dense gas tracers as well as likely shock tracers such as methanol and formaldehyde. Shock dominated positions are W1 and E2, both with high velocity red-shifted wings and a lack of dense gas tracers, while being bright in methanol and formaldehyde. These two peaks exhibit basically the same number and type of observed species. The only difference is that W1 has CS emission, while E2 does not, although the CS maps suggest that a contribution from the envelope around A/B sources can affect W1.

\subsection{On the nature of W2}\label{sec:W2}
The emission at the position of the W2 peak was already reported in early studies of the L1689N region~\citep[e.g.,][]{mizuno1990}. However, the origin of this emission has not yet been completely clarified. On the one hand, \citet{hirano2001} interpreted the peak as northern boundary of the western red lobe of the E--W outflow where the outflowing gas encounters the ambient medium. On the other hand, \citet{castets2001} suggested that W2 could be either an additional outflow of IRAS16293 A/B with a deeply embedded and therefore invisible counterpart, or an embedded source itself. 

As can be seen in Fig.~\ref{fig:cont_maps}d, the SCUBA-2 map at $\SI{450}{\micro\meter}$ reveals long wavelength continuum emission at the W2 emission peak, which suggests the presence of a cold dust source at this position. This conjecture is supported by the line profiles at this position (see Fig.~\ref{fig:spec_E2}), which show moderate line wings alongside self-absorption. Compared to the shock dominated regions W1 and E2, this region therefore seems to be relatively quiescent while the high optical depth suggests the presence of dense gas. Based on the H$_2$ densities given in Table~\ref{tab:h2}, the embedded object is likely to be less massive then the IRAS\,16293 A/B cores. 

Comparing the occurring species at this position, we note a similarity between the chemistry of W1 and W2 (see Sect.~\ref{sec:chemistry}). While we detect emission from a large number of species toward W2, the distribution of this emission (see Fig.~\ref{fig:offsets} and Appendix~\ref{app:maps}) only shows a clear peak for H$_2$CO and CH$_3$OH, while other species show rather broadly distributed weak emission in the region. Emission from other species observed at W2 therefore may be attributed to the extended envelope of IRAS\,16293 A/B instead of the cold source embedded at W2. 

It still remains unclear if the embedded object seen at W2 interacts with the outflows of IRAS\,16293 A. The CO~(3$-$2) maps suggest that W2 indeed is located at the northern boundaries of the E--W outflow, while the narrow line profiles at W2 do not suggest a strong interaction, as it is the case at E1. There may be a scenario in which the small scale NW-SE outflow, which~\citet{girart2014} observe to be redirected by IRAS\,16293~B, extends on larger scales up to W2. Nevertheless, since we do not observe a southern counterpart of this outflow with a spacing comparable to W2, it is unlikely that this outflow extends to larger scales.

\subsection{Molecular abundances in L1689N as compared to the literature}
Adapting a canonical $\U{CO}/ \U{H}_2$ ratio of $10^{-4}$, the derived column densities of HCO$^+$ and DCO$^+$ species at the position of IRAS\,16293 A/B (see Sect.~\ref{sec:cd} and Table~\ref{app:cd_co}) are comparable to the estimates from \citet{stark2004}, who derive $\U{HCO}^+/ \U{H}_2$ and $\U{DCO}^+/ \U{H}_2$ ratios on the order of $1\times 10^{-9}$ and $2\times 10^{-11}$. In contrast, the ratio of $\U{N}_2\U{H}^+/\U{H}_2 = \num{2.2e-10}$ at IRAS\,16293 A/B is about a magnitude larger than the value of $3\times 10^{-11}$ estimated by \citet{stark2004}, although it is in agreement with the value of $1.4\times 10^{-10}$ found by \citet{jorgensen2004}.

Also, the SiO column density at the position E1 is about an order of magnitude smaller than expected based on the results of~\citet{castets2001}, who derived a value of $5.8\times 10^{13}$\,\si{\per\centi\meter\squared} for the total SiO column density toward this position. This deviation originates in the different analyzed transitions, as \citet{castets2001} observe the SiO ($2-1$), SiO ($3-2$), and SiO ($5-4$) transitions, which have much smaller critical densities than the SiO ($8-7$) transition that we use in our estimations.

In general, most of the derived abundances at IRAS\,16293 A/B and E are comparable with the mean values for Class 0 protostars and prestellar objects found in the sample of~\citet{jorgensen2004}. Major differences to the sample of \citet{jorgensen2004} exist for N$_2$H$^+$, which is an order of magnitude less abundant in IRAS\,16293 A/B then in other Class 0 protostars, where they observe a N$_2$H$^+/\U{H}_2$ ratio of $2.5\times 10^{-9}$. Also the observed DCO$^+$ abundance of $9.8\times 10^{-11}$ with respect to H$_2$ is almost an order of magnitude larger in 16293E than in the compared prestellar objects, that show abundances of $2.3\times 10^{-11}$. The latter might be caused by the computation of our abundances relying on CO, while other authors used millimeter and submillimeter dust continuum emission to derive H$_2$ column densities. In cold regions such as 16293E the CO might be frozen out onto grains which would enhance molecular abundances relative to CO.

For Class 0 protostars in the Orion region specifically, \citet{johnstone2003astrochemistry} derived H$_2$CO and CH$_3$OH column densities of $4\times10^{13}\si{\per\centi\meter\squared}$ and $2\times 10^{13}\si{\per\centi\meter\squared}$ respectively. While these values are in agreement with values we derive for the outer layers of the protostar envelope in Appendix~\ref{app:twolayers}, our derived single-layer LTE column densities in Table~\ref{tab:cd} are about an order of magnitude larger. Therefore, the observed transitions are likely to include emission originating from the hot corino of the protostar.

Besides that, the $^{32}\U{SO}/^{34}\U{SO}$ ratio at IRAS\,16293 A/B suggest an $^{32}\U{S}/^{34}\U{S}$ isotope ratio of $\num{13.9}$, only about half of the value $24.4 \pm 5.0$ derived by \citet{Chin1996} which is commonly cited for comparable sources. Interestingly, this value is closer to the $^{32}\U{S}/^{34}\U{S}$ isotope ratio of $16.3^{+2.1}_{-1.7}$ observed toward the Galactic center \citep{Humire2020}. While the SO line profile appears mostly Gaussian, a reason for the deviation might be a high opacity of the considered SO transitions even without showing features of major self-absorption.

Finally, the deuterium fractionation as derived from the $\U{DCO}^+/\U{HCO}^+$ ratio is enhanced at 16293E, where a fractionation of $0.05$ is detected. The prestellar objects considered by~\citet{jorgensen2004} show on average deuterium fraction of $0.03$. However, the high deuterium fraction at 16293E is in agreement with the findings of~\citet{lis2002}, who observed a D/H ratio of about 10\% in this source. In contrast, a smaller fractionation of $0.006$ is observed at IRAS\,16293 A/B, which is in agreement with the average deuterium fractionation of $0.007$ in Class 0 protostars, as found in the sample of \citet{jorgensen2004}. The high deuterium fractionation in addition to the presence of several deuterated species make 16293E an interesting target for deeper integrating observations with the goal of investigating the variety of deuterated species in prestellar objects.

\section{Summary}\label{ch:summary}

In this work, we have studied the large-scale spatial distribution of the molecular content in the environment of IRAS\,16293--2422. This work fills an important information gap, since many previous studies toward IRAS\,16293--2422 and its environment were based on pointed observations, therefore missing the complex spatial molecular morphology. The analysis is based on LAsMA and FLASH+ APEX observations covering frequency ranges between 277--375\,$\si{\giga\hertz}$ and 476--493\,$\si{\giga\hertz}$ respectively. The results of this study are summarized as follows:

\begin{itemize}

    \item[$\bullet$] We identify a total of 144 transitions from 36 different molecular species in the observations. This is the first time that maps of the emission of such a large number of molecules have been presented for this region, which brings new information on the morphology of the molecular environment of IRAS\,16293--2422. \\
    
    \item[$\bullet$] The produced large-scale molecular maps of L1689N show the two cores of IRAS\,16293--2422 A/B and 16293E well separated and embedded in the molecular cloud, with extended envelopes surrounding them.~Also, the emissions trace the two known outflows driven by the Class 0 protostar of IRAS\,16293--2422 A. Outflow velocities up to $\SI{20}{\kilo\meter\per\second}$ are detected in the northeastern outflow of IRAS\,16293--2422 A.\\
    
    \item[$\bullet$] To easily compare and visualize the chemical wealth of each of the emission peaks, we have produced Venn diagrams, which allow us to directly analyze the molecular content. We find that the emission peaks related with the outflows show emission from other molecules besides the typical shock tracers (SiO, CO) such as HCN, CS, and NO. In addition, emission from deuterated species also is observed at these positions.\\

    \item[$\bullet$] A large-scale velocity gradient is observed for some species. Part of this gradient might be associated with the envelope around the A/B sources as it was suggested in previous studies, although we cannot disentangle the pure envelope rotation motion from the large-scale cloud kinematics.\\

    \item[$\bullet$] We have estimated new kinetic temperatures from para-H$_2$CO line ratios in L1689N at all considered positions, except for the prestellar core 16293E which shows only very weak H$_2$CO emission. The derived temperatures range from $\SI{31.3}{\kelvin}$ at the W2 position to $\SI{62.4}{\kelvin}$ as measured at IRAS\,16293 A/B. These values are in agreement with previous H$_2$CO temperature estimations in the literature.\\

    \item[$\bullet$] We have computed new column densities for all the detected species in all emission peaks by producing synthetic spectra using the LTE radiative transfer CLASS module weeds. Upper limits were computed where appropriate and the corresponding abundances relative to CO were calculated.\\
    
    \item[$\bullet$] We have produced a complex non-LTE radiative transfer model with multiple physical components to reproduce the self-absorption H$_2$CO line profiles observed toward all sources, except for those were the wings produced by the outflows have an important contribution. We find that the results from this RADEX model are in agreement with the average values derived from our LTE modeling.\\
    
    \item[$\bullet$] The derived H$_2$CO temperatures were used in conjunction with additional line ratios of H$_2$CO lines with different $J_\U{up}$ to estimate the H$_2$ volume densities in L1689N, resulting in values up to $\num{5}\times 10^6\si{\per\centi\meter\cubed}$ at IRAS\,16293--2422 A/B and values on the order of $\num{1}\times 10^6\si{\per\centi\meter\cubed}$ for the other considered positions.\\
    
    \item[$\bullet$] We tested the proposed scenario of an interaction of one of the outflows arising from IRAS\,16293--2422 A/B with the prestellar core 16293E. In this process, we were able to confirm the velocity offset between deuterated and nondeuterated species reported in literature and the trend of increasing velocities and line widths along the NE--SW axis across 16293E. These results, in combination with the high number of molecular species observed in the eastern cloud region, confirm the presence of an interaction between prestellar core and outflow.\\
    
    \item[$\bullet$] We show dust continuum maps obtained with Spitzer, Herschel, and JCMT telescopes. These maps reveal emission at the position W2 north of IRAS\,16293 A/B, which appears to be unrelated to the outflow-structure of the protostars. A comparison between the continuum and molecular maps suggests that the origin of the emission at W2 might be due to a colder dust source embedded in the L1698N cloud, as major dust-continuum emission is observed for long wavelengths at this position. It is unclear if the embedded source is influenced by the outflows of IRAS\,16293--2422 A.\\
    
    \end{itemize}
    In conclusion, combining all the results obtained in this work allows us to give a more complete and detailed view on the complex molecular distribution in the environment of IRAS\,16293--2422. We have have been able to determine new values of physical parameters such as kinetic temperatures, column densities, and volume densities, and present a detailed description of the chemical content found at each of the cloud cores and emission peaks, including the interaction region between IRAS\,16293--2422 and 16293E.

\begin{acknowledgements}
The authors are grateful to the anonymous referee for the useful comments and suggestions that helped improve this paper. We thank Arnaud Belloche for an early reading of the manuscript and for valuable suggestions.
This publication is based on data acquired with the Atacama Pathfinder Experiment (APEX) under programme ID [M-0102.F-9519A-2018]. APEX is a collaboration between the Max-Planck-Institut f\"{u}r Radioastronomie, the European Southern Observatory, and the Onsala Space Observatory. This work was partially funded
by the Collaborative Research Council 956 "Conditions and impact
of star formation" funded by the Deutsche Forschungsgemeinschaft (DFG). 
\end{acknowledgements}

\bibliographystyle{aa} % style aa.bst
\bibliography{ref} 

\begin{appendix}
\section{Observed frequency-bands}\label{app:obs_bands}
\begin{table}[H]
	\caption{Summary of the setups that were used during each observation run with the LAsMA receiver.}
	\label{tab:freq_int}
	\centering
	\begin{tabular}{c|ll}
		\hline
		\hline
		Observation Date & LSB / GHz  & USB / GHz \\ \hline
		07-19-2019           & 277.8    \, \, -- \, \,   281.8    & 289.8    \, \, -- \, \,   293.8    \\
		                     & 332.0    \, \, -- \, \,   336.0    & 344.0   \, \, -- \, \,    348.0 \\
		07-20-2019            & 292.8   \, \, -- \, \,    296.8   & 304.8   \, \, -- \, \,    308.8   \\ 
		07-21-2019            & 292.8   \, \, -- \, \,    296.8   & 304.8    \, \, -- \, \,   308.8     \\
		                      & 332.0   \, \, -- \, \,     336.0  & 344.0   \, \, -- \, \,    348.0\\ 
		07-22-2019            & 277.8   \, \, -- \, \,    281.8   & 289.8    \, \, -- \, \,   293.8   \\ 
		07-23-2019            & 342.2   \, \, -- \, \,    346.2   & 354.2  \, \, -- \, \,     358.2 \\ 
		07-25-2019            & 276.5   \, \, -- \, \,     280.5 \, F & 288.9  \, \, -- \, \,    292.9 \, F   \\
		                      & 476.8  \, \, -- \, \,    481.8 \, F & 489.2   \, \, -- \, \,   493.2 \, F \\
		07-26-2019            & 276.5   \, \, -- \, \,    280.5 \, F & 288.9    \, \, -- \, \,   292.9 \, F   \\
		                      & 335.9   \, \, -- \, \,    340.4 &  348.0  \, \, -- \, \,   352.4  \\ 
		                      &476.8    \, \, -- \, \,    480.8 &  489.2 \, \, -- \, \, 493.2 \\
		07-27-2019            & 359.0   \, \, -- \, \,    363.0   & 371.0    \, \, -- \, \,   375.0  \\ 
		07-28-2019            & 359.0   \, \, -- \, \,    363.0   & 371.0   \, \, -- \, \,    375.0   \\ \hline
	\end{tabular}
	\tablefoot{Bands observed only with the FLASH+ receiver are marked with an F.}
\end{table}

\section{Positions of the studied emission peaks}\label{app:positions}

\begin{table}[htb]
    \caption{Position of the emission peaks identified in our maps.}
    \label{tab:pos_marker}
	\centering
	\begin{tabular}{ccc}
		\hline
		\hline
		Position & RA (J2000) & DEC (J2000) \\ \hline
        IRAS\,16293 A/B & $16^\U{h}32^\U{m}22.8^\U{s}$ & $-24^\circ28\si{\arcminute}38.7\si{\arcsecond}$ \\
        16293E & $16^\U{h}32^\U{m}28.5^\U{s}$ & $-24^\circ29\si{\arcminute}02.0\si{\arcsecond}$ \\
        E1 & $16^\U{h}32^\U{m}26.7^\U{s}$ & $-24^\circ28\si{\arcminute}21.0\si{\arcsecond}$\\
        E2 & $16^\U{h}32^\U{m}28.4^\U{s}$ & $-24^\circ27\si{\arcminute}46.0\si{\arcsecond}$ \\
        W1 & $16^\U{h}32^\U{m}20.6^\U{s}$ & $-24^\circ28\si{\arcminute}19.0\si{\arcsecond}$ \\
        W2 & $16^\U{h}32^\U{m}21.4^\U{s}$ & $-24^\circ27\si{\arcminute}44.0\si{\arcsecond}$\\
        HE2& $16^\U{h}32^\U{m}27.8^\U{s}$ & $-24^\circ29\si{\arcminute}42.0\si{\arcsecond}$ \\ \hline
        SW & $16^\U{h}32^\U{m}12.3^\U{s}$ & $-24^\circ29\si{\arcminute}48.7\si{\arcsecond}$\\
        \hline
	\end{tabular}
	\tablefoot{The nomenclature for these sources is the same as in \citet{hirano2001} and \citet{castets2001}, although we do not use the exact same coordinates. The emission peak SW is covered only by our maps of the CO~($6-5$) transition.}
\end{table}

\section{Full list of the detected molecular line transitions}\label{app:molecules}
Table~\ref{tab:idlines} provides a full list of the spectral lines that were identified in our data. The list includes the name of the species, the quantum numbers of the transition, the frequency in MHz, the upper level energy, E$_\U{up}$, in K and the Einstein A coefficient, A$_{ij}$, in s$^{-1}$.

The second column describes the transition according to the values stated by the CDMS (Usually $J$ or $J_F$ for linear molecules, $J_K$ for symmetric tops and $J_{K_a, K_c}$ for asymmetric tops). An exception is methanol, for which we give the symmetry state and, as subscript, $K$ for the E-type and $K$ for the A-type species. NO, CN, and C$_2$H are described as $N_{J,F}$, where we additionally give the parity of NO in the subscript.

The sixth column shows which of the receivers used in our observations cover the frequency of the respective line, where it should be noted that OTF-maps were only observed with LAsMA. The RMS noise is given in the seventh column of Table~\ref{tab:idlines}, the values correspond to the RMS of the respective spectrum at A/B (in Kelvin) or the average RMS over each map (in K\,km\,s$^{-1}$) for mapped transitions indicated with an L in the sixth column. The analyzed LAsMA spectra and cubes have velocity resolutions between 0.14 and 0.10\,$\si{\kilo\meter\per\second}$, which correspond to the lowest and highest observed frequencies of 277 and 375\,$\si{\giga\hertz}$. The spectra taken with FLASH+ were smoothed to a velocity resolution of 0.08\,$\si{\kilo\meter\per\second}$.

The last column lists which of the considered positions show line emission with a signal to noise ratio above three. In order to determine the signal to noise ratio of a spectral line, we averaged all spectra in a $\SI{10}{\arcsecond}$ radius around the respected position, motivated by the beam size of the telescope. 

\begin{table*}[htbp]
    \caption{Overview of the identified transitions in the data and their basic parameters.}
    \label{tab:idlines}
    \centering
    \begin{tabular}[tb]{cccccccc}
        \hline
        \hline
			Molecule & Transition & Frequency  & E$_\U{up}$ & $\U{A}_\U{ij}$ & Receiver & RMS & Positions \\
			         &            & (MHz)      &  (K)         &  (s$^{-1}$)  &           & (K)$^{x)}$ &           \\ \hline
			$^{13}$CS		    & $6-5 $		 		 &  277455.405(30)  & 46.6  &$\num{4.40E-04 }$  & F     & 0.018         & -   \\ 
			$\U{CH}_3\U{OH}$-E	& $9_{-1}-8_{0}    $ &  278304.512(12)  & 110.0 &$\num{7.69E-05 }$  & F, L  & 0.061  & A/B \\ 
			$\U{CH}_3\U{OH}$-E  & $2_{-2} - 3_{-1}$  &  278342.261(11)  & 32.9  &$\num{1.65E-05 }$  & F, L*  & 0.014     & -  \\
			H$_2$CS             & $8_{1,7}-7_{1,6}$ 	 &  278887.661(50)  & 73.4  &$\num{3.18E-04 }$  & F, L  &  0.032 & A/B         \\ 
			$\U{CH}_3\U{OH}$-A$^-$& $11_{2}-10_{3}  $    &  279351.887(12)  & 190.9 &$\num{3.45E-05 }$  & F, L* &  0.014 & -    \\
			N$_2$H$^+$     		& $3-2$    				 &  279511.749(3)   & 26.8  &$\num{1.26E-03 }$  & F, L  & 0.028  & A/B, E, E1, E2, W1, W2, HE2   \\ 
    		OCS                 & $23-22$                &  279685.296(  2) & 161.1 & $\num{ 0.64E-04 }$& F, L    & 0.064  & A/B  \\
			H$_2$CO  			& $4_{1,4}-3_{1,3}     $ &  281526.929(10)  & 45.6  &$\num{5.88E-04 }$  & F, L     & 0.060  & A/B, E, E1, E2, W1, W2, HE2  \\  
	        SO$_2$              & $15_{1,15}-14_{0,14}$  &   281762.600(  1)& 107.4 &$\num{ 2.61E-04 }$ & L    & 0.055 & A/B \\
			C$^{34}$S  			& $6-5  $ 			  	 &  289209.068(1)   & 48.6  &$\num{4.98E-04 }$  & F     & 0.015 & -      \\
			DCN					& $4 - 3$				 &  289644.917(1)   & 34.8  &$\num{1.12E-03 }$  & F     & 0.014 & -   \\  
			$\U{CH}_3\U{OH}$-E	& $6_{0}-5_{0}    $      &  289939.377(5)   & 61.8  &$\num{1.06E-04 }$  & F, L*     & 0.056 & A/B, E1, W2 \\ 
			$\U{CH}_3\U{OH}$-E	& $6_{-1}-5_{-1}  $      &  290069.747(5)   & 54.3  &$\num{1.03E-04 }$  & F, L     & 0.053 & A/B, E1 \\
			$\U{CH}_3\U{OH}$-A$^+$	& $6_{0}-5_{0}  $    &  290110.637(5)   & 48.7  &$\num{1.06E-04 }$  & F, L     & 0.054 & A/B, E, E1, E2, W2, HE2    \\
			$\U{CH}_3\U{OH}$-A$^-$	$^\U{a)}$& $6_{4}-5_{4}  $  &  290161.348(5)   & 129.1  &$\num{5.90E-05 }$  & F, L*     & 0.013 & -    \\
			$\U{CH}_3\U{OH}$-A$^+$	$^\U{a)}$& $6_{4}-5_{4}  $  &  290161.352(5)   & 129.1  &$\num{5.90E-05 }$  & F, L*     & 0.013 & -    \\
			$\U{CH}_3\U{OH}$-E	    $^\U{a)}$& $6_{-4}-5_{-4}$  &  290162.356(5)   & 136.6  &$\num{5.89E-05}$   & F, L*     & 0.013 & -   \\
			$\U{CH}_3\U{OH}$-E	    $^\U{b)}$& $6_{4}-5_{4}  $  &  290183.289(5)  & 144.7 &$\num{5.93E-05}$ & F, L*     & 0.013 & -   \\
			$\U{CH}_3\U{OH}$-A$^-$	$^\U{b)}$& $6_{2}-5_{2}  $  & 290184.674(5)   & 86.5 &$\num{9.49E-05}$  & F, L*     & 0.013 & -   \\
			$\U{CH}_3\U{OH}$-A$^+$	$^\U{b)}$& $6_{3}-5_{3}  $  & 290189.515(5)   & 98.5 &$\num{7.95E-05}$  & F, L*     & 0.013 & -   \\
			$\U{CH}_3\U{OH}$-A$^-$	$^\U{b)}$& $6_{3}-5_{3}  $  & 290190.549(5)   & 98.5 &$\num{7.95E-05}$  & F, L*     & 0.013 & -   \\
			$\U{CH}_3\U{OH}$-E	$^\U{c)}$& $6_{-3}-5_{-3}$  & 290209.695(5)   & 111.5 &$\num{8.01E-05}$  & F, L*     & 0.017 & -   \\
			$\U{CH}_3\U{OH}$-E	$^\U{c)}$& $6_{3}-5_{3}  $  & 290213.180(5)   & 96.5  &$\num{7.97E-05}$  & F, L*     & 0.017 & -   \\
			$\U{CH}_3\U{OH}$-E	& $6_{1}-5_{1}    $      &  290248.685(5)   & 69.8  &$\num{1.06E-04 }$  & F, L     & 0.053 & A/B \\
			$\U{CH}_3\U{OH}$-A$^+$	& $6_{2}-5_{2}   $   &  290264.068(5)   & 86.5  &$\num{9.50E-05 }$  & F, L     & 0.030 & A/B  \\   
			$\U{CH}_3\U{OH}$-E	$^\U{d)}$& $6_{-2}-5_{-2}  $  &  290307.281(5)   & 74.7  &$\num{9.46E-05 }$  & F, L     & 0.036 & A/B, E1, W2 \\
			$\U{CH}_3\U{OH}$-E	$^\U{d)}$& $6_{2}-5_{2}  $    &  290307.738(5)   & 71.0  &$\num{9.35E-05 }$  & F, L     & 0.036 & A/B, E1, W2 \\
			$^{34}$SO           & $7_{ 6} - 6_{ 5}$ 	 &  290562.238(30)  & 63.8  &$\num{3.04E-04 }$  & F, L     & 0.009 & A/B        \\ 
			H$_2$CO				& $4_{0,4}-3_{0,3}    $  &  290623.405(10)  & 34.9  &$\num{6.90E-04 }$  & F, L     & 0.032 & A/B, E, E1, E2, W1, W2, HE2  \\
			H$_2$CO $^\U{e)}$			& $4_{2,3}-3_{2,2}    $  &  291237.766(2)   & 82.1  &$\num{5.21E-04 }$  & F, L     & 0.010 & A/B, E1, E2, W1, W2, HE2   \\ 
			H$_2$CO	$^\U{e)}$			& $4_{3,2}-3_{3,1}   $   &  291380.442(2)   & 140.9 &$\num{3.04E-04 }$  & F, L     & 0.012 & A/B, E1, E2, W1, W2 \\  
			H$_2$CO				& $4_{3,1}-3_{3,0}   $   &  291384.361(2)   & 140.9 &$\num{3.04E-04 }$  & F, L     & 0.012 & A/B, E1, E2, W1, W2 \\ 
			C$^{33}$S   		& $6 - 5$				 &  291485.935(30)  & 49.0  &$\num{5.10E-04 }$  & F, L     & 0.065 & A/B     \\
			D$_2$CO             & $5_{2,4}-4_{2,3}$      &  291745.747(  1) & 63.6 & $\num{ 0.60E-03 }$ & F, L     & 0.050 & A/B \\
			OCS                 & $24-23$                &  291839.653(  1) & 175.1 & $\num{ 0.72E-04 }$& F, L     & 0.051 & A/B \\
			H$_2$CO  			& $4_{2,2}-3_{2,1}   $   &  291948.067(2)   & 82.1  &$\num{5.25E-04 }$  & F, L     & 0.012 & A/B, E1, E2, W1, W2   \\
			$\U{CH}_3\U{OH}$-A$^-$	& $6_{1}-5_{1}   $ &  292672.889(5)   & 63.7  &$\num{1.06E-04 }$  & F, L     & 0.053 & A/B, E1 \\
			H$_2^{13}$CO 		& $4_{1,3}-3_{1,2} $	 &  293126.515(46)  & 47.0  &$\num{6.64E-04 }$  & L     & 0.031 & A/B \\ 
			$\U{CH}_3\U{OH}$-A$^+$	& $3_{2}-4_{1}   $   &  293464.055(13)  & 51.6  &$\num{2.86E-05 }$  & L*     & 0.025 & - \\
			CS					& $6-5$ 				 &  293912.086(3)   & 49.4  &$\num{5.23E-04 }$  & L     & 0.037 & A/B, E, E1, W1, W2, HE2    \\ 
			SO					& $7_{ 6} - 6_{ 5} $ 	 &  296550.064(30)  & 64.9  &$\num{3.23E-04 }$  & L     & 0.008 & A/B, E1, W1, W2, HE2       \\    
			DNC                 & $4-3$                  &  305206.219(30)  & 36.6  &$\num{1.37E-03 }$  & L     & 0.057 & A/B, E, E1, HE2 \\
			$\U{CH}_3\U{OH}$-A$^{-+}$	& $3_{1}-3_{0}$  &  305473.491(10)  & 28.6  &$\num{3.26E-04 }$  & L     & 0.009 & A/B, E, E1, E2, W1, W2, HE2   \\	
			$\U{CH}_3\U{OH}$-A$^{-+}$	& $4_{1}-4_{0}$  &  307165.924(10)  & 38.0  &$\num{3.31E-04 }$  & L     & 0.047 & A/B, E1, W2, HE2    \\
			HDCO                & $5_{1,5}-4_{1,4} $ 	 &  308418.200(100) & 52.4  &$\num{8.09E-04 }$  & L     & 0.008 & A/B, E1     \\
			N$_2$D$^+$          & $4-3$  				 &  308422.267(30)  & 37.0  &$\num{1.75E-03 }$  & L     & 0.006 & A/B, E, E1, HE2         \\
			SO$_2$				& $4_{3,1}-3_{2,2}$  	 &  332505.242(2)   & 31.3  &$\num{3.29E-04 }$  & L     & 0.085 & A/B, E1       \\
			NH$_2$D             & $1_{0,1,1}-0_{0,0,1}$  &  332781.890(100) & 16.6  &$\num{7.82E-06 }$  & L     & 0.078 & A/B, E, HE2\\
            NH$_2$D             & $1_{0,1,1}-0_{0,0,0}$  &  332822.510(100) & 16.0  &$\num{7.29e-06 }$  & L     & 0.075 & E \\
			SO$_2$				& $8_{2,6}-7_{1,7}$   	 &  334673.353(2)   & 43.1  &$\num{1.27E-04 }$  & L     & 0.063 & A/B, E1  \\ 
			HDCO                & $5_{1,4}-4_{1,3}$      &  335096.783(  2) & 56.2  &$\num{ 0.10E-02 }$ & L*     & 0.073 & A/B\\
			\hline
    \end{tabular}
    \tablefoot{Blended lines are indicated with small letters a)-l). The frequency uncertainty for each transition (obtained from the CDMS) is given in parenthesis in units of kHz. The sixth column describes the receivers that cover the respective frequencies, where it is differentiated between FLASH+ (F), LAsMA (L), and spectral lines covered by LAsMA but for which the maps are not displayed here (L*). x) The RMS corresponds to the spectrum RMS at A/B (in Kelvin) or the average RMS over each map (in K\,km\,s$^{-1}$) for mapped transitions. The last column gives information about which positions show the respective spectral lines with a signal to noise ratio of three or higher.}
\end{table*}
			
\begin{table*}
\centering
\caption{continued.}
		\begin{tabular}[tb]{cccccccc}
			\hline
			\hline
			Molecule & Transition & Frequency  & E$_\U{up}$ & $\U{A}_\U{ij}$ & Receiver & RMS & Positions \\
			         &            & (MHz)      &  (K)         &  (s$^{-1}$)  &           & (K)$^{x)}$ &           \\ \hline
			 $\U{CH}_3\U{OH}$-A$^-$	& $2_{2}-3_{1}   $   & 335133.570(13)   & 44.7   &$\num{2.69E-05 }$  & L*    & 0.037 & - \\
			NHD$_2$             & $1_{1,1,1}-0_{0,0,1}$  &  335446.321(3)   & 16.3  &$\num{1.47e-05 }$  & L     & 0.076 & E \\
            NHD$_2$             & $1_{1,1,0}-0_{0,0,0}$  & 335513.793(3)    & 16.1  &$\num{1.29e-05 }$  & L     & 0.082 & E \\
            CH$_3$OH-A$^+$      & $7_{1}-6_{1}$          &  335582.017(5)   & 79.0  & $\num{ 1.63E-04 }$ & L* & 0.040 & - \\
            CH$_3$OH-A$^{-+}$   & $12_{1}-12_{0}$        &  336865.149( 12) & 197.1 & $\num{ 4.07E-04 }$ & L* & 0.037 & - \\
			C$^{17}$O           & $3-2 $				 &  337061.214(2)   & 32.4  &$\num{1.81E-06 }$  & L     & 0.033 & A/B, E, E1, E2, W1, W2, HE2      \\ 
			$\U{CH}_3\U{OH}$-E	& $3_{3}-4_{2}   $       &  337135.853(14)  & 61.6  &$\num{1.58E-05 }$  & L*    & 0.037 & - \\
			C$^{34}$S 			& $7-6   $ 				 &  337396.459(1)   & 64.8  &$\num{8.00E-04 }$  & L     & 0.054 & A/B    \\ 
			H$_2$CS             & $10_{1,10}-9_{1,9}$    &  338083.195(50)  & 102.4 & $\num{ 0.58E-03 }$& L     & 0.047 & A/B \\
			$\U{CH}_3\U{OH}$-E	& $7_{0}-6_{0}$          &  338124.488(5)   & 78.1  &$\num{1.70E-04 }$  & L     & 0.061 & A/B, E1, HE2    \\
			SO$_2$              & $18_{4,14}-18_{3,15}$  &  338305.993(1)   & 196.8 & $\num{ 3.27E-04 }$& L     & 0.058 & A/B\\
			$\U{CH}_3\U{OH}$-E	& $7_{-1}-6_{-1}    $    &  338344.588(5)   & 70.6  &$\num{1.67E-04 }$  & L     & 0.054 & A/B, E1, E2, W2, HE2 \\
			$\U{CH}_3\U{OH}$-A$^+$	& $7_{0}-6_{0}    $  &  338408.698(5)   & 65.0  &$\num{1.70E-04 }$  & L     & 0.044 & A/B, E1, E2, W2, HE2    \\ 
			$\U{CH}_3\U{OH}$-E	$^\U{f)}$& $7_{-4}-6_{-4}    $    &  338504.065(5)   & 152.9  &$\num{1.15E-04 }$  & L*    & 0.036 & -   \\
			$\U{CH}_3\U{OH}$-A$^-$	$^\U{f)}$& $7_{4}-6_{4}    $  &  338512.632(5)   & 145.3  &$\num{1.15E-04 }$  & L*    & 0.036 & -   \\ 
			$\U{CH}_3\U{OH}$-A$^+$	$^\U{f)}$& $7_{4}-6_{4}    $  &  338512.644(5)   & 145.3  &$\num{1.15E-04 }$  & L*    & 0.036 & -   \\ 
			$\U{CH}_3\U{OH}$-A$^-$	$^\U{f)}$& $7_{2}-6_{2}    $  &  338512.853(5)   & 102.7  &$\num{1.57E-04 }$  & L*    & 0.036 & -   \\ 
			$\U{CH}_3\U{OH}$-A$^+$	$^\U{g)}$& $7_{3}-6_{3}    $  &  338540.826(5)   & 114.8  &$\num{1.39E-04 }$  & L*    & 0.036 & -   \\
			$\U{CH}_3\U{OH}$-A$^-$	$^\U{g)}$& $7_{3}-6_{3}    $  &  338543.152(5)   & 114.8  &$\num{1.39E-04 }$  & L*    & 0.036 & -   \\
			$\U{CH}_3\U{OH}$-E	& $7_{-3}-6_{-3}    $  &  338559.963(5)   & 127.7  &$\num{1.40E-04 }$  & L*    & 0.036 & -   \\
			$\U{CH}_3\U{OH}$-E	& $7_{3}-6_{3}    $    &  338583.216(5)   & 112.7  &$\num{1.39E-04 }$  & L*    & 0.036 & -   \\
			$\U{CH}_3\U{OH}$-E	& $7_{1}-6_{1}    $    &  338614.936(5)   & 86.1   &$\num{1.71E-04 }$  & L*    & 0.034 & -   \\
			$\U{CH}_3\U{OH}$-A$^+$	& $7_{2}-6_{2}    $    &  338639.802(5)   & 102.7   &$\num{1.58E-04 }$  & L*    & 0.034 & -   \\
			$\U{CH}_3\U{OH}$-E	$^\U{h)}$& $7_{2}-6_{2}     $   &  338721.693(5)   & 87.3  &$\num{1.55E-04 }$  & L     & 0.039 & A/B, E1   \\ 
			$\U{CH}_3\U{OH}$-E	$^\U{h)}$& $7_{-2}-6_{-2}     $ &   338722.898(5)  & 90.9  &$\num{1.57E-04 }$  & L     & 0.039 & A/B, E1   \\ 
			SO  				& $3_{ 3 }- 2_{3}$ 		 &  339341.459(70)  & 25.5  &$\num{1.45E-05 }$  & L     & 0.047 & A/B, E, E1, E2, W2 \\
			$^{34}$SO			& $8_{ 9} - 7_{ 8}$		 &  339857.269(14)  & 77.3  &$\num{5.08E-04 }$  & L     & 0.048 & A/B   \\  
			CN                  &$3_{5/2,7/2}-2_{3/2,5/2}$   & 340031.549(3)    & 32.6  &$\num{3.85E-04 }$  & L     & 0.036 & A/B, E, E1 \\   
            CN                  &$3_{5/2,5/2}-2_{3/2,3/2}$ & 340035.408(50) & 32.6  &$\num{3.23E-04 }$  & L     & 0.043 & A/B, E, HE2 \\
            C$^{33}$S           & $7_0-6_0$              &  340052.575(  0) & 65.3  & $\num{ 0.82E-03 }$ & L     & 0.071 & A/B\\
            $\U{CH}_3\U{OH}$-A$^+$	& $2_{2}-3_{1}    $  &  340141.143(13)  & 44.7  &$\num{2.78E-05 }$  & L*    & 0.059 & -   \\
            CN $^\U{i)}$           &$3_{7/2,9/2} - 2_{5/2,7/2}$     & 340247.770(50)   & 32.7  &$\num{4.13E-04 }$  & L     & 0.064 & A/B, E, E1, HE2\\
            CN $^\U{i)}$           &$3_{7/2,5/2} - 2_{5/2,3/2}$   & 340248.544(4)    & 32.7  &$\num{3.67E-04 }$  & L     & 0.064 & A/B, E, E1, HE2\\
            $\U{CH}_3\U{OH}$-A$^{-+}$	& $13_{1}-13_{0}  $  &  342729.796(13)  & 227.5 &$\num{4.23E-04 }$  & L*    & 0.054 & -   \\
			CS					& $7-6$  				 &  342882.850(5)   & 65.8  &$\num{8.39E-04 }$  & L     & 0.092 & A/B, E1, W1    \\ 
			SO					& $8_{ 8} - 7_{ 7} $ 	 &  344310.612(70)  & 87.5  &$\num{5.19E-04 }$  & L     & 0.006 & A/B, E1, W1    \\ 
		    SO$_2$ $^\U{j)}$    & $13_{2,12}-12_{1,11}$  &  345338.538(2)   & 93.0  &$\num{2.38E-04 }$  & L*     & 0.041 & A/B       \\
			H$^{13}$CN $^\U{j)}$& $4-3 $  				 &  345339.769(1)   & 41.4  &$\num{1.90E-03 }$  & L     & 0.041 & A/B       \\ 
			CO    			    &		 $3-2 $    		 &  345795.990(1)   & 33.2  &$\num{2.50E-06 }$  & L     & 0.046 & A/B, E, E1, E2, W1, W2, HE2 \\
			$\U{CH}_3\U{OH}$-A$^-$	$^\U{k)}$& $5_{4}-6_{3}    $  &  346202.719(15)  & 115.2 &$\num{2.18E-05 }$  & L*    & 0.038 & -   \\
			$\U{CH}_3\U{OH}$-A$^+$	$^\U{k)}$& $5_{4}-6_{3}    $  &  346204.271(15)  & 115.2 &$\num{2.18E-05 }$  & L*    & 0.038 & -   \\
			SO					& $8_{ 9} - 7_{8}$   	 &  346528.481(70)  & 78.8  &$\num{5.38E-04 }$  & L     & 0.084 & A/B, E1, W1, W2 \\
			SO$_2$				& $19_{1,19} - 18_{0,18}$ & 346652.169(1)   & 168.1 &$\num{5.22E-04 }$  & L     & 0.074 & A/B \\
			H$^{13}$CO$^+$      & $4-3$  				 &  346998.344(12)  & 41.6  &$\num{3.29E-03 }$  & L     & 0.013 & A/B, E, E1, HE2        \\ 
			SiO                 & $8-7$   				 &  347330.581(7)   & 75.0  &$\num{2.20E-03 }$  & L     & 0.011 & A/B, E1       \\ 
	    	H$_2$CS             & $10_{1,9}-9_{1,8}$     &  348534.365(50)  & 105.2 &$\num{ 0.63E-03 }$ & L     & 0.031 & A/B \\
			$\U{C}_2\U{H}$	    & $4_{9/2,4} - 3_{7/2,3}$ 	 &  349337.706(22)  & 41.9  &$\num{1.31E-04 }$  & L     & 0.012 & A/B, E1     \\ 
			$\U{C}_2\U{H}$      & $4_{7/2,3}-3_{5/2,2}$  	 &  349399.276(21)  & 41.9  &$\num{1.25E-04 }$  & L     & 0.012 & A/B, E1 \\
			CH$_3$OH-E $^\U{l)}$ & $4_{0}-3_{-1} $        &  350687.662(13)  & 36.3  &$\num{8.67e-05 }$  & L     & 0.048 & A/B, E, E1, E2, W2, HE2\\
			NO  $^\U{l)}$        & $4_{-,7/2,9/2}-3_{+,5/2,7/2}$ &  350689.494(20)  & 36.1  &$\num{5.42e-06 }$  & L     & 0.053 & A/B, E, E1, HE2\\
    		NO  $^\U{l)}$        & $4_{-,7/2,7/2}-3_{+,5/2,5/2}$ & 350690.766(20)   & 36.1  &$\num{4.98e-06 }$  & L     & 0.053 & A/B, E, HE2\\
    		CH$_3$OH-A$^+$      & $1_{1}-0_{0}$          & 350905.100(11)   & 16.8  &$\num{3.32e-04 }$  & L     & 0.054 & A/B, E, E1, E2, W2, HE2\\
            NO                  & $4_{+,7/2,9/2}-3_{-,5/2,7/2}$ & 351043.524(10)   & 36.1  &$\num{5.43e-06 }$  & L     & 0.040 & A/B, E, E1, HE2\\
            NO                  & $4_{+,7/2,7/2}-3_{-,5/2,5/2}$ &  351051.705(10)  & 36.1  &$\num{4.99e-06 }$ & L     & 0.040 & A/B, E, E1, HE2\\ 
            			SO$_2$				& $5_{3,3}-4_{2,2}$ 	 &  351257.223(2)   & 35.9  &$\num{3.36E-04 }$  & L     & 0.059 & A/B, E1  \\ 
			\hline

\end{tabular}
\end{table*}

\begin{table*} 
\centering
\caption{continued.}
		\begin{tabular}[tb]{cccccccc}
			\hline
			\hline
			Molecule & Transition & Frequency  & E$_\U{up}$ & $\U{A}_\U{ij}$ & Receiver & RMS & Positions \\
			         &            & (MHz)      &  (K)         &  (s$^{-1}$)  &           & (K)$^{x)}$ &           \\ \hline
			HNCO                & $16_{0,16}-15_{0,15}$  &  351633.257(10)  & 143.5 &$\num{6.12E-04 }$  & L    & 0.056 & A/B \\
			SO$_2$				& $14_{4,10}-14_{3,11}$  &  351873.873(1)   & 135.9  &$\num{3.43E-04 }$ & L*     & 0.038 & -  \\
			H$_2$CO 			& $5_{1,5}-4_{1,4} $     &  351768.645(30)  & 62.5  &$\num{1.20E-03 }$  & L     & 0.005 & A/B, E, E1, E2, W1, W2, HE2 \\
			HCN                 & $4-3$       			 &  354505.477(1)   & 42.5  &$\num{2.05E-03 }$  & L     & 0.017 & A/B, E, E1, E2, W1, W2, HE2   \\ 
			SO$_2$              & $12_{4,8}-12_{3,9}$    &  355045.517(1)   & 111.0 & $\num{ 3.40E-04 }$& L     & 0.180 & A/B \\
			$\U{CH}_3\U{OH}$-A$^+$	& $13_{0}-12_{1}$    &  355602.945(12)  & 211.0 &$\num{2.53E-04 }$  & L*    & 0.061 & -   \\
			HCO$^+$             & $4-3$        			 &  356734.223(2)   & 42.8  &$\num{3.57E-03 }$  & L     & 0.015 & A/B, E, E1, E2, W1, W2, HE2   \\
			SO$_2$              & $13_{4,10}-13_{3,11}$  &  357165.390(1)   & 123.0 & $\num{ 3.51E-04 }$& L*    & 0.033 & - \\
			SO$_2$              & $11_{4,8}-11_{3,9}$    &  357387.579(1)   & 100.0 & $\num{ 3.38E-04 }$& L     & 0.033 & - \\
			SO$_2$              & $8_{4,4}-8_{3,5}$      &  357581.449(1)   & 72.4  & $\num{ 3.06E-04 }$& L     & 0.239 & A/B \\
			SO$_2$              & $7_{4,4}-7_{3,5}$      &  357892.442(1)   & 65.0  & $\num{ 2.87E-04 }$& L*    & 0.033 & - \\
			SO$_2$              & $6_{4,2}-6_{3,3}$      &  357925.848(1)   & 58.6  & $\num{ 2.60E-04 }$& L*    & 0.033 & - \\
			SO$_2$              & $5_{4,2}-5_{3,3}$      &  358013.154(1)   & 53.1  & $\num{ 2.18E-04 }$& L*    & 0.033 & - \\
			DCO$^+$             & $5-4$ 			 	 &  360169.778(5)   & 51.9  &$\num{3.76E-03 }$  & L     & 0.010 & A/B, E, E1, HE2     \\ 
			$\U{CH}_3\U{OH}$-E	& $11_{0}-10_{1} $       &  360848.946(11)  & 166.0 &$\num{1.21E-04 }$  & L*    & 0.033 & -  \\
			$\U{CH}_3\U{OH}$-E	& $8_{1}-7_{2}     $     &  361852.195(13)  & 104.6 &$\num{7.72E-05 }$  & L*    & 0.035 & -  \\ 
			DCN					& $5-4$    				 &  362045.753(1)   & 52.1  &$\num{2.25E-03 }$  & L     & 0.010 & A/B       \\
			HNC                 & $4-3 $  				 &  362630.303(20)  & 43.5  &$\num{2.30E-03 }$  & L     & 0.009 & A/B, E, E1, E2, W1, W2, HE2\\
			H$_2$CO 			& $5_{0,5}-4_{0,4}  $    &  362736.048(30)  & 52.3  &$\num{1.37E-03 }$  & L     & 0.061 & A/B, E, E1, E2, W1, W2, HE2 \\
			SO$_2$				& $6_{3,3}-5_{2,4}$  	 &  371172.451(2)   & 41.4  &$\num{3.55E-04 }$  & L     & 0.117 & A/B      \\
			H$_2$CS             & $11_{1,11}-10_{1,10}$  &  371847.395( 50) & 120.3 & $\num{ 0.77E-03 }$& L*     & 0.130 & A/B\\
			H$_2$D$^+$          & $1_{1,0}-1_{1,1}$ 	 &  372421.356(3)   & 104.2 &$\num{1.08E-04 }$  & L     & 0.053 & E     \\ 
			N$_2$H$^+$			& $4-3$        		     &  372672.481(3)   & 44.7  &$\num{3.10E-03 }$  & L     & 0.024 & A/B, E, E1, HE2 \\ 
			SiO					& $11-10$  				 &  477504.635(4)   & 137.5 &$\num{5.82E-03 }$  & F & 0.040 & -  \\ 
			$\U{CH}_3\U{OH}$-A$^{+-}$ &$ 4_{2}-4_{1}  $    &  478633.216(12)  & 60.9  &$\num{5.28E-04 }$  & F & 0.036 & -   \\ 
			CH$_3$OH-A$^+$      & $10_{1}-9_{1}$         &  479215.536(6)   & 141.1 & $\num{ 4.89E-04 }$& F & 0.035 & -  \\
			H$_2^{13}$CO        & $7_{1,7}-6_{1,6}$      &  480194.687(20)  & 104.1 & $\num{ 0.32E-02 }$& F & 0.041 & -  \\
			$\U{CH}_3\U{OH}$-A$^{+-}$	& $3_{2}-3_{1}  $    &  480269.256(13)  & 51.6  &$\num{4.92E-04 }$  & F & 0.041 & -     \\ 
			CS					& $10-9 $ 			 	 &  489750.921(4)   & 129.3 &$\num{2.50E-03 }$  & F & 0.065 & -       \\
			HDO                 & $2_{0,2}-1_{1,1}$   	 &  490596.640(58)  & 66.4  &$\num{5.25E-04 }$  & F & 0.056 & -        \\ 
			$\U{CH}_3\U{OH}$-A$^{-+}$ 	& $6_{2}-6_{1} $ &  491550.719(12)  & 86.5  &$\num{5.87E-04 }$  & F & 0.044 & -      \\
			SO$_2$				& $7_{4,4}-6_{3,3}$  	 &  491934.721(2)   & 65.0  &$\num{9.49E-04 }$  & F & 0.043 & -       \\ 
            HDCO                & $8_{1,8}-7_{1,7}$      &  491936.965(  2) & 114.4 & $\num{ 0.35E-02 }$& F & 0.043 & -  \\
			H$_2$CO 			& $7_{1,7}-6_{1,6}$      &  491968.369(3)   & 106.3 &$\num{3.44E-03 }$  & F & 0.043 & -    \\
			C                   & $1-0$    		 		 &  492160.651(55)  & 23.6  &$\num{7.99E-08 }$  & F & 0.043 & -         \\
			$\U{CH}_3\U{OH}$-A$^+$ 	& $ 4_{1}-3_{0}  $   &  492278.692(11)  & 37.6  &$\num{7.65E-04 }$  & F & 0.040 & -      \\ 
			$\U{CH}_3\U{OH}$-A$^+$ 	& $5_{3}-4_{2}  $    &  493699.112(15)  & 84.6  &$\num{6.62E-04 }$  & F & 0.359 & - \\
			$\U{CH}_3\U{OH}$-A$^-$ 	& $5_{3}-4_{2}  $    &  493733.687(15)  & 84.6  &$\num{6.62E-04 }$  & F & 0.359 & - \\
			\hline

\end{tabular}
\end{table*}

%------------------------------------------------------------------------------
\section{Determination of the H$_2$CO excitation temperatures under LTE conditions}\label{app:temp_derivation}
In the LTE approximation, the population between two energy levels $n_\U{u}$ and $n_\U{l} $ is described by the Boltzmann distribution 
\begin{equation}\label{eq:boltzmann}
	\frac{n_\U{u}}{n_\U{l}} = \frac{g_\U{u}}{g_\U{l}} \exp\left(-\frac{E_\U{u}-E_\U{l}}{k  T_\U{ex}}\right),
\end{equation}
where $g_\U{u}$ and $g_\U{l}$ are the degeneracies of the upper and lower levels, respectively, $E_\U{u}$ and $E_\U{l}$ are their energies and $T_\U{ex}$ is the excitation temperature; $k$ is the Boltzmann constant. The degeneracy $g_\U{u}$ of an upper level can be computed as $g_\U{u} = g_\U{J} \times g_\U{K}\times g_\U{I}$, where $g_\U{J} = 2J +1$ is the rotational degeneracy and $g_\U{K} =g_\U{I} = 1$ applies as we consider H$_2$CO transitions in the same symmetry state~\citep{mangum2015calculate}.

Assuming optically thin emission, the column density $N_\U{u}$ of a transitions upper level is proportional to the integrated line intensity $I$ of this transition \citep[e.g., see][]{goldsmith1999population} and is given by
\begin{equation}\label{eq:cd_thin}
	N_\U{u} = \frac{8 \pi k \nu^2 }{h c^3 A_\U{ul}} I,
\end{equation}
where $\nu$ is the frequency, $c$ is the speed of light, $A_\U{ul}$ is the spontaneous emission Einstein (A) coefficient of a respective transition. $I$ is computed by integrating the main beam temperature $T_\U{mb}$ over the considered velocities according to
\begin{equation}
	I = \int T_\U{mb} \, \U{d}\U{v}.
\end{equation}
For this analysis, the ratio of column densities $N_\U{u}/N_\U{l}$ as derived from the integrated main beam temperatures is used in Eq.~\ref{eq:boltzmann} instead of the relative level populations. The excitation temperature can therefore be written as
\begin{equation}
    T_\U{ex} = - \frac{E_\U{u}-E_\U{l}}{k}\left[\ln \left(\frac{N_\U{u}}{N_\U{l}}\frac{g_\U{l}}{g_\U{u}} \right)\right]^{-1}.
\end{equation}

%-----------------------------------------------------------------------------
The observed formaldehyde transitions allow the derivation of excitation temperatures for ortho- and para-H$_2$CO separately. Since only three transitions per species are available, these temperatures were obtained from the respective line ratios. For para-H$_2$CO was therefore possible to use the line ratios \mbox{($4_{2,2} - 3_{2,1}$)/($4_{0,4} - 3_{0,3}$)} and ($4_{2,3} - 3_{2,2}$)/($4_{0,4} - 3_{0,3}$). In case of ortho-H$_2$CO, the ($4_{3,2} - 3_{3,1}$) and ($4_{3,1} - 3_{3,0}$) lines are blended. For this reason, the ratio $[(4_{3,2} - 3_{3,1})+(4_{3,1} - 3_{3,0})]/[2 \times (4_{1,4}-3_{1,3})]$ was considered. Table~\ref{tab:h2coparams} lists the applied parameters for these calculations.

\begin{table*}[hbt]
    \caption{List of line parameters for the formaldehyde transitions considered for the temperature derivation.}
    \label{tab:h2coparams}
	\centering
	\begin{tabular}{cccccc}
		\hline
		\hline
		Transition & o/p & $\nu$ & $E_\U{up}$ & $A_\U{ul}$ & $g_\U{up}$\\ 
			&       &  (MHz) & (K) & ($s^{-1}$) & \\ \hline
	$4_{1,4} - 3_{1,3}$ &  o & 281526.929 & 45.6 & 5.88$\times 10^{-4}$ & 27 \\
	$4_{0,4} - 3_{0,3}$ &  p & 290623.405 & 34.9 & 6.90$\times 10^{-4}$ & 9 \\
	$4_{2,3} - 3_{2,2}$ &  p & 291237.780 & 82.1 & 5.21$\times 10^{-4}$ & 9  \\
	$4_{3,2} - 3_{3,1}$ &  o & 291380.442 & 140.9& 3.04$\times 10^{-4}$ & 27  \\
	$4_{3,1} - 3_{3,0}$ &  o & 291384.361 & 140.9& 3.04$\times 10^{-4}$ & 27  \\
	$4_{2,2} - 3_{2,1}$ &  p & 291948.060 & 82.1 & 5.24$\times 10^{-4}$ & 9  \\
	$5_{1,5} - 4_{1,4}$ &  o & 351768.645 & 62.5 & 1.20$\times 10^{-3}$ & 33 \\ 
	$5_{0,5} - 4_{0,4}$ &  p & 362736.048 & 52.3 & 1.37$\times 10^{-3}$ & 11 \\ \hline

	\end{tabular}
	\tablefoot{Frequencies $\nu$, upper level energies $E_\U{up}$, Einstein coefficients $A_\U{ul}$ and degeneracies $g_\U{up}$ are adopted from the CDMS and JPL catalogs. The second column indicates if the transition correspond to the ortho (o) or para (p) form of H$_2$CO respectively.}
\end{table*}

\clearpage
\clearpage
%-----------------------------------------------------------------------------

\section{Temperatures from RADEX non-LTE models}\label{app:RADEX_temp}
In order to test the validity of the assumption of LTE conditions, the correlation between calculated temperatures and line ratios from ortho- and para-H$_2$CO are compared with non-LTE models that were computed with the RADEX radiative transfer code \citep{van2007computer} considering different H$_2$ volume densities $n$ and using collision rates from \citet{wiesenfeld2013rotational}.

\begin{figure}[tbph]
	\centering
	\subfigure[]{\includegraphics[width=0.49\textwidth]{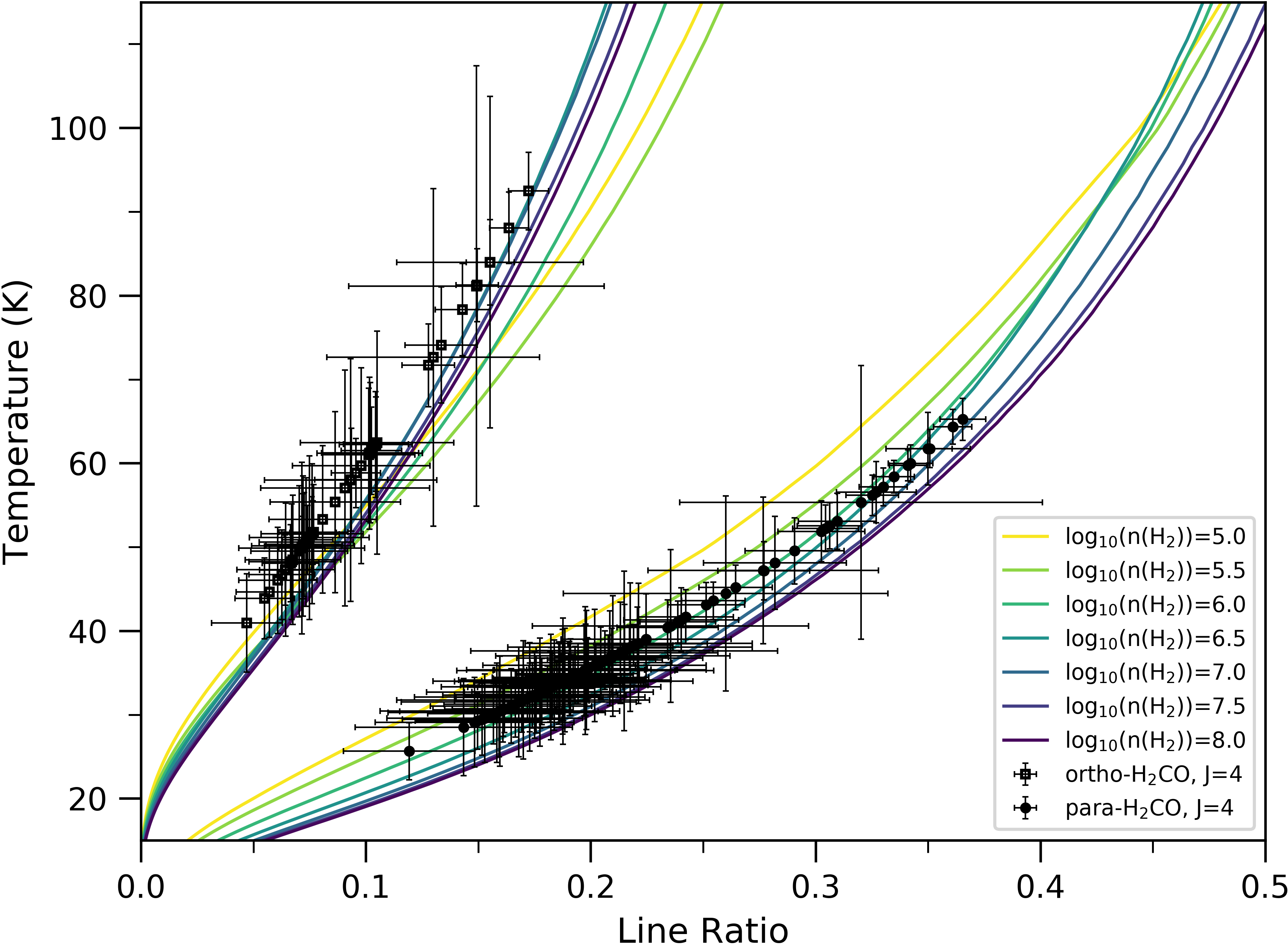}}\\
	\subfigure[]{\includegraphics[width=0.49\textwidth]{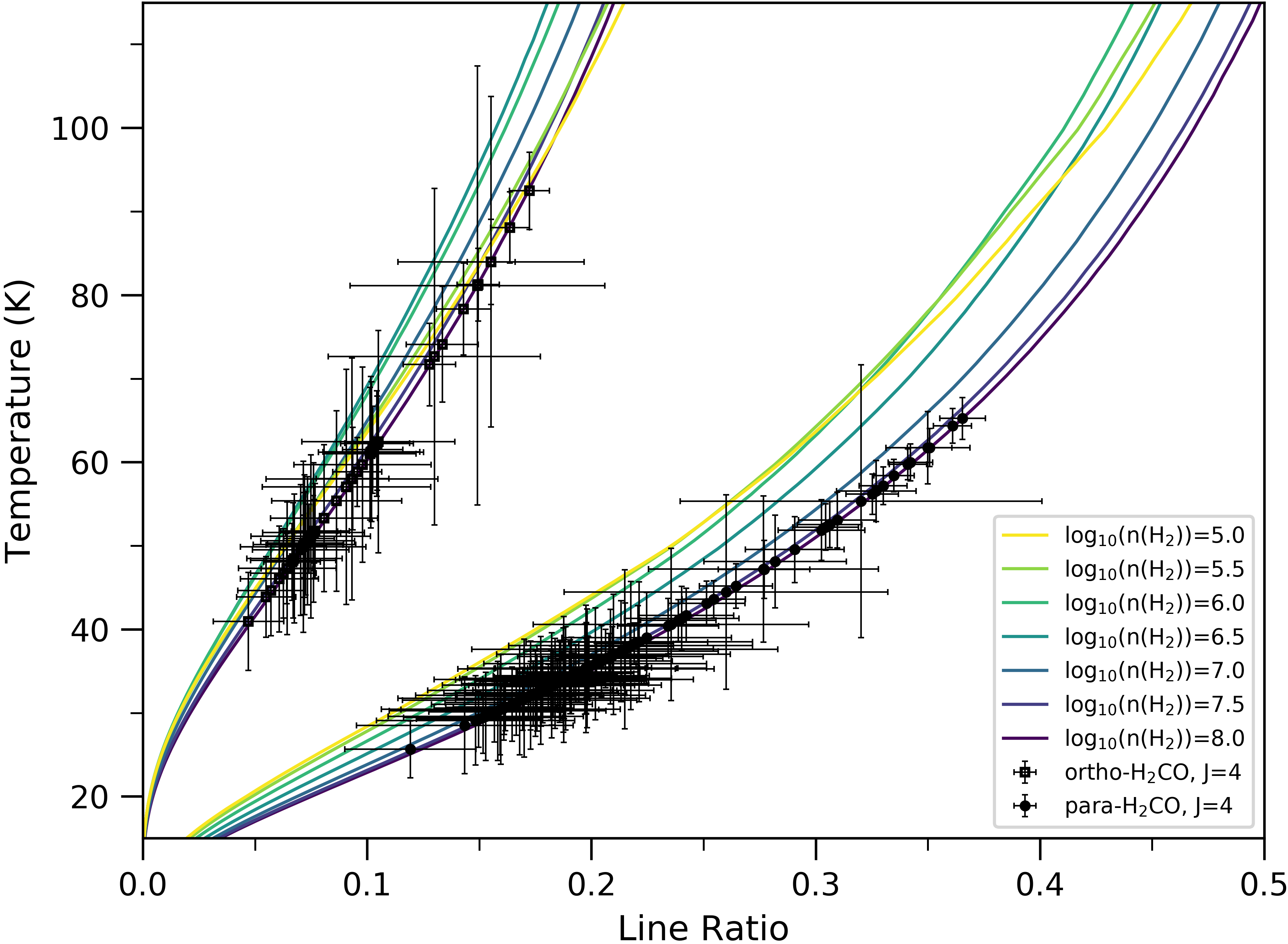}}
	\caption{Temperature (in Kelvin) as function of line ratio of the H$_2$CO transitions with $J_\U{up} = 4$. The datapoints show the calculated excitation temperatures from the individual pixels of the maps shown in Fig.~\ref{fig:temp_ratio}. Square markers in the upper left side of the figure correspond to the ortho-H$_2$CO transitions, while circle markers in the lower right side correspond to the para-H$_2$CO transitions. The solid color lines indicate the RADEX models of kinetic temperature as function of line ratio for different values of H$_2$ volume density as indicated in the lower right panel. These models were computed using H$_2$CO column densities of (a) $\num{3.8}\times \num{10}^{14}\si{\per\centi\meter\squared}$ and (b) $\num{5.2}\times \num{10}^{13}\si{\per\centi\meter\squared}$, which correspond to the column densities at A/B and the average on the other positions, respectively.}
	\label{fig:radex_1}
\end{figure}

Figure~\ref{fig:radex_1} shows several RADEX models for the kinetic temperature as function of line ratio alongside datapoints from our maps, which correspond to the individual pixels displayed in Fig.~\ref{fig:temp_ratio}. The uncertainties on these points were calculated by Gaussian error propagation of the mean RMS from the associated maps.

The RADEX models were computed assuming a formaldehyde line width of $\SI{3.0}{\kilo\meter\per\second}$ and considering only the cosmic microwave background for the background temperature of $\SI{2.73}{\kelvin}$. Based on the computations in Sect.~\ref{sec:cd}, a H$_2$CO column density of $\num{3.8}\times\num{10}^{14}\si{\per\centi\meter\squared}$ was derived at at the position of IRAS\,16293 A/B and $\num{5.2}\times\num{10}^{13}\si{\per\centi\meter\squared}$ on average at the other positions. Since the \textit{weeds} computations did not distinguish between ortho- and para-formaldehyde, the individual ortho- and para-column densities for the RADEX modeling are determined using a ratio of $\U{ortho}/\U{para}=3$ \citep{kahane1984measurement}.

A comparison between the results from the LTE temperature derivation and the low column density non-LTE RADEX models in Fig.~\ref{fig:radex_1}b shows that the excitation temperatures computed from the data agree best with RADEX models for high volume densities, as the assumption of LTE holds in these cases. In contrast, the temperatures are underestimated for lower volume-densities by up to 20\%. High column densities of H$_2$CO lead to an increase in optical depth for the observed transitions, which affects the ($4_{0,4} - 3_{0,3}$) and ($4_{1,4}-3_{1,3}$) transitions more than the fainter upper energy state transitions. This effectively leads to an increase of the observed line ratios and therefore to an overestimation of the derived excitation temperatures when assuming optically thin emission. In contrast, RADEX accounts for these optical depth effects, resulting in the temperature offset for high volume densities seen in Fig.~\ref{fig:radex_1}a. Due to this, temperatures derived at A/B are likely to be in better agreement with non-LTE temperatures than values derived at the other emission peaks.

%------------------------------------------------------------------------------

\section{Derived parameters from the radiative transfer modeling with the CLASS module \textit{weeds}}\label{app:cd1}
\begin{figure}[bhtp]
	\centering
	\includegraphics[width=0.49\textwidth]{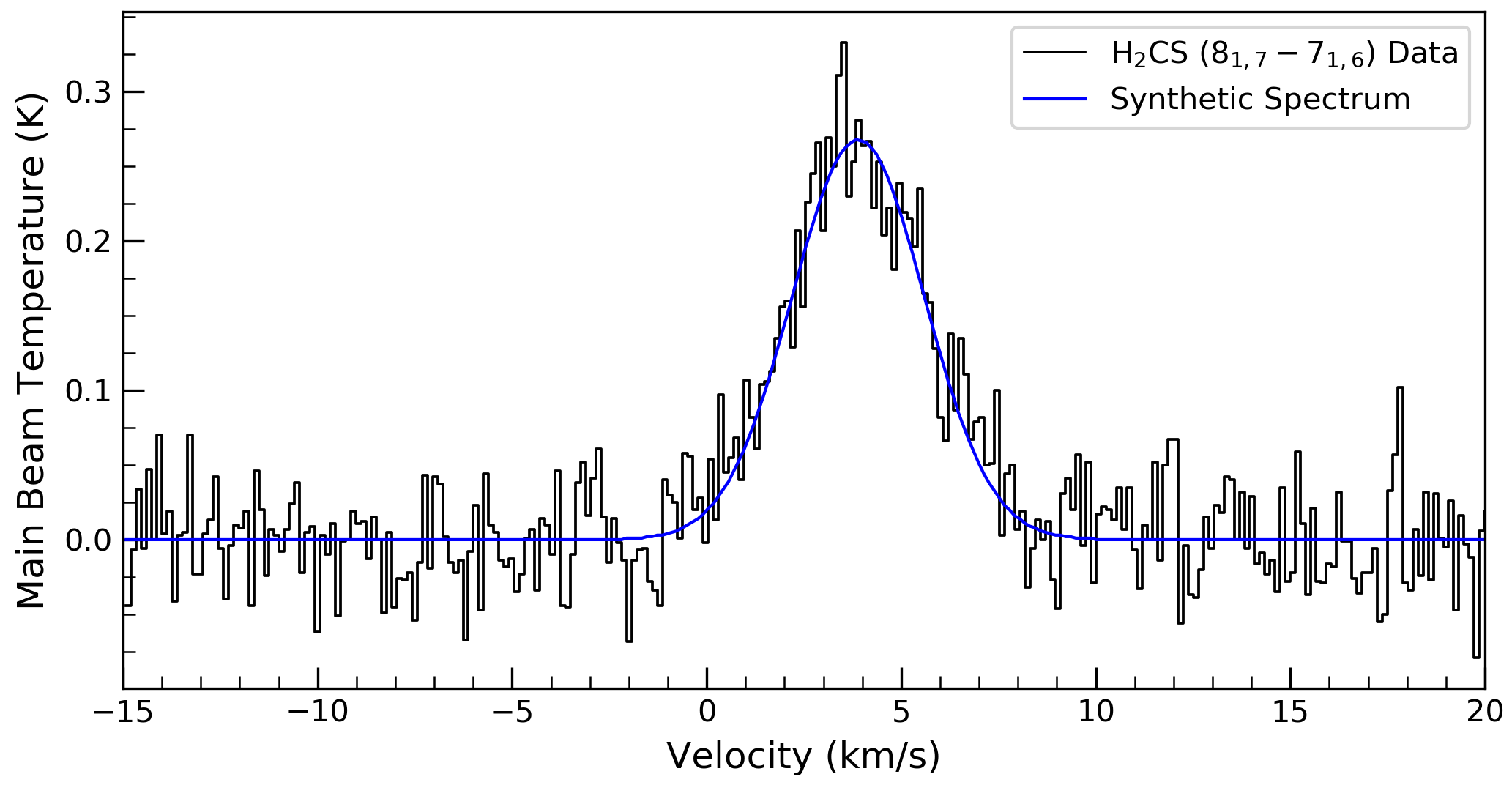}
	\caption{H$_2$CS ($8_{1,7} - 7_{1,6}$) transition at $\SI{278.9}{\giga\hertz}$ detected at the position of IRAS\,16293--2422 A/B. The data is shown in black color, while the LTE radiative transfer model computed with \textit{weeds} is shown in blue color.}
	\label{fig:weeds}
\end{figure}
We performed a number of LTE radiative transfer models with the CLASS-\textit{weeds} module in order the reproduce the observed molecular line profiles on each of the individual sources. The derived parameters from such models are shown in this Appendix. The derived values are the column density in cm$^{-2}$, the full width at half maximum (FWHM) in $\si{\kilo\meter\per\second}$ and the offset ($\delta$) in $\si{\kilo\meter\per\second}$ with respect to the local standard of rest (LSR) velocity of the source ($\SI{4}{\kilo\meter\per\second}$). We have used the derived temperatures reported in Table \ref{tab:temps} for each source. In addition to these values, a constant source size of $\SI{20}{\arcsecond}$ was assumed during the modeling and the velocities derived in Sect.~\ref{subsec:temperatures} were applied.

In Fig.~\ref{fig:weeds} we show an example of a synthetic spectrum obtained with \textit{weeds} for the H$_2$CS ($8_{1,7} - 7_{1,6}$) transition detected in IRAS\,16293--2422 A/B. Parameters used for the computation of simulated line profiles can be found in Table~\ref{tab:cdp0}-\ref{tab:cdp6}. Based on a conservative estimate on the calibration uncertainty of about 20\%, it can be expected that this uncertainty is transferred to the column density values. 

An overview of the derived column densities and upper limits is presented in Table~\ref{tab:cd}, which states values derived from self-absorbed or distorted line profiles in parenthesis. Column densities that are based on optically thin isotopes are given in the last three rows of the table. 

For comparing these values more easily with common literature, abundances of all detected transitions relative to CO are given in Table~\ref{app:cd_co}. For this, the CO column density was derived from the C$^{17}$O column density by assuming a ratio of $^{16}\U{O}/^{17}\U{O} = 1790$ based the values suggested by \citet{wilson1994abmilam}.
%-----------------------------------------------------------------------------
%------------------------------------------------------------------------------

\begin{table}[hbt]
    \caption{Derived parameters from the LTE CLASS-\textit{weeds} simulation of line profiles at the position of IRAS\,16293 A/B.}
    \label{tab:cdp0}
	\centering
	\begin{tabular}{cccc}
		\hline
		\hline
		Molecule & Column Density  & FWHM                          & $\delta$  \\ 
		& (cm$^{-2}$)     & ($\si{\kilo\meter\per\second}$) & (\si{\kilo\meter\per\second}) \\ \hline
        H$_2$CO & $\num{3.8e+14}$ & $\num{2.5}$ & $\num{0.0}$ \\ 
        H$_2^{13}$CO & $\num{1.0e+13}$ & $\num{3.5}$ & $\num{0.0}$ \\ 
        HDCO & $\num{3.6e+13}$ & $\num{3.0}$ & $\num{0.0}$ \\ 
        D$_2$CO & $\num{1.5e+13}$ & $\num{2.6}$ & $\num{0.0}$ \\ 
        C$_2$H & $\num{7.5e+13}$ & $\num{1.2}$ & $\num{-0.4}$ \\ 
        CH$_3$OH & $\num{6e+14}$ & $\num{3.4}$ & $\num{0.0}$ \\ 
        C$^{17}$O & $\num{7.6e+15}$ & $\num{2.6}$ & $\num{-0.2}$ \\ 
        CN & $\num{2.8e+13}$ & $\num{2.3}$ & $\num{0.0}$ \\ 
        CS & $\num{8.8e+13}$ & $\num{2.3}$ & $\num{0.0}$ \\ 
        C$^{33}$S & $\num{4.0e+12}$ & $\num{3.5}$ & $\num{0.0}$ \\ 
        C$^{34}$S & $\num{1.2e+13}$ & $\num{3.0}$ & $\num{-0.2}$ \\ 
        H$^{13}$CO$^+$ & $\num{4e+12}$ & $\num{2.5}$ & $\num{0.0}$ \\ 
        DCO$^+$ & $\num{1.6e+12}$ & $\num{2.0}$ & $\num{0.0}$ \\ 
        H$_2$CS & $\num{4.5e+13}$ & $\num{4.0}$ & $\num{-0.1}$ \\ 
        H$^{13}$CN & $\num{1.8e+12}$ & $\num{4.0}$ & $\num{0.2}$ \\ 
        DCN & $\num{1.4e+12}$ & $\num{3.0}$ & $\num{0.0}$ \\ 
        H$_2$D$^+$ & $<\num{1.5e+13}$ & $\num{3.0}$ & $\num{0.0}$ \\ 
        HNC & $\num{1.3e+13}$ & $\num{2.3}$ & $\num{0.0}$ \\ 
        DNC & $\num{1.4e+12}$ & $\num{2.0}$ & $\num{0.0}$ \\ 
        HNCO & $\num{3e+13}$ & $\num{4.5}$ & $\num{-0.2}$ \\ 
        NH$_2$D & $\num{4.0e+14}$ & $\num{1.2}$ & $\num{0.3}$ \\ 
        NHD$_2$ & $<\num{5.3e+14}$ & $\num{3.0}$ & $\num{0.0}$ \\ 
        N$_2$H$^+$ & $\num{3.0e+13}$ & $\num{1.0}$ & $\num{0.0}$ \\ 
        N$_2$D$^+$ & $\num{7.6e+11}$ & $\num{0.7}$ & $\num{0.3}$ \\ 
        NO & $\num{1.2e+15}$ & $\num{2.6}$ & $\num{0.0}$ \\ 
        OCS & $\num{6.2e+14}$ & $\num{4.0}$ & $\num{-0.4}$ \\ 
        SiO & $\num{7.9e+12}$ & $\num{5.5}$ & $\num{0.2}$ \\ 
        SO & $\num{5.0e+14}$ & $\num{3.8}$ & $\num{0.0}$ \\ 
        $^{34}$SO & $\num{3.6e+13}$ & $\num{4.0}$ & $\num{0.0}$ \\ 
        SO$_2$ & $\num{2.4e+14}$ & $\num{3.5}$ & $\num{0.0}$ \\ 
		\hline

	\end{tabular}
	\tablefoot{A temperature of $\SI{62.4}{\kelvin}$ and a beam size of $\SI{20}{\arcsecond}$ are assumed for the computation of the synthetic spectra.}
\end{table}

\begin{table}[hbt]
    \caption{Derived parameters from the LTE CLASS-\textit{weeds} simulation of line profiles at the position of 16293E.}
	\centering
	\begin{tabular}{cccc}
		\hline
		\hline
		Molecule & Column Density  & FWHM                          & $\delta$  \\ 
		& (cm$^{-2}$)     & ($\si{\kilo\meter\per\second}$) & (\si{\kilo\meter\per\second}) \\ \hline
        H$_2$CO & $\num{2.4e+13}$ & $\num{0.8}$ & $\num{-0.2}$ \\ 
        H$_2^{13}$CO & $<\num{1.4e+12}$ & $\num{0.7}$ & $\num{0.0}$ \\ 
        HDCO & $<\num{5.2e+12}$ & $\num{0.7}$ & $\num{0.0}$ \\ 
        D$_2$CO & $<\num{1.5e+13}$ & $\num{0.7}$ & $\num{0.0}$ \\ 
        C$_2$H & $<\num{2.2e+13}$ & $\num{0.7}$ & $\num{0.0}$ \\ 
        CH$_3$OH & $\num{8e+12}$ & $\num{0.6}$ & $\num{-0.2}$ \\ 
        C$^{17}$O & $\num{3.3e+15}$ & $\num{0.8}$ & $\num{-0.2}$ \\ 
        CN & $\num{1.5e+13}$ & $\num{0.7}$ & $\num{-0.4}$ \\ 
        CS & $\num{1.5e+13}$ & $\num{1.2}$ & $\num{0.0}$ \\ 
        C$^{33}$S & $<\num{2.8e+12}$ & $\num{0.7}$ & $\num{0.0}$ \\ 
        C$^{34}$S & $<\num{6.9e+12}$ & $\num{0.7}$ & $\num{0.0}$ \\ 
        H$^{13}$CO$^+$ & $\num{1.8e+12}$ & $\num{0.8}$ & $\num{-0.2}$ \\ 
        DCO$^+$ & $\num{5.8e+12}$ & $\num{0.5}$ & $\num{-0.3}$ \\ 
        H$_2$CS & $<\num{2.5e+13}$ & $\num{0.7}$ & $\num{0.0}$ \\ 
        H$^{13}$CN & $<\num{3.8e+11}$ & $\num{0.7}$ & $\num{0.0}$ \\ 
        DCN & $<\num{7.6e+11}$ & $\num{0.7}$ & $\num{0.0}$ \\ 
        H$_2$D$^+$ & $\num{1.1e+15}$ & $\num{0.5}$ & $\num{-0.4}$ \\ 
        HNC & $\num{1.1e+13}$ & $\num{0.8}$ & $\num{-0.2}$ \\ 
        DNC & $\num{2.1e+12}$ & $\num{0.6}$ & $\num{-0.4}$ \\ 
        HNCO & $<\num{6.1e+15}$ & $\num{0.7}$ & $\num{0.0}$ \\ 
        NH$_2$D & $\num{4.1e+14}$ & $\num{0.8}$ & $\num{-0.4}$ \\ 
        NHD$_2$ & $\num{7.5e+13}$ & $\num{0.7}$ & $\num{-0.4}$ \\ 
        N$_2$H$^+$ & $\num{9.1e+13}$ & $\num{0.6}$ & $\num{-0.3}$ \\ 
        N$_2$D$^+$ & $\num{9.6e+12}$ & $\num{0.5}$ & $\num{-0.4}$ \\ 
        NO & $\num{5.9e+14}$ & $\num{0.6}$ & $\num{-0.3}$ \\ 
        OCS & $<\num{2.4e+17}$ & $\num{0.7}$ & $\num{0.0}$ \\ 
        SiO & $<\num{7.2e+12}$ & $\num{0.7}$ & $\num{0.0}$ \\ 
        SO & $\num{9.7e+13}$ & $\num{0.7}$ & $\num{-0.2}$ \\ 
        $^{34}$SO & $<\num{3.0e+13}$ & $\num{0.7}$ & $\num{0.0}$ \\ 
        SO$_2$ & $<\num{1.1e+13}$ & $\num{0.7}$ & $\num{0.0}$ \\ 
		\hline

	\end{tabular}
	\tablefoot{A temperature of $\SI{12}{\kelvin}$ and a beam size of $\SI{20}{\arcsecond}$ are assumed for the computation of the synthetic spectra.}
	\label{tab:cdp1}
\end{table}

\begin{table}[hbt]
    \caption{Derived parameters from the LTE CLASS-\textit{weeds} simulation of line profiles at the outflow position E1.}
	\centering
	\begin{tabular}{cccc}
		\hline
		\hline
		Molecule & Column Density  & FWHM                          & $\delta$  \\ 
		& (cm$^{-2}$)     & ($\si{\kilo\meter\per\second}$) & (\si{\kilo\meter\per\second}) \\ \hline
        H$_2$CO & $\num{1.0e+14}$ & $\num{2.0}$ & $\num{0.0}$ \\ 
        H$_2^{13}$CO & $<\num{9.5e+11}$ & $\num{1.0}$ & $\num{0.0}$ \\ 
        HDCO & $\num{4.0e+12}$ & $\num{1.9}$ & $\num{0.1}$ \\ 
        D$_2$CO & $<\num{3.6e+12}$ & $\num{1.0}$ & $\num{0.0}$ \\ 
        C$_2$H & $\num{9.5e+12}$ & $\num{1.2}$ & $\num{0.0}$ \\ 
        CH$_3$OH & $\num{2.7e+14}$ & $\num{3.2}$ & $\num{1.0}$ \\ 
        C$^{17}$O & $\num{1.6e+15}$ & $\num{1.0}$ & $\num{-0.2}$ \\ 
        CN & $\num{3.9e+12}$ & $\num{0.8}$ & $\num{0.0}$ \\ 
        CS & $\num{1.2e+13}$ & $\num{1.5}$ & $\num{0.3}$ \\ 
        C$^{33}$S & $<\num{9.6e+11}$ & $\num{1.0}$ & $\num{0.0}$ \\ 
        C$^{34}$S & $<\num{8.3e+11}$ & $\num{1.0}$ & $\num{0.0}$ \\ 
        H$^{13}$CO$^+$ & $\num{3.2e+11}$ & $\num{1.2}$ & $\num{0.0}$ \\ 
        DCO$^+$ & $\num{1.4e+11}$ & $\num{0.5}$ & $\num{0.0}$ \\ 
        H$_2$CS & $<\num{4.4e+12}$ & $\num{1.0}$ & $\num{0.0}$ \\ 
        H$^{13}$CN & $<\num{1.6e+11}$ & $\num{1.0}$ & $\num{0.0}$ \\ 
        DCN & $<\num{1.8e+11}$ & $\num{1.0}$ & $\num{0.0}$ \\ 
        H$_2$D$^+$ & $<\num{7.0e+12}$ & $\num{1.0}$ & $\num{0.0}$ \\ 
        HNC & $\num{1.5e+12}$ & $\num{1.3}$ & $\num{0.0}$ \\ 
        DNC & $\num{2.3e+11}$ & $\num{0.7}$ & $\num{-0.3}$ \\ 
        HNCO & $<\num{1.3e+13}$ & $\num{1.0}$ & $\num{0.0}$ \\ 
        NH$_2$D & $<\num{1.2e+14}$ & $\num{1.0}$ & $\num{0.0}$ \\ 
        NHD$_2$ & $<\num{8.6e+13}$ & $\num{1.0}$ & $\num{0.0}$ \\ 
        N$_2$H$^+$ & $\num{8.0e+12}$ & $\num{1.0}$ & $\num{0.0}$ \\ 
        N$_2$D$^+$ & $\num{1.5e+11}$ & $\num{0.5}$ & $\num{-0.2}$ \\ 
        NO & $\num{2.7e+14}$ & $\num{0.7}$ & $\num{0.0}$ \\ 
        OCS & $<\num{1.7e+14}$ & $\num{1.0}$ & $\num{0.0}$ \\ 
        SiO & $\num{1.0e+12}$ & $\num{1.0}$ & $\num{0.0}$ \\ 
        SO & $\num{7.6e+13}$ & $\num{1.5}$ & $\num{0.0}$ \\ 
        $^{34}$SO & $<\num{3.5e+12}$ & $\num{1.0}$ & $\num{0.0}$ \\ 
        SO$_2$ & $\num{2.5e+13}$ & $\num{1.0}$ & $\num{0.2}$ \\ 
		\hline
	\end{tabular}
	\tablefoot{A temperature of $\SI{36.4}{\kelvin}$ and a beam size of $\SI{20}{\arcsecond}$ are assumed for the computation of the synthetic spectra.}
	\label{tab:cdp2}
\end{table}

\begin{table}[hbt]
    \caption{Derived parameters from the LTE CLASS-\textit{weeds} simulation of line profiles at the outflow position E2.}
	\centering
	\begin{tabular}{cccc}
		\hline
		\hline
		Molecule & Column Density  & FWHM                          & $\delta$  \\ 
		& (cm$^{-2}$)     & ($\si{\kilo\meter\per\second}$) & (\si{\kilo\meter\per\second}) \\ \hline
        H$_2$CO & $\num{2.8e+13}$ & $\num{2.0}$ & $\num{0.0}$ \\ 
        H$_2^{13}$CO & $<\num{7.7e+11}$ & $\num{0.7}$ & $\num{0.0}$ \\ 
        HDCO & $<\num{1.9e+12}$ & $\num{0.7}$ & $\num{0.0}$ \\ 
        D$_2$CO & $<\num{2.5e+12}$ & $\num{0.7}$ & $\num{0.0}$ \\ 
        C$_2$H & $<\num{7.9e+12}$ & $\num{0.7}$ & $\num{0.0}$ \\ 
        CH$_3$OH & $\num{7.1e+13}$ & $\num{2.0}$ & $\num{0.5}$ \\ 
        C$^{17}$O & $\num{8.2e+14}$ & $\num{0.6}$ & $\num{-0.2}$ \\ 
        CN & $<\num{1.8e+12}$ & $\num{0.7}$ & $\num{0.0}$ \\ 
        CS & $<\num{1.1e+12}$ & $\num{0.7}$ & $\num{0.0}$ \\ 
        C$^{33}$S & $<\num{6.7e+11}$ & $\num{0.7}$ & $\num{0.0}$ \\ 
        C$^{34}$S & $<\num{7.4e+11}$ & $\num{0.7}$ & $\num{0.0}$ \\ 
        H$^{13}$CO$^+$ & $<\num{8.1e+10}$ & $\num{0.7}$ & $\num{0.0}$ \\ 
        DCO$^+$ & $<\num{1.2e+11}$ & $\num{0.7}$ & $\num{0.0}$ \\ 
        H$_2$CS & $<\num{4.7e+12}$ & $\num{0.7}$ & $\num{0.0}$ \\ 
        H$^{13}$CN & $<\num{1.3e+11}$ & $\num{0.7}$ & $\num{0.0}$ \\ 
        DCN & $<\num{2.0e+11}$ & $\num{0.7}$ & $\num{0.0}$ \\ 
        H$_2$D$^+$ & $<\num{7.9e+12}$ & $\num{0.7}$ & $\num{0.0}$ \\ 
        HNC & $\num{3.7e+11}$ & $\num{0.8}$ & $\num{0.0}$ \\ 
        DNC & $<\num{2.2e+11}$ & $\num{0.7}$ & $\num{0.0}$ \\ 
        HNCO & $<\num{1.4e+13}$ & $\num{0.7}$ & $\num{0.0}$ \\ 
        NH$_2$D & $<\num{8.2e+13}$ & $\num{0.7}$ & $\num{0.0}$ \\ 
        NHD$_2$ & $<\num{6.6e+13}$ & $\num{0.7}$ & $\num{0.0}$ \\ 
        N$_2$H$^+$ & $\num{5.3e+11}$ & $\num{0.5}$ & $\num{-0.1}$ \\ 
        N$_2$D$^+$ & $<\num{1.7e+11}$ & $\num{0.7}$ & $\num{0.0}$ \\ 
        NO & $<\num{2.1e+14}$ & $\num{0.7}$ & $\num{0.0}$ \\ 
        OCS & $<\num{1.9e+14}$ & $\num{0.7}$ & $\num{0.0}$ \\ 
        SiO & $<\num{3.4e+11}$ & $\num{0.7}$ & $\num{0.0}$ \\ 
        SO & $\num{7.0e+13}$ & $\num{0.3}$ & $\num{-0.2}$ \\ 
        $^{34}$SO & $<\num{3.2e+12}$ & $\num{0.7}$ & $\num{0.0}$ \\ 
        SO$_2$ & $<\num{9.3e+12}$ & $\num{0.7}$ & $\num{0.0}$ \\ 
		\hline

	\end{tabular}
	\tablefoot{A temperature of $\SI{34.1}{\kelvin}$ and a beam size of $\SI{20}{\arcsecond}$ are assumed for the computation of the synthetic spectra.}
	
	\label{tab:cdp3}
\end{table}

\begin{table}[hbt]
	\caption{Derived parameters from the LTE CLASS-\textit{weeds} simulation of line profiles at the outflow position W1.}
	\centering
	\begin{tabular}{cccc}
		\hline
		\hline
		Molecule & Column Density  & FWHM                          & $\delta$  \\ 
		& (cm$^{-2}$)     & ($\si{\kilo\meter\per\second}$) & (\si{\kilo\meter\per\second}) \\ \hline
        H$_2$CO & $\num{5.7e+13}$ & $\num{3.0}$ & $\num{0.5}$ \\ 
        H$_2^{13}$CO & $<\num{1.9e+12}$ & $\num{1.3}$ & $\num{0.0}$ \\ 
        HDCO & $<\num{7.4e+12}$ & $\num{1.3}$ & $\num{0.0}$ \\ 
        D$_2$CO & $<\num{9.7e+12}$ & $\num{1.3}$ & $\num{0.0}$ \\ 
        C$_2$H & $<\num{2.3e+13}$ & $\num{1.3}$ & $\num{0.0}$ \\ 
        CH$_3$OH & $\num{1.0e+14}$ & $\num{4.0}$ & $\num{1.0}$ \\ 
        C$^{17}$O & $\num{1.9e+15}$ & $\num{0.8}$ & $\num{0.0}$ \\ 
        CN & $<\num{6.3e+12}$ & $\num{1.3}$ & $\num{0.0}$ \\ 
        CS & $\num{8.3e+12}$ & $\num{1.3}$ & $\num{-0.4}$ \\ 
        C$^{33}$S & $<\num{2.6e+12}$ & $\num{1.3}$ & $\num{0.0}$ \\ 
        C$^{34}$S & $<\num{2.3e+12}$ & $\num{1.3}$ & $\num{0.0}$ \\ 
        H$^{13}$CO$^+$ & $<\num{2.8e+11}$ & $\num{1.3}$ & $\num{0.0}$ \\ 
        DCO$^+$ & $<\num{3.1e+11}$ & $\num{1.3}$ & $\num{0.0}$ \\ 
        H$_2$CS & $<\num{1.2e+13}$ & $\num{1.3}$ & $\num{0.0}$ \\ 
        H$^{13}$CN & $<\num{5.6e+11}$ & $\num{1.3}$ & $\num{0.0}$ \\ 
        DCN & $<\num{5.6e+11}$ & $\num{1.3}$ & $\num{0.0}$ \\ 
        H$_2$D$^+$ & $<\num{2.5e+13}$ & $\num{1.3}$ & $\num{0.0}$ \\ 
        HNC & $\num{1.3e+12}$ & $\num{1.5}$ & $\num{-0.4}$ \\ 
        DNC & $<\num{1.1e+12}$ & $\num{1.3}$ & $\num{0.0}$ \\ 
        HNCO & $<\num{4.5e+13}$ & $\num{1.3}$ & $\num{0.0}$ \\ 
        NH$_2$D & $<\num{3.5e+14}$ & $\num{1.3}$ & $\num{0.0}$ \\ 
        NHD$_2$ & $<\num{2.7e+14}$ & $\num{1.3}$ & $\num{0.0}$ \\ 
        N$_2$H$^+$ & $\num{8.1e+11}$ & $\num{0.7}$ & $\num{0.2}$ \\ 
        N$_2$D$^+$ & $<\num{6.0e+11}$ & $\num{1.3}$ & $\num{0.0}$ \\ 
        NO & $<\num{7.3e+14}$ & $\num{1.3}$ & $\num{0.0}$ \\ 
        OCS & $<\num{4.6e+14}$ & $\num{1.3}$ & $\num{0.0}$ \\ 
        SiO & $<\num{1.5e+12}$ & $\num{1.3}$ & $\num{0.0}$ \\ 
        SO & $\num{5.6e+13}$ & $\num{0.6}$ & $\num{-0.7}$ \\ 
        $^{34}$SO & $<\num{1.1e+13}$ & $\num{1.3}$ & $\num{0.0}$ \\ 
        SO$_2$ & $<\num{3.2e+13}$ & $\num{1.3}$ & $\num{0.0}$ \\ 
		\hline
	\end{tabular}
	\tablefoot{A temperature of $\SI{34.8}{\kelvin}$ and a beam size of $\SI{20}{\arcsecond}$ are assumed for the computation of the synthetic spectra.}
	
	\label{tab:cdp4}
\end{table}

\begin{table}[hbt]
    \caption{Derived parameters from the LTE CLASS-\textit{weeds} simulation of line profiles at the position W2.}
	\centering
	\begin{tabular}{cccc}
		\hline
		\hline
		Molecule & Column Density  & FWHM                          & $\delta$  \\ 
		& (cm$^{-2}$)     & ($\si{\kilo\meter\per\second}$) & (\si{\kilo\meter\per\second}) \\ \hline
        H$_2$CO & $\num{5.6e+13}$ & $\num{1.6}$ & $\num{-0.1}$ \\ 
        H$_2^{13}$CO & $<\num{1.7e+12}$ & $\num{1.2}$ & $\num{0.0}$ \\ 
        HDCO & $<\num{6.7e+12}$ & $\num{1.2}$ & $\num{0.0}$ \\ 
        D$_2$CO & $<\num{7.9e+12}$ & $\num{1.2}$ & $\num{0.0}$ \\ 
        C$_2$H & $<\num{2.3e+13}$ & $\num{1.2}$ & $\num{0.0}$ \\ 
        CH$_3$OH & $\num{1.0e+14}$ & $\num{1.5}$ & $\num{0.0}$ \\ 
        C$^{17}$O & $\num{1.9e+15}$ & $\num{0.9}$ & $\num{-0.1}$ \\ 
        CN & $<\num{6.3e+12}$ & $\num{1.2}$ & $\num{0.0}$ \\ 
        CS & $\num{6.2e+12}$ & $\num{1.4}$ & $\num{0.0}$ \\ 
        C$^{33}$S & $<\num{1.8e+12}$ & $\num{1.2}$ & $\num{0.0}$ \\ 
        C$^{34}$S & $<\num{2.4e+12}$ & $\num{1.2}$ & $\num{0.0}$ \\ 
        H$^{13}$CO$^+$ & $<\num{2.5e+11}$ & $\num{1.2}$ & $\num{0.0}$ \\ 
        DCO$^+$ & $<\num{2.9e+11}$ & $\num{1.2}$ & $\num{0.0}$ \\ 
        H$_2$CS & $<\num{1.1e+13}$ & $\num{1.2}$ & $\num{0.0}$ \\ 
        H$^{13}$CN & $<\num{4.4e+11}$ & $\num{1.2}$ & $\num{0.0}$ \\ 
        DCN & $<\num{5.7e+11}$ & $\num{1.2}$ & $\num{0.0}$ \\ 
        H$_2$D$^+$ & $<\num{2.3e+13}$ & $\num{1.2}$ & $\num{0.0}$ \\ 
        HNC & $\num{5.4e+11}$ & $\num{0.9}$ & $\num{-0.1}$ \\ 
        DNC & $<\num{6.3e+11}$ & $\num{1.2}$ & $\num{0.0}$ \\ 
        HNCO & $<\num{5.6e+13}$ & $\num{1.2}$ & $\num{0.0}$ \\ 
        NH$_2$D & $<\num{2.0e+14}$ & $\num{1.2}$ & $\num{0.0}$ \\ 
        NHD$_2$ & $<\num{1.5e+14}$ & $\num{1.2}$ & $\num{0.0}$ \\ 
        N$_2$H$^+$ & $\num{1.3e+12}$ & $\num{0.8}$ & $\num{-0.2}$ \\ 
        N$_2$D$^+$ & $<\num{6.3e+11}$ & $\num{1.2}$ & $\num{0.0}$ \\ 
        NO & $<\num{6.4e+14}$ & $\num{1.2}$ & $\num{0.0}$ \\ 
        OCS & $<\num{5.9e+14}$ & $\num{1.2}$ & $\num{0.0}$ \\ 
        SiO & $<\num{1.3e+12}$ & $\num{1.2}$ & $\num{0.0}$ \\ 
        SO & $\num{2.3e+13}$ & $\num{1.2}$ & $\num{-0.3}$ \\ 
        $^{34}$SO & $<\num{1.4e+13}$ & $\num{1.2}$ & $\num{0.0}$ \\ 
        SO$_2$ & $<\num{2.6e+13}$ & $\num{1.2}$ & $\num{0.0}$ \\ 
        \hline
	\end{tabular}
	\tablefoot{A temperature of $\SI{31.3}{\kelvin}$ and a beam size of $\SI{20}{\arcsecond}$ are assumed for the computation of the synthetic spectra.}
	
	\label{tab:cdp5}
\end{table}

\begin{table}[hbt]
    \caption{Derived parameters from the LTE CLASS-\textit{weeds} simulation of line profiles at the outflow position HE2.}
	\centering
	\begin{tabular}{cccc}
		\hline
		\hline
		Molecule & Column Density  & FWHM                          & $\delta$  \\ 
		& (cm$^{-2}$)     & ($\si{\kilo\meter\per\second}$) & (\si{\kilo\meter\per\second}) \\ \hline
        H$_2$CO & $\num{4.9e+13}$ & $\num{2.6}$ & $\num{-1.7}$ \\ 
        H$_2^{13}$CO & $<\num{6.4e+11}$ & $\num{0.7}$ & $\num{0.0}$ \\ 
        HDCO & $<\num{2.1e+12}$ & $\num{0.7}$ & $\num{0.0}$ \\ 
        D$_2$CO & $<\num{2.7e+12}$ & $\num{0.7}$ & $\num{0.0}$ \\ 
        C$_2$H & $<\num{7.3e+12}$ & $\num{0.7}$ & $\num{0.0}$ \\ 
        CH$_3$OH & $\num{9.4e+13}$ & $\num{2.3}$ & $\num{-2.0}$ \\ 
        C$^{17}$O & $\num{1.6e+15}$ & $\num{0.8}$ & $\num{-0.1}$ \\ 
        CN & $\num{3.8e+12}$ & $\num{0.6}$ & $\num{0.0}$ \\ 
        CS & $\num{8.1e+12}$ & $\num{1.3}$ & $\num{-0.2}$ \\ 
        C$^{33}$S & $<\num{6.4e+11}$ & $\num{0.7}$ & $\num{0.0}$ \\ 
        C$^{34}$S & $<\num{8.4e+11}$ & $\num{0.7}$ & $\num{0.0}$ \\ 
        H$^{13}$CO$^+$ & $\num{2.3e+11}$ & $\num{0.5}$ & $\num{-0.1}$ \\ 
        DCO$^+$ & $\num{1.8e+11}$ & $\num{0.3}$ & $\num{-0.1}$ \\ 
        H$_2$CS & $<\num{3.5e+12}$ & $\num{0.7}$ & $\num{0.0}$ \\ 
        H$^{13}$CN & $<\num{1.6e+11}$ & $\num{0.7}$ & $\num{0.0}$ \\ 
        DCN & $<\num{1.7e+11}$ & $\num{0.7}$ & $\num{0.0}$ \\ 
        H$_2$D$^+$ & $<\num{5.4e+12}$ & $\num{0.7}$ & $\num{0.0}$ \\ 
        HNC & $\num{8.6e+11}$ & $\num{0.7}$ & $\num{0.0}$ \\ 
        DNC & $\num{1.8e+11}$ & $\num{0.4}$ & $\num{-0.2}$ \\ 
        HNCO & $<\num{1.5e+13}$ & $\num{0.7}$ & $\num{0.0}$ \\ 
        NH$_2$D & $\num{9.7e+13}$ & $\num{0.7}$ & $\num{0.0}$ \\ 
        NHD$_2$ & $<\num{7.0e+13}$ & $\num{0.7}$ & $\num{0.0}$ \\ 
        N$_2$H$^+$ & $\num{1.0e+13}$ & $\num{0.4}$ & $\num{-0.2}$ \\ 
        N$_2$D$^+$ & $\num{2.2e+11}$ & $\num{0.3}$ & $\num{-0.2}$ \\ 
        NO & $\num{2.5e+14}$ & $\num{0.7}$ & $\num{0.0}$ \\ 
        OCS & $<\num{1.4e+14}$ & $\num{0.7}$ & $\num{0.0}$ \\ 
        SiO & $<\num{4.5e+11}$ & $\num{0.7}$ & $\num{0.0}$ \\ 
        SO & $\num{1.9e+13}$ & $\num{1.3}$ & $\num{-1.5}$ \\ 
        $^{34}$SO & $<\num{3.5e+12}$ & $\num{0.7}$ & $\num{0.0}$ \\ 
        SO$_2$ & $<\num{8.6e+12}$ & $\num{0.7}$ & $\num{0.0}$ \\ 
		\hline
	\end{tabular}
	\tablefoot{A temperature of $\SI{34.0}{\kelvin}$ and a beam size of $\SI{20}{\arcsecond}$ are assumed for the computation of the synthetic spectra.}
	
	\label{tab:cdp6}
\end{table}

%%%%%%%%%%%%%%%%%%%%%%%%%%%%%%%%%%%%%%%%%%%%%%%%%%%%%%%%%%%%%%%%%%%%%%%%%%%%%%%%%%%%%%%%%%%%%%%%%
\begin{table*}[p]
    \caption{Column densities of the observed molecules in units of cm$^{-2}$ derived with the CLASS-Module \textit{weeds}.}
    \label{tab:cd}
	\centering
	\begin{tabular}{crrrrrrr}
		\hline
		\hline
		Molecule & A/B \,\, & E\,\, & E1\,\, & E2\,\, & W1\,\, & W2\,\, & HE2\,\, \\ \hline
        H$_2$CO & ($\num{3.8e+14}$) & $\num{2.4e+13}$ & ($\num{1.0e+14}$) & ($\num{2.8e+13}$) & ($\num{5.7e+13}$) & $\num{5.6e+13}$ & ($\num{4.9e+13}$) \\ 
        H$_2^{13}$CO & $\num{1.0e+13}$ & $<\num{1.4e+12}$ & $<\num{9.5e+11}$ & $<\num{7.7e+11}$ & $<\num{1.9e+12}$ & $<\num{1.7e+12}$ & $<\num{6.4e+11}$ \\ 
        HDCO & $\num{3.6e+13}$ & $<\num{5.2e+12}$ & ($\num{4.0e+12}$) & $<\num{1.9e+12}$ & $<\num{7.4e+12}$ & $<\num{6.7e+12}$ & $<\num{2.1e+12}$ \\ 
        D$_2$CO & ($\num{1.5e+13}$) & $<\num{1.5e+13}$ & $<\num{3.6e+12}$ & $<\num{2.5e+12}$ & $<\num{9.7e+12}$ & $<\num{7.9e+12}$ & $<\num{2.7e+12}$ \\ 
        C$_2$H & $\num{7.5e+13}$ & $<\num{2.2e+13}$ & $\num{9.5e+12}$ & $<\num{7.9e+12}$ & $<\num{2.3e+13}$ & $<\num{2.3e+13}$ & $<\num{7.3e+12}$ \\ 
        CH$_3$OH & ($\num{6.0e+14}$) & ($\num{8.0e+12}$) & ($\num{2.7e+14}$) & ($\num{7.1e+13}$) & ($\num{1.0e+14}$) & ($\num{1.0e+14}$) & ($\num{9.4e+13}$) \\ 
        C$^{17}$O & $\num{7.6e+15}$ & $\num{3.3e+15}$ & $\num{1.6e+15}$ & $\num{8.2e+14}$ & ($\num{1.9e+15}$) & $\num{1.9e+15}$ & $\num{1.6e+15}$ \\ 
        CN & ($\num{2.8e+13}$) & $\num{1.5e+13}$ & $\num{3.9e+12}$ & $<\num{1.8e+12}$ & $<\num{6.3e+12}$ & $<\num{6.3e+12}$ & $\num{3.8e+12}$ \\ 
        CS & ($\num{8.8e+13}$) & ($\num{1.5e+13}$) & ($\num{1.2e+13}$) & $<\num{1.1e+12}$ & ($\num{8.3e+12}$) & $\num{6.2e+12}$ & ($\num{8.1e+12}$) \\ 
        C$^{33}$S & ($\num{4.0e+12}$) & $<\num{2.8e+12}$ & $<\num{9.6e+11}$ & $<\num{6.7e+11}$ & $<\num{2.6e+12}$ & $<\num{1.8e+12}$ & $<\num{6.4e+11}$ \\ 
        C$^{34}$S & $\num{1.2e+13}$ & $<\num{6.9e+12}$ & $<\num{8.3e+11}$ & $<\num{7.4e+11}$ & $<\num{2.3e+12}$ & $<\num{2.4e+12}$ & $<\num{8.4e+11}$ \\ 
        H$^{13}$CO$^+$ & $\num{4.0e+12}$ & $\num{1.8e+12}$ & $\num{3.2e+11}$ & $<\num{8.1e+10}$ & $<\num{2.8e+11}$ & $<\num{2.5e+11}$ & $\num{2.3e+11}$ \\ 
        DCO$^+$ & ($\num{1.6e+12}$) & $\num{5.8e+12}$ & $\num{1.4e+11}$ & $<\num{1.2e+11}$ & $<\num{3.1e+11}$ & $<\num{2.9e+11}$ & $\num{1.8e+11}$ \\ 
        H$_2$CS & $\num{4.5e+13}$ & $<\num{2.5e+13}$ & $<\num{4.4e+12}$ & $<\num{4.7e+12}$ & $<\num{1.2e+13}$ & $<\num{1.1e+13}$ & $<\num{3.5e+12}$ \\ 
        H$^{13}$CN & $\num{1.8e+12}$ & $<\num{3.8e+11}$ & $<\num{1.6e+11}$ & $<\num{1.3e+11}$ & $<\num{5.6e+11}$ & $<\num{4.4e+11}$ & $<\num{1.6e+11}$ \\ 
        DCN & $\num{1.4e+12}$ & $<\num{7.6e+11}$ & $<\num{1.8e+11}$ & $<\num{2.0e+11}$ & $<\num{5.6e+11}$ & $<\num{5.7e+11}$ & $<\num{1.7e+11}$ \\ 
        H$_2$D$^+$ & $<\num{1.5e+13}$ & $\num{1.1e+15}$ & $<\num{7.0e+12}$ & $<\num{7.9e+12}$ & $<\num{2.5e+13}$ & $<\num{2.3e+13}$ & $<\num{5.4e+12}$ \\ 
        HNC & ($\num{1.3e+13}$) & ($\num{1.1e+13}$) & $\num{1.5e+12}$ & ($\num{3.7e+11}$) & ($\num{1.3e+12}$) & ($\num{5.4e+11}$) & $\num{8.6e+11}$ \\ 
        DNC & ($\num{1.4e+12}$) & $\num{2.1e+12}$ & ($\num{2.3e+11}$) & $<\num{2.2e+11}$ & $<\num{1.1e+12}$ & $<\num{6.3e+11}$ & $\num{1.8e+11}$ \\ 
        HNCO & $\num{3.0e+13}$ & $<\num{6.1e+15}$ & $<\num{1.3e+13}$ & $<\num{1.4e+13}$ & $<\num{4.5e+13}$ & $<\num{5.6e+13}$ & $<\num{1.5e+13}$ \\ 
        NH$_2$D & $\num{4.0e+14}$ & $\num{4.1e+14}$ & $<\num{1.2e+14}$ & $<\num{8.2e+13}$ & $<\num{3.5e+14}$ & $<\num{2.0e+14}$ & $\num{9.7e+13}$ \\ 
        NHD$_2$ & $<\num{5.3e+14}$ & $\num{7.5e+13}$ & $<\num{8.6e+13}$ & $<\num{6.6e+13}$ & $<\num{2.7e+14}$ & $<\num{1.5e+14}$ & $<\num{7.0e+13}$ \\ 
        N$_2$H$^+$ & ($\num{3.0e+13}$) & $\num{9.1e+13}$ & $\num{8.0e+12}$ & $\num{5.3e+11}$ & $\num{8.1e+11}$ & ($\num{1.3e+12}$) & $\num{1.0e+13}$ \\ 
        N$_2$D$^+$ & $\num{7.6e+11}$ & $\num{9.6e+12}$ & $\num{1.5e+11}$ & $<\num{1.7e+11}$ & $<\num{6.0e+11}$ & $<\num{6.3e+11}$ & $\num{2.2e+11}$ \\ 
        NO & $\num{1.2e+15}$ & $\num{5.9e+14}$ & $\num{2.7e+14}$ & $<\num{2.1e+14}$ & $<\num{7.3e+14}$ & $<\num{6.4e+14}$ & $\num{2.5e+14}$ \\ 
        OCS & ($\num{6.2e+14}$) & $<\num{2.4e+17}$ & $<\num{1.7e+14}$ & $<\num{1.9e+14}$ & $<\num{4.6e+14}$ & $<\num{5.9e+14}$ & $<\num{1.4e+14}$ \\ 
        SiO & $\num{7.9e+12}$ & $<\num{7.2e+12}$ & $\num{1.0e+12}$ & $<\num{3.4e+11}$ & $<\num{1.5e+12}$ & $<\num{1.3e+12}$ & $<\num{4.5e+11}$ \\ 
        SO & $\num{5.0e+14}$ & ($\num{9.7e+13}$) & ($\num{7.6e+13}$) & ($\num{7.0e+13}$) & ($\num{5.6e+13}$) & ($\num{2.3e+13}$) & ($\num{1.9e+13}$) \\ 
        $^{34}$SO & $\num{3.6e+13}$ & $<\num{3.0e+13}$ & $<\num{3.5e+12}$ & $<\num{3.2e+12}$ & $<\num{1.1e+13}$ & $<\num{1.4e+13}$ & $<\num{3.5e+12}$ \\ 
        SO$_2$ & $\num{2.4e+14}$ & $<\num{1.1e+13}$ & $\num{2.5e+13}$ & $<\num{9.3e+12}$ & $<\num{3.2e+13}$ & $<\num{2.6e+13}$ & $<\num{8.6e+12}$ \\ \hline
        CO & $\num{1.4e+19}$ & $\num{5.9e+18}$ & $\num{2.9e+18}$ & $\num{1.5e+18}$ & ($\num{3.4e+18}$) & $\num{3.4e+18}$ & $\num{2.9e+18}$ \\ 
        HCN & $\num{1.2e+14}$ & $<\num{2.6e+13}$ & $<\num{1.1e+13}$ & $<\num{8.8e+12}$ & $<\num{3.8e+13}$ & $<\num{3.0e+13}$ & $<\num{1.1e+13}$ \\ 
        HCO$^+$ & $\num{2.7e+14}$ & $\num{1.2e+14}$ & $\num{2.2e+13}$ & $<\num{5.5e+12}$ & $<\num{1.9e+13}$ & $<\num{1.7e+13}$ & $\num{1.6e+13}$ \\  \hline
		\hline
	\end{tabular}
	\tablefoot{Upper limits have been derived for positions that do not show significant emission. Values derived from distorted line profiles are written in parenthesis. Uncertainties are estimated to be about 20\% from the radiative transfer modeling. The last three rows contain values derived from the column densities of the less abundant isotopologues C$^{17}$O, H$^{13}$CN, and H$^{13}$CO$^+$. Note that the main isotopologues of these species are detected at all considered positions.}
\end{table*}

%------------------------------------------------------------------------------------

\begin{table*}[p]
    \caption{Column densities and upper limits for the observed molecules in units of the CO column density at the respective positions.}
    \label{app:cd_co}
	\centering
	\begin{tabular}{crrrrrrr}
		\hline
		\hline
		Molecule &A/B & E & E1 & E2 & W1 & W2 & HE2 \\ \hline
        H$_2$CO & ($\num{2.8e-05}$) & $\num{4.1e-06}$ & ($\num{3.5e-05}$) & ($\num{1.9e-05}$) & ($\num{1.7e-05}$) & $\num{1.6e-05}$ & ($\num{1.7e-05}$) \\ 
        H$_2^{13}$CO & $\num{7.4e-07}$ & $<\num{2.4e-07}$ & $<\num{3.3e-07}$ & $<\num{5.2e-07}$ & $<\num{5.6e-07}$ & $<\num{5.0e-07}$ & $<\num{2.2e-07}$ \\ 
        HDCO & $\num{2.6e-06}$ & $<\num{8.8e-07}$ & ($\num{1.4e-06}$) & $<\num{1.3e-06}$ & $<\num{2.2e-06}$ & $<\num{2.0e-06}$ & $<\num{7.3e-07}$ \\ 
        D$_2$CO & ($\num{1.1e-06}$) & $<\num{2.5e-06}$ & $<\num{1.3e-06}$ & $<\num{1.7e-06}$ & $<\num{2.9e-06}$ & $<\num{2.3e-06}$ & $<\num{9.4e-07}$ \\ 
        C$_2$H & $\num{5.5e-06}$ & $<\num{3.7e-06}$ & $\num{3.3e-06}$ & $<\num{5.4e-06}$ & $<\num{6.8e-06}$ & $<\num{6.8e-06}$ & $<\num{2.5e-06}$ \\ 
        CH$_3$OH & ($\num{4.4e-05}$) & ($\num{1.4e-06}$) & ($\num{9.4e-05}$) & ($\num{4.8e-05}$) & ($\num{2.9e-05}$) & ($\num{2.9e-05}$) & ($\num{3.3e-05}$) \\ 
        CN & ($\num{2.1e-06}$) & $\num{2.5e-06}$ & $\num{1.4e-06}$ & $<\num{1.2e-06}$ & $<\num{1.9e-06}$ & $<\num{1.9e-06}$ & $\num{1.3e-06}$ \\ 
        CS & ($\num{6.5e-06}$) & ($\num{2.5e-06}$) & ($\num{4.2e-06}$) & $<\num{7.5e-07}$ & ($\num{2.4e-06}$) & $\num{1.8e-06}$ & ($\num{2.8e-06}$) \\ 
        C$^{33}$S & ($\num{2.9e-07}$) & $<\num{4.7e-07}$ & $<\num{3.4e-07}$ & $<\num{4.6e-07}$ & $<\num{7.6e-07}$ & $<\num{5.3e-07}$ & $<\num{2.2e-07}$ \\ 
        C$^{34}$S & $\num{8.8e-07}$ & $<\num{1.2e-06}$ & $<\num{2.9e-07}$ & $<\num{5.0e-07}$ & $<\num{6.8e-07}$ & $<\num{7.1e-07}$ & $<\num{2.9e-07}$ \\ 
        H$^{13}$CO$^+$ & $\num{2.9e-07}$ & $\num{3.0e-07}$ & $\num{1.1e-07}$ & $<\num{5.5e-08}$ & $<\num{8.2e-08}$ & $<\num{7.4e-08}$ & $\num{8.0e-08}$ \\ 
        DCO$^+$ & ($\num{1.2e-07}$) & $\num{9.8e-07}$ & $\num{4.9e-08}$ & $<\num{8.2e-08}$ & $<\num{9.1e-08}$ & $<\num{8.5e-08}$ & $\num{6.3e-08}$ \\ 
        H$_2$CS & $\num{3.3e-06}$ & $<\num{4.2e-06}$ & $<\num{1.5e-06}$ & $<\num{3.2e-06}$ & $<\num{3.5e-06}$ & $<\num{3.2e-06}$ & $<\num{1.2e-06}$ \\ 
        H$^{13}$CN & $\num{1.3e-07}$ & $<\num{6.4e-08}$ & $<\num{5.6e-08}$ & $<\num{8.9e-08}$ & $<\num{1.6e-07}$ & $<\num{1.3e-07}$ & $<\num{5.6e-08}$ \\ 
        DCN & $\num{1.0e-07}$ & $<\num{1.3e-07}$ & $<\num{6.3e-08}$ & $<\num{1.4e-07}$ & $<\num{1.6e-07}$ & $<\num{1.7e-07}$ & $<\num{5.9e-08}$ \\ 
        H$_2$D$^+$ & $<\num{1.1e-06}$ & $\num{1.9e-04}$ & $<\num{2.4e-06}$ & $<\num{5.4e-06}$ & $<\num{7.4e-06}$ & $<\num{6.8e-06}$ & $<\num{1.9e-06}$ \\ 
        HNC & ($\num{9.6e-07}$) & ($\num{1.9e-06}$) & $\num{5.2e-07}$ & ($\num{2.5e-07}$) & ($\num{3.8e-07}$) & ($\num{1.6e-07}$) & $\num{3.0e-07}$ \\ 
        DNC & ($\num{1.0e-07}$) & $\num{3.6e-07}$ & ($\num{8.0e-08}$) & $<\num{1.5e-07}$ & $<\num{3.2e-07}$ & $<\num{1.9e-07}$ & $\num{6.3e-08}$ \\ 
        HNCO & $\num{2.2e-06}$ & $<\num{1.0e-03}$ & $<\num{4.5e-06}$ & $<\num{9.5e-06}$ & $<\num{1.3e-05}$ & $<\num{1.6e-05}$ & $<\num{5.2e-06}$ \\ 
        NH$_2$D & $\num{2.9e-05}$ & $\num{6.9e-05}$ & $<\num{4.2e-05}$ & $<\num{5.6e-05}$ & $<\num{1.0e-04}$ & $<\num{5.9e-05}$ & $\num{3.4e-05}$ \\ 
        NHD$_2$ & $<\num{3.9e-05}$ & $\num{1.3e-05}$ & $<\num{3.0e-05}$ & $<\num{4.5e-05}$ & $<\num{7.9e-05}$ & $<\num{4.4e-05}$ & $<\num{2.4e-05}$ \\ 
        N$_2$H$^+$ & ($\num{2.2e-06}$) & $\num{1.5e-05}$ & $\num{2.8e-06}$ & $\num{3.6e-07}$ & $\num{2.4e-07}$ & ($\num{3.8e-07}$) & $\num{3.5e-06}$ \\ 
        N$_2$D$^+$ & $\num{5.6e-08}$ & $\num{1.6e-06}$ & $\num{5.2e-08}$ & $<\num{1.2e-07}$ & $<\num{1.8e-07}$ & $<\num{1.9e-07}$ & $\num{7.7e-08}$ \\ 
        NO & $\num{8.8e-05}$ & $\num{1.0e-04}$ & $\num{9.4e-05}$ & $<\num{1.4e-04}$ & $<\num{2.1e-04}$ & $<\num{1.9e-04}$ & $\num{8.7e-05}$ \\ 
        OCS & ($\num{4.6e-05}$) & $<\num{4.1e-02}$ & $<\num{5.9e-05}$ & $<\num{1.3e-04}$ & $<\num{1.4e-04}$ & $<\num{1.7e-04}$ & $<\num{4.9e-05}$ \\ 
        SiO & $\num{5.8e-07}$ & $<\num{1.2e-06}$ & $\num{3.5e-07}$ & $<\num{2.3e-07}$ & $<\num{4.4e-07}$ & $<\num{3.8e-07}$ & $<\num{1.6e-07}$ \\ 
        SO & $\num{3.7e-05}$ & ($\num{1.6e-05}$) & ($\num{2.7e-05}$) & ($\num{4.8e-05}$) & ($\num{1.6e-05}$) & ($\num{6.8e-06}$) & ($\num{6.6e-06}$) \\ 
        $^{34}$SO & $\num{2.6e-06}$ & $<\num{5.1e-06}$ & $<\num{1.2e-06}$ & $<\num{2.2e-06}$ & $<\num{3.2e-06}$ & $<\num{4.1e-06}$ & $<\num{1.2e-06}$ \\ 
        SO$_2$ & $\num{1.8e-05}$ & $<\num{1.9e-06}$ & $\num{8.7e-06}$ & $<\num{6.3e-06}$ & $<\num{9.4e-06}$ & $<\num{7.6e-06}$ & $<\num{3.0e-06}$ \\ 
        HCN & $\num{9.0e-06}$ & $<\num{4.4e-06}$ & $<\num{3.8e-06}$ & $<\num{6.0e-06}$ & $<\num{1.1e-05}$ & $<\num{8.8e-06}$ & $<\num{3.8e-06}$ \\ 
        HCO$^+$ & $\num{2.0e-05}$ & $\num{2.1e-05}$ & $\num{7.6e-06}$ & $<\num{3.8e-06}$ & $<\num{5.6e-06}$ & $<\num{5.0e-06}$ & $\num{5.5e-06}$ \\ 
        \hline
	\end{tabular}
	\tablefoot{Values which involve emission from distorted line profiles are written in parenthesis.}
\end{table*}

%%%%%%%%%%%%%%%%%%%%%%%%%%%%%%%%%%%%%%%%%%%%%%%%%%%%%%%%%%%%%%%%%%%%%%%%%%%%%%%%%%%%%%%%%%%%%%%%%
\clearpage
\clearpage

\section{Multilayer non-LTE radiative transfer model for H$_2$CO}\label{app:twolayers}

In this part of the Appendix we describe in detail the multicomponent line modeling of H$_2$CO toward IRAS\,16293--2422. The H$_2$CO temperatures and column densities derived with CLASS-weeds and RADEX in the main body of this work already describe very well most of the line profiles. Nevertheless, some of the H$_2$CO lines present self-absorption features. To model these lines properly, a more sophisticated model is needed. For that reason, we have used a non-LTE multilayer radiative transfer model to reproduce such line profiles. In this case, we have used the CASSIS\footnote{\url{http://cassis.irap.omp.eu}} software \citep{Vastel2015} to perform such models. We have also used the collision rates for the ortho- and para-H$_2$CO -- H$_2$ system derived by \citet{wiesenfeld2013rotational}.

In CASSIS, it is possible to create a model that consists of any number of physical components. Each of these physical components is defined by six parameters: Column density, temperature, linewidths (FWHM), LSR velocity, source size and H$_2$ volume density. CASSIS makes use of the Monte Carlo Markov Chain (MCMC) method \citep{Hastings1970} to explore the space of parameters and to find the best combination between them to reproduce the line profiles. This is done by means of a $\chi^2$ minimization.

Not all the emission peaks could be modeled, since some of them (E1, E2, W1) present strong outflow wings in the line profiles (see Fig.~\ref{fig:spec_E2}). In contrast, the sources IRAS\,16293--2422 A/B, 16293E, W2, and HE2 exhibit (approximately) Gaussian line profiles, although in some cases it is clear that there is contribution from the outflows. 

\subsection{IRAS\,16293--2422 A/B}
\begin{figure}
\centering
\setlength\tabcolsep{3.7pt}
\begin{tabular}{c c c}
\includegraphics[width=0.315\textwidth, trim= 0 0 0 0, clip]{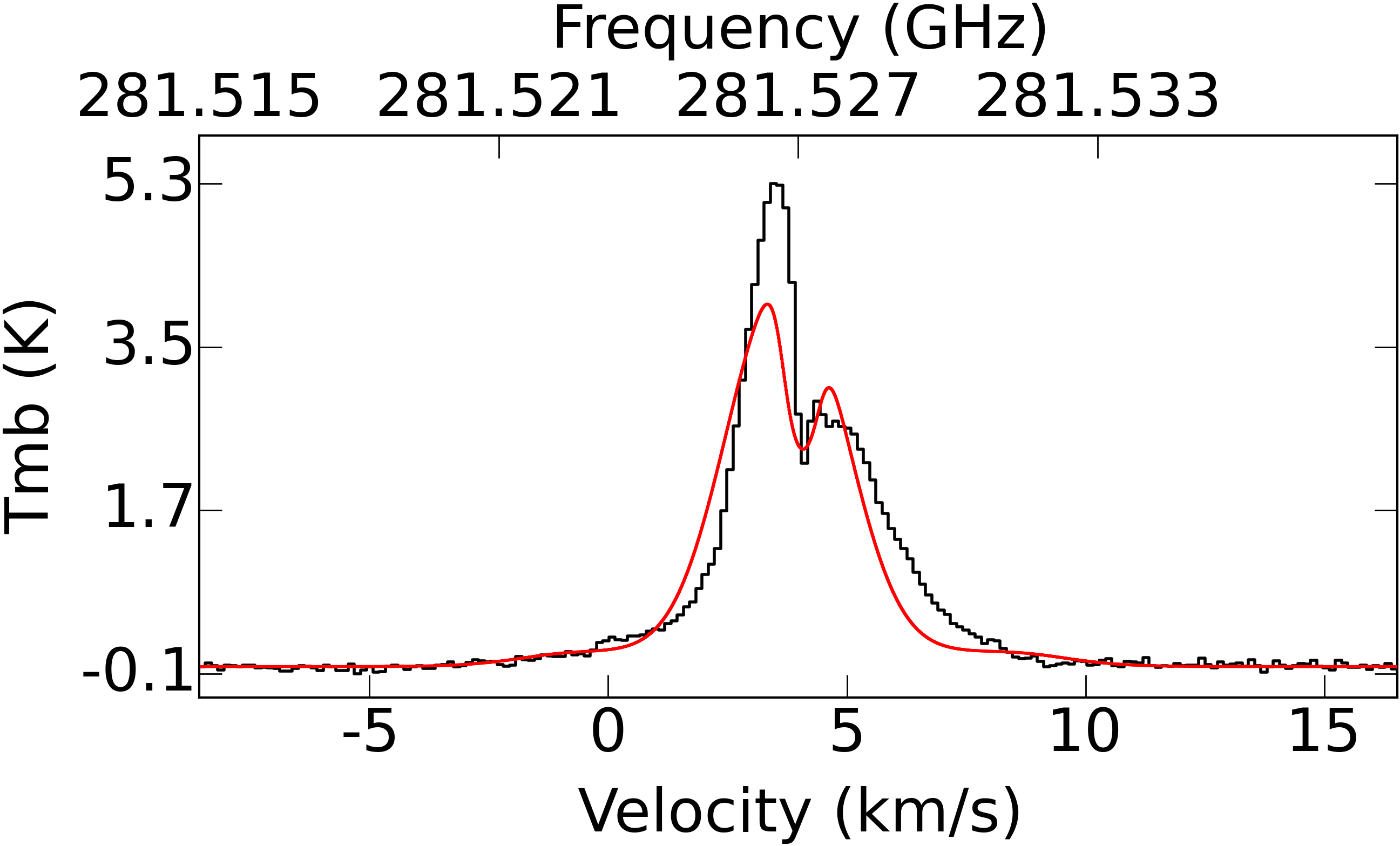} &\includegraphics[width=0.315\textwidth,trim = 0 0 0 0,clip]{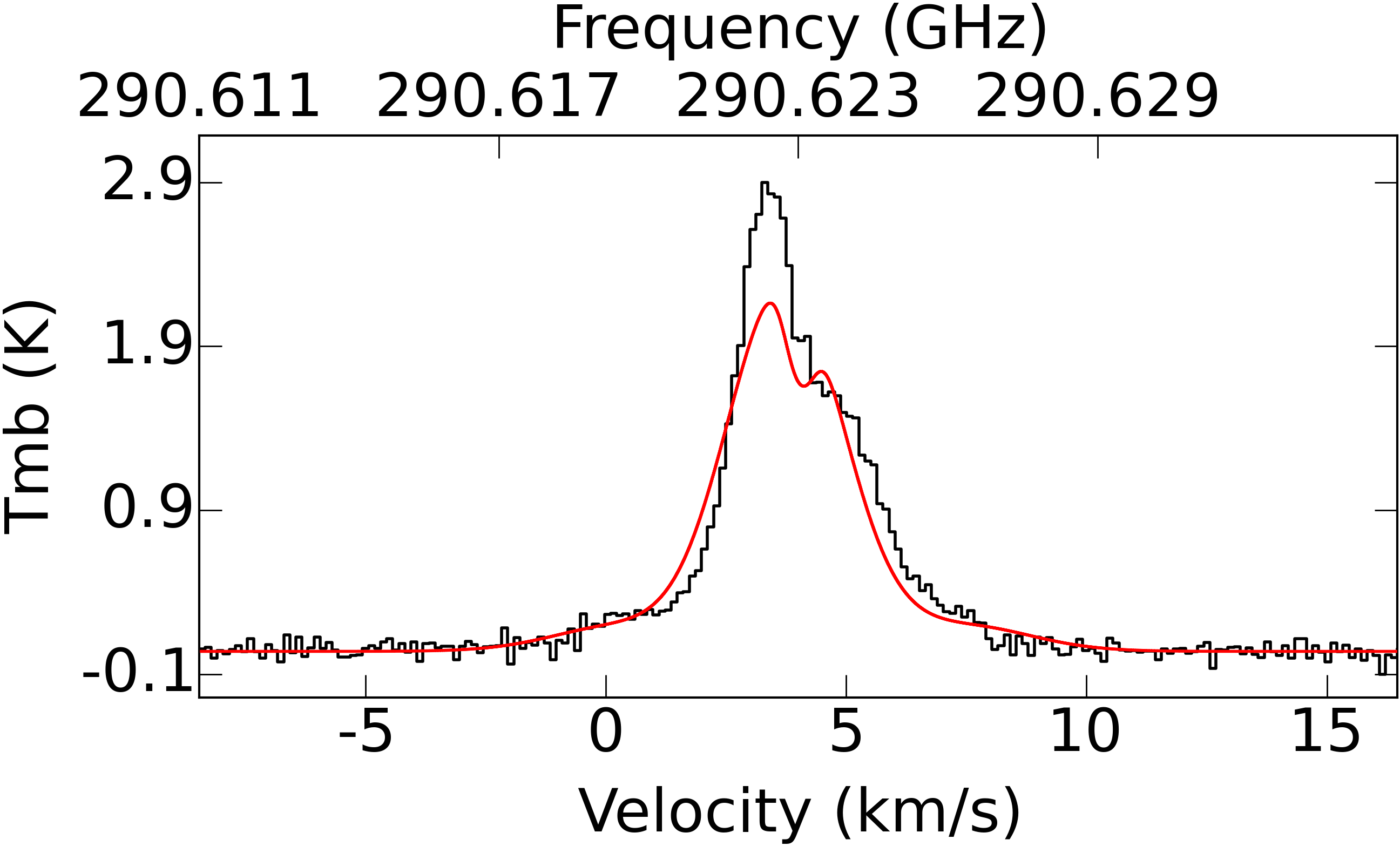}  &\includegraphics[width=0.315\textwidth,trim = 0 0 0 0,clip]{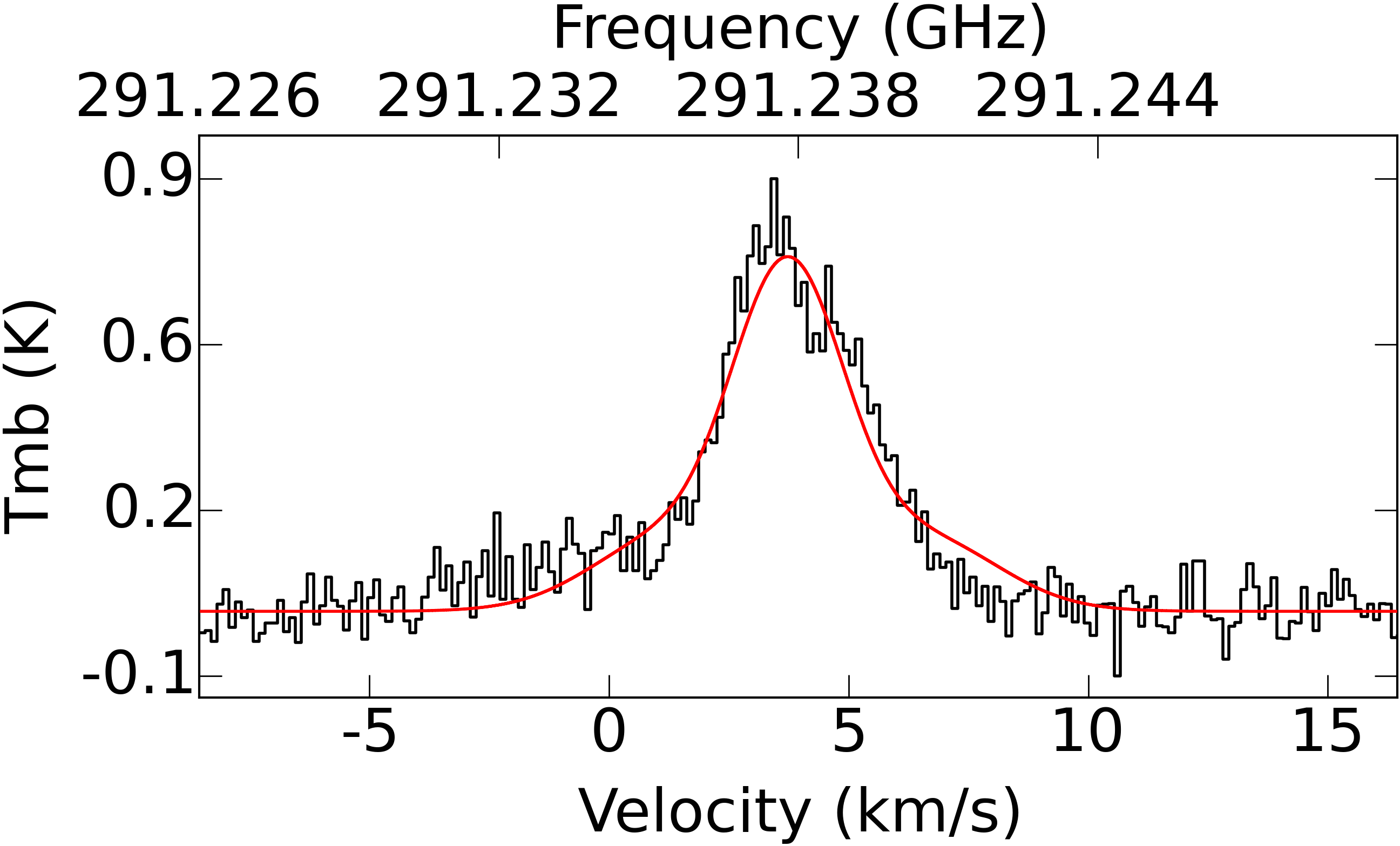}  \\

\includegraphics[width=0.315\textwidth, trim= 0 0 0 0, clip]{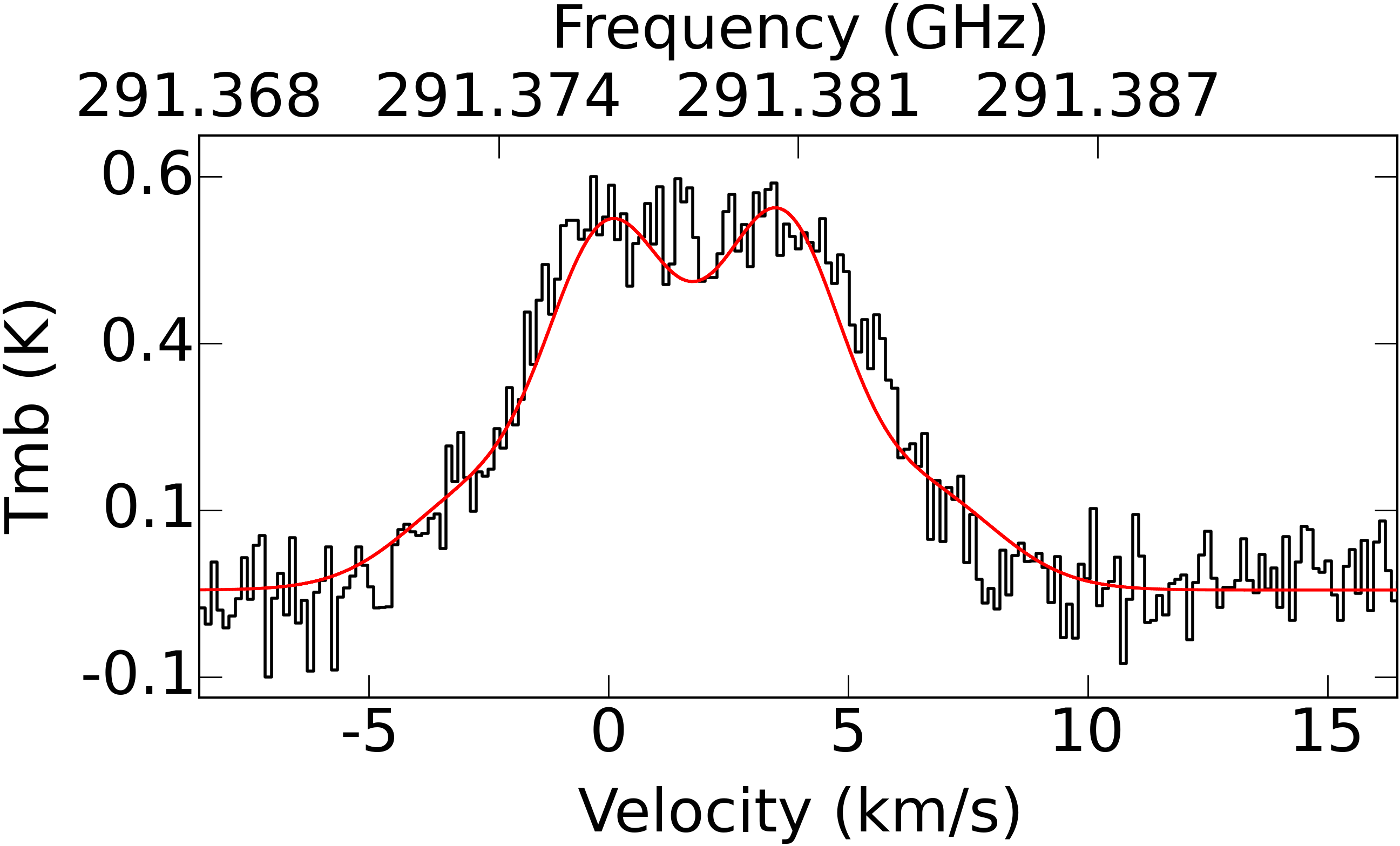} &\includegraphics[width=0.315\textwidth,trim = 0 0 0 0,clip]{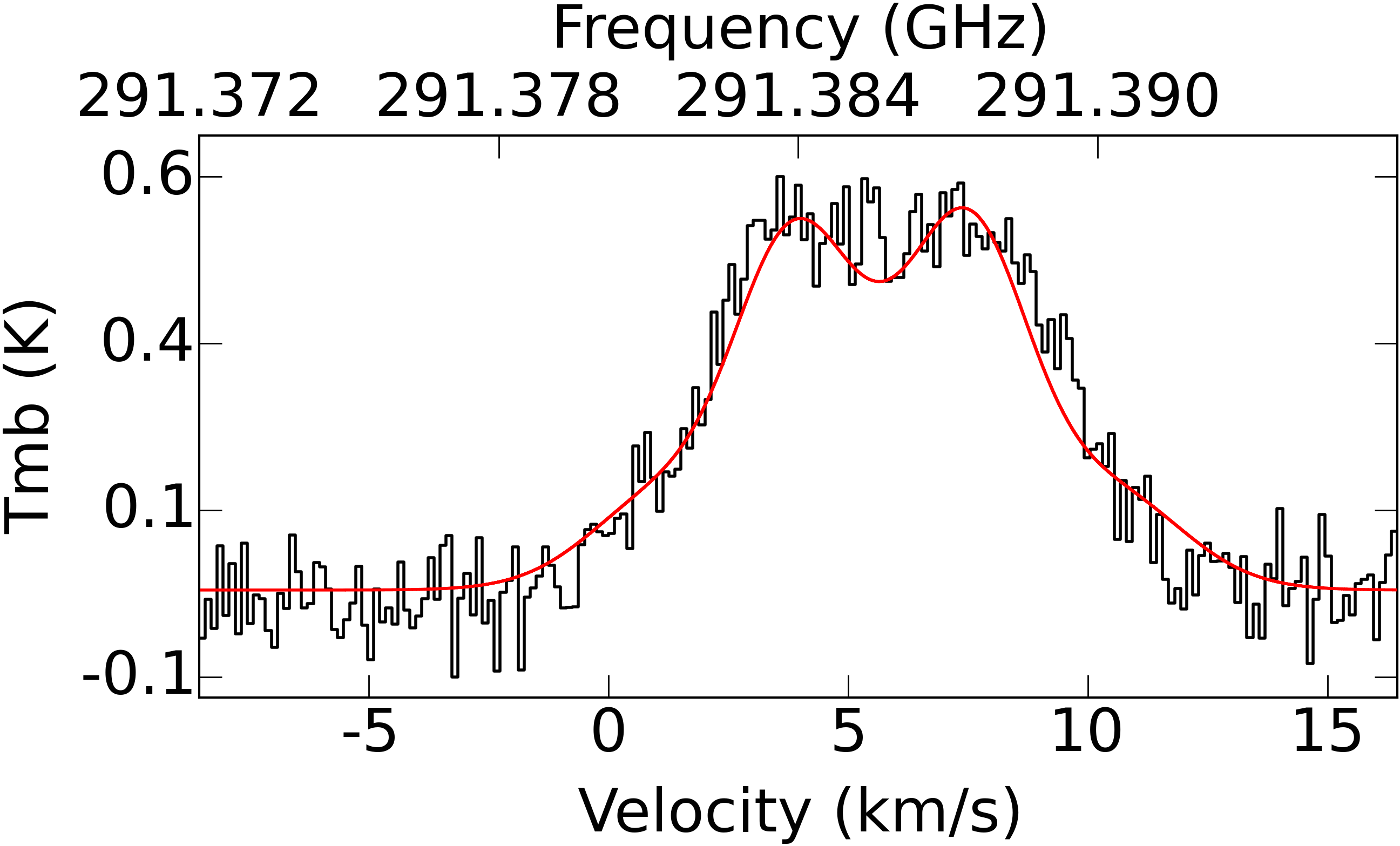}  &\includegraphics[width=0.315\textwidth,trim = 0 0 0 0,clip]{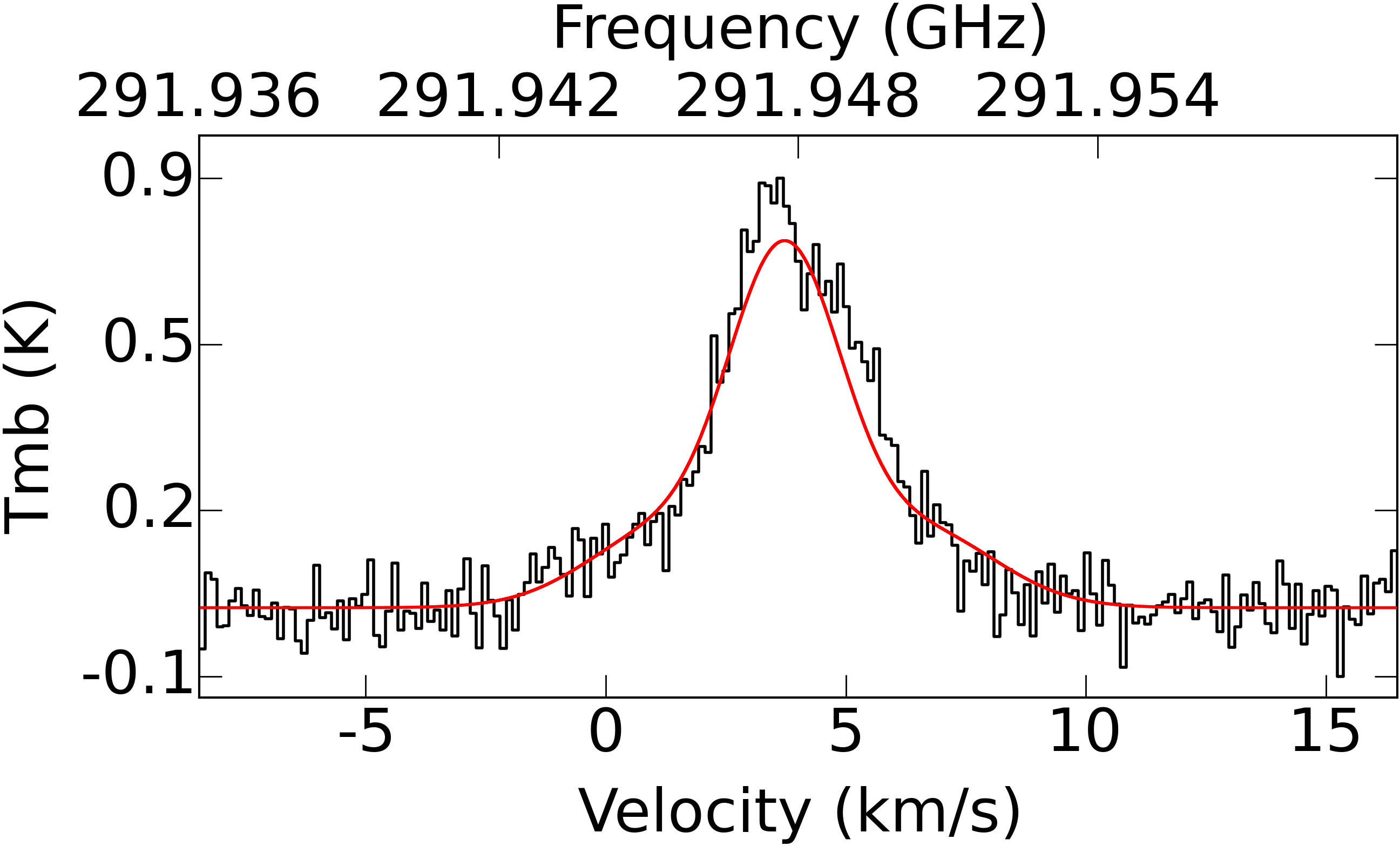}  \\

\includegraphics[width=0.315\textwidth, trim= 0 0 0 0, clip]{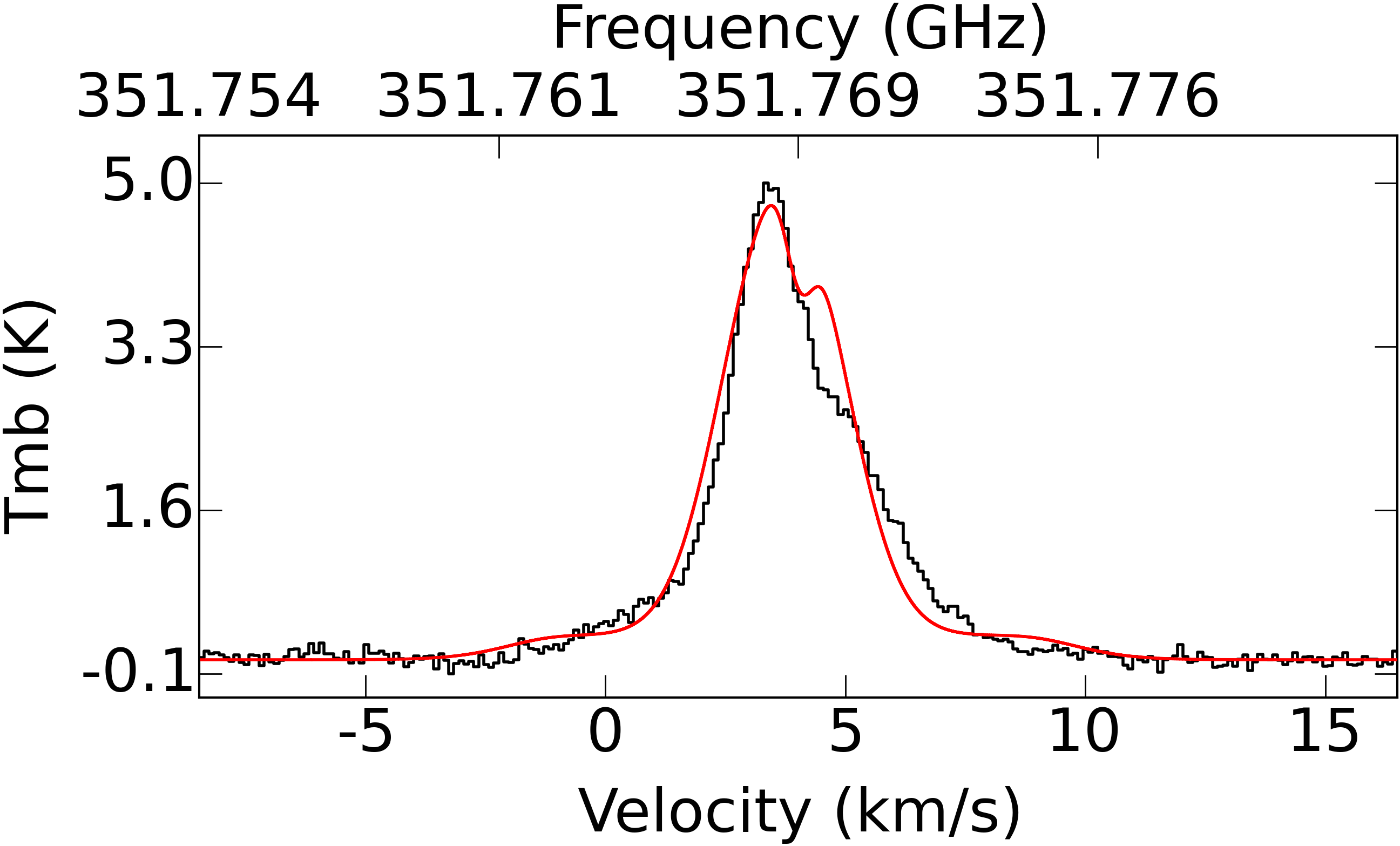} &\includegraphics[width=0.315\textwidth,trim = 0 0 0 0,clip]{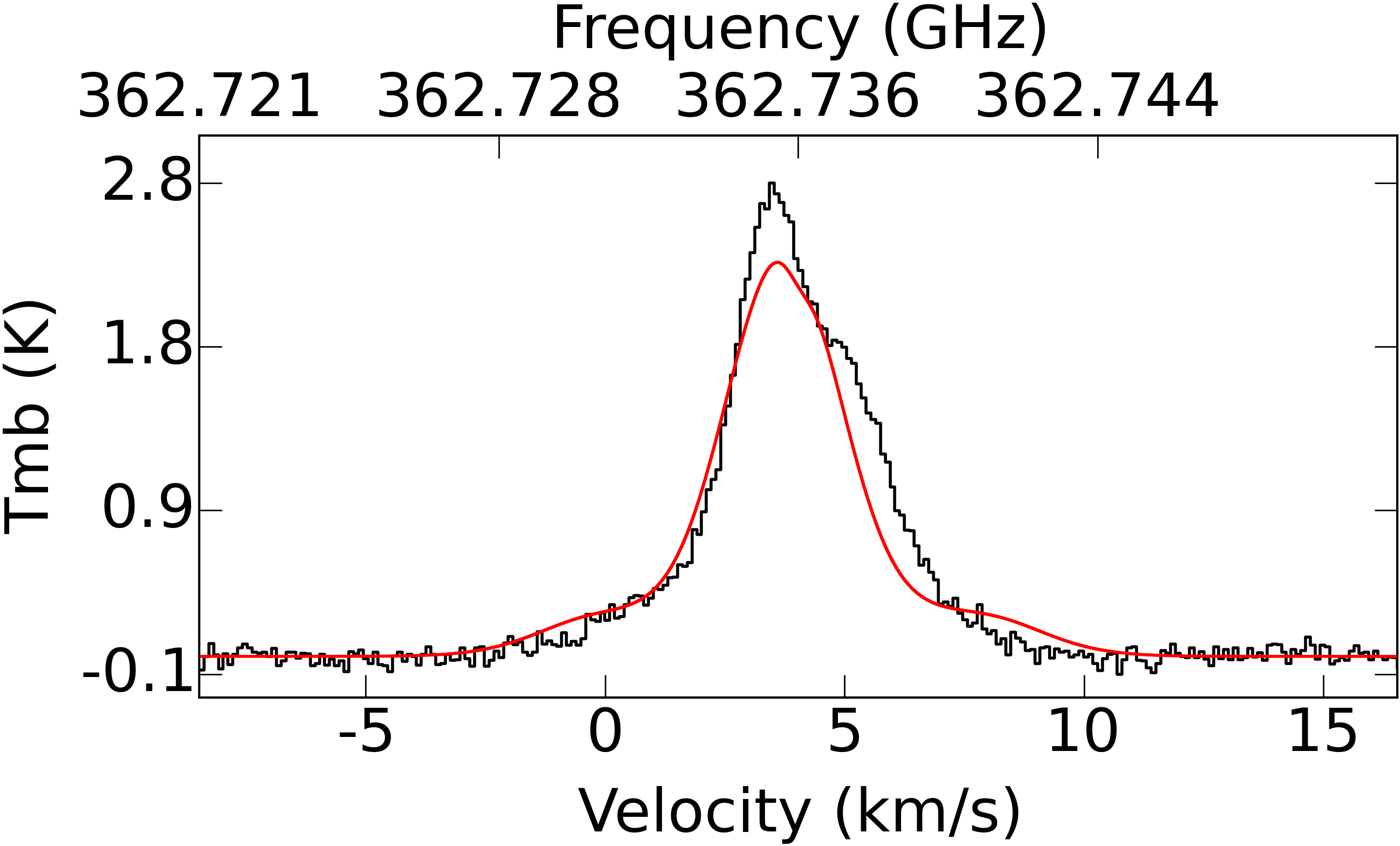}  &  \\

\end{tabular}
\caption{Best model of the line profiles at IRAS\,16293--2422 A/B. In black, we show the H$_2$CO lines observed toward IRAS\,16293--2422 A/B, while in red we show the non-LTE radiative transfer model. Three physical components were needed to reproduce the line profiles for this source. The parameters of each component are summarized in Table \ref{tab:cassis_ab}.}
\label{fig:cassis_ab}
\end{figure}

Some of the H$_2$CO lines (such as the $4_{1,4} - 3_{1, 3}$ transition) in this source present self-absorption features (see Fig.~\ref{fig:cassis_ab}). Indeed, a number of molecules detected toward this source have been reported to have self-absorption features. In some cases, this feature is associated with the presence of an absorbing foreground cloud at a velocity of 4.2\,km\,s$^{-1}$ \citep[e.g.,][]{coutens2012, bottinelli2014}. To reproduce such absorption in our model, we need an extended and cold component. For our full radiative transfer model, we have considered a total of three physical components. The first component is compact and warm, associated with the hot corino in this source. The second component is associated with the warm envelope, while the third component correspond to the outermost extended and cold layer. The presence of these 3 physical components in IRAS\,16293--2422 have been also identified in other molecules such as HNCO \citep{hernandez-gomez2019-2}. 
Some of the parameters were fixed during the modeling. First, the size of the hot corino has been fixed to 1$''$ as the individual sources A and B have angular sizes below this limit. Since our observations cannot resolve between A and B, we assume that the bulk of the emission associated with the hot corino comes from a single warm source with this size. The temperature of the hot corino has been fixed to 90\,K motivated by the results from the H$_2$CO line ratios presented in the main body of this work. 
For the warm envelope, we assume that the emission fills the beam, and therefore its size was fixed to 20$''$. The temperature was also fixed to 50\,K as preliminar models show that this values does not change significantly during the modeling. 
Since the extended component is assumed to be very extended, we fixed its size to 100$''$. We have noted that is not easy to reproduce the absorption observed in the $4_{1,4} - 3_{1, 3}$ transition, unless a very low temperature is considered. When having this parameter as free during the modeling, the absorption is easily destroyed during the modeling when increasing the temperature. For this reason, we have fixed the temperature of this cold component to 10\,K.
To determine the best choice for the velocity of each component, we have made some Gaussian fits to the line profiles and choose the best values based on them. In particular, we need to fix the velocities of all sources to reproduce the self-absorption feature correctly.

Finally, we have used the H$_2$ density radial profile determined by \citet{crimier2010} to choose a suitable H$_2$ average value for each of the components depending on their sizes. The best parameters derived with CASSIS-RADEX are shown in Table \ref{tab:cassis_ab}, while the produced synthetic spectrum is shown in Fig.~\ref{fig:cassis_ab}. As it can be seen, the non-LTE radiative transfer model is in good agreement with the observations.
Interestingly, the velocity of the absorption is seen at 4.05\,km\,s$^{-1}$, closer to the systemic velocity of the source. 

A very useful utility of CASSIS-RADEX is that both forms of ortho-para H$_2$CO can be minimized at the same time during the modeling by using an extra parameter called iso, which corresponds to the ortho/para ratio. As a first approach, we fixed the ortho/para ratio to 3.0. However, we have noticed that changing this value between $1-3$ does not affect the final result. 

It is worth mentioning that \citet{Ceccarelli2000} studied the H$_2$CO emission in IRAS\,16293--2422 and proposed a two components model that consisted of an evaporation region that corresponds to the hot corino in this source and a cold outer region. Their derived H$_2$CO abundances in these components are lower by a factor of about $2-4$ compared with our model. These differences might be due to the different radial H$_2$ density profiles used for the modeling. In addition, we have use different number of components and different component sizes, which can also affect the computed abundance values.  

\subsection{16293E}
\begin{figure*}[htbp]
\centering
\setlength\tabcolsep{3.7pt}
\begin{tabular}{c c c}
\includegraphics[width=0.315\textwidth, trim= 0 0 0 0, clip]{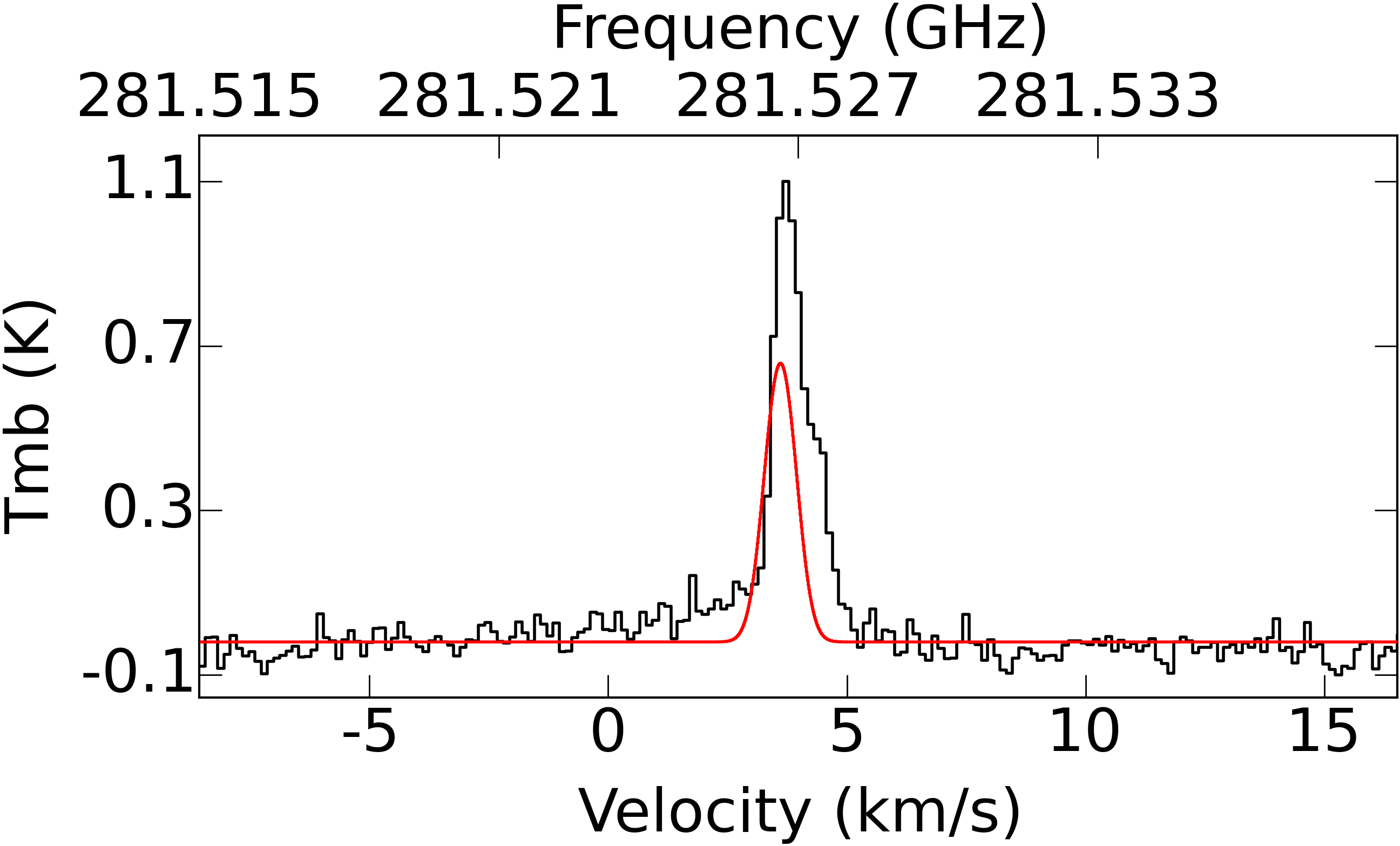} &\includegraphics[width=0.315\textwidth,trim = 0 0 0 0,clip]{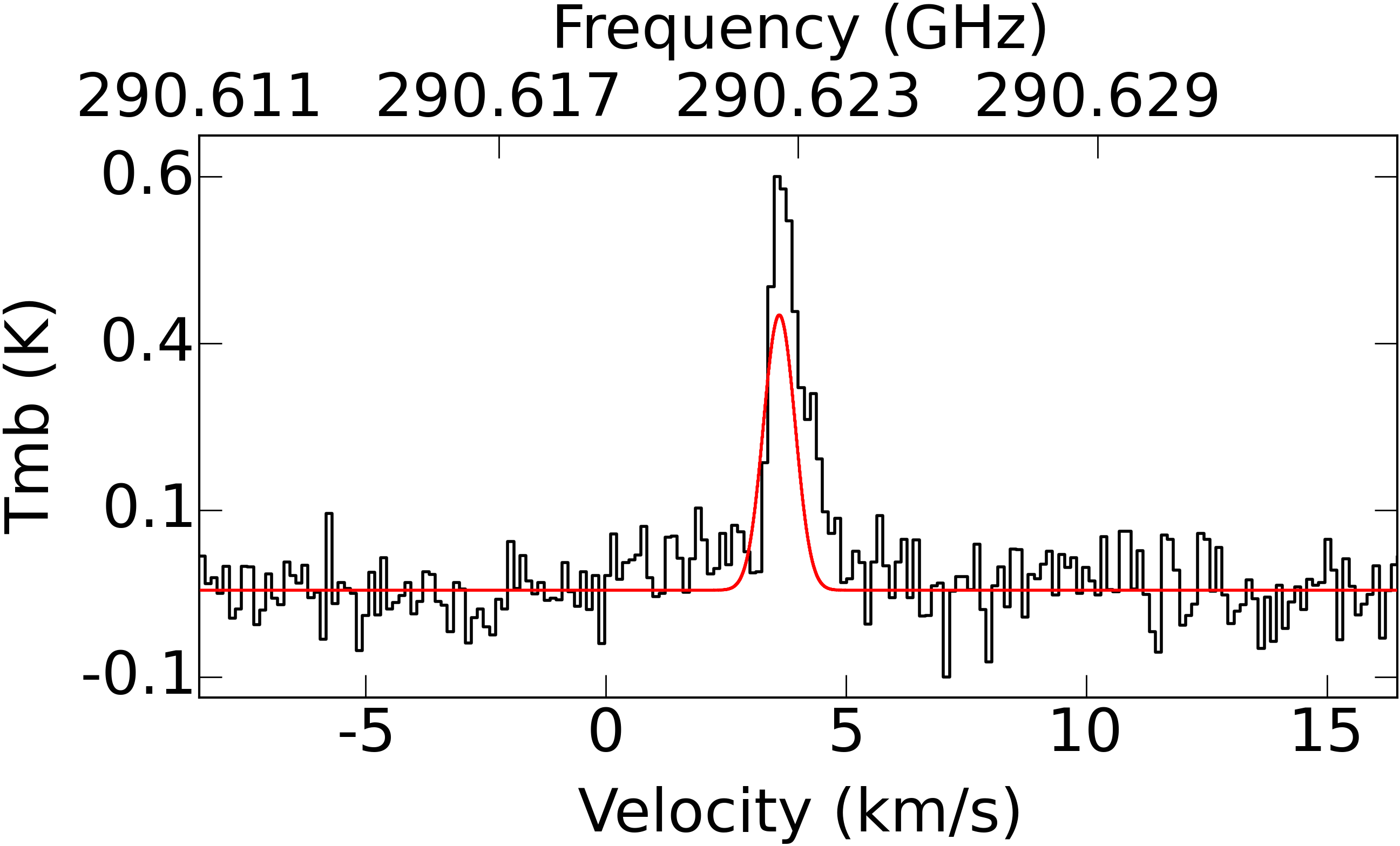}  &\includegraphics[width=0.315\textwidth,trim = 0 0 0 0,clip]{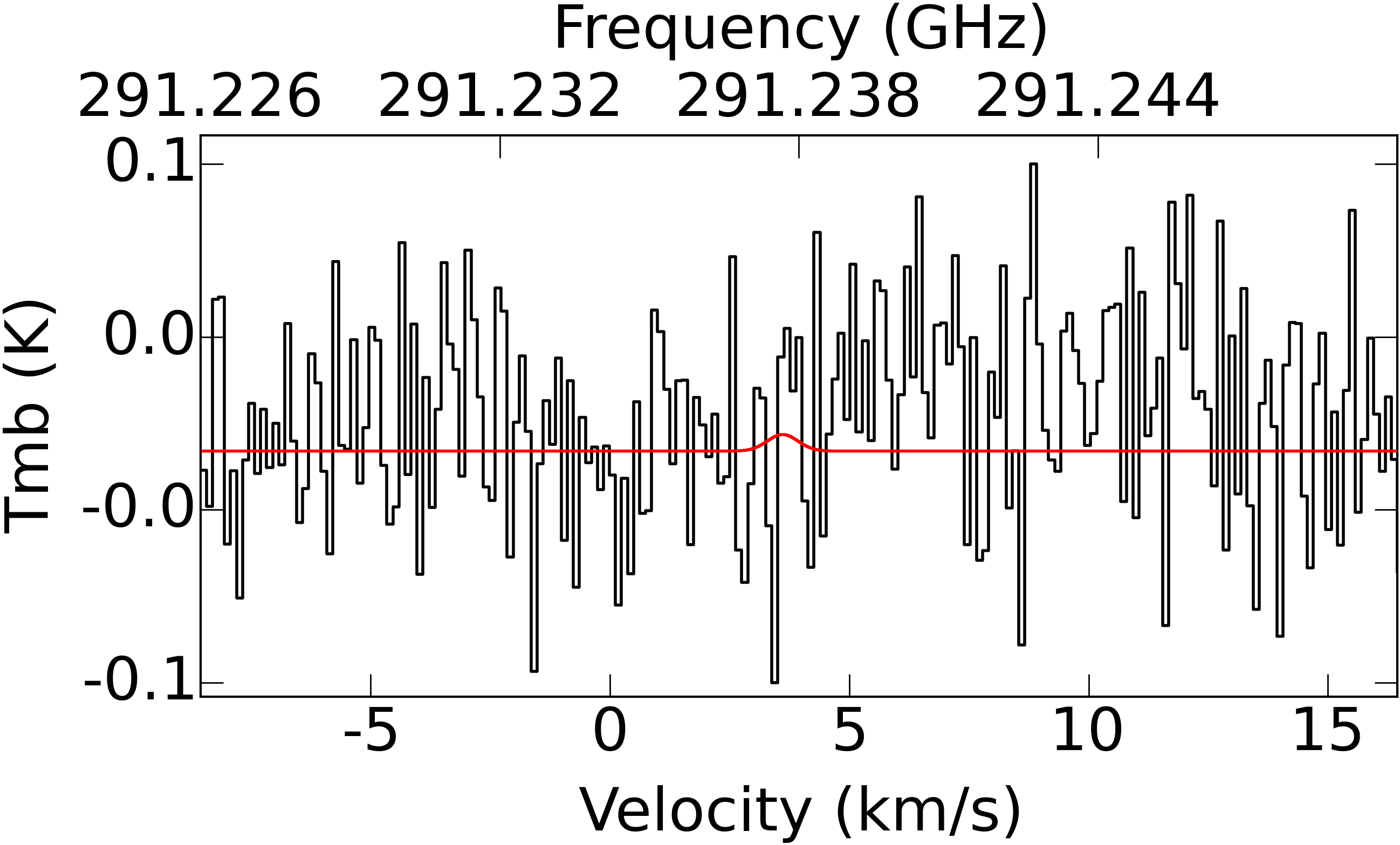}  \\

\includegraphics[width=0.315\textwidth, trim= 0 0 0 0, clip]{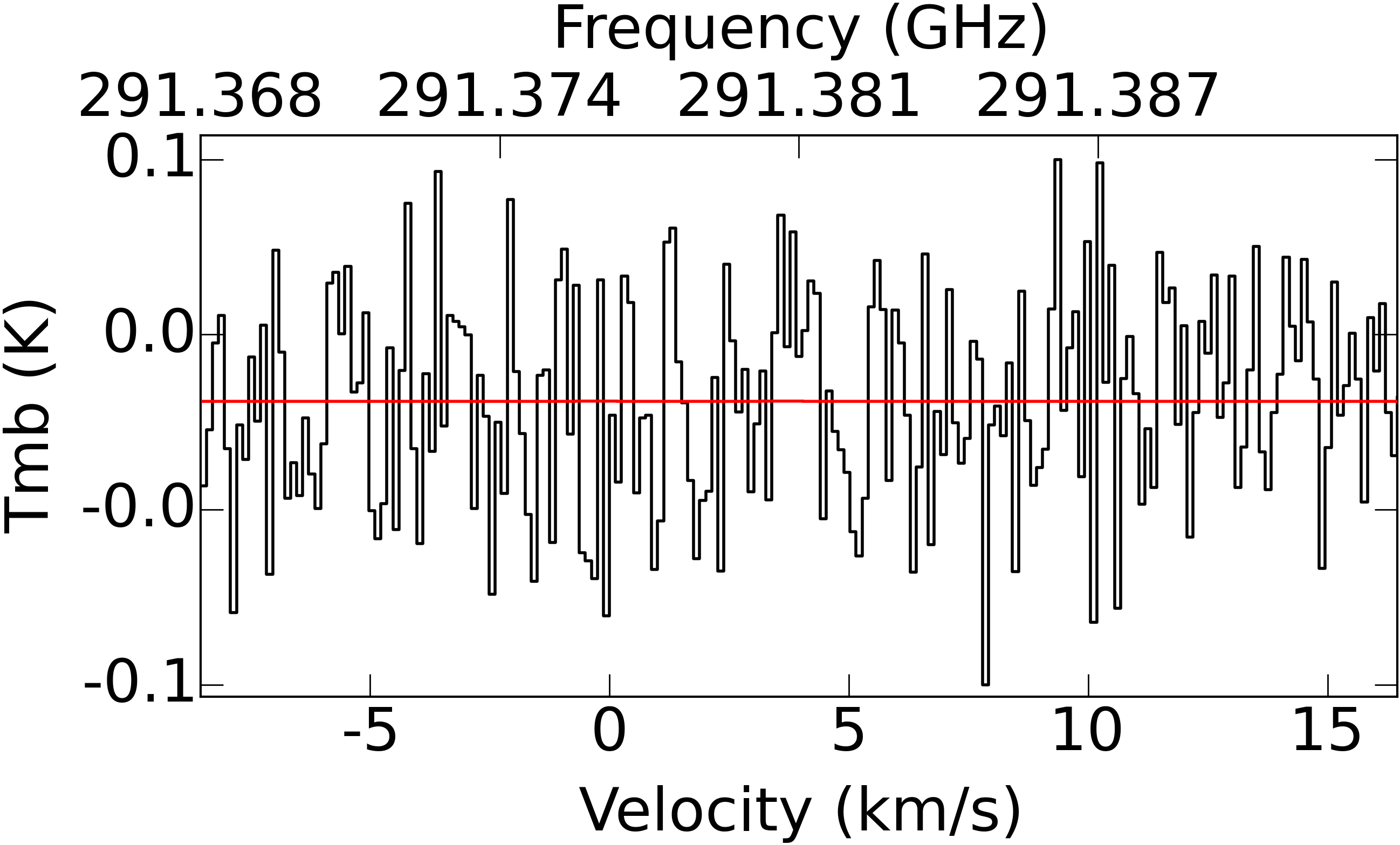} &\includegraphics[width=0.315\textwidth,trim = 0 0 0 0,clip]{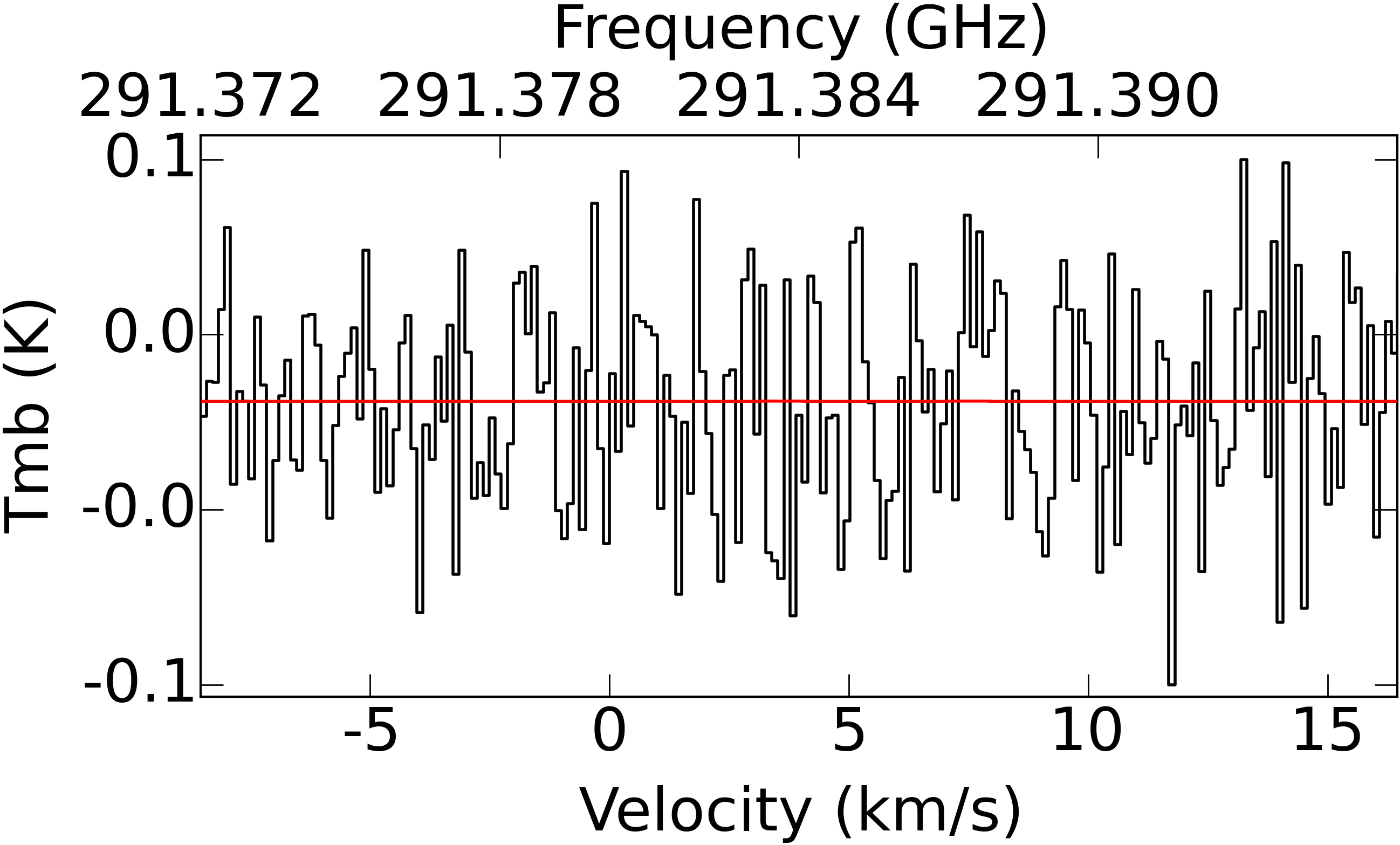}  &\includegraphics[width=0.315\textwidth,trim = 0 0 0 0,clip]{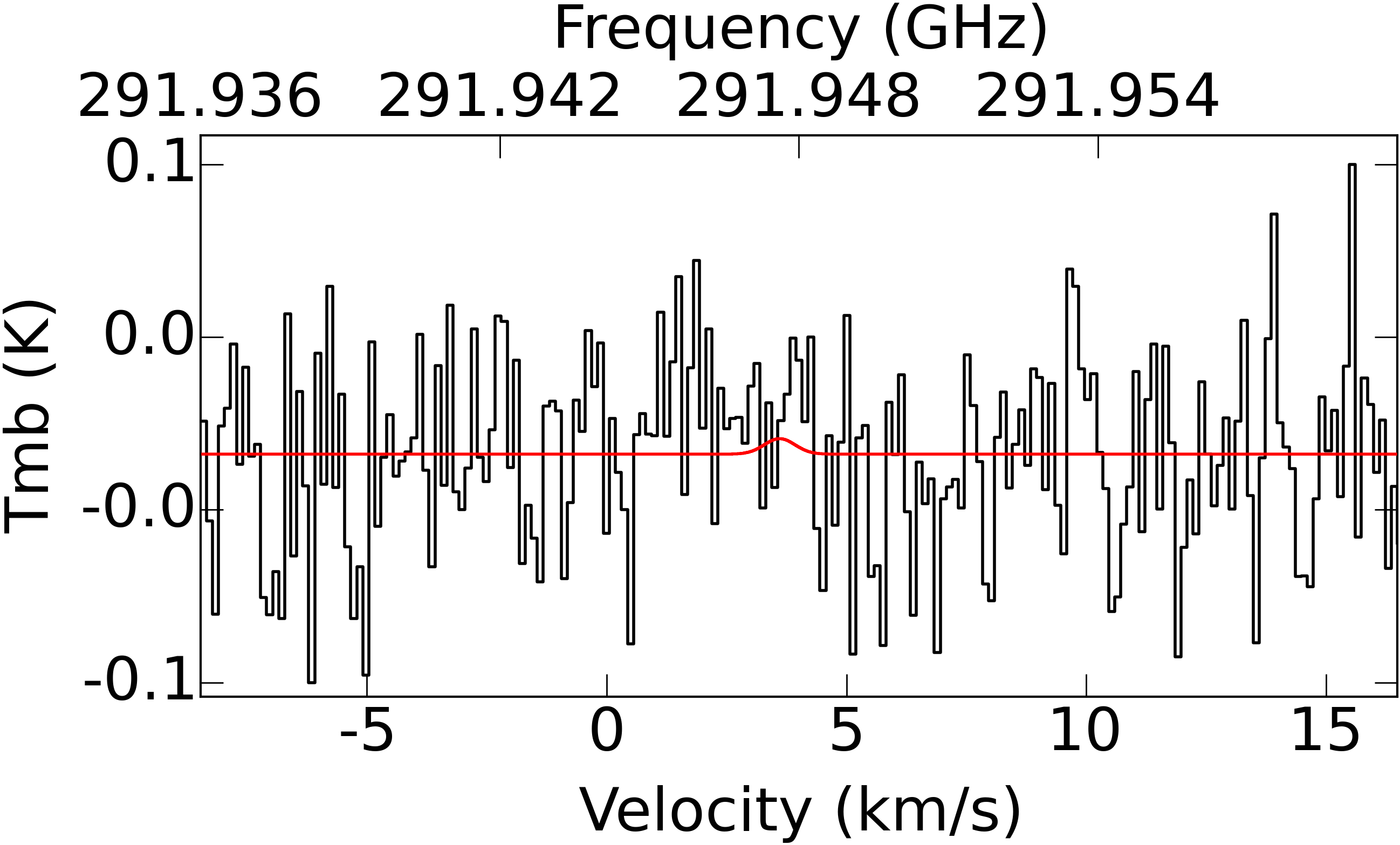}  \\

\includegraphics[width=0.315\textwidth, trim= 0 0 0 0, clip]{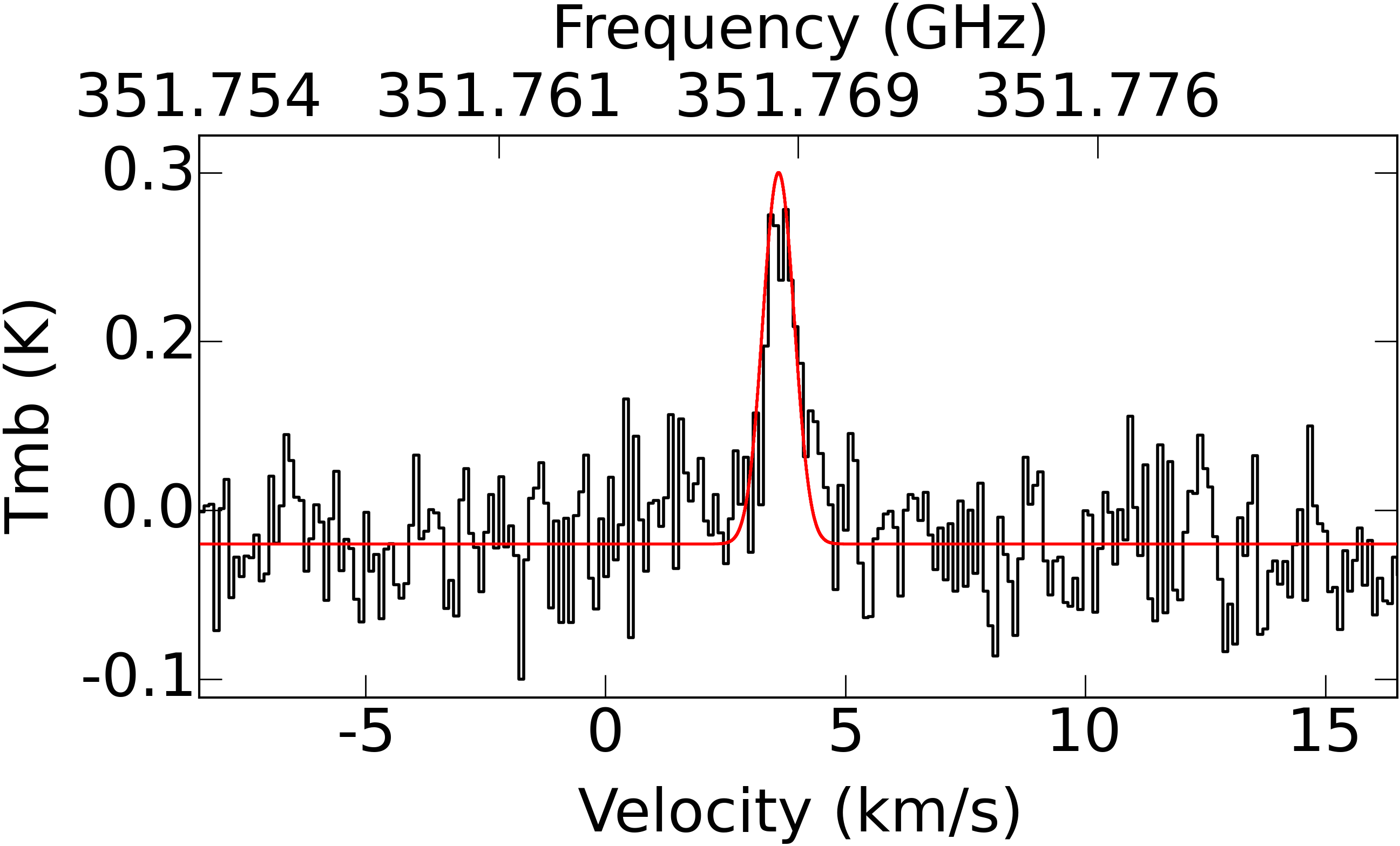} &\includegraphics[width=0.315\textwidth,trim = 0 0 0 0,clip]{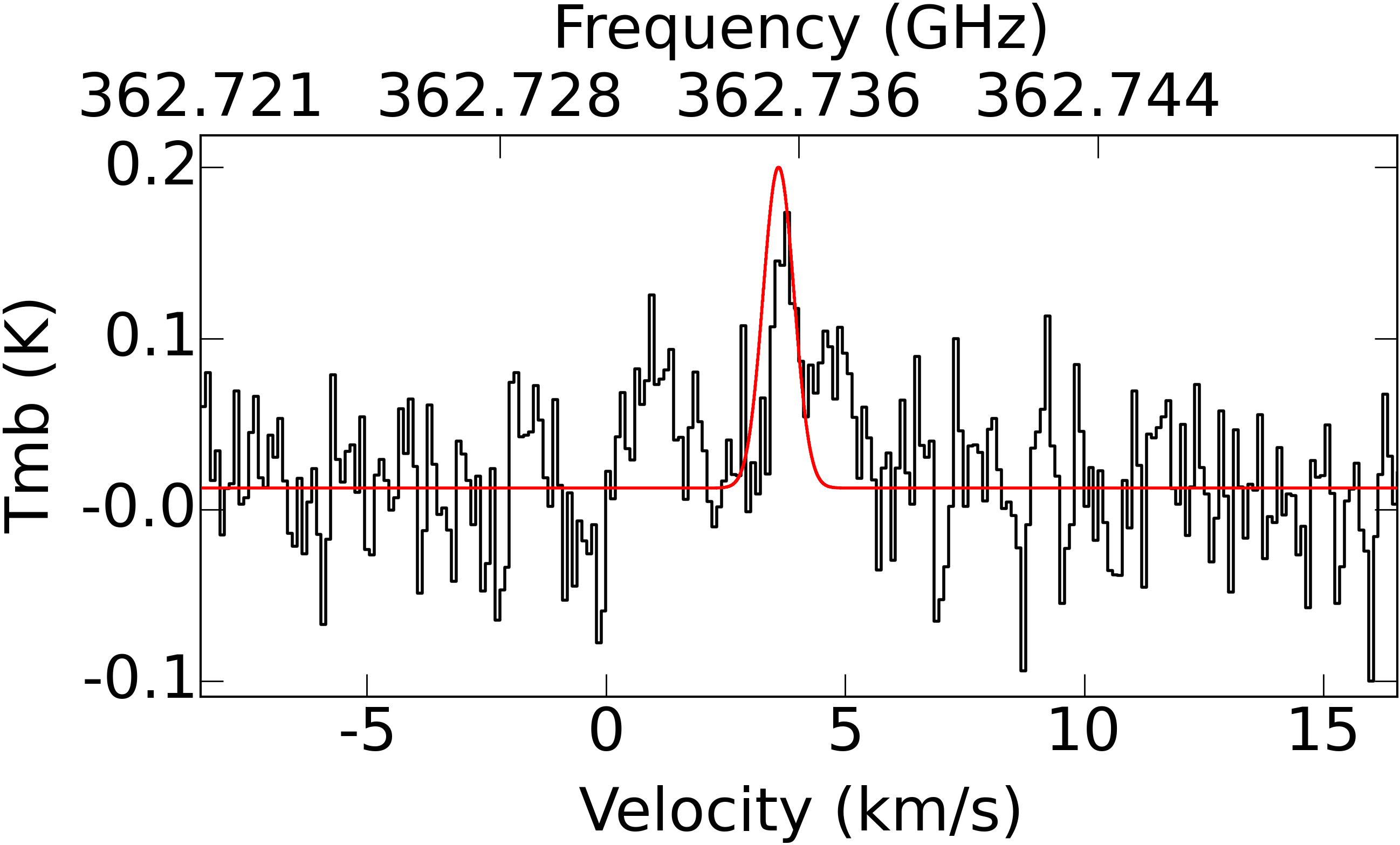}  &  \\

\end{tabular}
\caption{Best model of the line profiles at 16293E. In black, we show the H$_2$CO lines observed toward 16293E, while in red we show the non-LTE radiative transfer model. One physical component was needed to reproduce the line profiles for this source. The derived parameters from the model are summarized in Table \ref{tab:cassis_e}.}
\label{fig:cassis_e}
\end{figure*}

The prestellar core 16293E exhibits narrow Gaussian H$_2$CO line profiles. A total of four formaldehyde transitions are detected toward this source (see Fig.~\ref{fig:cassis_e}). Since these line profiles show a simpler morphology compared with IRAS\,16293--2422 A/B, only one physical component was used during the modeling. Previous works have estimated the temperature in this source (12\,K for the gas \citep{lis2002}; 16-20\,K for the dust \citep{stark2004}; 11\,K for the dust \citep{bacmann2016}). Also, the H$_2$ density has been recently estimated in different ranges: $3.3\times 10^7$\,cm$^{-3}$ \citep{pattle2015}; (1.1$-$1.9)$\times 10^7$\,cm$^{-3}$ \citep{lis2016}; $1.4\times 10^7$\,cm$^{-3}$ \citep{bacmann2016}. For our model, we have left the temperature to be a free parameter varying between $12-20$\,K. The H$_2$ density was varying also between $1\times 10^6 - 1\times 10^8$\,cm$^{-3}$. The ortho/para ratio was fixed to 3.0. We verified that varying this ratio did not change the final results. The best parameters obtained with CASSI-RADEX are shown in Table \ref{tab:cassis_e} and Fig.~\ref{fig:cassis_e}, respectively. A temperature of about 12\,K and an approximate H$_2$ density of $3 \times 10^7$\,cm$^{-3}$ resulted from our model. We note that although these values give a better reduced $\chi^2$ value during the minimization, there is some emission lacking for the $4_{1,4} - 3_{1,3}$ and $4_{0,4} - 3_{0,3}$ lines in our model.
\clearpage
\onecolumn
%AB-----------------------------------------------------------------------------------
\vfill
   
\begin{table}
\centering
\caption{Best physical parameters derived from the $\chi^2$ minimization for IRAS\,16293--2422 A/B.}
\label{tab:cassis_ab}
\scalebox{1.0}{
\begin{tabular}{lccccccc}

\hline
Component 	    & N 						        & T$_{\rm kin}$ 	& FWHM 		    & V$_{\rm LSR}$	& Size 		    & $n$(H$_2$)        & ortho/para	\\
		 	    &(cm$^{-2}$)				        & (K)			    & (km s$^{-1}$) & (km s$^{-1}$)	& ($''$)		& (cm$^{-3}$)	    &   \\
\hline
Hot corino 	    &$(6.95 \pm 0.03 )\times 10^{16}$	& 90$^{*}$          & 5.0$^{*}$         & 3.88$^*$		& 1.0$^{*}$	    & 1.0$\times 10^{8*}$ & 3.0$^{*}$	\\
Warm envelope 	&$(8.55\pm 0.12)\times 10^{13}$	    & 50$^{*}$          & $2.64\pm 0.03$	& 3.72$^{*}$	& 20.0$^{*}$    & 1.0$\times 10^{6*}$ & 3.0$^{*}$ \\
Cold envelope 	&$(4.11\pm 0.01)\times 10^{13}$	    & 10$^*$    	    & 0.65$^{*}$        & 4.05$^{*}$	& 100.0$^{*}$	& 1.0$\times 10^{5*}$ & 3.0$^{*}$\\
\hline
\end{tabular}
}
\tablefoot{The values marked with a $^*$ symbol were fixed during the modeling. The derived column densities correspond to the ortho form of H$_2$CO. The column density of the para-form can be computed by multiplying by the ortho/para ratio $=3$.}
\end{table}

\vfill
%AB-----------------------------------------------------------------------------------
\clearpage
%E-----------------------------------------------------------------------------------
\vfill
   
\begin{table*}[htbp]
\centering
\caption{Best physical parameters derived from the $\chi^2$ minimization for 16293E.}
\label{tab:cassis_e}
\scalebox{1.0}{
\begin{tabular}{cccccccc}

\hline
Component 	    & N 						   & T$_{\rm kin}$ 	& FWHM 		    & V$_{\rm LSR}$	        & Size 		    & $n$(H$_2$)        & ortho/para	\\
		 	    &(cm$^{-2}$)				   & (K)			 & (km s$^{-1}$) & (km s$^{-1}$)	    & ($''$)		& (cm$^{-3}$)	    &   \\
\hline
1 	    &$(8.29 \pm 0.8 )\times 10^{12}$	   & $12.03\pm 0.05$   & $0.75\pm 0.02$	& $3.59\pm 0.01$	& $20.04\pm 0.4$	& 2.98$\times 10^{7}$ & 3.0$^*$	\\

  \\
\hline
\end{tabular}
}
\tablefoot{The values marked with a $^*$ symbol were fixed during the modeling.}
\end{table*}

\vspace{5cm}
%E-----------------------------------------------------------------------------------

\clearpage

\subsection{W2}
\begin{figure*}[htbp]
\centering
\setlength\tabcolsep{3.7pt}
\begin{tabular}{c c c}
\includegraphics[width=0.315\textwidth, trim= 0 0 0 0, clip]{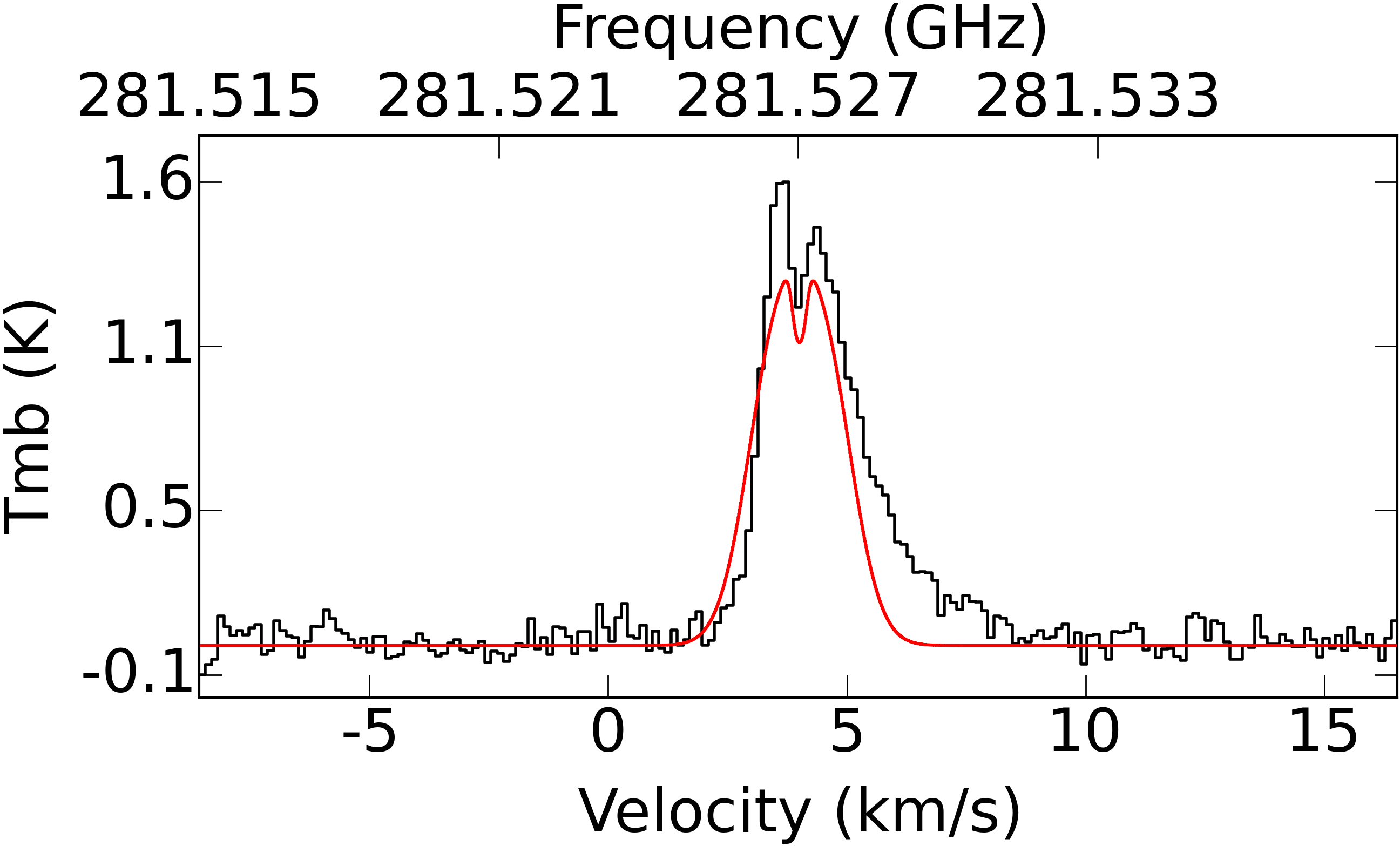} &\includegraphics[width=0.315\textwidth,trim = 0 0 0 0,clip]{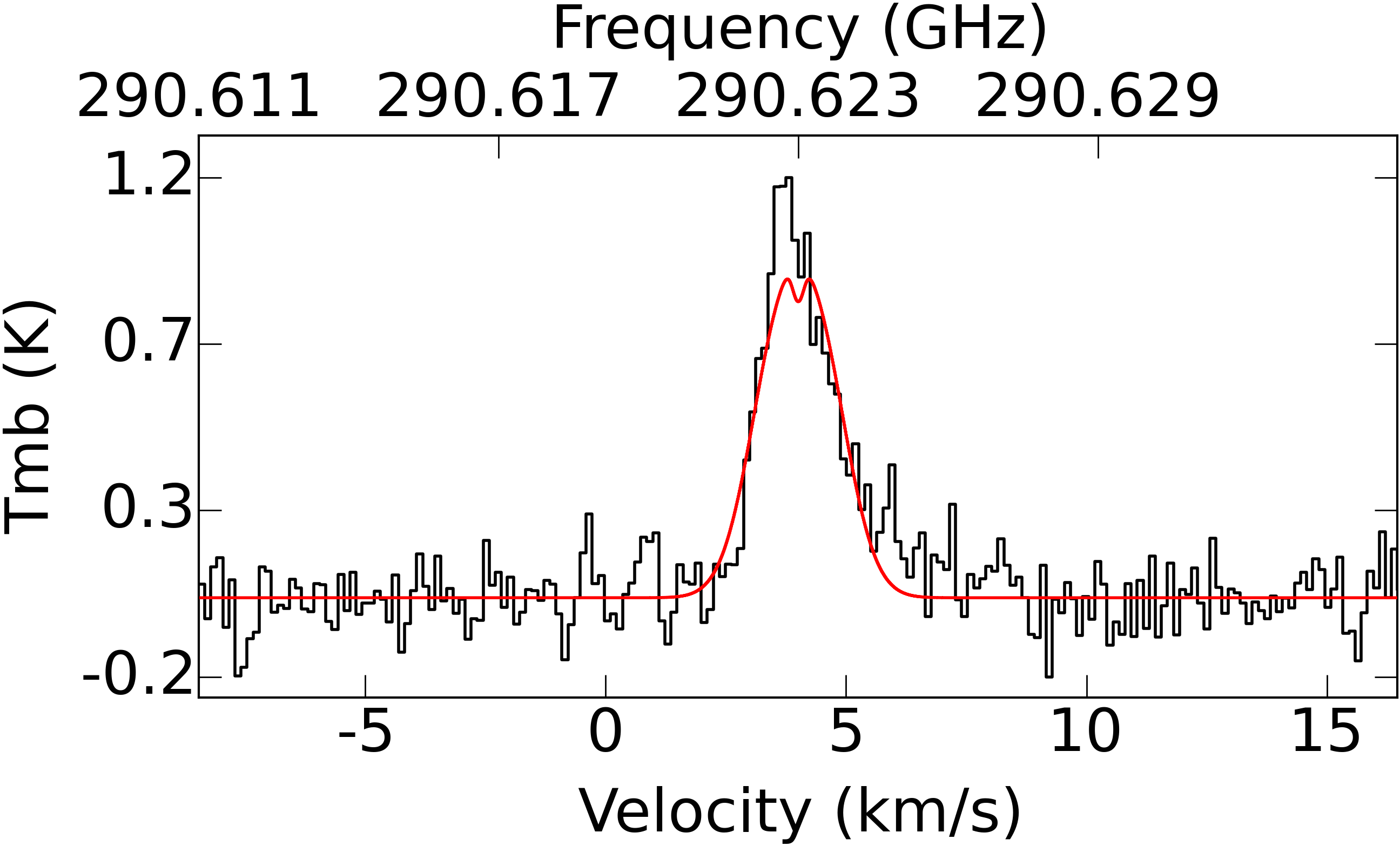}  &\includegraphics[width=0.315\textwidth,trim = 0 0 0 0,clip]{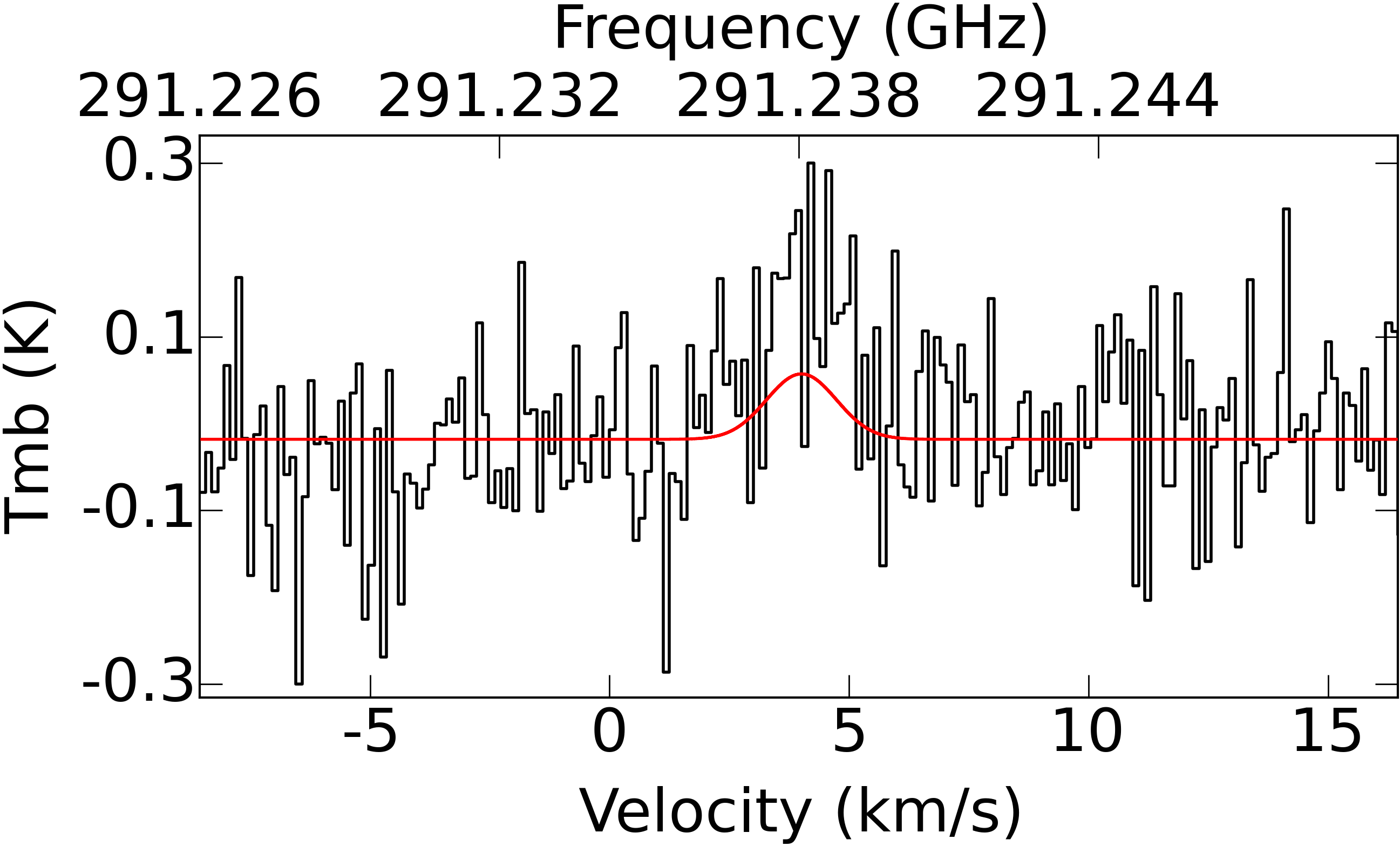}  \\

\includegraphics[width=0.315\textwidth, trim= 0 0 0 0, clip]{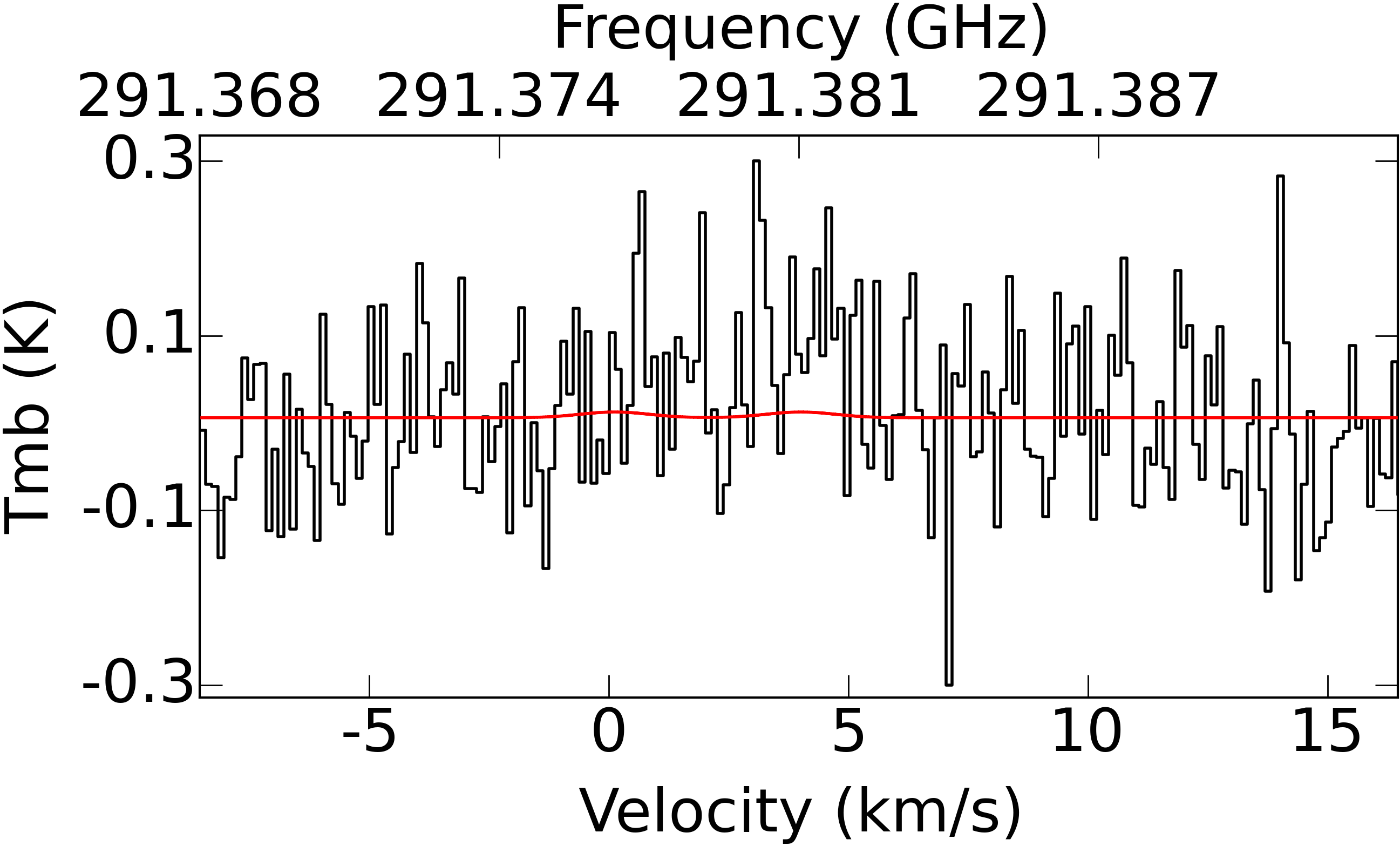} &\includegraphics[width=0.315\textwidth,trim = 0 0 0 0,clip]{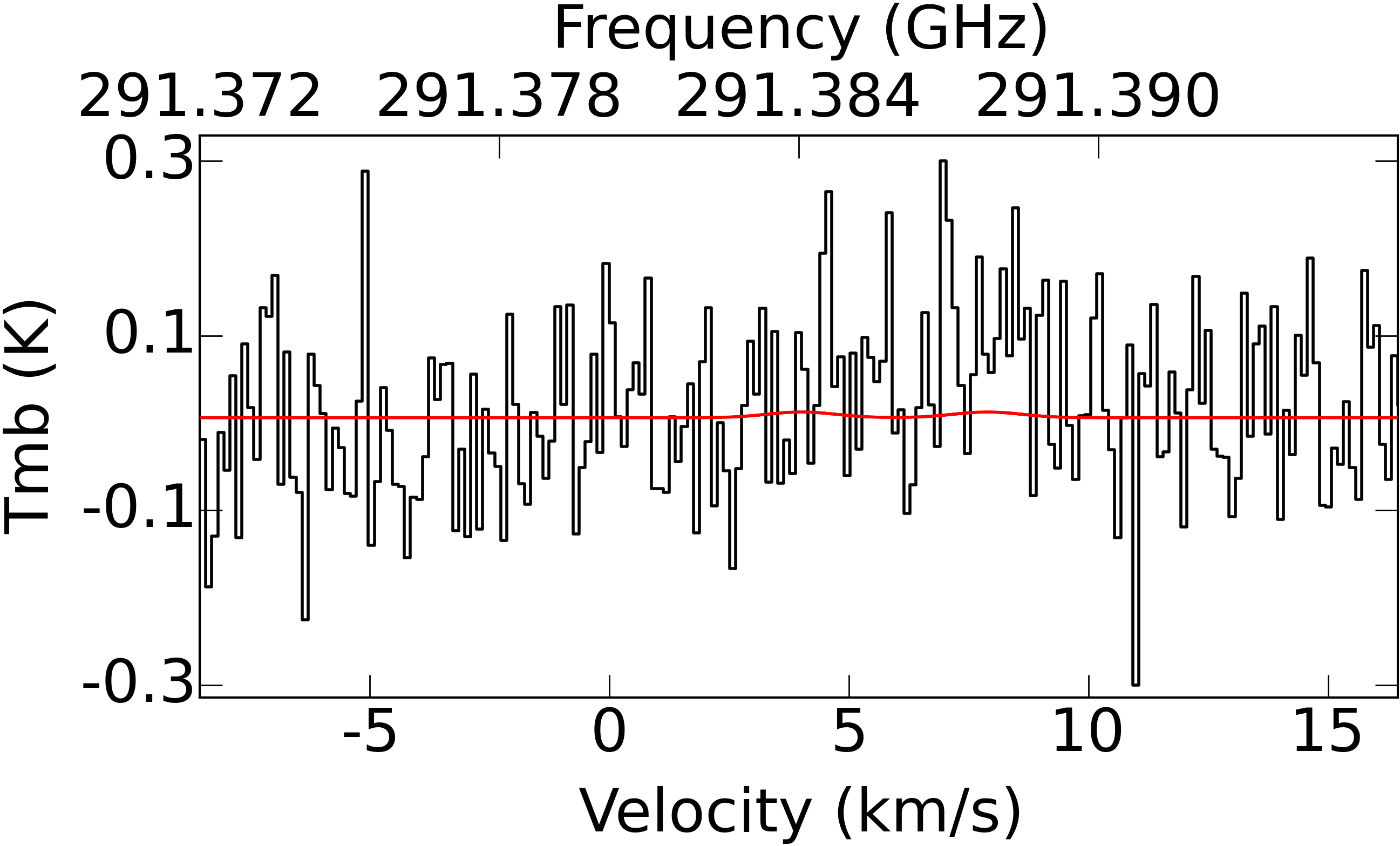}  &\includegraphics[width=0.315\textwidth,trim = 0 0 0 0,clip]{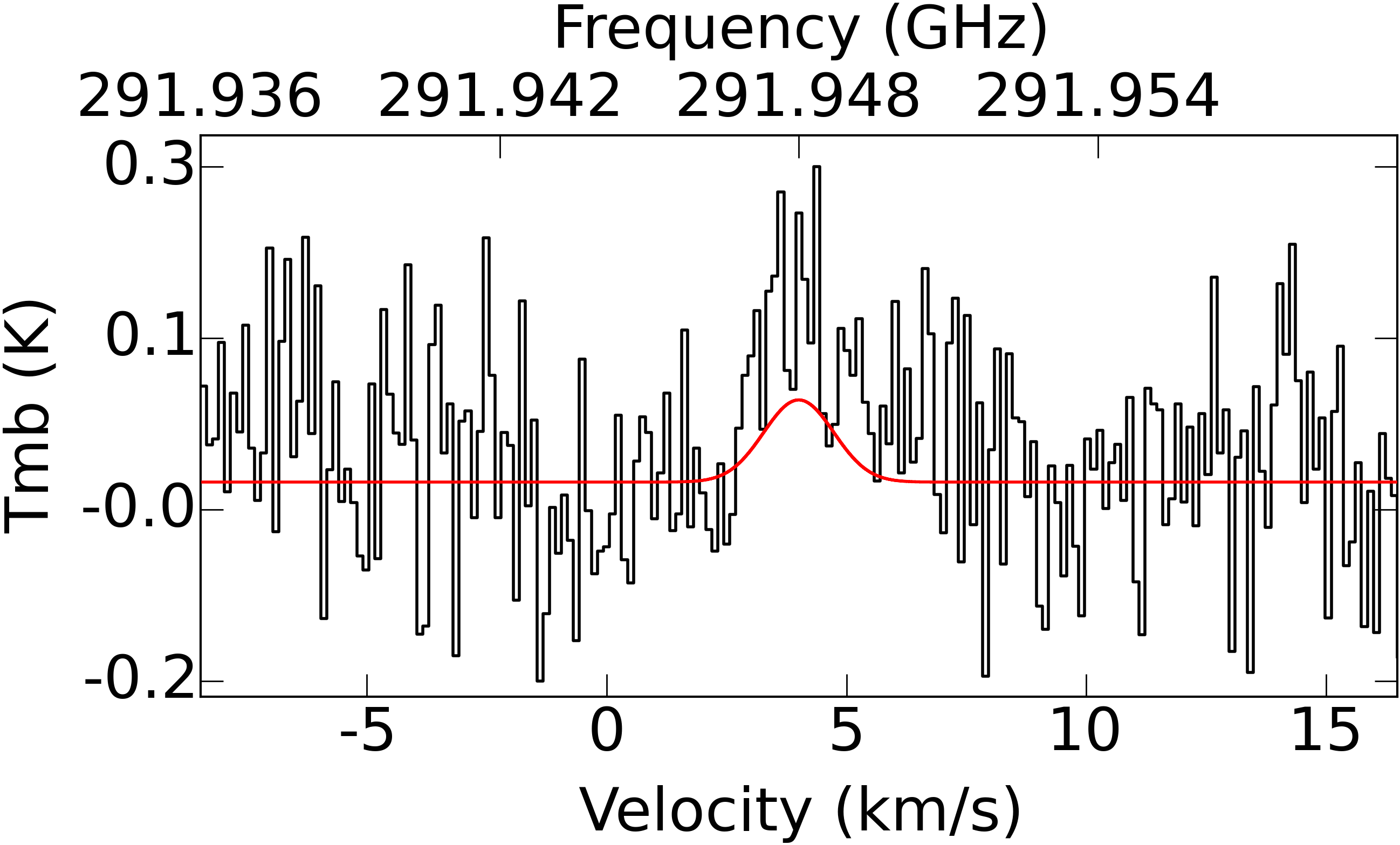}  \\

\includegraphics[width=0.315\textwidth, trim= 0 0 0 0, clip]{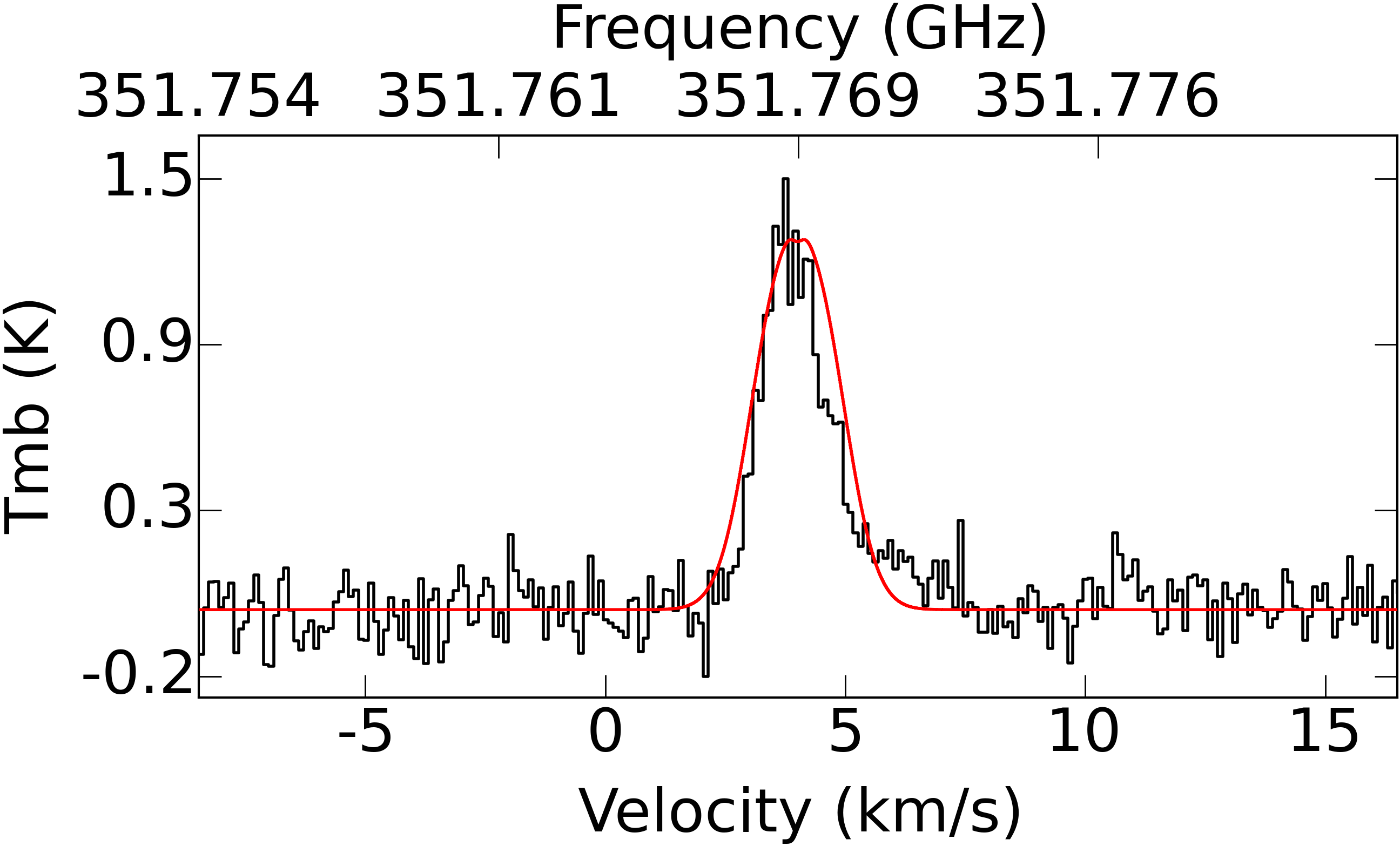} &\includegraphics[width=0.315\textwidth,trim = 0 0 0 0,clip]{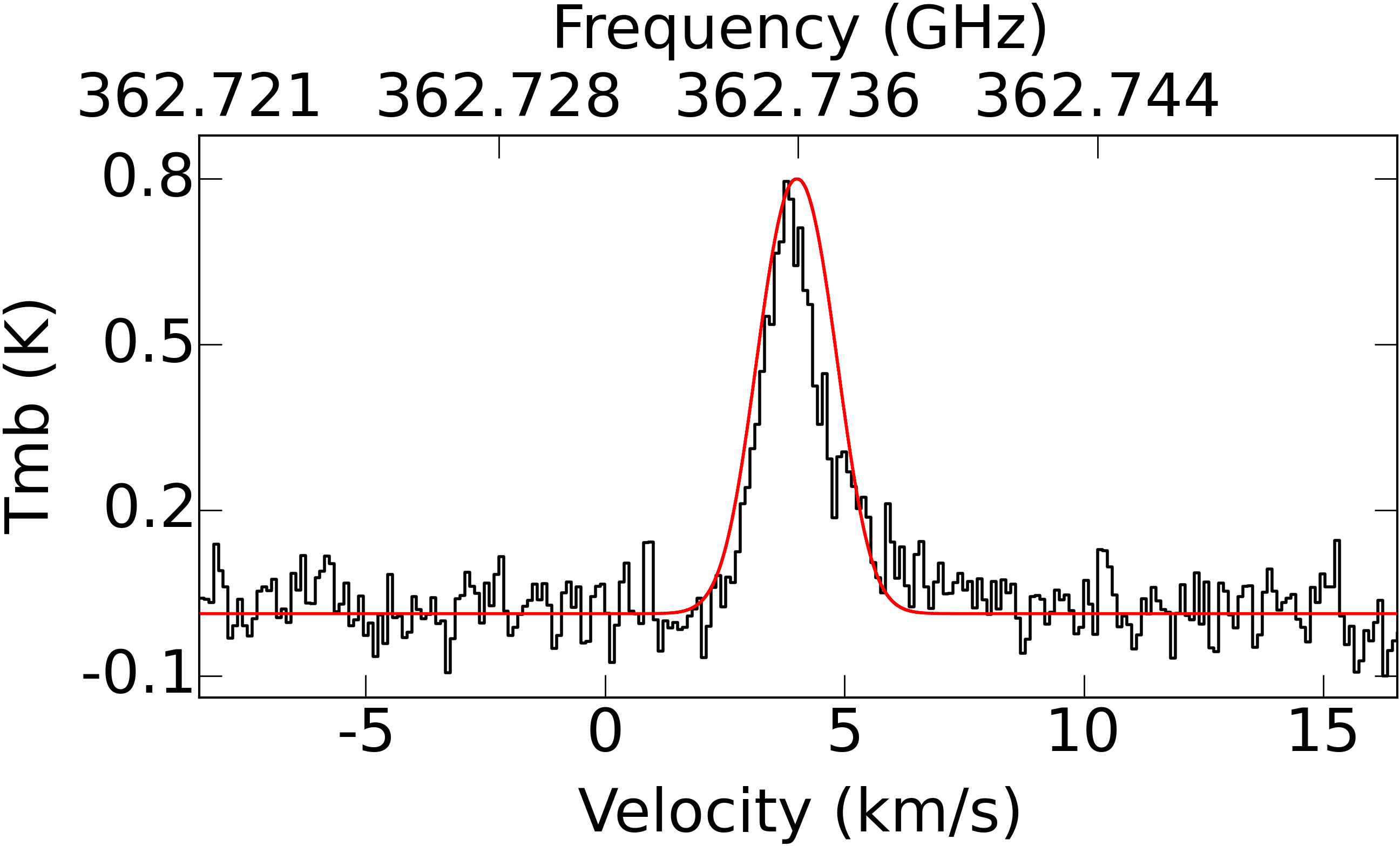}  &  \\

\end{tabular}
\caption{Best model of the line profiles at W2. In black, we show the H$_2$CO lines observed toward W2, while in red we show the non-LTE radiative transfer model. Two physical components were needed to reproduce the line profiles for this source. The parameters of each component are summarized in Table \ref{tab:cassis_w2}.}
\label{fig:cassis_w2}
\end{figure*}

\begin{multicols}{2}
\noindent
For the emission peak W2, four H$_2$CO transitions are detected in the data. In particular, the line at 281\,GHz ($4_{1,4} - 3_{1, 3}$) presents self-absorption features. In this case two physical components are need to reproduce the data. It is worth mentioning that some of these lines are clearly contaminated by the outflows. Nevertheless, the radiative transfer model can satisfactorily reproduce most of the line profiles. The results from such modeling are presented in Table \ref{tab:cassis_w2} and Fig.~\ref{fig:cassis_w2} respectively. Note that in the case of the self-absorption, some emission is lacking, even when the absorption is well reproduced. Compared with the temperatures derived from the H$_2$CO line ratios, we need much smaller values. Such discrepancy might be due to the degeneracy of our model. This is because we have seen that different combinations of the parameters can give similar results, although in most of these cases the self-absorption is destroyed. Therefore they are not representative of the physical conditions in this emission peak that seems to be associated with a cold dust source. While we require an extended and cold component as in the case of IRAS\,16293--2422, lower values for the H$_2$ density in this component are needed.  
\end{multicols}
%W2-----------------------------------------------------------------------------------
\begin{table*}[htbp]
\centering
\caption{Best physical parameters derived from the $\chi^2$ minimization for source W2.}
\label{tab:cassis_w2}
\scalebox{1.0}{
\begin{tabular}{cccccccc}

\hline
Component 	    & N 						   & T$_{\rm kin}$ 	& FWHM 		    & V$_{\rm LSR}$	        & Size 		    & $n$(H$_2$)        & ortho/para	\\
		 	    &(cm$^{-2}$)				   & (K)			 & (km s$^{-1}$) & (km s$^{-1}$)	    & ($''$)		& (cm$^{-3}$)	    &   \\
\hline
1 	    &$(1.12 \pm 0.05 )\times 10^{14}$	   & $20.0\pm 0.3$   & $1.7\pm 0.09$	& $4.0^*$      	& $9.0\pm 0.2$	& 1$\times 10^{6*}$ & 3.0$^*$	\\
2 	    &$(2.67 \pm 0.04 )\times 10^{12}$	   & $10.0\pm 1.3$   & $0.25\pm 0.02$	& $4.0^*$      	& $17\pm 0.4$	& 1$\times 10^{4*}$ & 3.0$^*$	\\
\hline
\end{tabular}
}
\tablefoot{The values marked with a $^*$ symbol were fixed during the modeling.}
\end{table*}

%W2-----------------------------------------------------------------------------------

\clearpage
\subsection{HE2}
\begin{figure*}[htbp]
\centering
\setlength\tabcolsep{3.7pt}
\begin{tabular}{c c c}
\includegraphics[width=0.315\textwidth, trim= 0 0 0 0, clip]{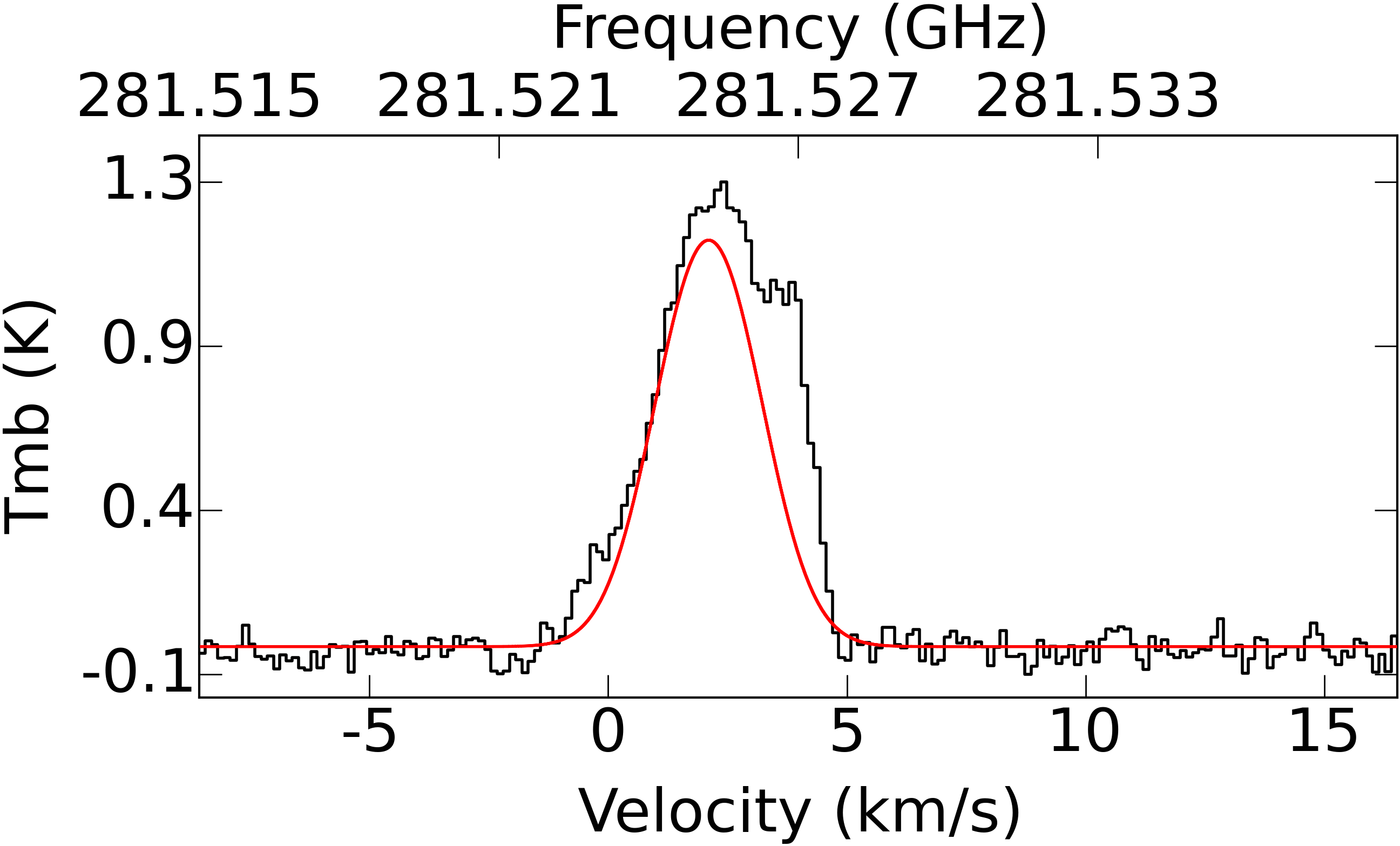} &\includegraphics[width=0.315\textwidth,trim = 0 0 0 0,clip]{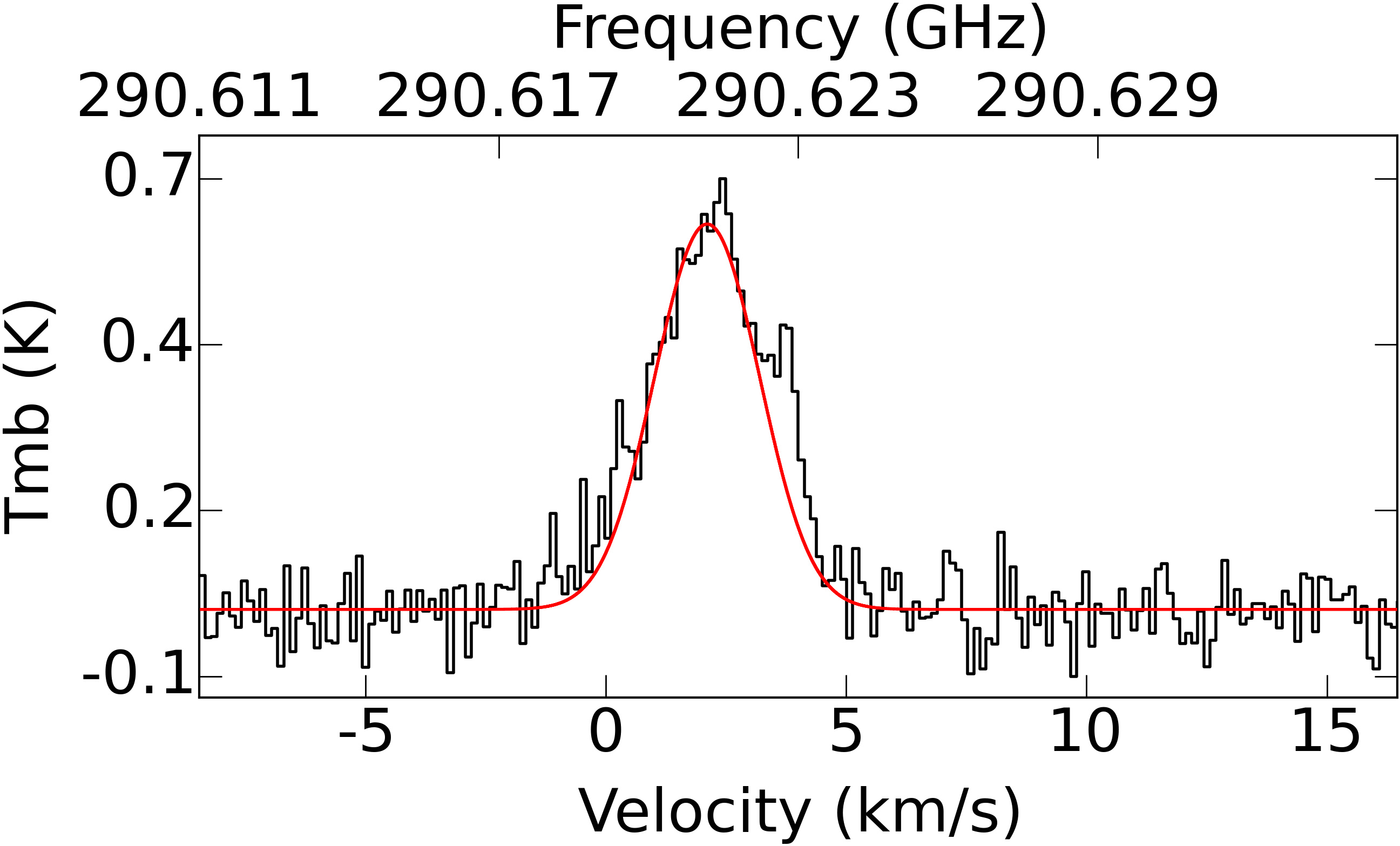}  &\includegraphics[width=0.315\textwidth,trim = 0 0 0 0,clip]{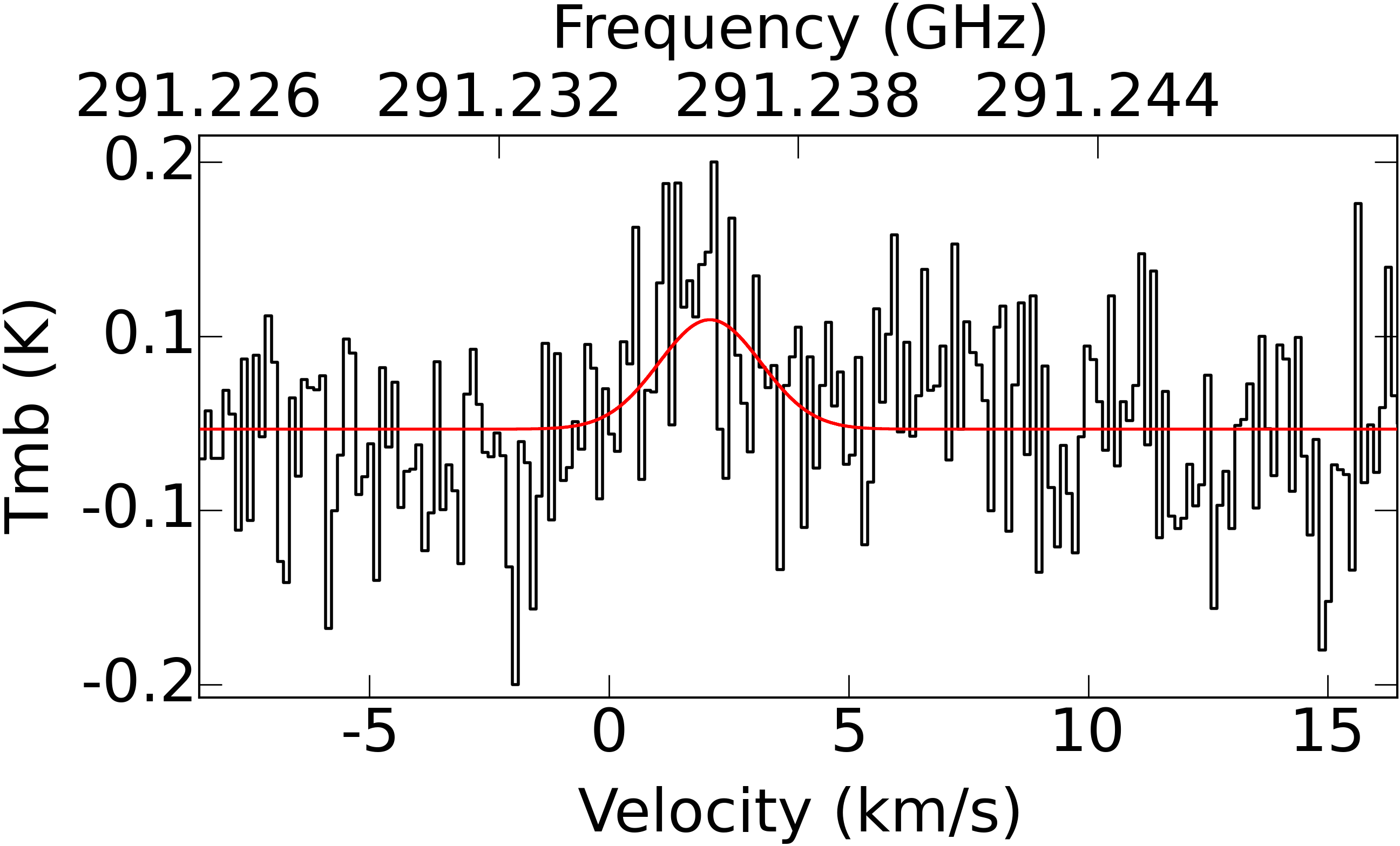}  \\

\includegraphics[width=0.315\textwidth, trim= 0 0 0 0, clip]{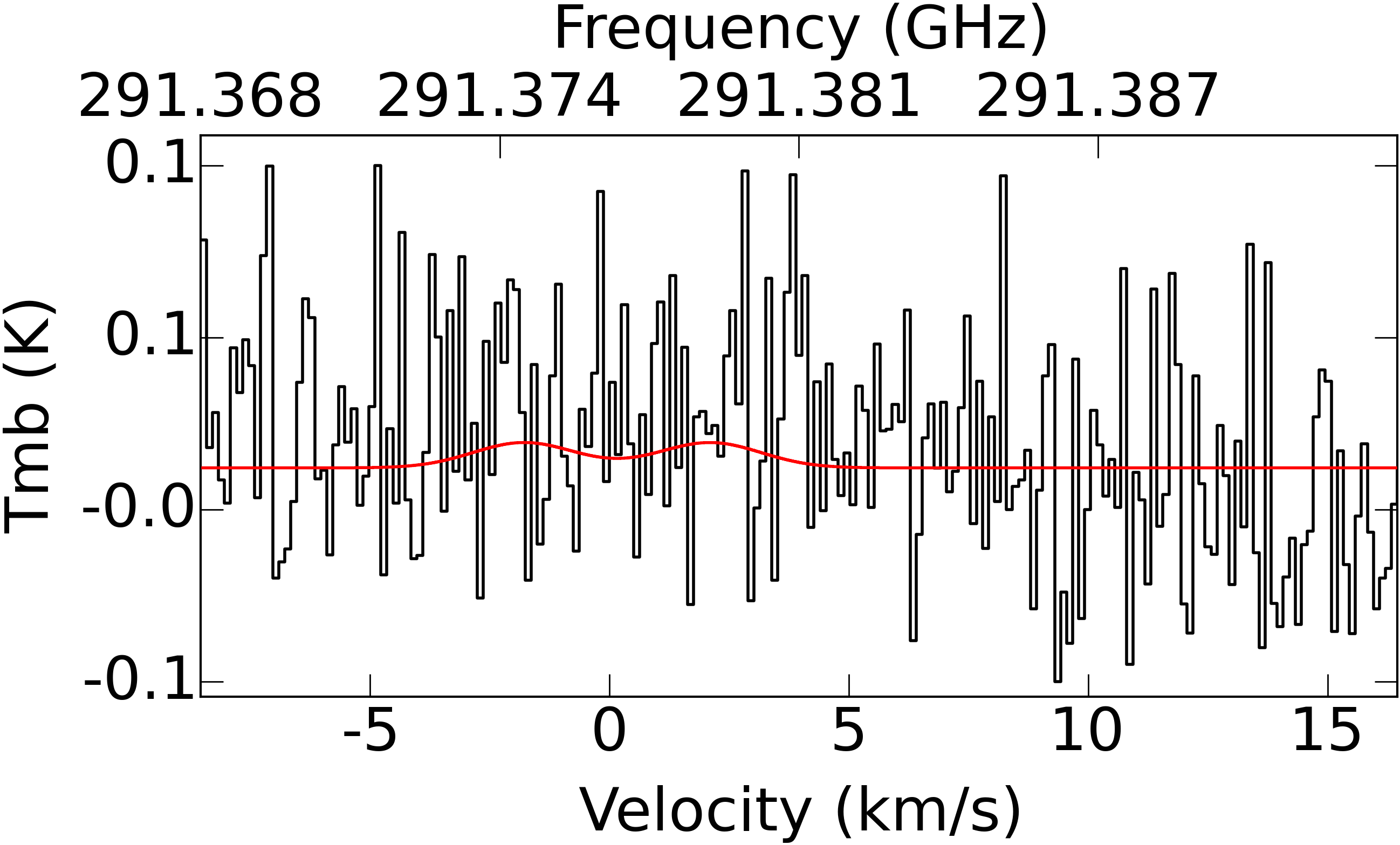} &\includegraphics[width=0.315\textwidth,trim = 0 0 0 0,clip]{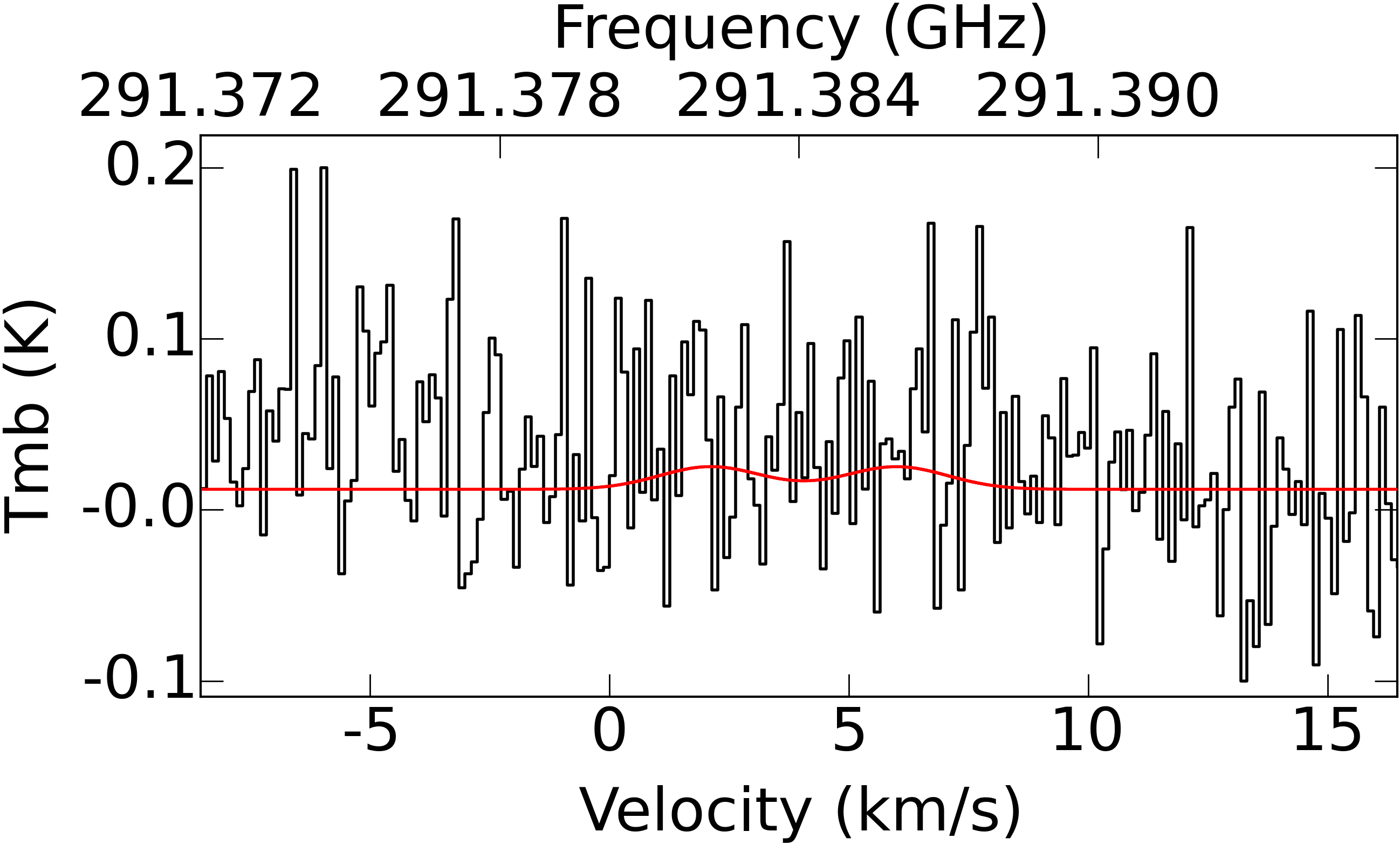}  &\includegraphics[width=0.315\textwidth,trim = 0 0 0 0,clip]{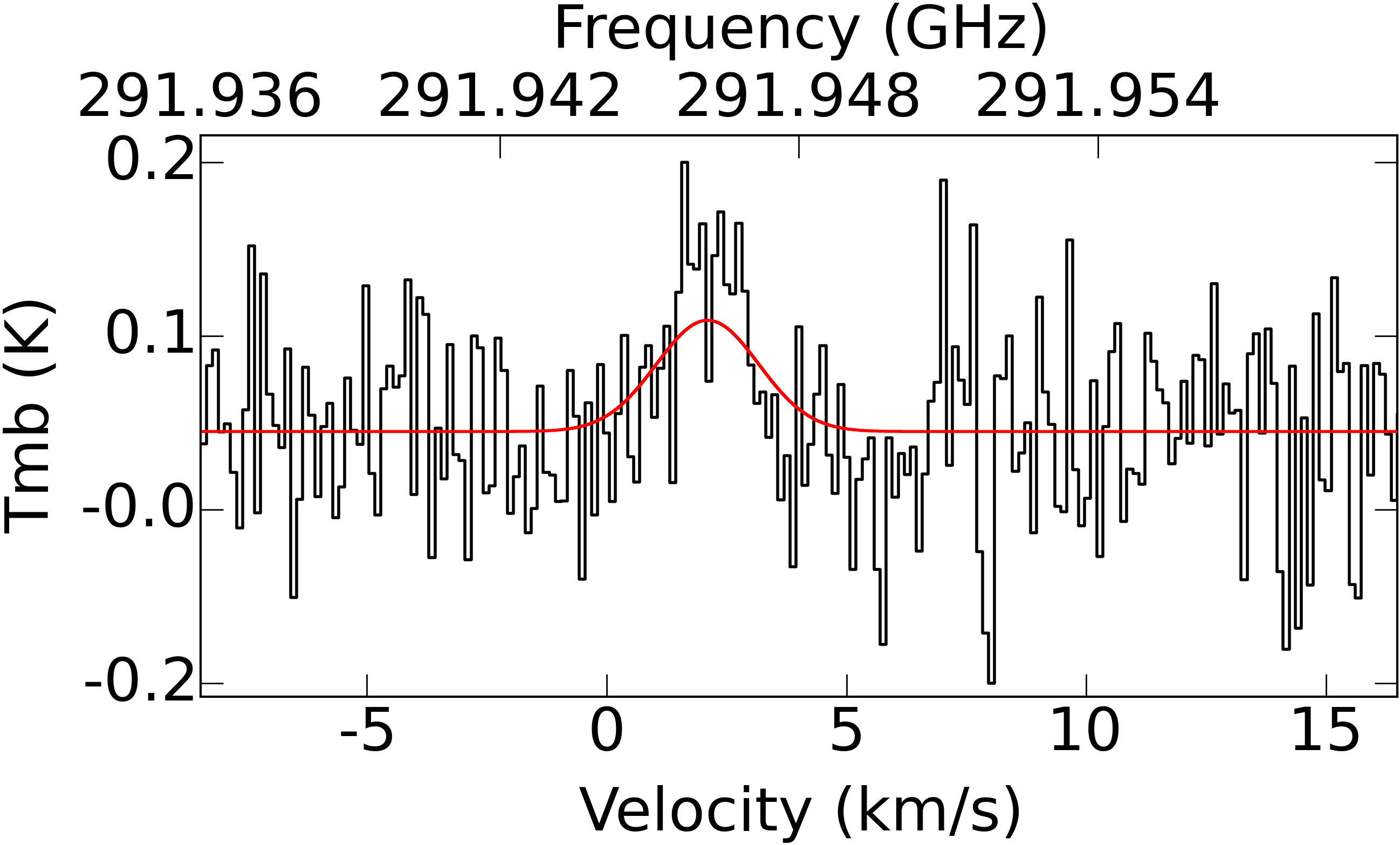}  \\

\includegraphics[width=0.315\textwidth, trim= 0 0 0 0, clip]{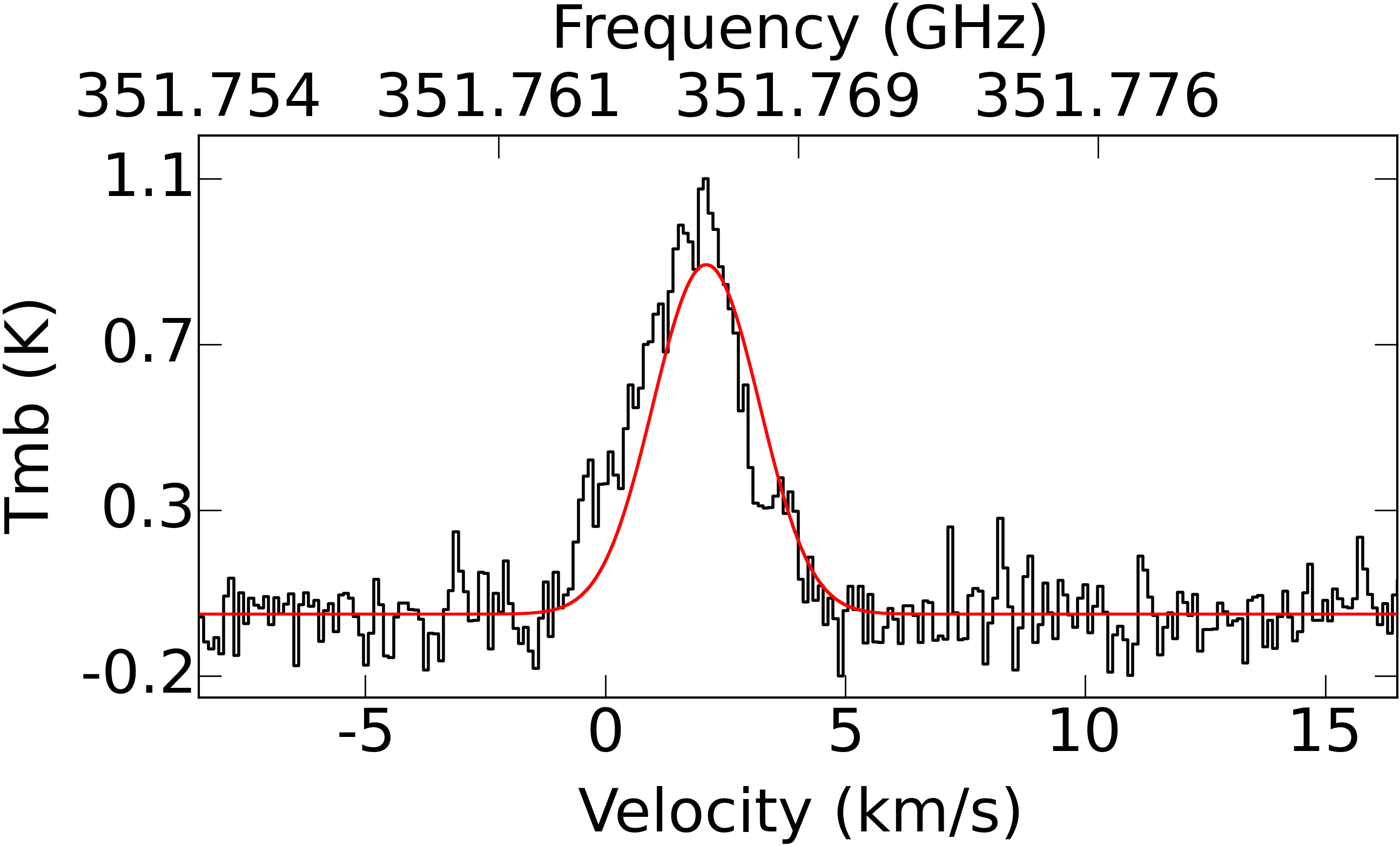} &\includegraphics[width=0.315\textwidth,trim = 0 0 0 0,clip]{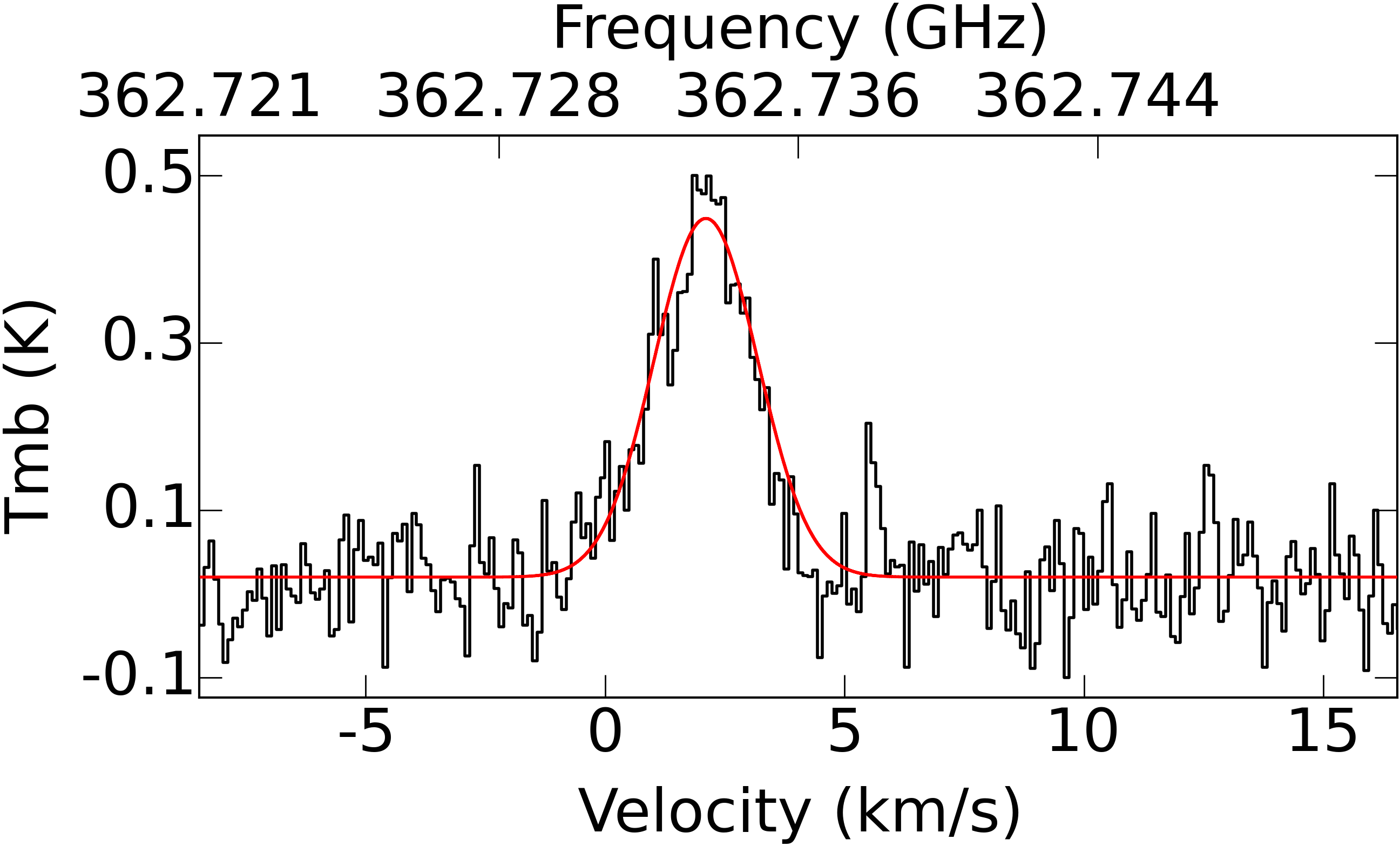}  &  \\

\end{tabular}
\caption{Best model of the line profiles at HE2. In black, we show the H$_2$CO lines observed toward HE2, while in red we show the non-LTE radiative transfer model. One physical component is needed to reproduce the line profiles for this source. The parameters derived from the model are summarized in Table \ref{tab:cassis_he2}.}
\label{fig:cassis_he2}
\end{figure*}

\begin{multicols}{2}
\noindent
The morphology of the H$_2$CO line profiles in the emission peak HE2 is not completely Gaussian. Nevertheless, they can still be modeled with CASSIS-RADEX. Four transitions are clearly detected in this source and since there is no clear self-absorption detection, we used only one physical component to model the lines. The results obtained with CASSIS-RADEX are summarized in Table \ref{tab:cassis_he2} and Fig.~\ref{fig:cassis_he2} respectively.
\end{multicols}
%HE2-----------------------------------------------------------------------------------
\begin{table*}[htbp]
\centering
\caption{Best physical parameters derived from the $\chi^2$ minimization for source HE2.}
\label{tab:cassis_he2}
\scalebox{1.0}{
\begin{tabular}{cccccccc}

\hline
Component 	    & N 						   & T$_{\rm kin}$ 	& FWHM 		    & V$_{\rm LSR}$	        & Size 		    & $n$(H$_2$)        & ortho/para	\\
		 	    &(cm$^{-2}$)				   & (K)			 & (km s$^{-1}$) & (km s$^{-1}$)	    & ($''$)		& (cm$^{-3}$)	    &   \\
\hline
1 	    &$(2.45 \pm 0.01 )\times 10^{13}$	   & $27.3\pm 2.12$   & $2.5\pm 0.25$	& $2.1\pm 0.01$	& 20.0$^*$	& 1$\times 10^{6*}$ & 3.0$^*$	\\
\hline
\end{tabular}
}
\tablefoot{The values marked with a $^*$ symbol were fixed during the modeling.}
\end{table*}

%HE2-----------------------------------------------------------------------------------
\clearpage
\begin{multicols}{2}
\section{Venn diagrams for the occurrence of molecules}\label{app:VennDiag}
In this section we show Venn diagrams to visually compare the occurrence of different molecules on each of the emission peaks E, E1. E2, W1, W2, and HE2. 
\end{multicols}
\begin{figure*}[htbp]
	\centering
    \subfigure[]{\includegraphics[width=0.42\textwidth]{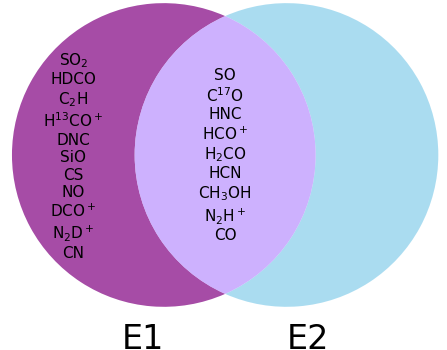}}\hspace{0.5cm}
    \subfigure[]{\includegraphics[width=0.42\textwidth]{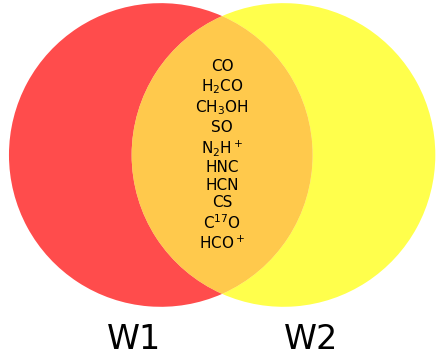}}\\
    \subfigure[]{\includegraphics[width=0.42\textwidth]{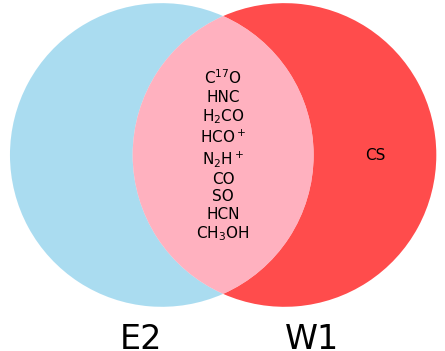}}\hspace{0.5cm}
    \subfigure[]{\includegraphics[width=0.42\textwidth]{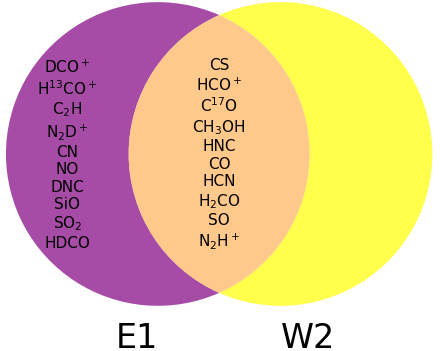}}\\
    \subfigure[]{\includegraphics[width=0.42\textwidth]{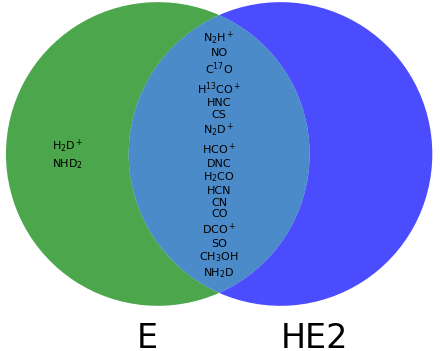}}
    \caption{Venn diagrams for the occurrence of molecules in different emission peaks in the IRAS\,16293--2422 environment.}
    \label{fig:venn}
\end{figure*}

\twocolumn
\section{Line profiles of deuterated species in the vicinity of 16293E}\label{app:gauss}
\begin{figure}[tbhp]
	\centering
	\includegraphics[width=0.49\textwidth]{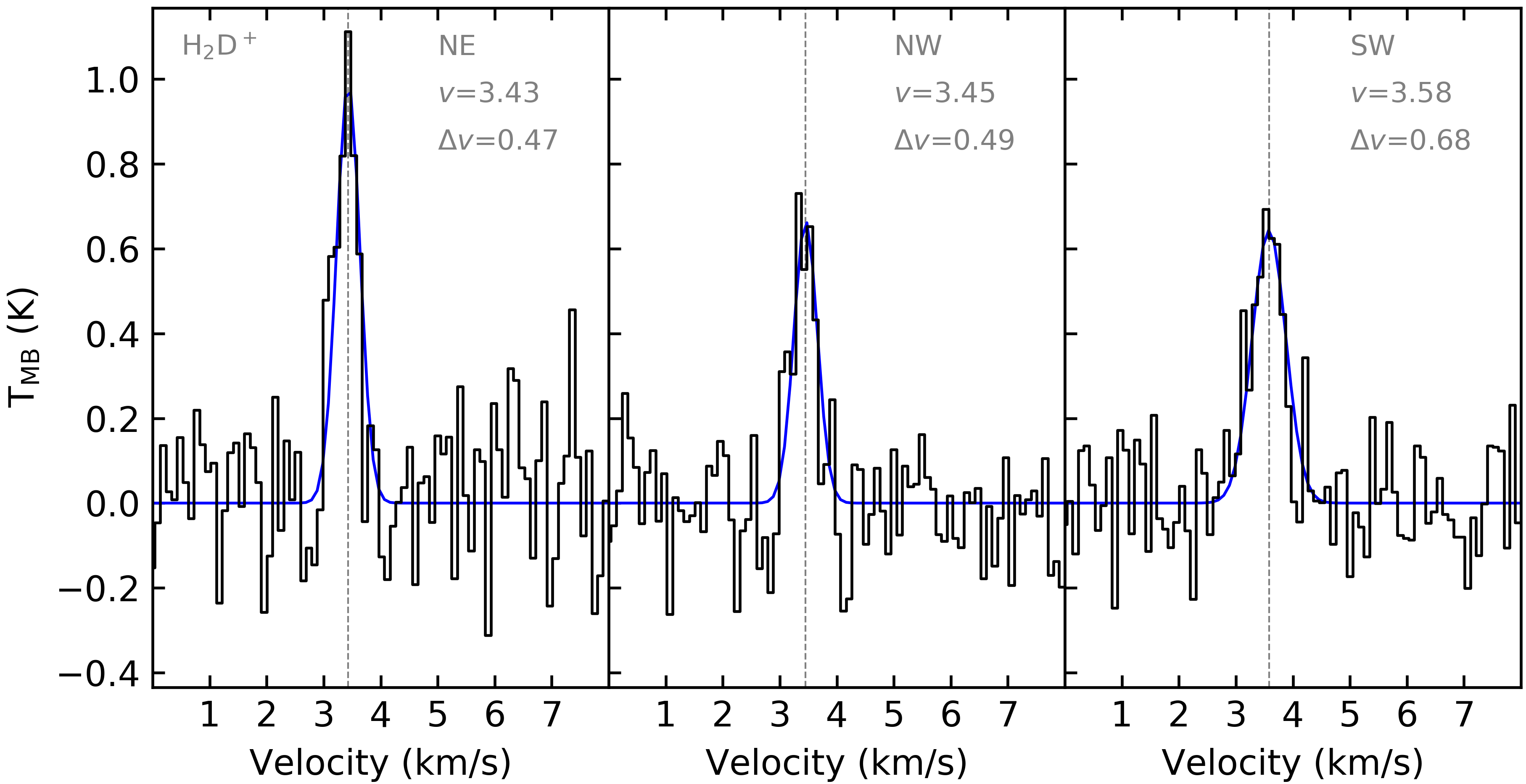}
	\caption{Line profiles of the H$_2$D$^+$ (372421.356\,$\si{\mega\hertz}$) transition in the vicinity of 16293E. The data is displayed in black while the blue line shows the computed Gaussian fits to the spectra. Gray vertical lines mark the position of the fit Gaussians.}
	\label{fig:gaussfits1}
\end{figure}

\begin{figure}[tbhp]
	\centering
	\includegraphics[width=0.49\textwidth]{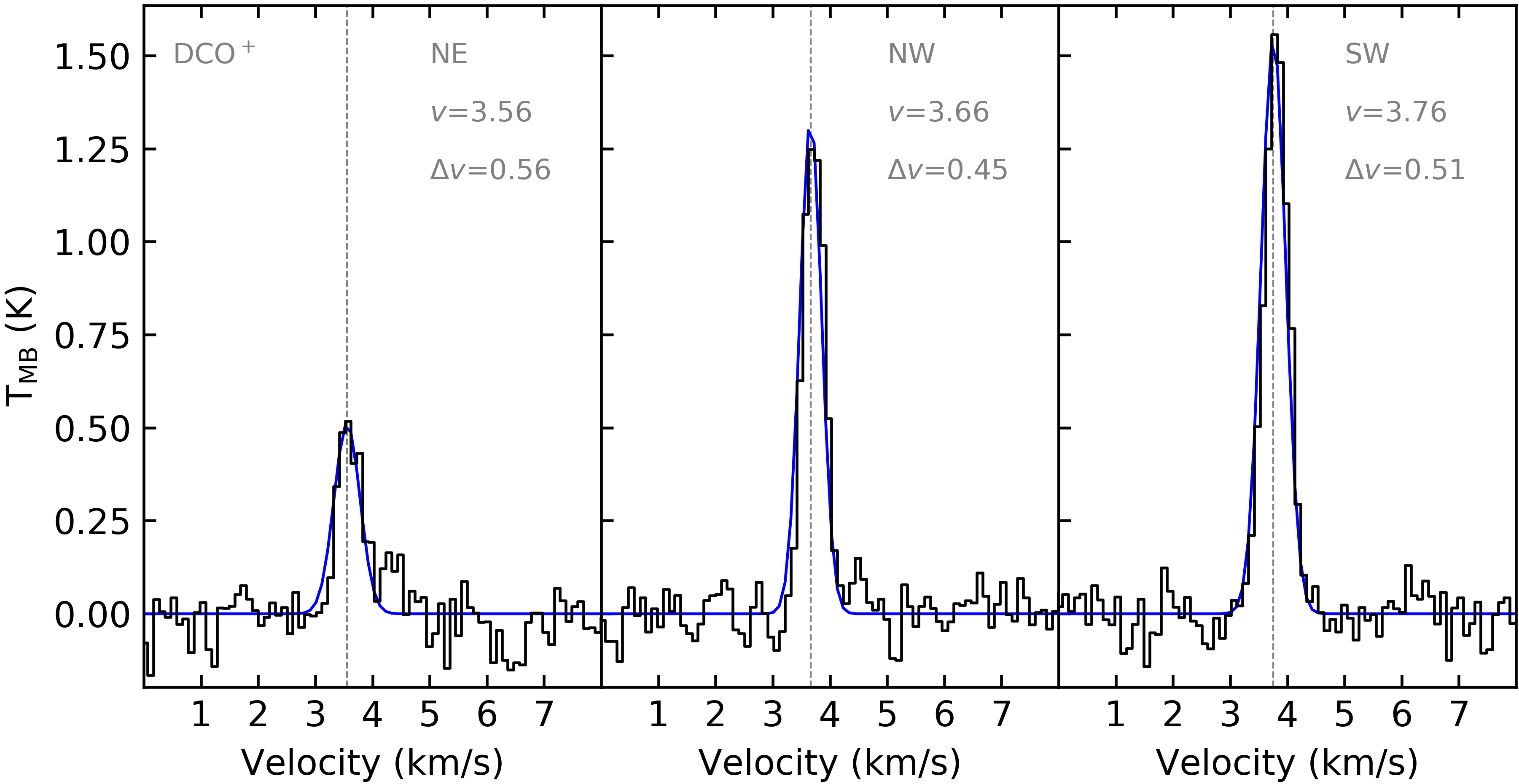}
	\caption{Line profiles of the DCO$^+$ (360169.778\,$\si{\mega\hertz}$) transition in the vicinity of 16293E. The data is displayed in black while the blue line shows the computed Gaussian fits to the spectra. Gray vertical lines mark the position of the fit Gaussians.}
	\label{fig:gaussfits2}
\end{figure}

\begin{figure}[tbhp]
	\centering
	\includegraphics[width=0.49\textwidth]{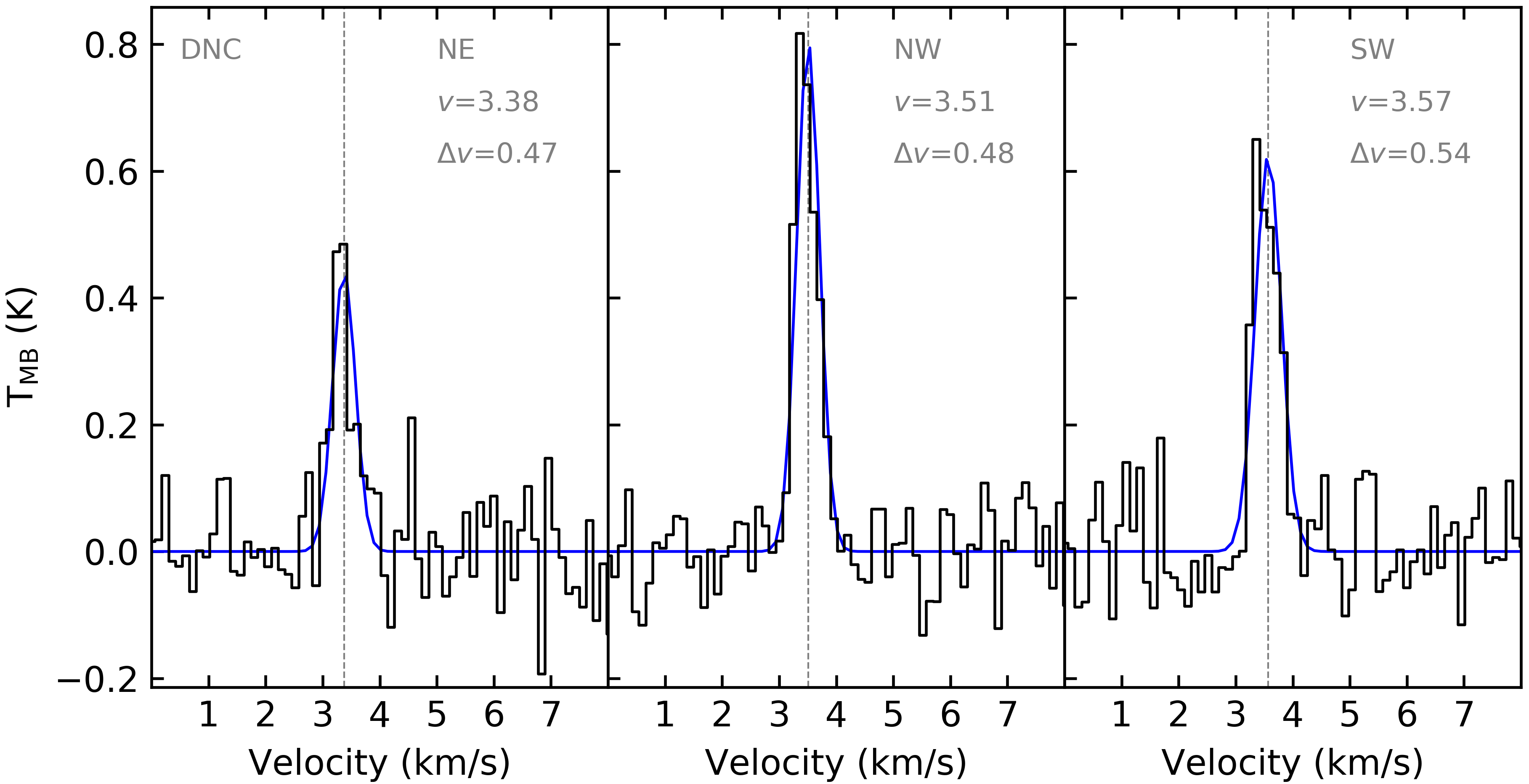}
	\caption{Line profiles of the DNC (305206.219\,$\si{\mega\hertz}$) transition in the vicinity of 16293E. The data is displayed in black while the blue line shows the computed Gaussian fits to the spectra. Gray vertical lines mark the position of the fit Gaussians.}
	\label{fig:gaussfits3}
\end{figure}

\begin{figure}[tbhp]
	\centering
	\includegraphics[width=0.49\textwidth]{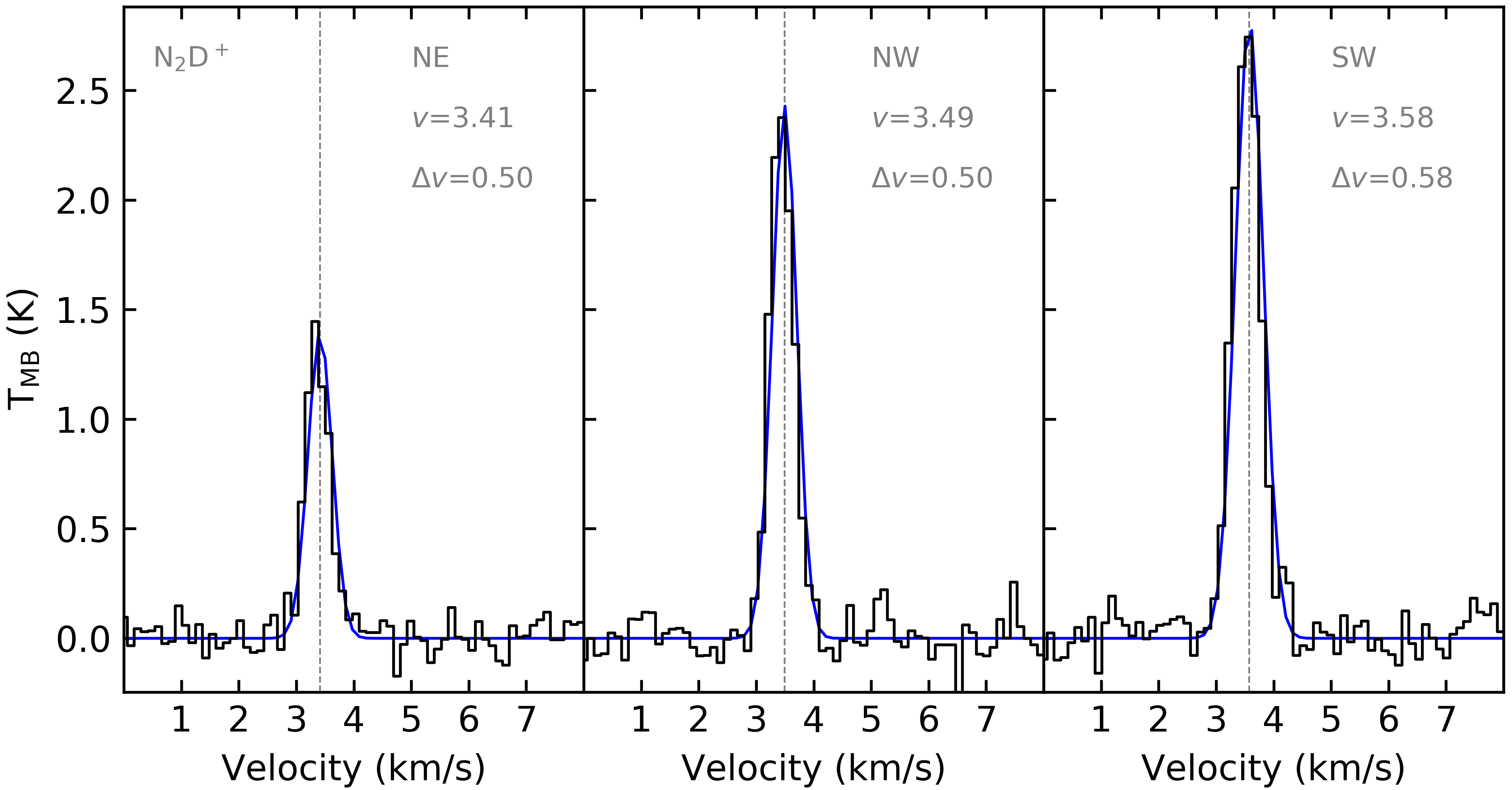}
	\caption{Line profiles of the N$_2$D$^+$ (308422.267\,$\si{\mega\hertz}$) transition in the vicinity of 16293E. The data is displayed in black while the blue line shows the computed Gaussian fits to the spectra. Gray vertical lines mark the position of the fit Gaussians.}
	\label{fig:gaussfits4}
\end{figure}

\begin{figure}[tbhp]
	\centering
	\includegraphics[width=0.49\textwidth]{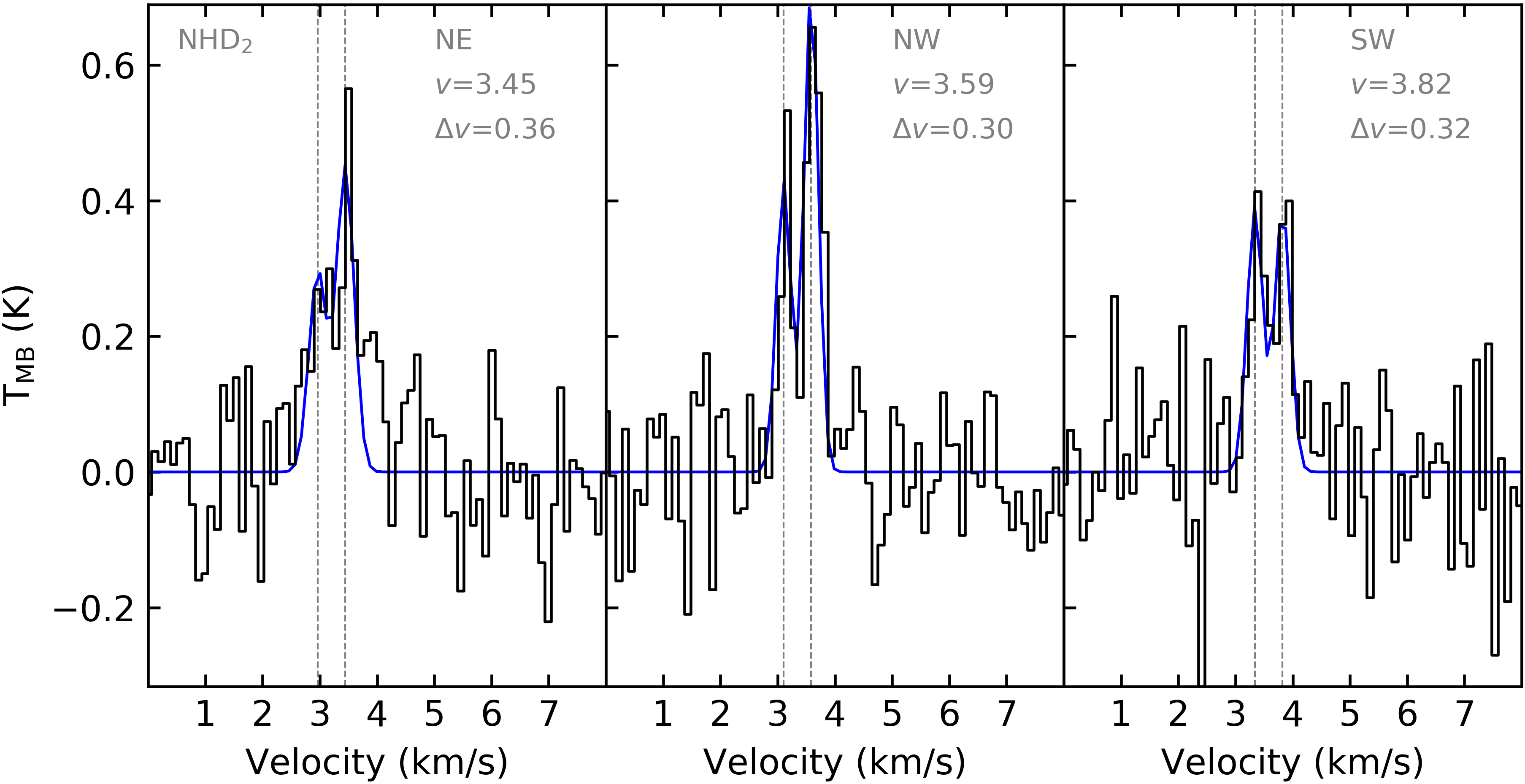}
	\caption{Line profiles of the NHD$_2$ (335513.793\,$\si{\mega\hertz}$) transition in the vicinity of 16293E. The data is displayed in black while the blue line shows the computed Gaussian fits to the spectra. Gray vertical lines mark the position of the fit Gaussians.}
	\label{fig:gaussfits5}
\end{figure}

We extracted spectra of three positions where we see clear emission for all deuterated species that peak near 16293E: H$_2$D$^+$, DCO$^+$, DNC, N$_2$D$^+$, NHD$_2$, and NH$_2$D. The exact coordinates are the following: NE: (16$^\U{h}$32$^\U{m}$30.6$^\U{s}$, -24$^\circ$28$'$46.7$''$), NW: (16$^\U{h}$32$^\U{m}$28.9$^\U{s}$, -24$^\circ$28$'$46.7$''$), SW: (16$^\U{h}$32$^\U{m}$28.9$^\U{s}$, -24$^\circ$29$'$13.7$''$).
In order to test the existence of a trend in line width and velocity shift of these lines, we conducted a Gaussian fit to these line profiles. The resulting fits to the spectra are shown in Figs.~\ref{fig:gaussfits1}-\ref{fig:gaussfits5}, the derived peak velocity and FWHM values are shown in Table \ref{tab:vel_offsets}. 
\begin{table}[hbt]
    \caption{Peak velocities and line widths (FWHM) for several transitions from deuterated species at the NE, NW, and SW positions shown in Fig.~\ref{fig:H2D+}.}
	\label{tab:vel_offsets}
	\centering
	\begin{tabular}{C{3.2cm}C{1.128cm}C{1.5cm}C{1.5cm}}
		\hline
		\hline
		Transition         & Position    & Peak velocity            & FWHM \\
		                &             &   (km s$^{-1}$)     & (km s$^{-1}$) \\ \hline
		                & NE           & $3.43 \pm 0.02$  & $0.47 \pm 0.05$\\ 
		H$_2$D$^+$ $(1_{1,\, 0} - 1_{1,\, 1})$   & NW           & $3.45 \pm 0.03$  & $0.49 \pm 0.06$\\ 
		                & SW           & $3.58 \pm 0.04$  & $0.68 \pm 0.10$\\ \hline
%---------------------------------------------------------------------------------------------
		 		       & NE           & $3.56 \pm 0.03$  & $0.56 \pm 0.07$\\ 
		DCO$^+$ $(5-4)$        & NW           & $3.66 \pm 0.01$  & $0.45 \pm 0.02$\\ 
		               & SW           & $3.76 \pm 0.01$  & $0.51 \pm 0.02$\\ \hline
%---------------------------------------------------------------------------------------------
		 		       & NE           & $3.38 \pm 0.03$  & $0.47 \pm 0.08$\\ 
		DNC $(4-3)$           & NW           & $3.51 \pm 0.01$  & $0.48 \pm 0.03$\\ 
		               & SW           & $3.57 \pm 0.02$  & $0.54 \pm 0.04$\\ \hline
%---------------------------------------------------------------------------------------------
		 		       & NE           & $3.41 \pm 0.01$  & $0.50 \pm 0.02$\\ 
		N$_2$D$^+$ $(4-3)$    & NW           & $3.49 \pm 0.01$  & $0.50 \pm 0.01$\\ 
		               & SW           & $3.58 \pm 0.01$  & $0.58 \pm 0.01$\\ \hline
%---------------------------------------------------------------------------------------------
		 		       & NE           & $3.45 \pm 0.03$  & $0.36 \pm 0.04$\\ 
		NHD$_2$  $(1_{1,\,1,\,0} - 0_{0,\, 0,\,0})$      & NW           & $3.59 \pm 0.02$  & $0.30 \pm 0.02$\\ 
		               & SW           & $3.82 \pm 0.03$  & $0.32 \pm 0.04$\\ \hline
%---------------------------------------------------------------------------------------------
		 		       & NE           & $3.49 \pm 0.01$  & $0.35 \pm 0.01$\\ 
		NH$_2$D $(1_{0,\,1,\,1} - 0_{0,\, 0,\,1})$       & NW           & $3.62 \pm 0.01$  & $0.43 \pm 0.01$\\ 
		               & SW           & $3.84 \pm 0.03$  & $0.49 \pm 0.02$\\ \hline
%---------------------------------------------------------------------------------------------
	\end{tabular}
	\tablefoot{A trend in the velocity is seen for all species as it increases closer to the SW position of 16293E where the outflow-core interaction is taking place. A similar trend is seen in the line widths, except for DCO$^+$ and NHD$_2$.}
\end{table}
\\[3.3cm]
\\[1.2cm]

\newpage

\section{Integrated intensity maps for the observed molecules}\label{app:maps}
In this part of the Appendix we show all the velocity-integrated intensity (moment 0) maps for all the detected molecules in the observations. The figures are sorted based on the frequencies of the corresponding transitions (see Table~\ref{tab:idlines}).

The name of the corresponding molecule and its transition are indicated in red color in the upper left side of each map, while the velocity range considered for the integration is given the lower right corner. In case of blended lines the integration range was chosen such that all lines are covered, the contributing transitions are given in the caption of the corresponding figures. The beam is shown in the lower left corner of the maps.
The color bar in the right side has units of $\si{\kelvin\kilo\meter\per\second}$. The contour levels are drawn in black color at $3 \sigma$, in steps of $2\sigma$ between $4\sigma$ and $10\sigma$ and in steps of $10\sigma$ afterwards. The markers show the position of the emission peaks introduced in Sect.~\ref{ch:results}. Given that some methanol transitions exhibit similar spatial morphologies, we have not mapped them all (see Table \ref{tab:idlines}). The intensity is given in terms of antenna temperatures $T_{\rm A}^*$. To convert to brightness temperature $T_{\rm MB}$, a multiplicative factor of $1/\num{0.7}$ needs to be considered.

Some transitions only show weak emission, such that the signal to noise ratio of a single pixel is not sufficient for a significant detection. In these cases, we averaged the spectra in a $\SI{10}{\arcsecond}$ radius around the respective emission peaks to confirm a detection. The averaged spectra are shown next to the associated maps. Some of these spectra are smoothed to a lower velocity resolution in order to improve the signal to noise ratio for the visualization.

%%%%%%%%%%%%%%%%%%%%%%%%%%%%%%%%%%%%%%%%%%%%%%%%%

\begin{figure}[th]
	\centering
    \subfigure[]{\includegraphics[width=0.42\textwidth, trim={0 0.65cm 5.4cm 1.18cm},clip]{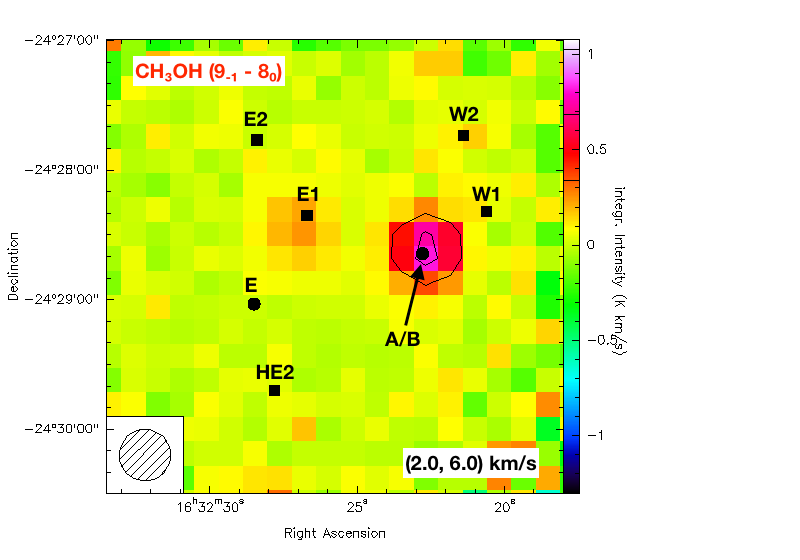}}\\   \subfigure[]{\includegraphics[width=0.42\textwidth, trim={0 0.65cm 0 1.18cm},clip]{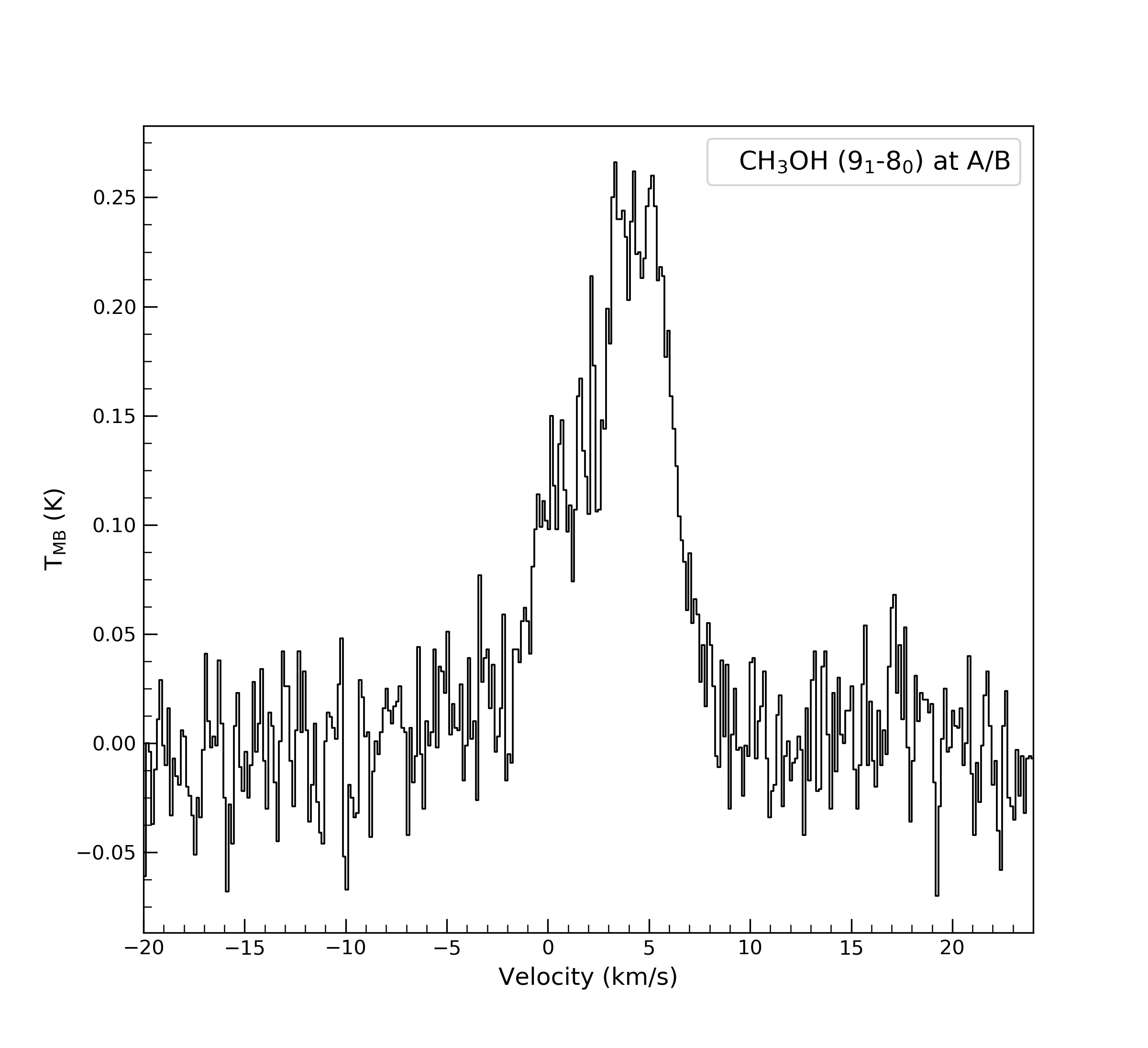}}
    \caption{(a) CH$_3$OH-E ($9_{-1}- 8_{0}$) transition at 278304.512\,MHz. Additional contours are drawn at 1$\sigma$ and 2$\sigma$. (b) Averaged spectrum of this transition in a $\SI{10}{\arcsecond}$ radius at the position of IRAS\,16293 A/B.}
    \label{fig:15}
\end{figure}
\clearpage
\begin{figure*}[ht]
	\centering
    \includegraphics[width=0.42\textwidth, trim={0 0.65cm 0 1.18cm},clip]{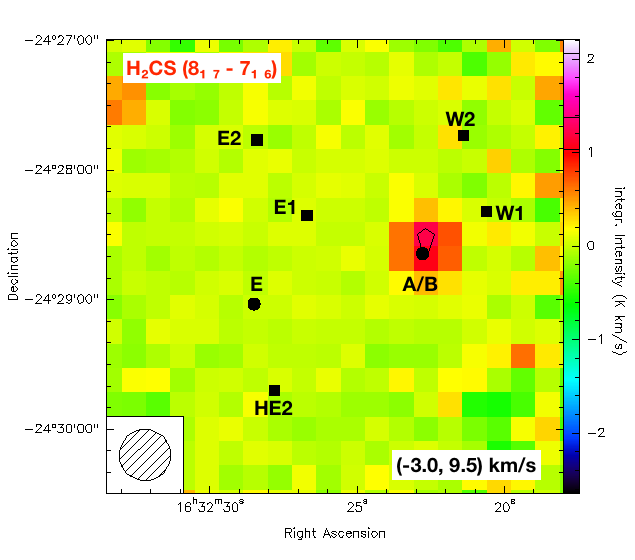}
    \caption{H$_2$CS ($8_{1,7} - 7_{1,6}$) transition at 278887.661\,MHz.}
    \label{fig:22}
\end{figure*}
\begin{figure*}[ht]
	\centering
    \subfigure[]{\includegraphics[width=0.42\textwidth, trim={0 0.65cm 0 1.18cm},clip]{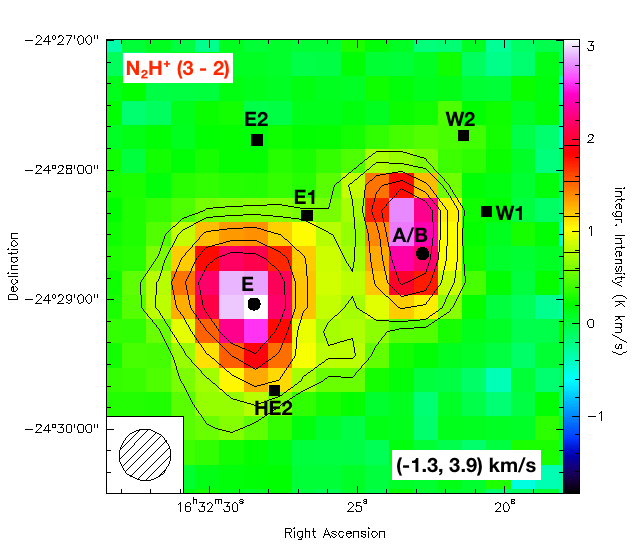}}
    \subfigure[]{\includegraphics[width=0.42\textwidth, trim={0 0.65cm 0 1.18cm},clip]{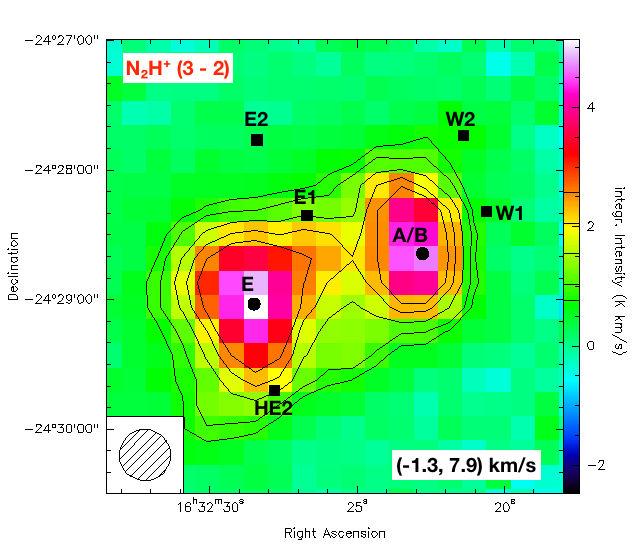}}
    \subfigure[]{\includegraphics[width=0.42\textwidth, trim={0 0.65cm 0 1.18cm},clip]{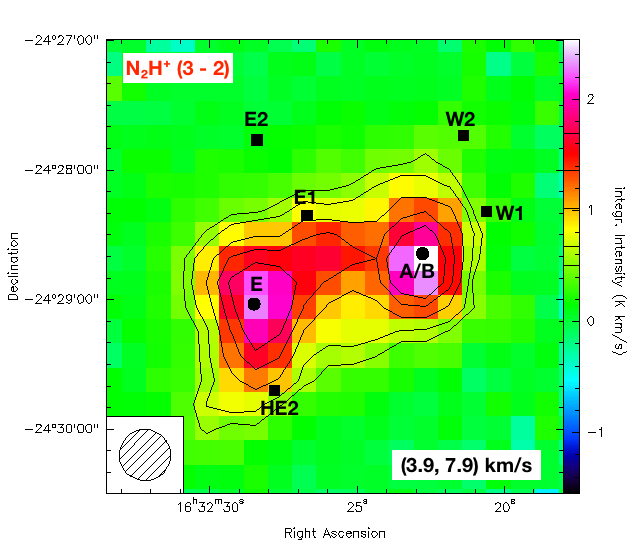}}
    \caption{N$_2$H$^+$ ($3 - 2 $) transition at 279511.749\,MHz.}
    \label{fig:38}
\end{figure*}

\begin{figure*}[ht]
	\centering
    \subfigure[]{\includegraphics[width=0.42\textwidth, trim={0 0.65cm 5.4cm 1.18cm},clip]{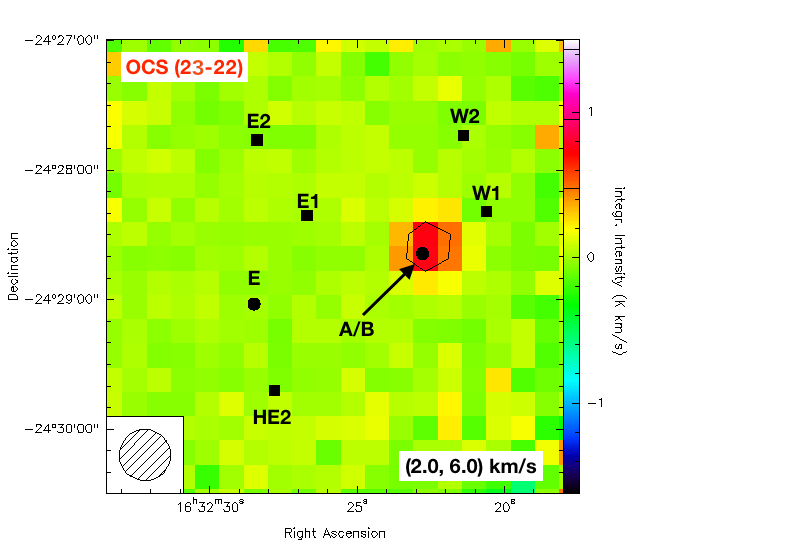}}
    \subfigure[]{\includegraphics[width=0.42\textwidth, trim={0 0.65cm 0 1.18cm},clip]{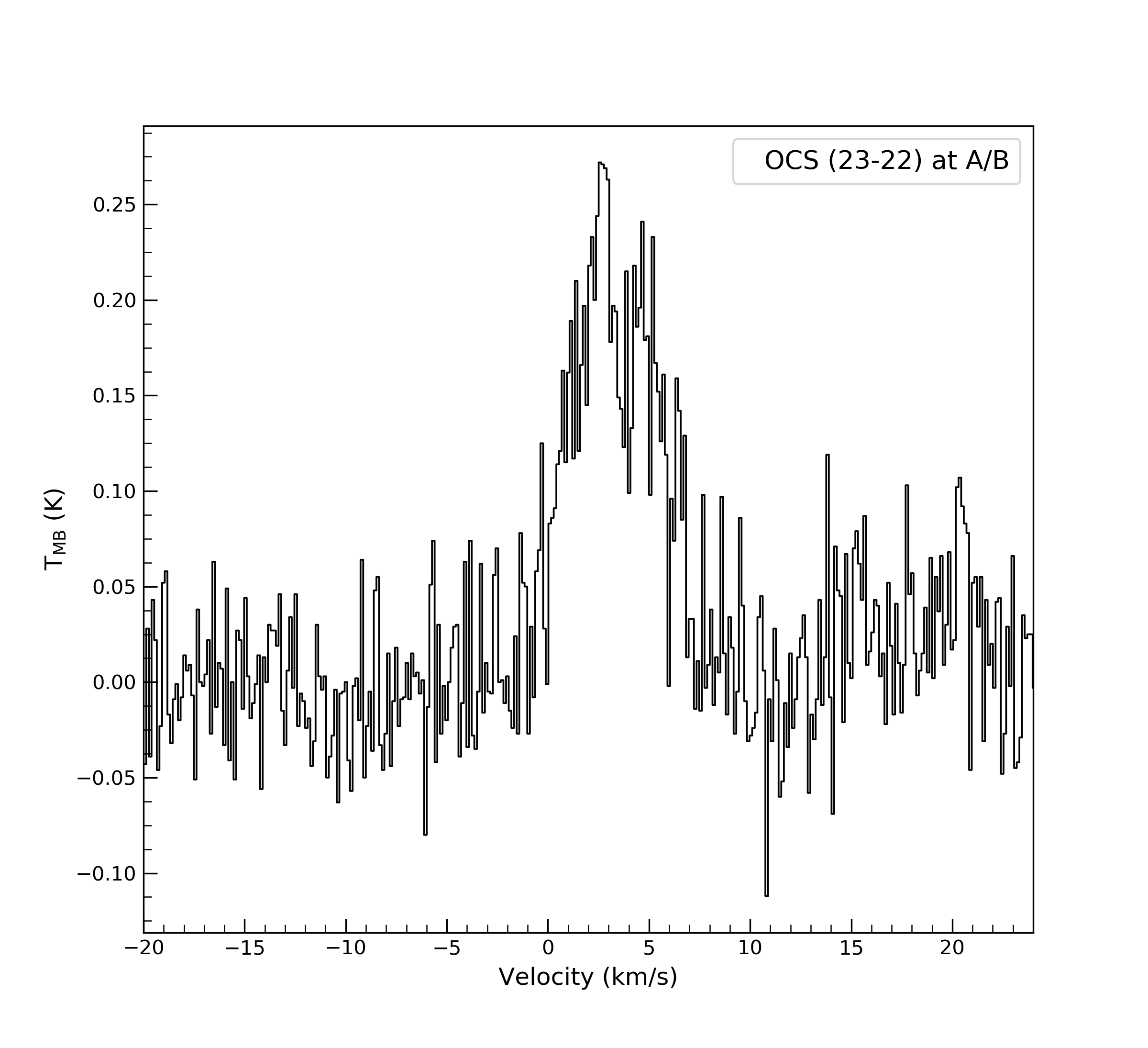}}
    \caption{(a) OCS $(23 - 22)$ transition at 279685.3\,MHz. Additional contours are drawn at 1$\sigma$. (b) Averaged spectrum of this transition in a $\SI{10}{\arcsecond}$ radius at the position of IRAS\,16293 A/B.}
    \label{fig:66}
\end{figure*}

\begin{figure*}[ht]
	\centering
    \subfigure[]{\includegraphics[width=0.42\textwidth, trim={0 0.65cm 5.4cm 1.18cm},clip]{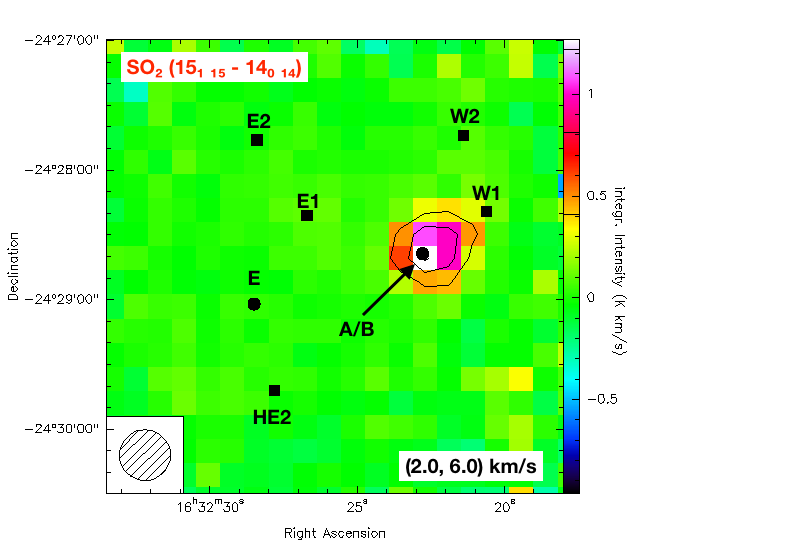}}
    \subfigure[]{\includegraphics[width=0.42\textwidth, trim={0 0.65cm 0 1.18cm},clip]{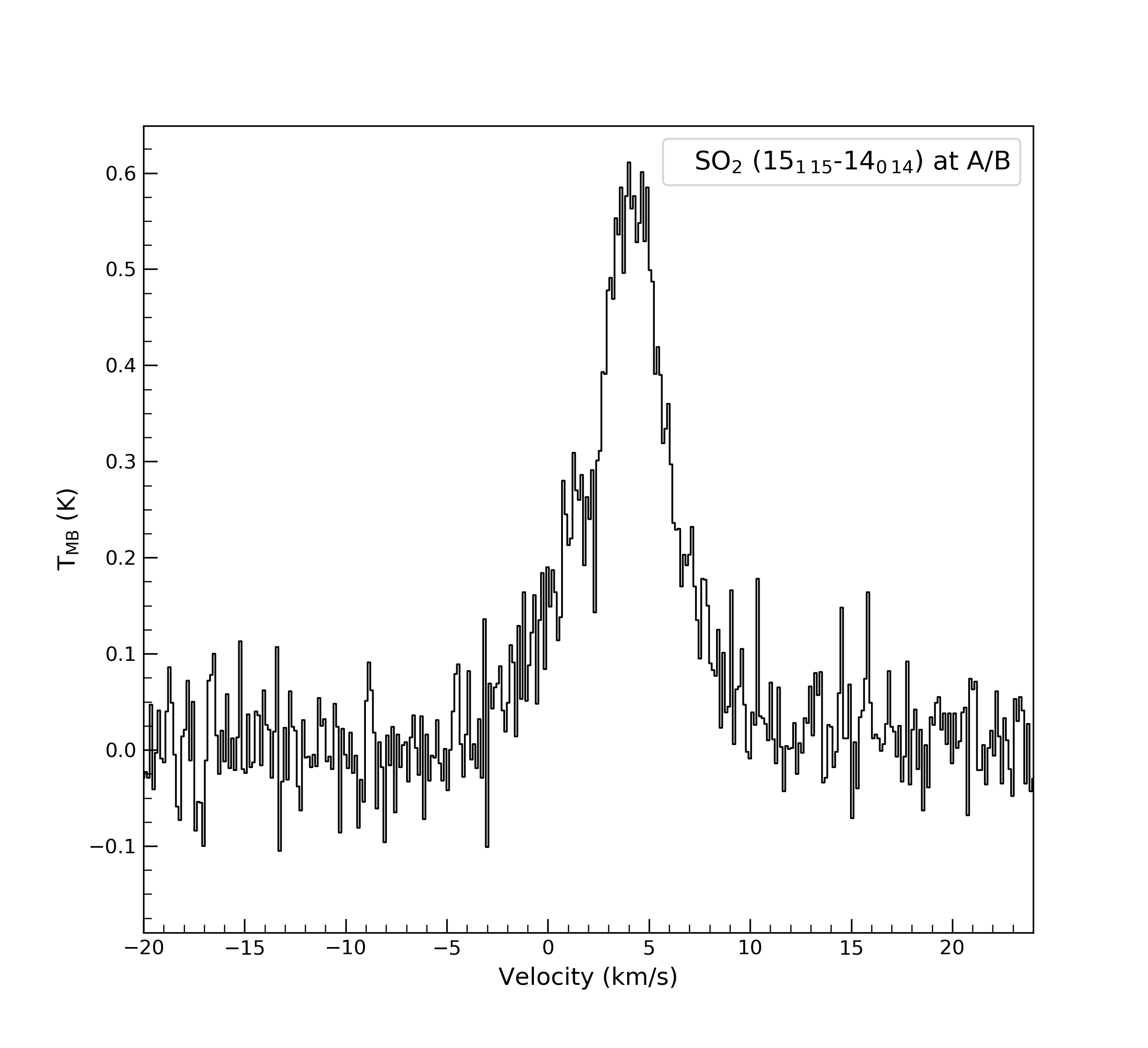}}
    \caption{(a) SO$_2$ $(15_{1, 15} - 14_{0, 14})$ transition at 281762.600\,MHz. Additional contours are drawn at 1$\sigma$ and 2$\sigma$. (b) Averaged spectrum of this transition in a $\SI{10}{\arcsecond}$ radius at the position of IRAS\,16293 A/B.}
    \label{fig:69}
\end{figure*}

\begin{figure*}[ht]
	\centering
    \subfigure[]{\includegraphics[width=0.42\textwidth, trim={0 0.65cm 5.4cm 1.18cm},clip]{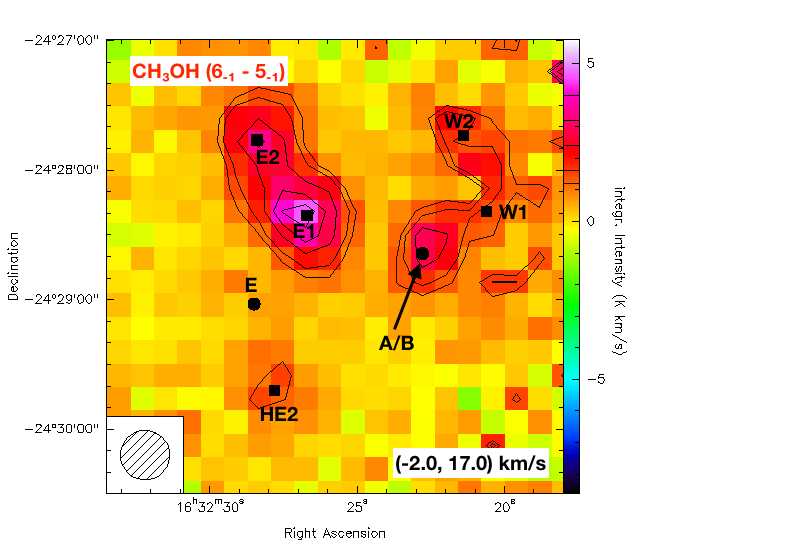}}
    \subfigure[]{\includegraphics[width=0.42\textwidth, trim={0 0.65cm 5.4cm 1.18cm},clip]{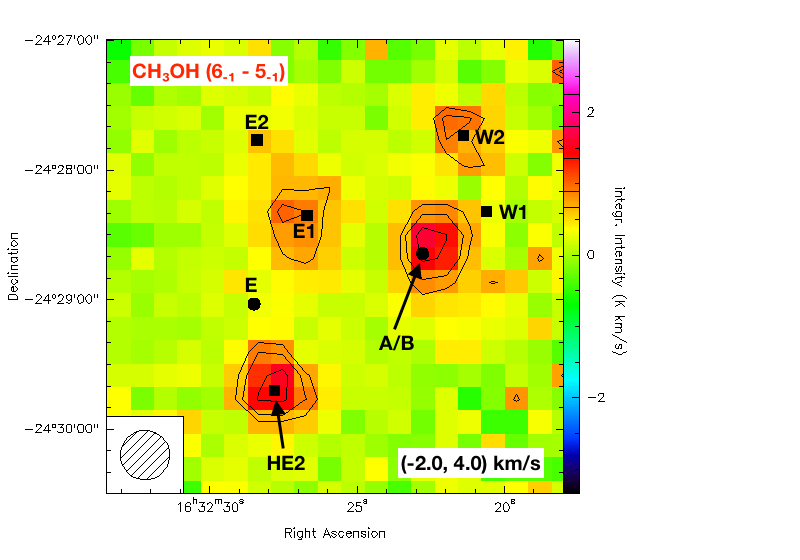}}
    \subfigure[]{\includegraphics[width=0.42\textwidth, trim={0 0.65cm 5.4cm 1.18cm},clip]{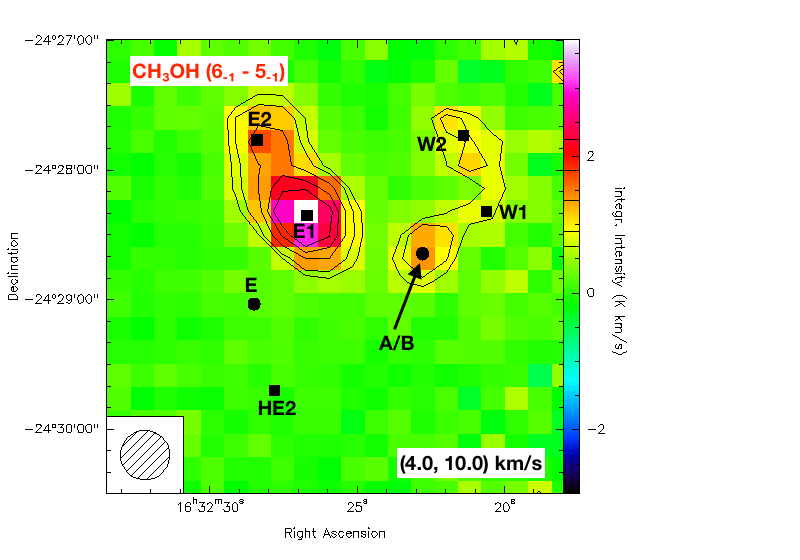}}
    \subfigure[]{\includegraphics[width=0.42\textwidth, trim={0 0.65cm 5.4cm 1.18cm},clip]{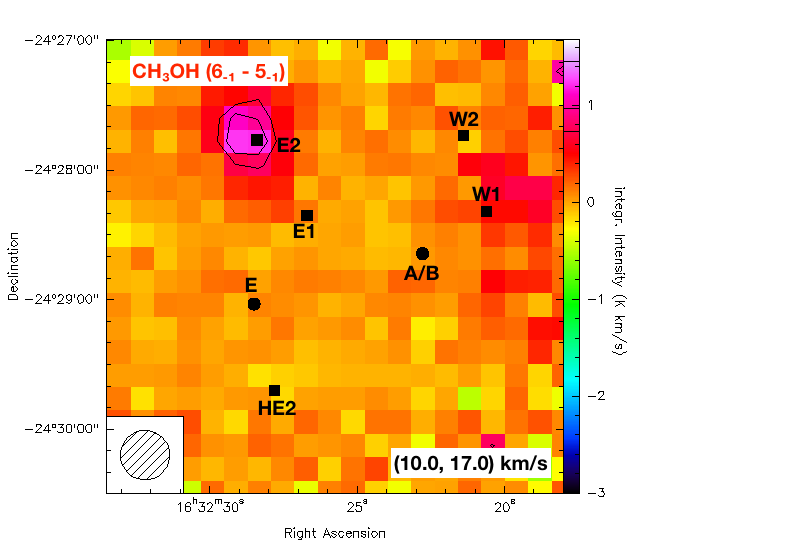}}
    \caption{CH$_3$OH-E ($6_{-1}- 5_{-1}$) transition at 290069.747\,MHz.}
    \label{fig:9}
\end{figure*}

\begin{figure*}[ht]
	\centering
    \subfigure[]{\includegraphics[width=0.42\textwidth, trim={0 0.65cm 0 1.18cm},clip]{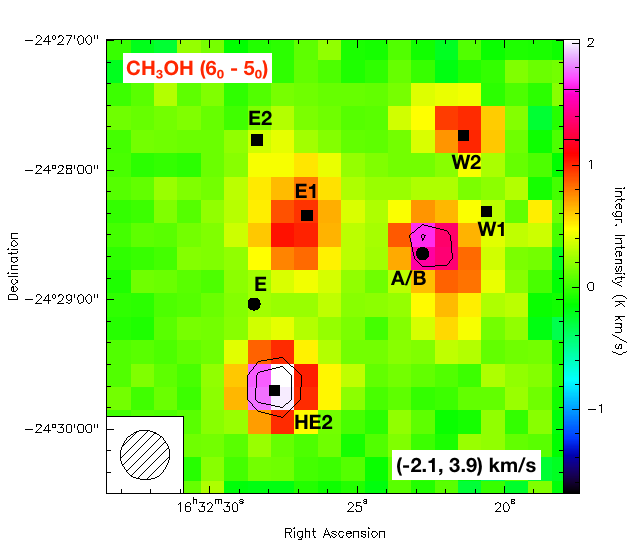}}
    \subfigure[]{\includegraphics[width=0.42\textwidth, trim={0 0.65cm 0 1.18cm},clip]{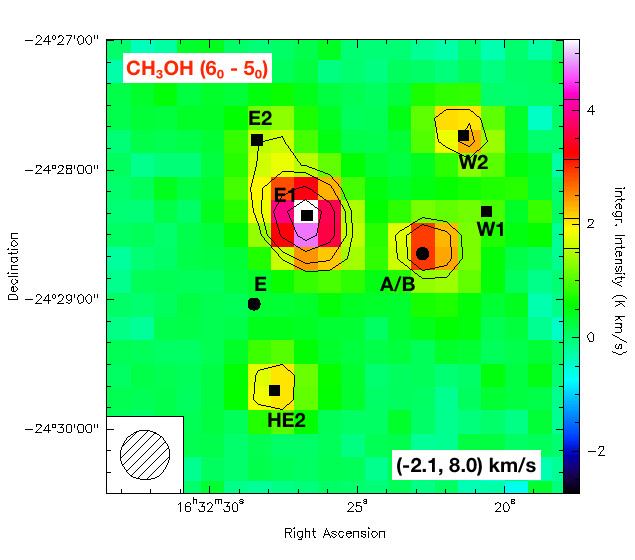}}
    \subfigure[]{\includegraphics[width=0.42\textwidth, trim={0 0.65cm 0 1.18cm},clip]{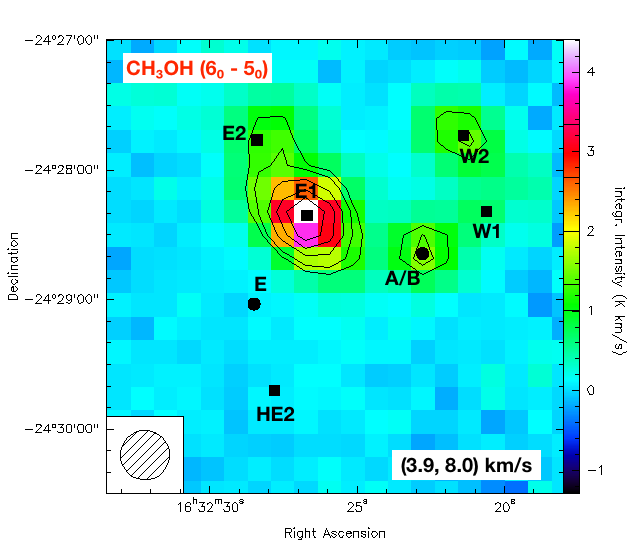}}
    \caption{CH$_3$OH-A$^{+}$ ($6_{0}- 5_{0}$) transition at 290110.637\,MHz.}
    \label{fig:5}
\end{figure*}

\begin{figure*}[ht]
	\centering
    \includegraphics[width=0.42\textwidth, trim={0 0.65cm 5.4cm 1.18cm},clip]{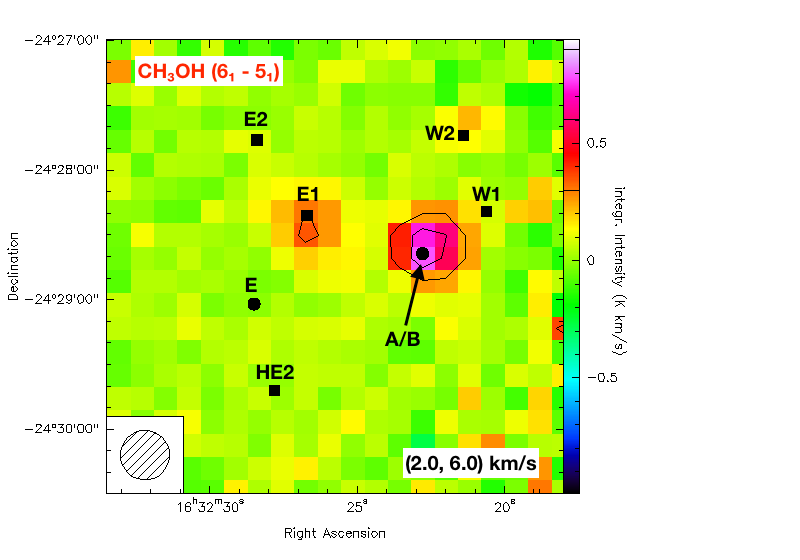}
    \caption{CH$_3$OH-E ($6_{1}- 5_{1}$) transition at 290248.685\,MHz. Additional contours are drawn at 1$\sigma$ and 2$\sigma$.}
    \label{fig:6}
\end{figure*}

\begin{figure*}[ht]
	\centering
    \includegraphics[width=0.42\textwidth, trim={0 0.65cm 5.4cm 1.18cm},clip]{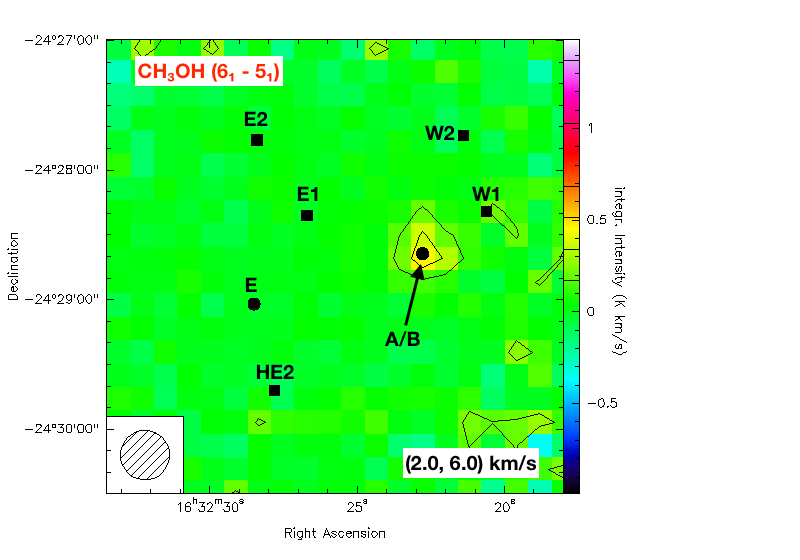}
    \caption{CH$_3$OH-A$^{+}$ ($6_{2}- 5_{2}$) transition at 290264.068\,MHz.}
    \label{fig:7}
\end{figure*}

\begin{figure*}[ht]
	\centering
    \includegraphics[width=0.42\textwidth, trim={0 0.65cm 5.4cm 1.18cm},clip]{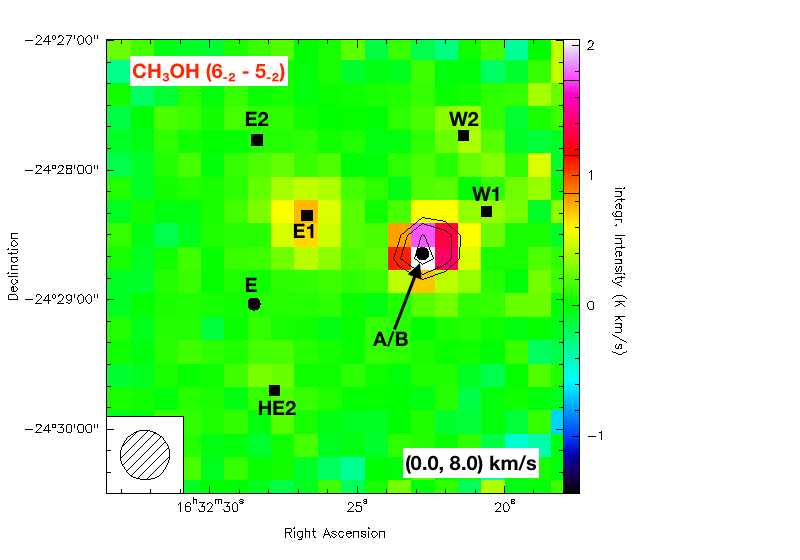}
    \caption{CH$_3$OH-E ($6_{-2}- 5_{-2}$) and CH$_3$OH-E ($6_{2}- 5_{2}$) transition at 290307.281\,MHz and 290307.738\,MHz. The velocity scale is calculated based on a rest frequency of 290307.281\,MHz.}
    \label{fig:10}
\end{figure*}

\begin{figure*}[ht]
	\centering
    \includegraphics[width=0.42\textwidth, trim={0 0.65cm 0 1.18cm},clip]{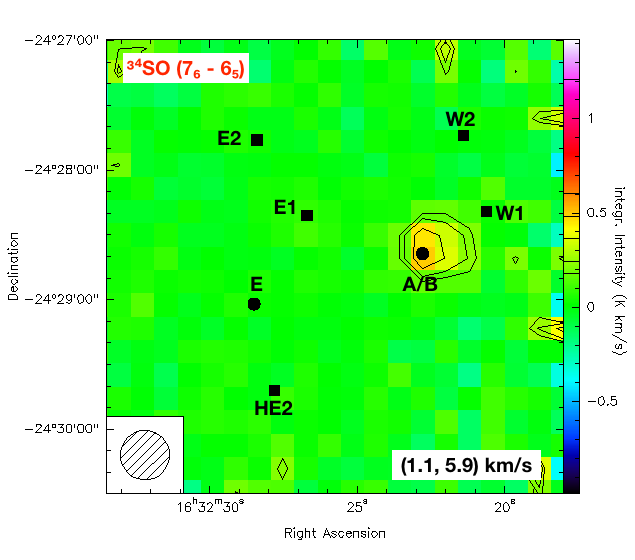}
    \caption{$^{34}$SO ($7_6 - 6_5$) transition at 290562.238\,MHz.}
    \label{fig:46}
\end{figure*}

\begin{figure*}[ht]
	\centering
    \subfigure[]{\includegraphics[width=0.42\textwidth, trim={0 0.65cm 0 1.18cm},clip]{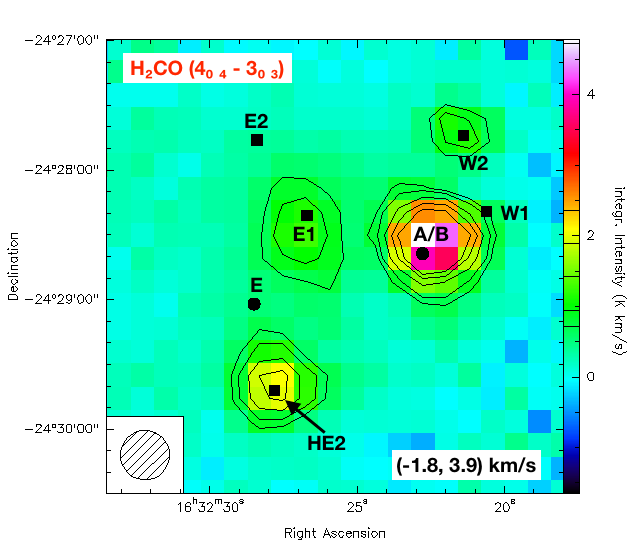}}
    \subfigure[]{\includegraphics[width=0.42\textwidth, trim={0 0.65cm 0 1.18cm},clip]{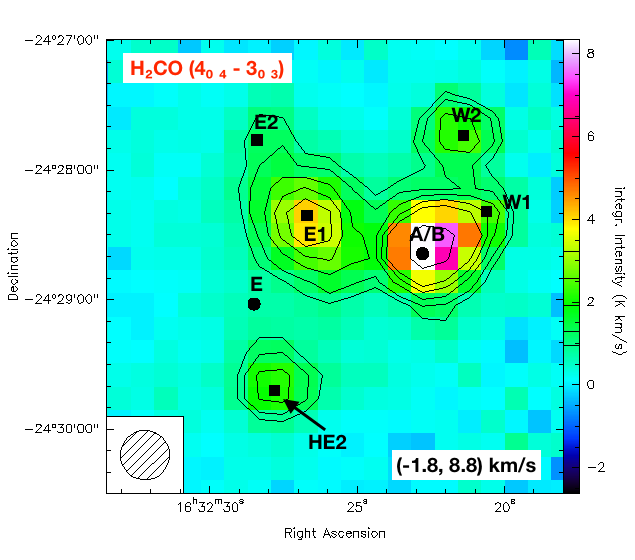}}
    \subfigure[]{\includegraphics[width=0.42\textwidth, trim={0 0.65cm 0 1.18cm},clip]{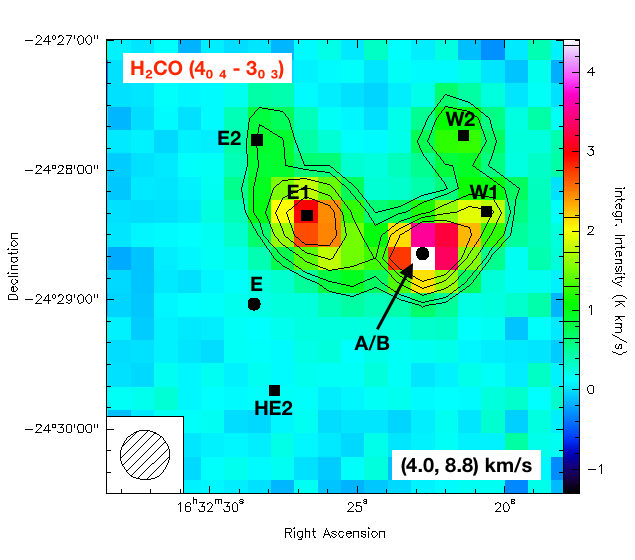}}
    \subfigure[]{\includegraphics[width=0.42\textwidth, trim={0 0.65cm 0 1.18cm},clip]{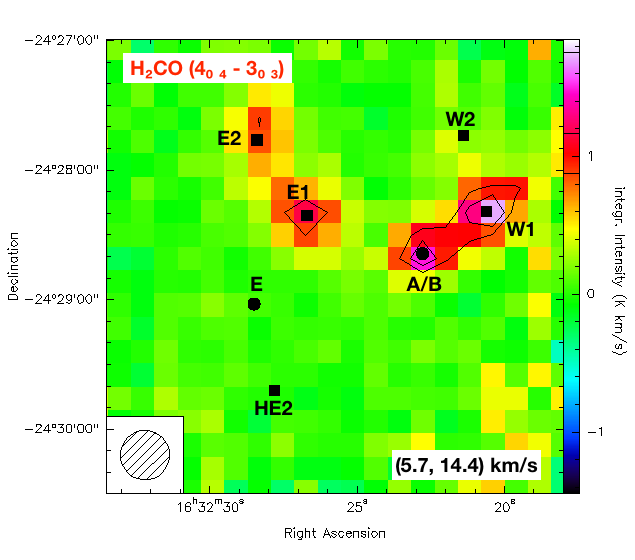}}
    \caption{H$_2$CO ($4_{0,4} - 3_{0,3} $) transition at 290623.405\,MHz.}
    \label{fig:29}
\end{figure*}

\begin{figure*}[ht]
	\centering
    \includegraphics[width=0.42\textwidth, trim={0 0.65cm 0 1.18cm},clip]{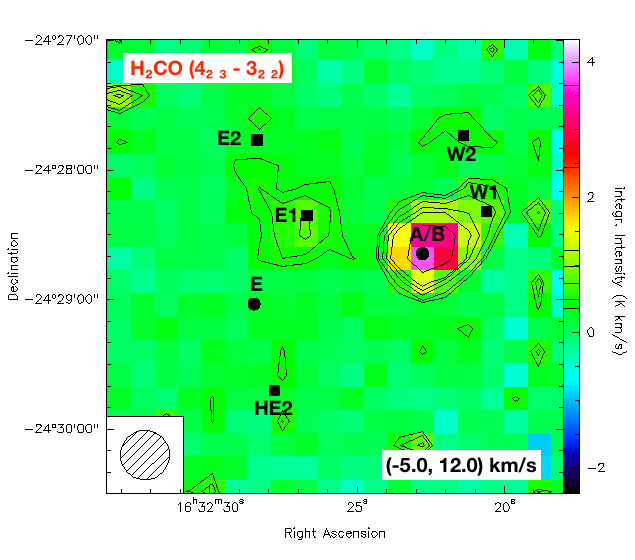}
    \caption{H$_2$CO ($4_{2,3} - 3_{2,2} $) transition at 291237.766\,MHz.}
    \label{fig:31}
\end{figure*}

\begin{figure*}[ht]
	\centering
    \subfigure[]{\includegraphics[width=0.40\textwidth, trim={0 0.65cm 0 1.18cm},clip]{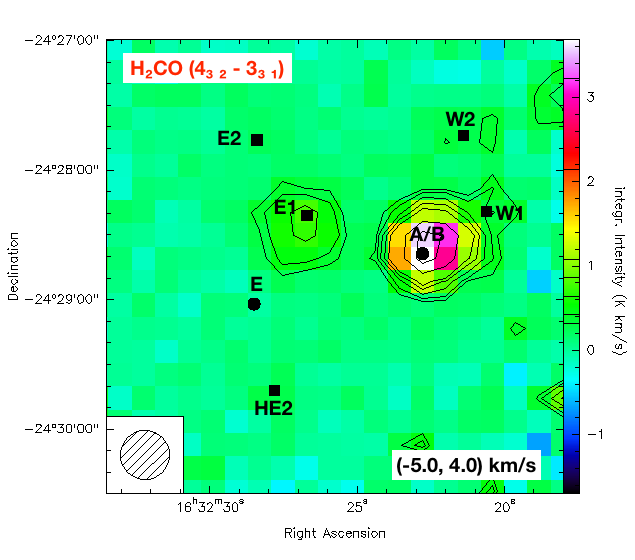}}
    \subfigure[]{\includegraphics[width=0.40\textwidth, trim={0 0.65cm 0 1.18cm},clip]{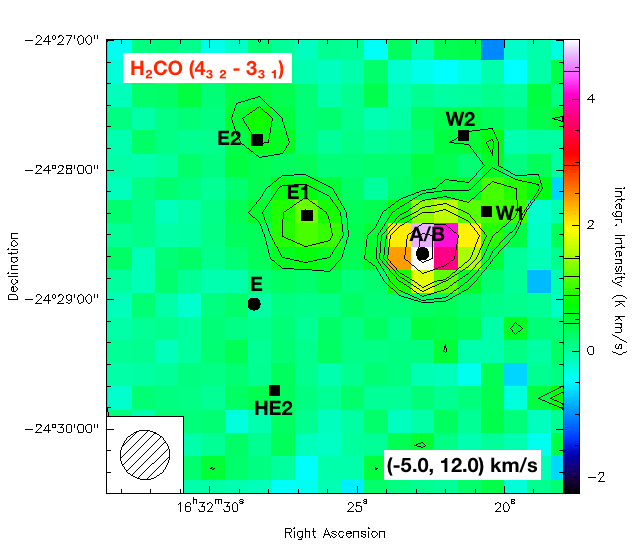}}
    \caption{H$_2$CO ($4_{3,2} - 3_{3,1} $) and H$_2$CO ($4_{3,1} - 3_{3,0} $) transitions at 291380.442\,MHz and 291384.361\,MHz. The velocity scale is calculated based on a rest frequency of 291380.442\,MHz}
    \label{fig:32}
\end{figure*}

\begin{figure*}[ht]
	\centering
    \subfigure[]{\includegraphics[width=0.40\textwidth, trim={0 0.65cm 5.4cm 1.18cm},clip]{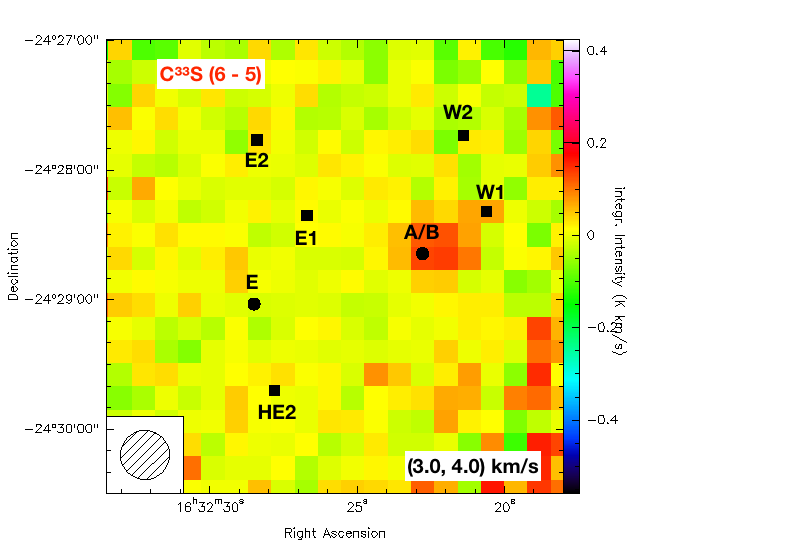}}
    \subfigure[]{\includegraphics[width=0.40\textwidth, trim={0 0.65cm 0 1.18cm},clip]{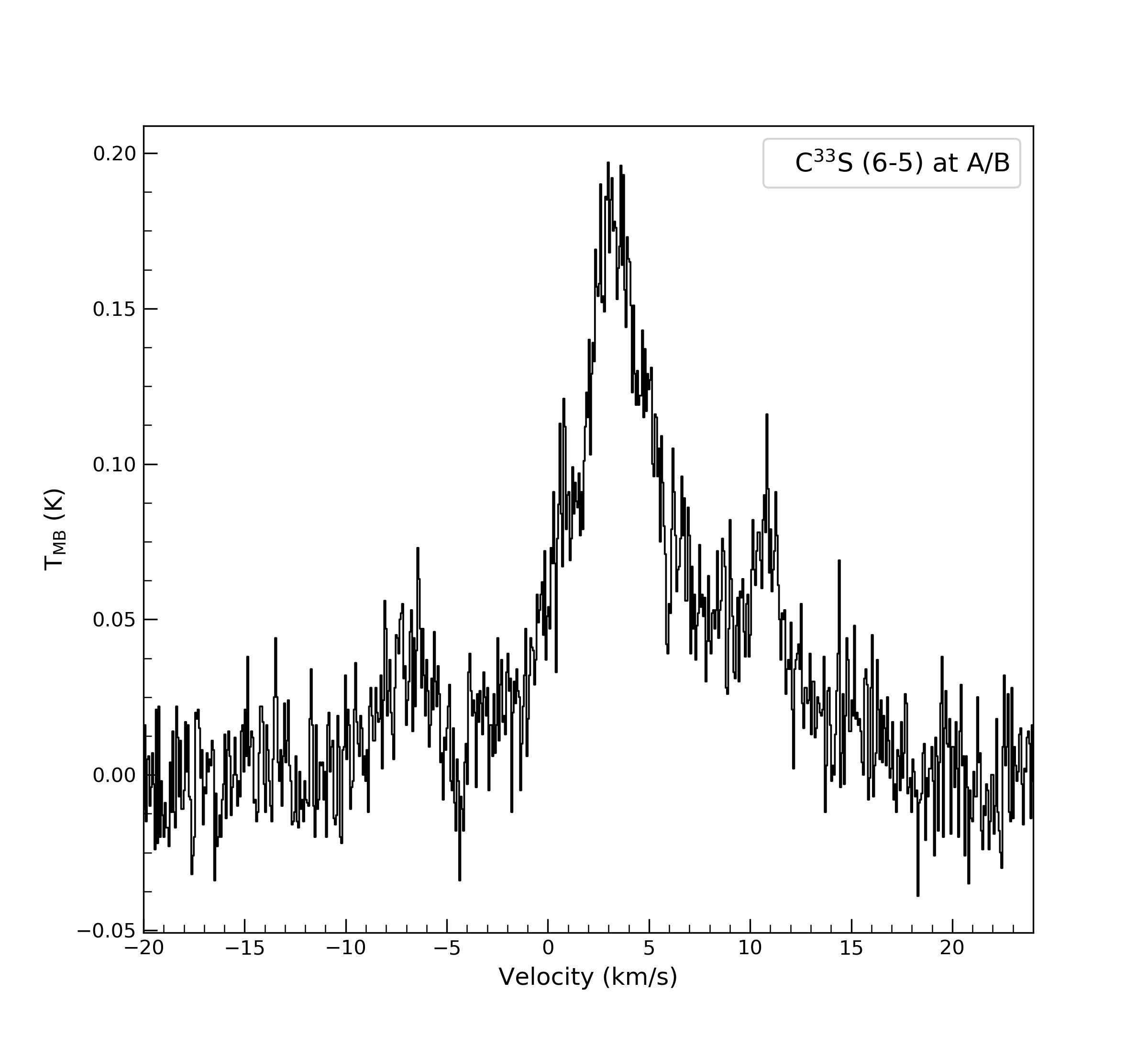}}
    \caption{(a) C$^{33}$S ($6-5$) transition at 291485.935\,MHz. (b) Averaged spectrum of this transition in a $\SI{10}{\arcsecond}$ radius at the position of IRAS\,16293 A/B, including data from LAsMA and FLASH+.}
    \label{fig:18}
\end{figure*}

\begin{figure*}[ht]
	\centering
    \subfigure[]{\includegraphics[width=0.40\textwidth, trim={0 0.65cm 5.4cm 1.18cm},clip]{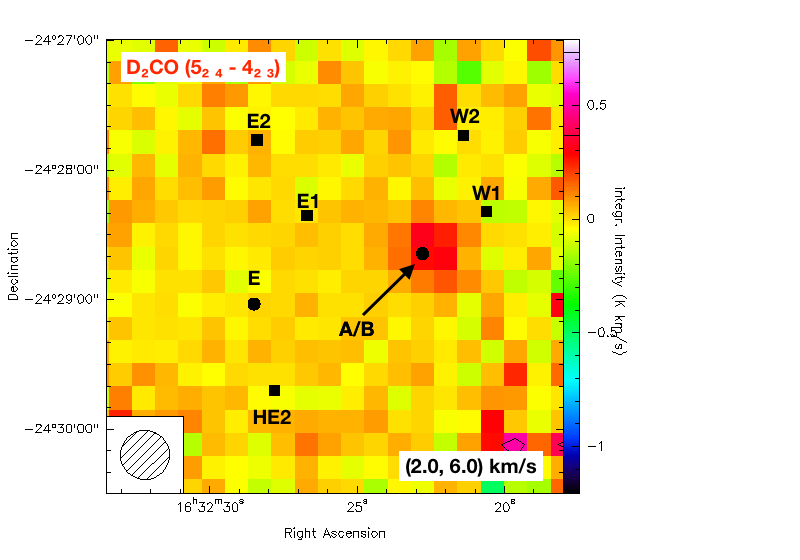}}
    \subfigure[]{\includegraphics[width=0.40\textwidth, trim={0 0.65cm 0 1.18cm},clip]{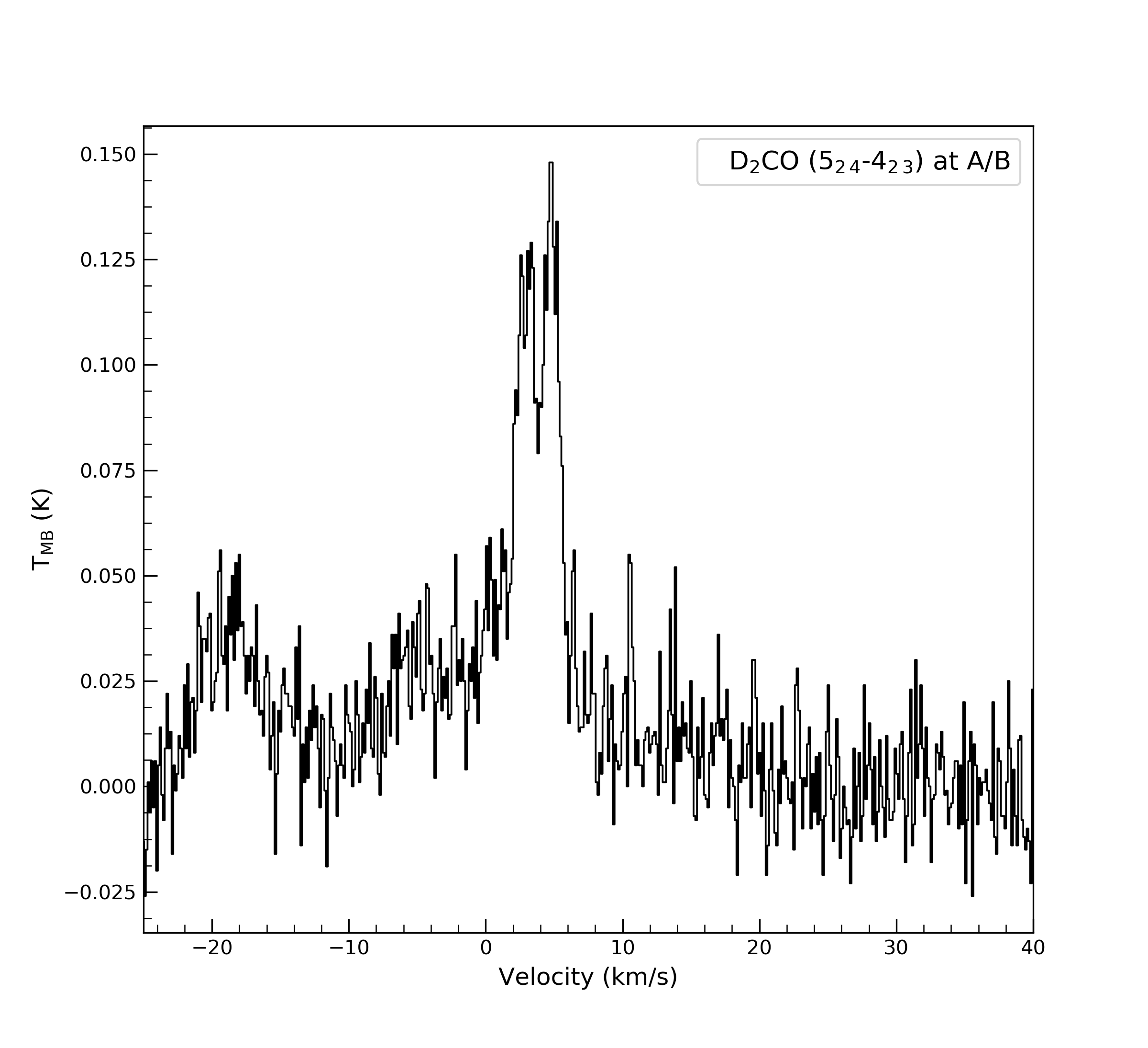}}
    \caption{(a) D$_2$CO $(5_{2, 4} - 4_{2, 3})$) transition at 291745.747\,MHz.  (b) Averaged spectrum of this transition in a $\SI{10}{\arcsecond}$ radius at the position of IRAS\,16293 A/B, including data from LAsMA and FLASH+.}
    \label{fig:63}
\end{figure*}

\begin{figure*}[ht]
	\centering
    \subfigure[]{\includegraphics[width=0.42\textwidth, trim={0 0.65cm 5.4cm 1.18cm},clip]{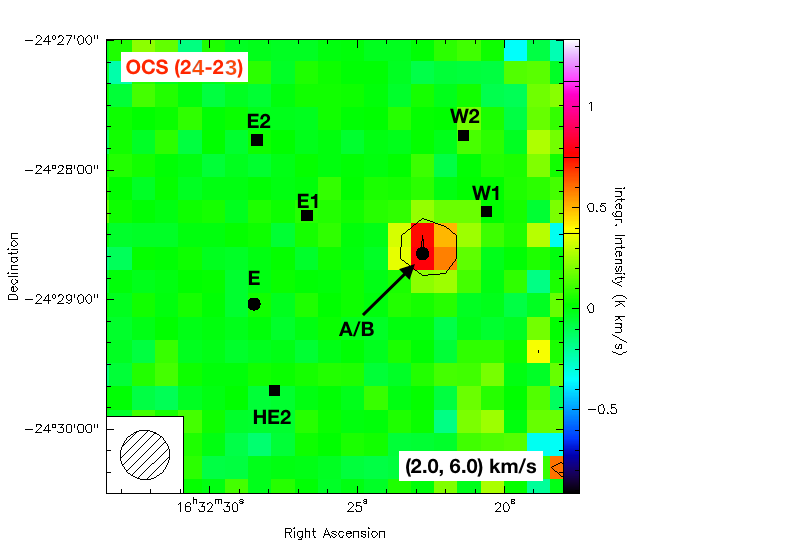}}
    \subfigure[]{\includegraphics[width=0.42\textwidth, trim={0 0.65cm 0 1.18cm},clip]{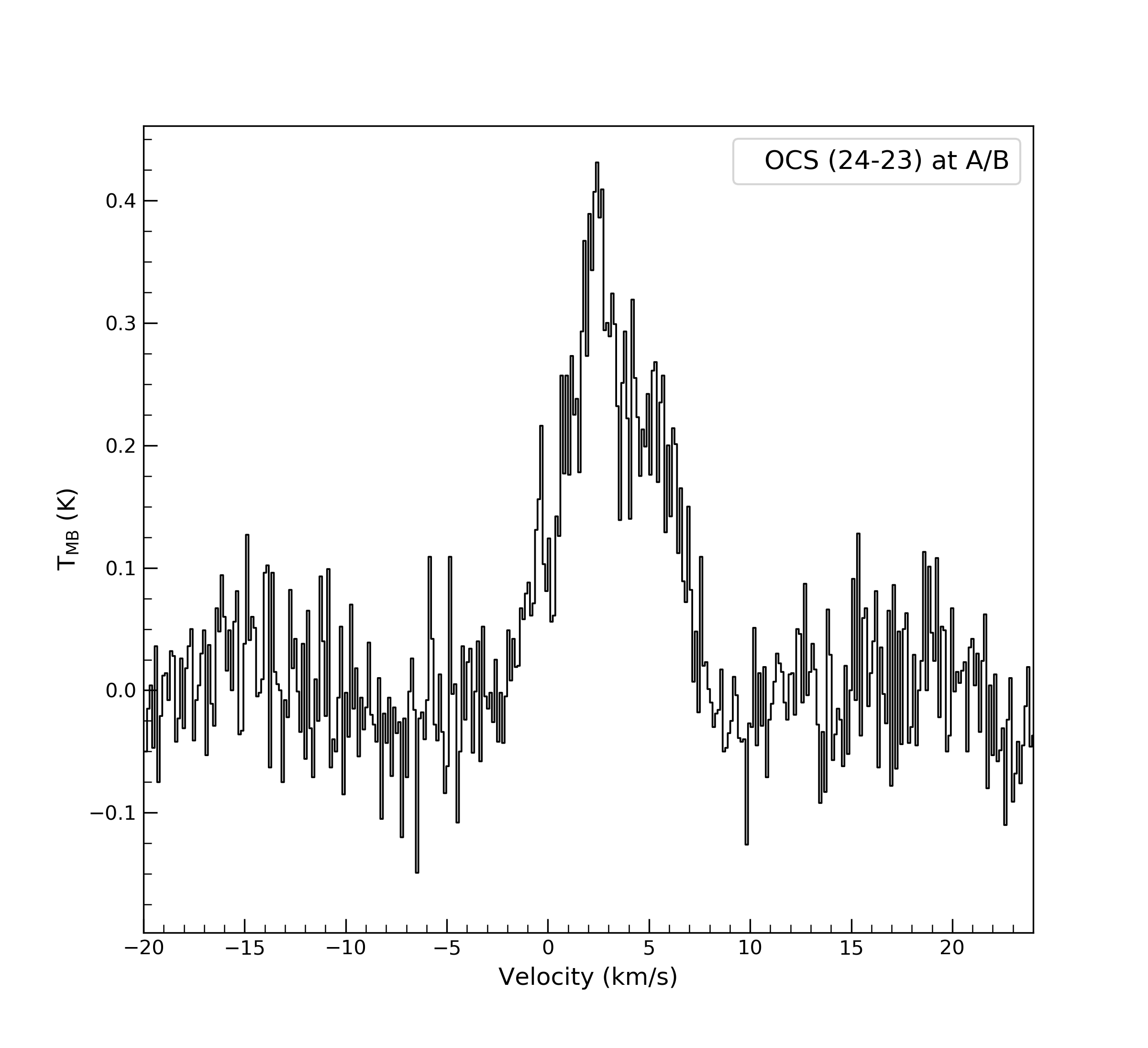}}
    \caption{(a) OCS $(24 - 23)$ transition at 291839.653\,MHz. Additional contours are drawn at 1$\sigma$ and 2$\sigma$. (b) Averaged spectrum of this transition in a $\SI{10}{\arcsecond}$ radius at the position of IRAS\,16293 A/B.}
    \label{fig:67}
\end{figure*}

\begin{figure*}[ht]
	\centering
    \includegraphics[width=0.42\textwidth, trim={0 0.65cm 0 1.18cm},clip]{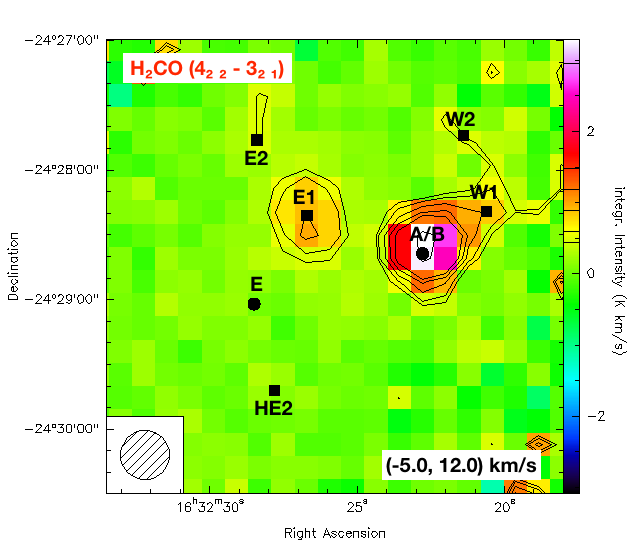}
    \caption{H$_2$CO ($4_{2,2} - 3_{2,1} $) transition at 291948.067\,MHz.}
    \label{fig:30}
\end{figure*}

\begin{figure*}[ht]
	\centering
    \includegraphics[width=0.42\textwidth, trim={0 0.65cm 5.4cm 1.18cm},clip]{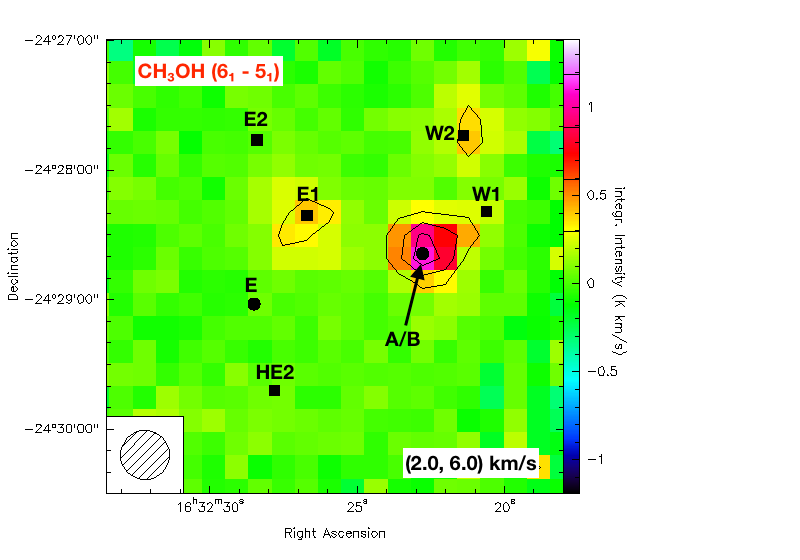}
    \caption{CH$_3$OH-A$^{-}$ ($6_{1}- 5_{1}$) transition at 292672.889\,MHz. Additional contours are drawn at 1$\sigma$ and 2$\sigma$.}
    \label{fig:8}
\end{figure*}

\begin{figure*}[ht]
	\centering
    \subfigure[]{\includegraphics[width=0.42\textwidth, trim={0 0.3cm 0 0.2cm},clip]{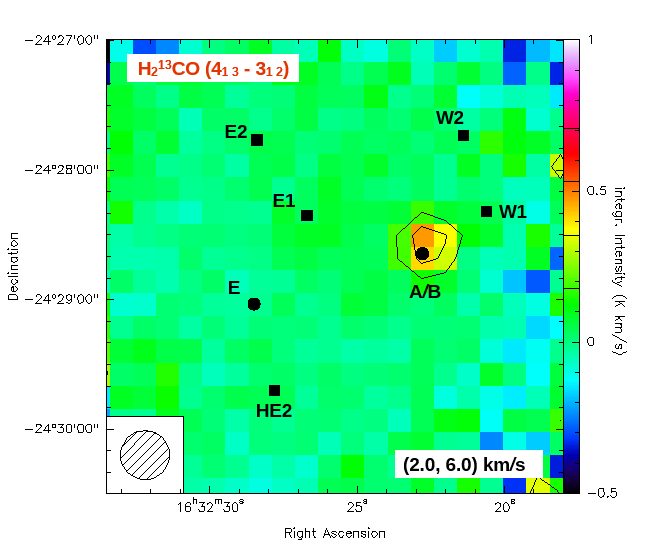}}
    \subfigure[]{\includegraphics[width=0.42\textwidth, trim={0 0.65cm 0 1.18cm},clip]{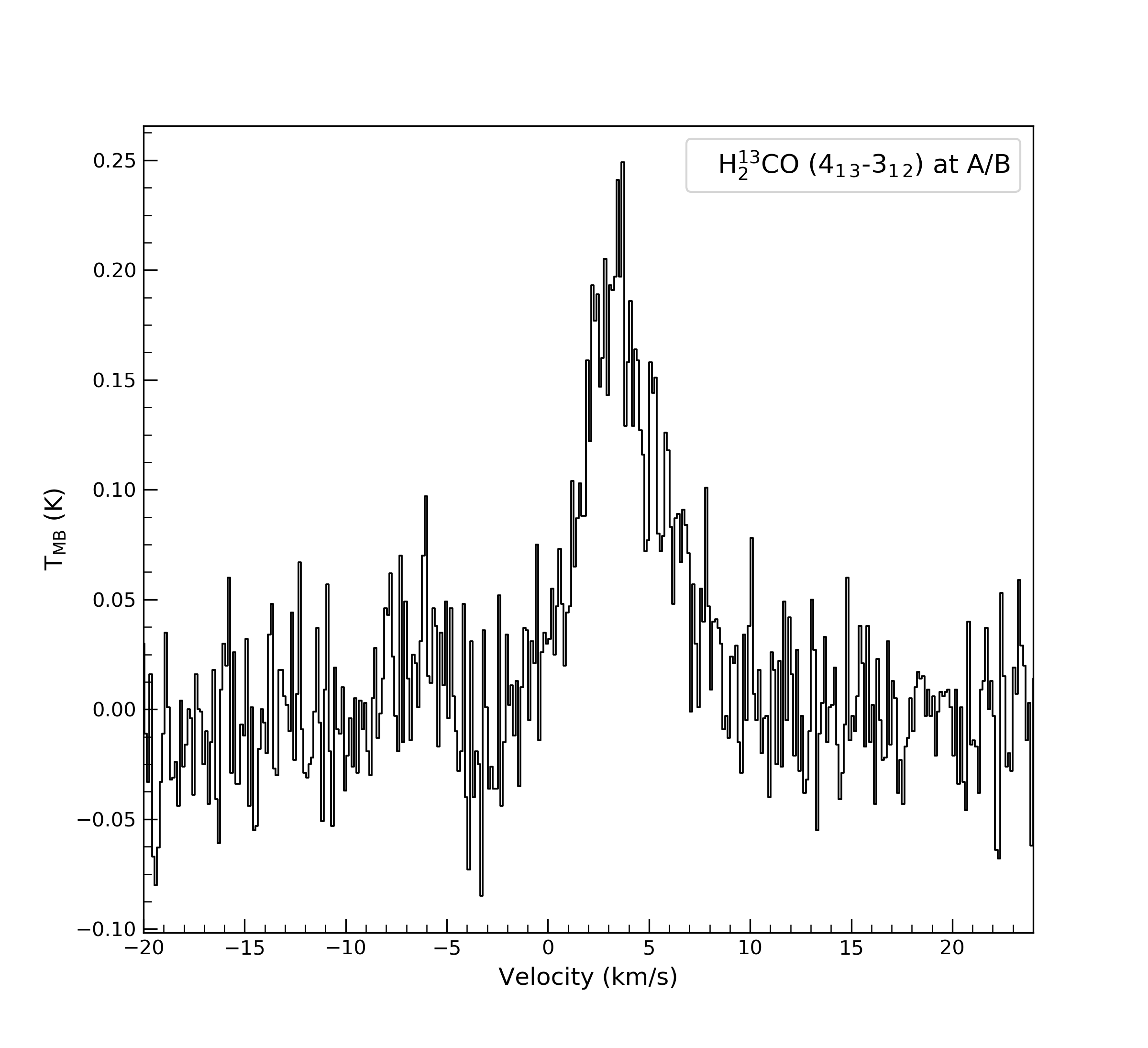}}
    \caption{(a) H$_2^{13}$CO ($4_{1,3} - 3_{1,2} $) transition at 293126.515\,MHz. Additional contours are drawn at 1$\sigma$ and 2$\sigma$. (b) Averaged spectrum of this transition in a $\SI{10}{\arcsecond}$ radius at the position of IRAS\,16293 A/B.}
    \label{fig:35}
\end{figure*}

\begin{figure*}[ht]
	\centering
    \includegraphics[width=0.42\textwidth, trim={0 0.65cm 0 1.18cm},clip]{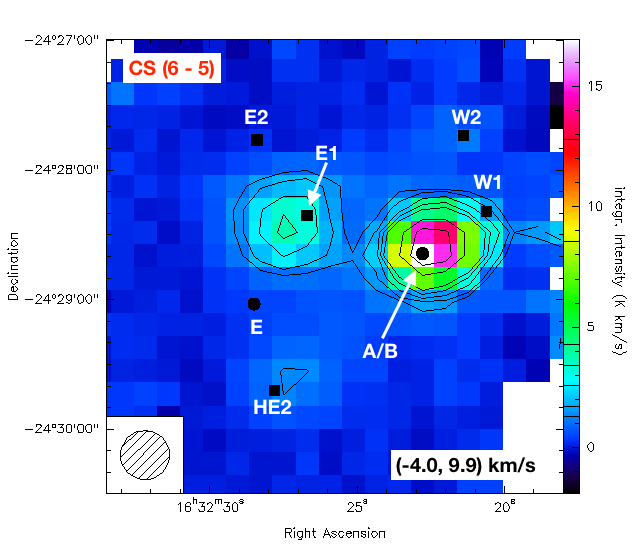}
    \caption{CS ($6 - 5$) transition at 293912.086\,MHz.}
    \label{fig:16}
\end{figure*}

\begin{figure*}[ht]
	\centering
	\subfigure[]{\includegraphics[width=0.42\textwidth, trim={0 0.65cm 0 1.18cm},clip]{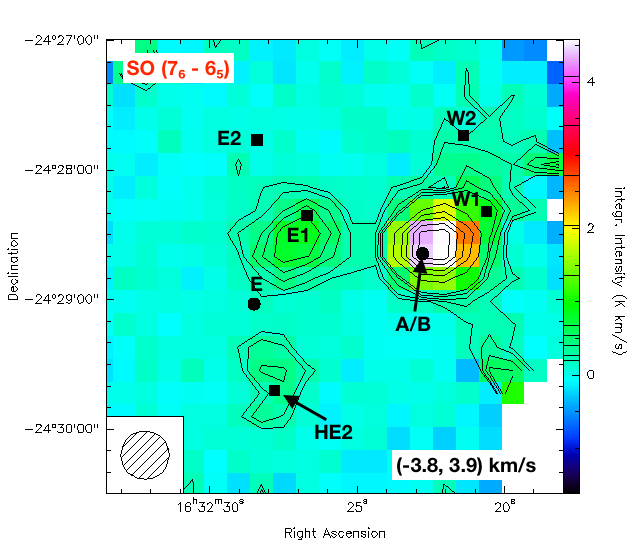}}
    \subfigure[]{\includegraphics[width=0.42\textwidth, trim={0 0.65cm 0 1.18cm},clip]{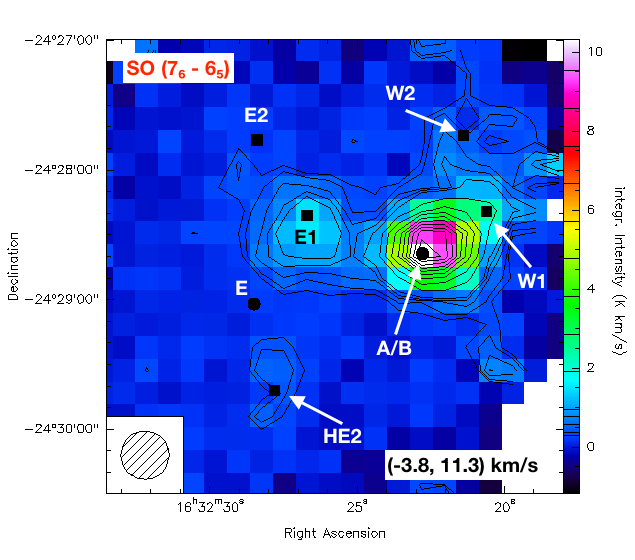}}
    \subfigure[]{\includegraphics[width=0.42\textwidth, trim={0 0.65cm 0 1.18cm},clip]{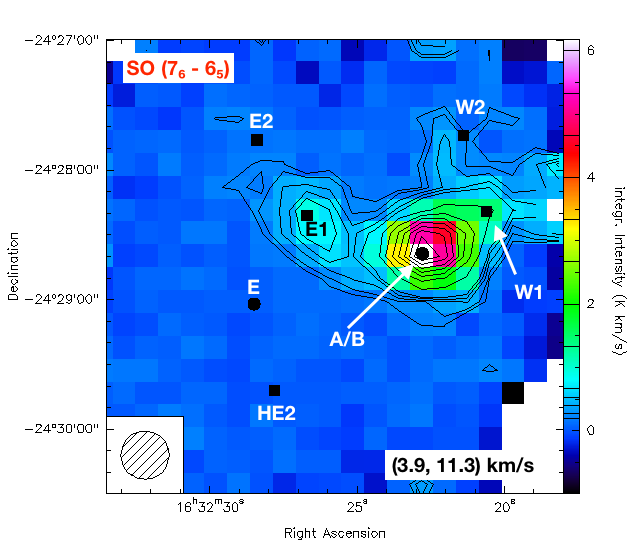}}
    \caption{SO ($7_6 - 6_5$) transition at 296550.064\,MHz.}
    \label{fig:43}
\end{figure*}

\begin{figure*}[ht]
	\centering
    \subfigure[]{\includegraphics[width=0.42\textwidth, trim={0 0.65cm 5.4cm 1.18cm},clip]{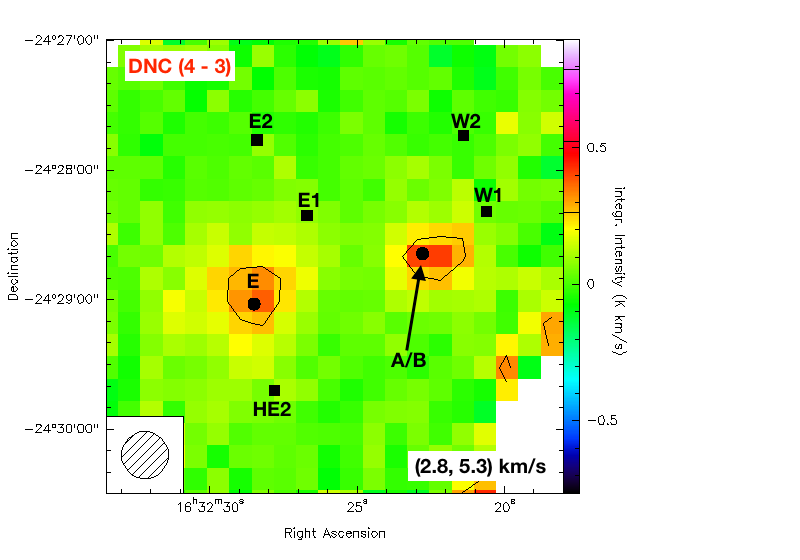}}
    \subfigure[]{\includegraphics[width=0.42\textwidth, trim={0 0.65cm 0 1.18cm},clip]{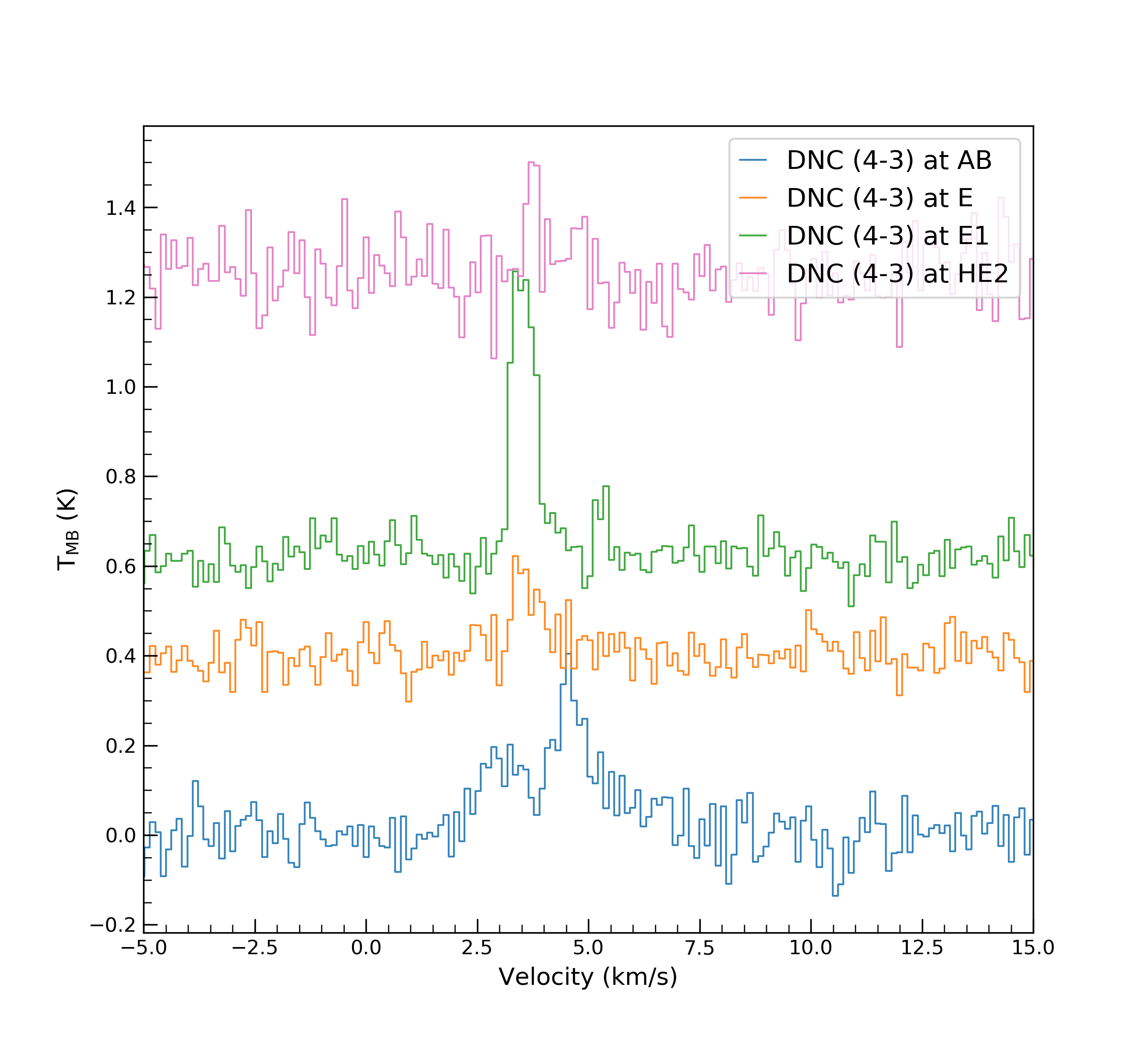}}
    \caption{DNC ($4 - 3$) transition at 305206.219\,MHz.Additional contours are drawn at 1$\sigma$. (b) Averaged spectra of this transition in a $\SI{10}{\arcsecond}$ radius at the positions with visible emission.}
    \label{fig:20_2}
\end{figure*}

\begin{figure*}[ht]
	\centering
    \includegraphics[width=0.42\textwidth, trim={0 0.65cm 0 1.18cm},clip]{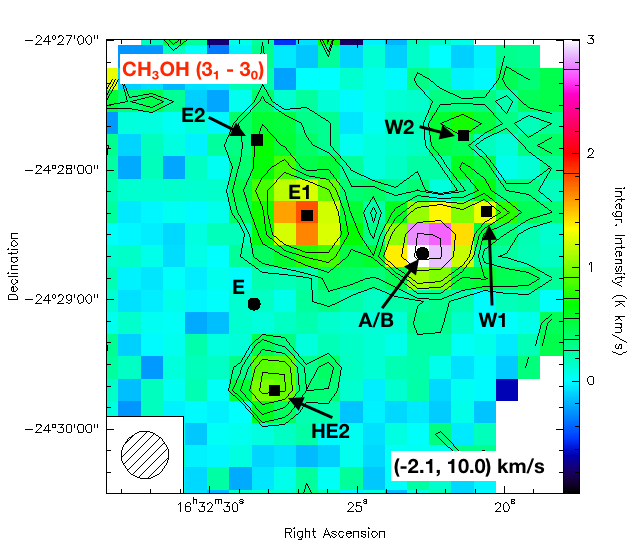}
    \caption{CH$_3$OH-A$^{-+}$ ($3_{1} - 3_{0}$) transition at 305473.491\,MHz.}
    \label{fig:3}
\end{figure*}

\begin{figure*}[ht]
	\centering
    \includegraphics[width=0.42\textwidth, trim={0 0.65cm 0 1.18cm},clip]{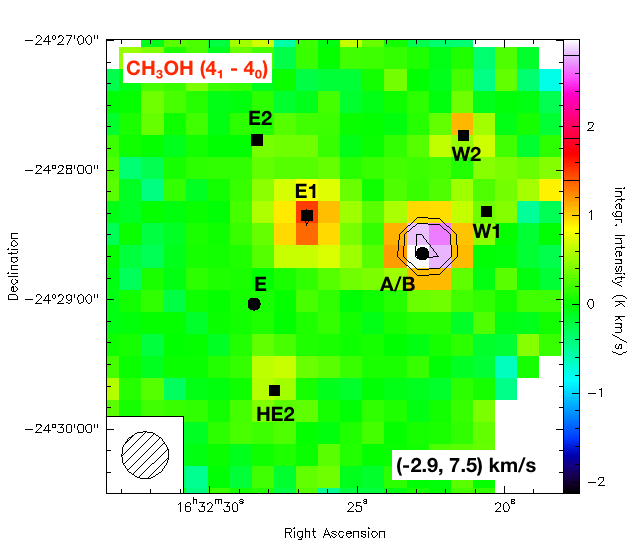}
    \caption{CH$_3$OH-A$^{-+}$ ($4_{1} - 4_{0}$) transition at 307165.924\,MHz.}
    \label{fig:4}
\end{figure*}

\begin{figure*}[ht]
	\centering
    \includegraphics[width=0.42\textwidth, trim={0 0.65cm 0 1.18cm},clip]{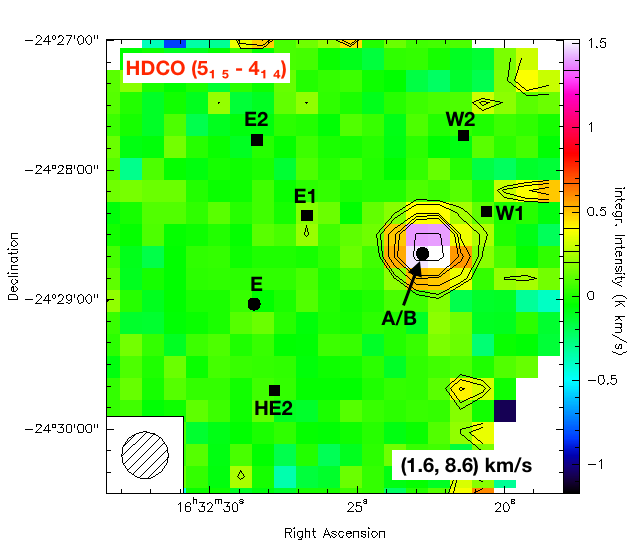}
    \caption{HDCO ($5_{1,5} - 4_{1,4} $) transition at 308418.200\,MHz.}
    \label{fig:28}
\end{figure*}

\begin{figure*}[ht]
	\centering
    \includegraphics[width=0.42\textwidth, trim={0 0.65cm 0 1.18cm},clip]{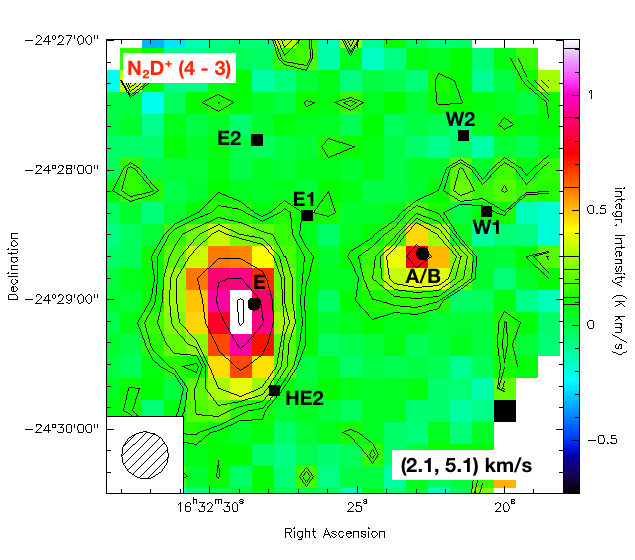}
    \caption{N$_2$D$^+$($4 - 3 $) transition at 308422.267\,MHz.}
    \label{fig:37}
\end{figure*}

\begin{figure*}[ht]
	\centering
    \includegraphics[width=0.42\textwidth, trim={0 0.65cm 0 1.18cm},clip]{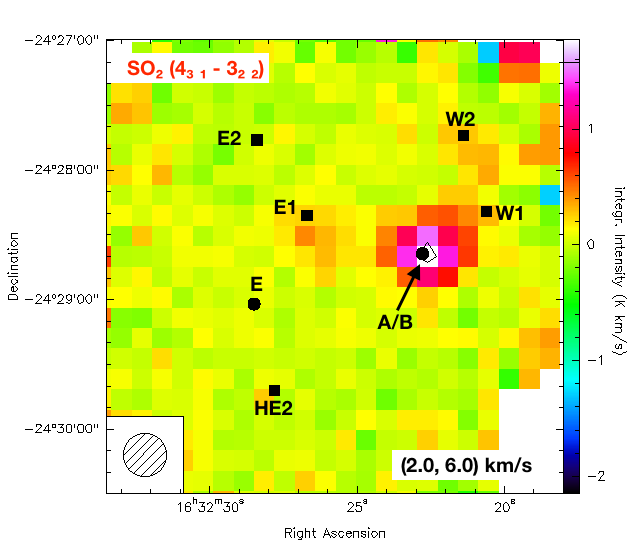}
    \caption{SO$_2$ ($4_{3,1} - 3_{2,2}$) transition at 332505.242\,MHz.}
    \label{fig:48}
\end{figure*}

\begin{figure*}[ht]
	\centering
    \subfigure[]{\includegraphics[width=0.42\textwidth, trim={0 0.65cm 5.4cm 1.18cm},clip]{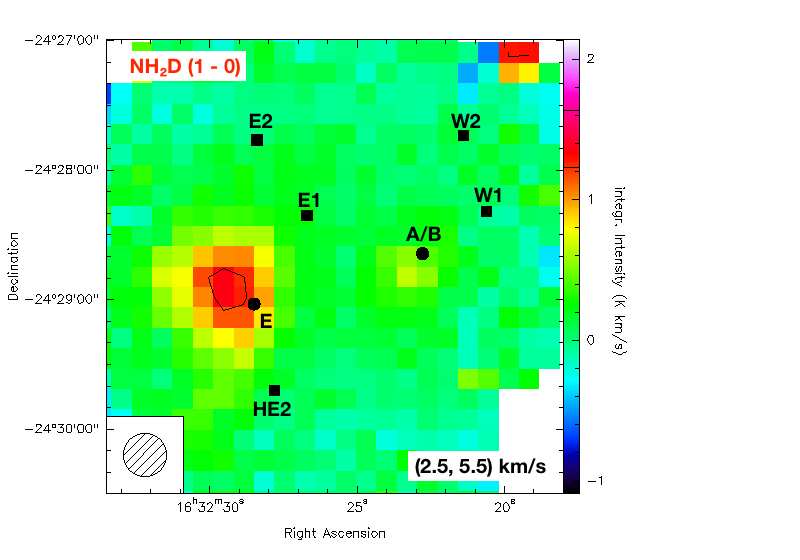}}
    \subfigure[]{\includegraphics[width=0.42\textwidth, trim={0 0.65cm 0 1.18cm},clip]{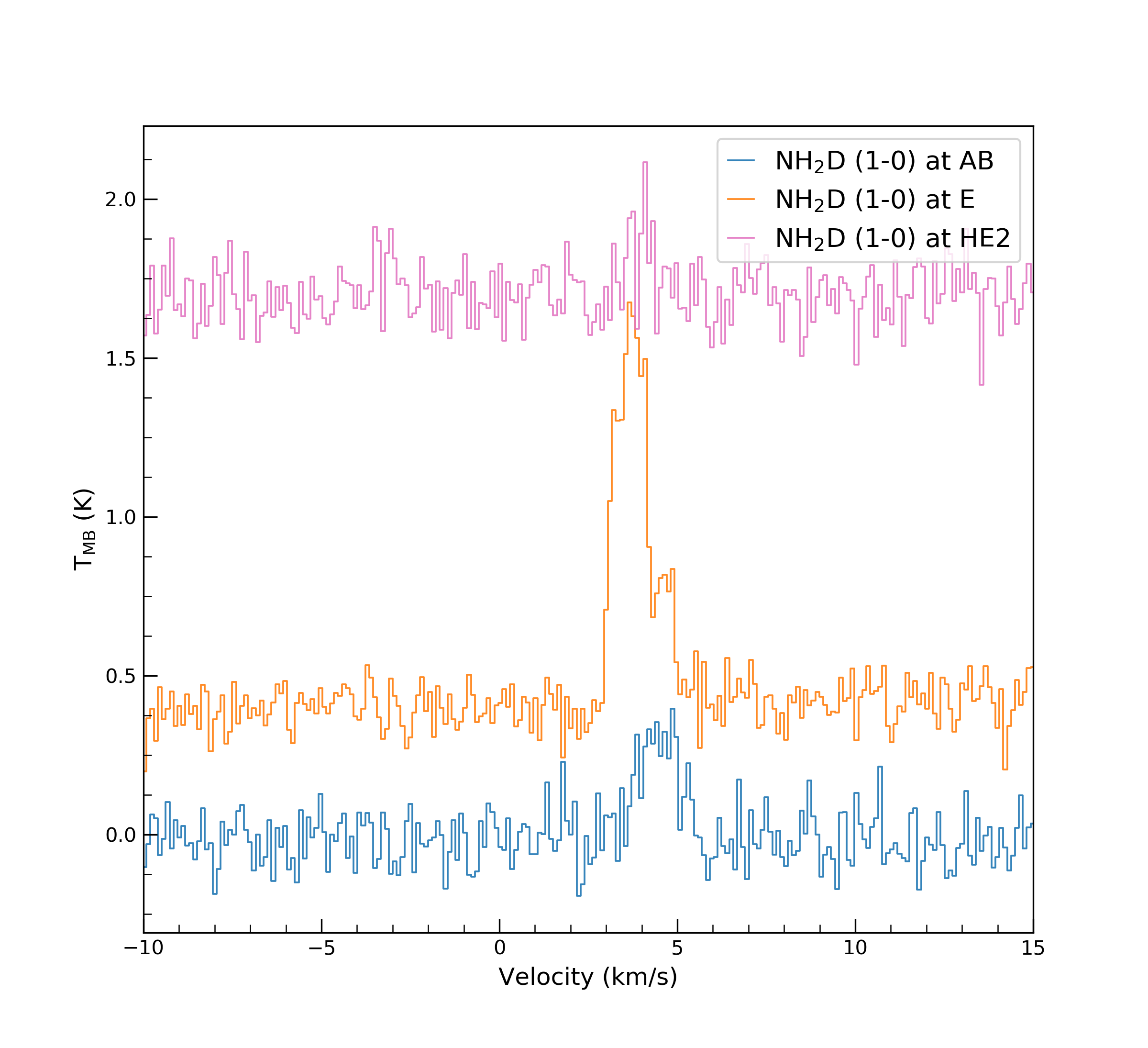}}
    \caption{(a) NH$_2$D ($1_{0, 1, 1, 2} - 0_{0, 0, 1, 1}$) transition at 332781.890\,MHz. Additional contours are drawn at 1$\sigma$. (b) Averaged spectrum of this transition in a $\SI{10}{\arcsecond}$ radius at the positions indicated in the upper right corner.}
    \label{fig:53}
\end{figure*}

\begin{figure*}[ht]
	\centering
    \subfigure[]{\includegraphics[width=0.42\textwidth, trim={0 0.65cm 5.4cm 1.18cm},clip]{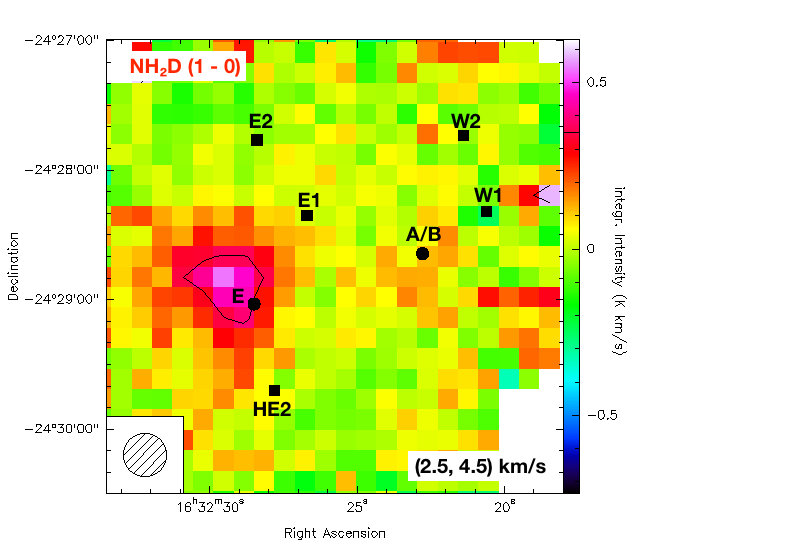}}
    \subfigure[]{\includegraphics[width=0.42\textwidth, trim={0 0.65cm 0 1.18cm},clip]{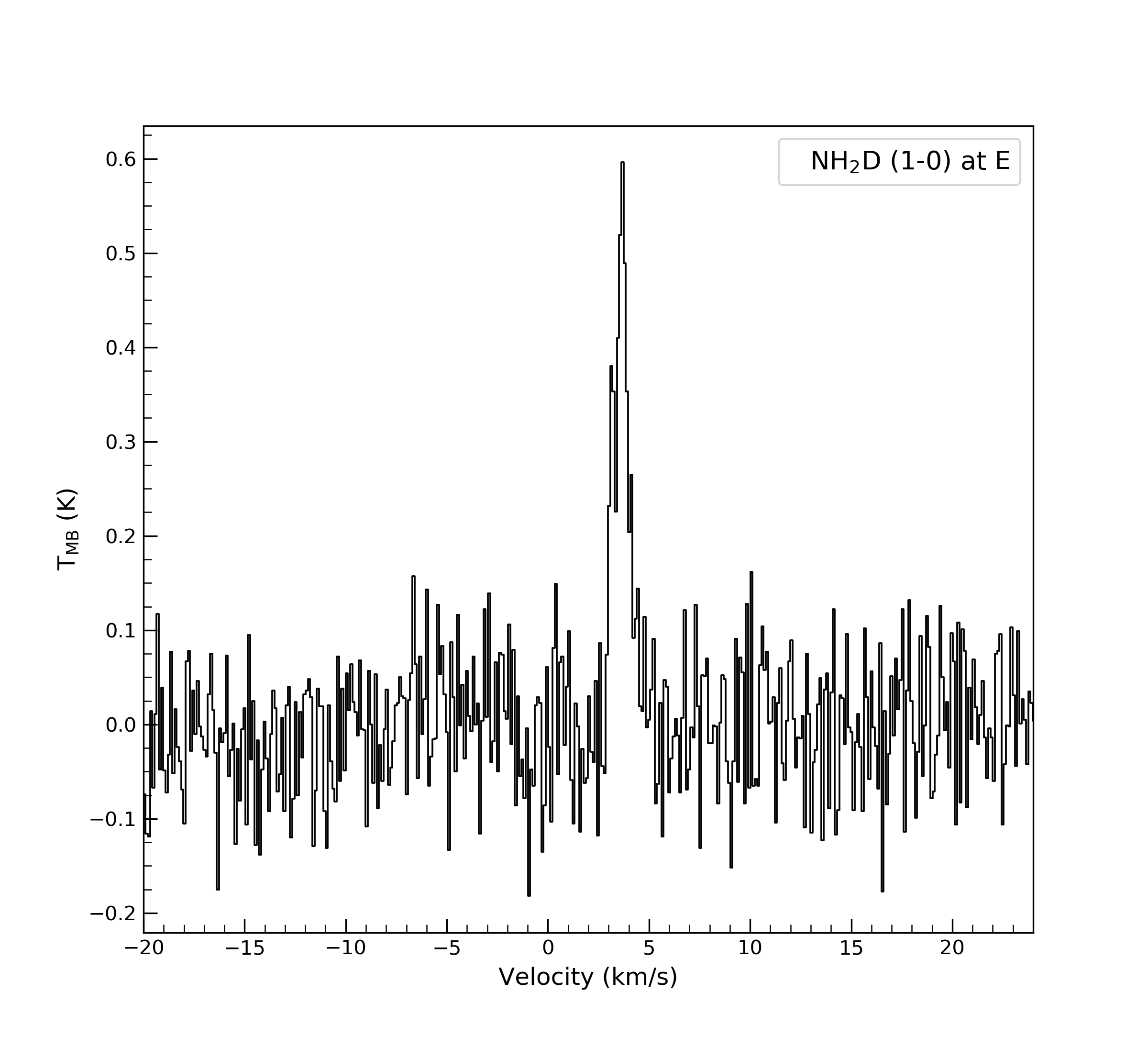}}
    \caption{(a) NH$_2$D ($1_{0, 1, 0, 2} - 0_{0, 0, 0, 1}$) transition at 332822.510\,MHz. Additional contours are drawn at 1$\sigma$. (b) Averaged spectrum of this transition in a $\SI{10}{\arcsecond}$ radius at the position of IRAS\,16293 A/B.}
    \label{fig:54a}
\end{figure*}

\begin{figure*}[ht]
	\centering
    \includegraphics[width=0.42\textwidth, trim={0 0.65cm 5.4cm 1.18cm},clip]{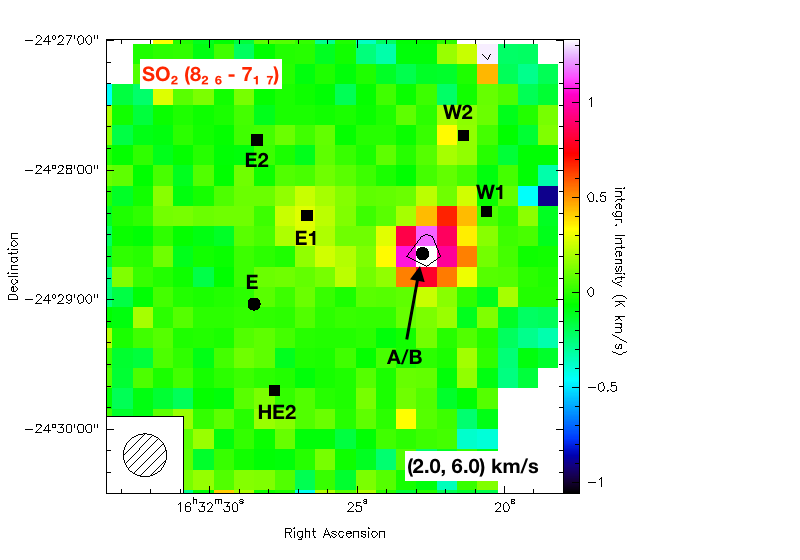}
    \caption{SO$_2$ ($8_{2,6}-7_{1,7}$) transition at 334673.353\,MHz.}
    \label{fig:51}
\end{figure*}

\begin{figure*}[ht]
	\centering
    \subfigure[]{\includegraphics[width=0.42\textwidth, trim={0 0.65cm 5.4cm 1.18cm},clip]{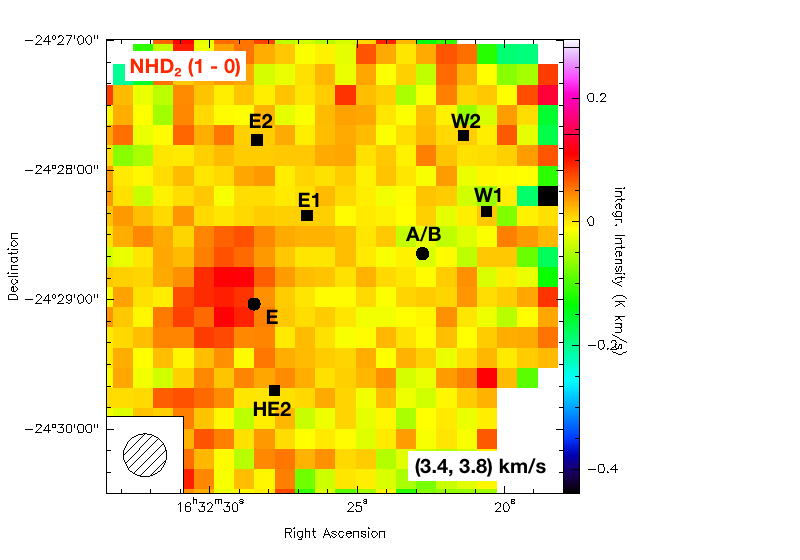}}
    \subfigure[]{\includegraphics[width=0.42\textwidth, trim={0 0.65cm 0 1.18cm},clip]{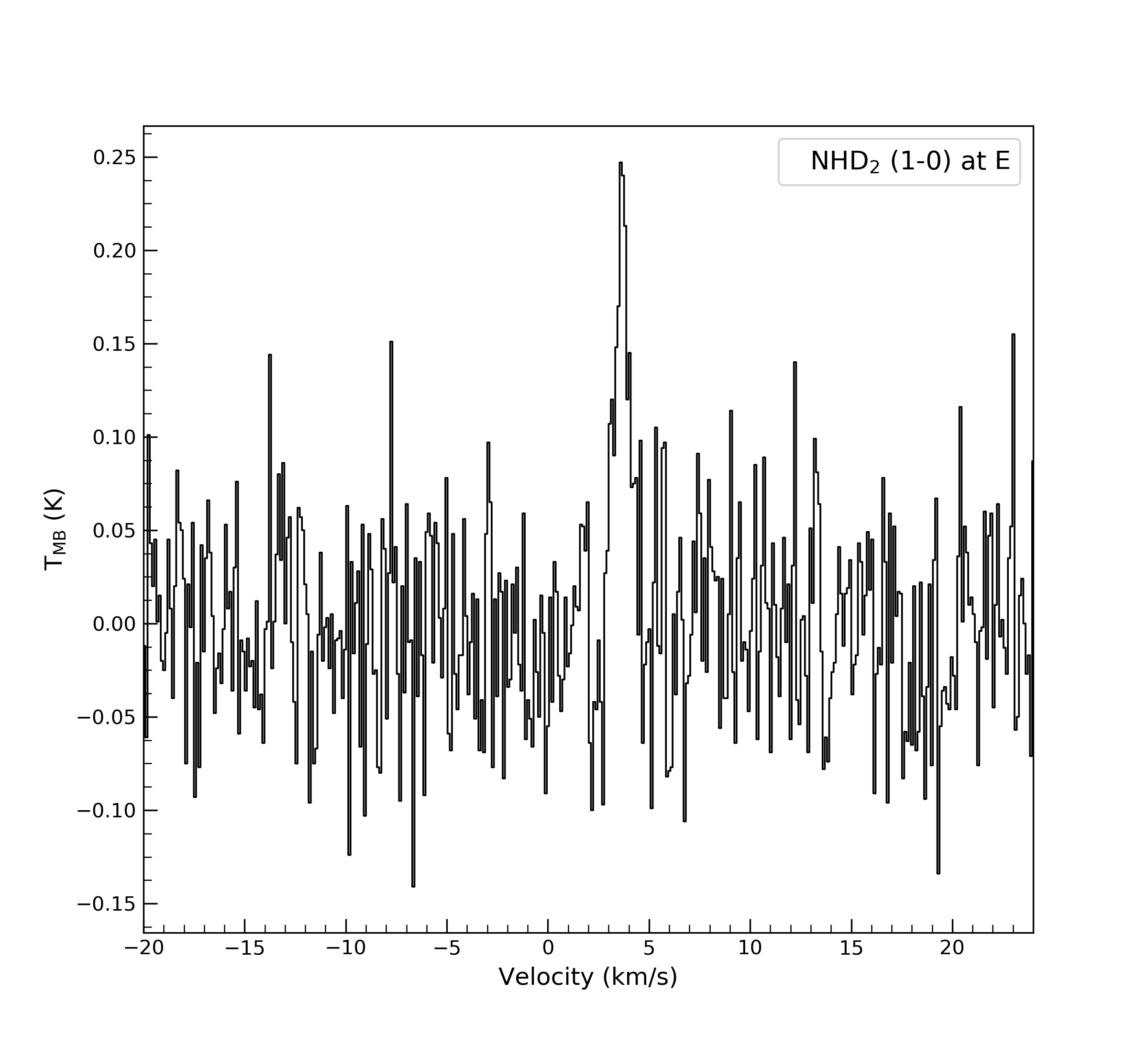}}
    \caption{(a) NHD$_2$ ($1_{1, 1, 1} - 0_{0, 0, 1}$) transition at 335446.321\,MHz. (b) Averaged spectrum of this transition in a $\SI{10}{\arcsecond}$ radius at the position of 16293E.}
    \label{fig:54b}
\end{figure*}

\begin{figure*}[ht]
	\centering
    \subfigure[]{\includegraphics[width=0.42\textwidth, trim={0 0.65cm 5.4cm 1.18cm},clip]{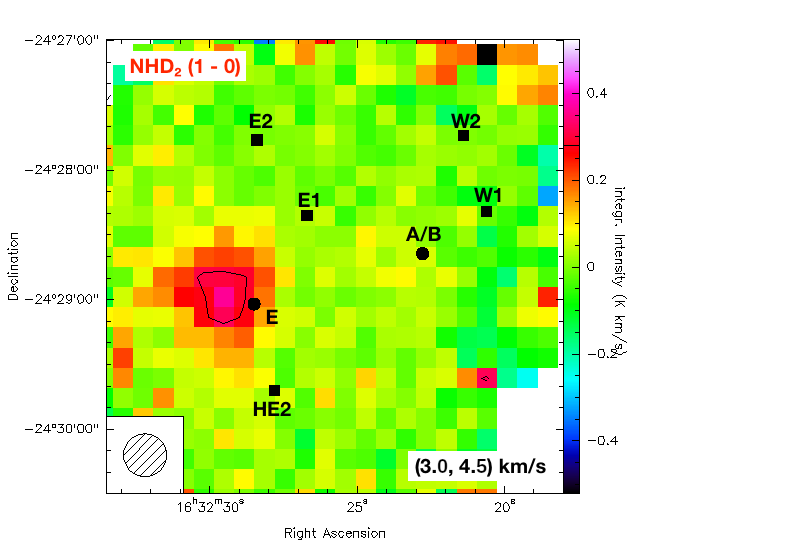}}
    \subfigure[]{\includegraphics[width=0.42\textwidth, trim={0 0.65cm 0 1.18cm},clip]{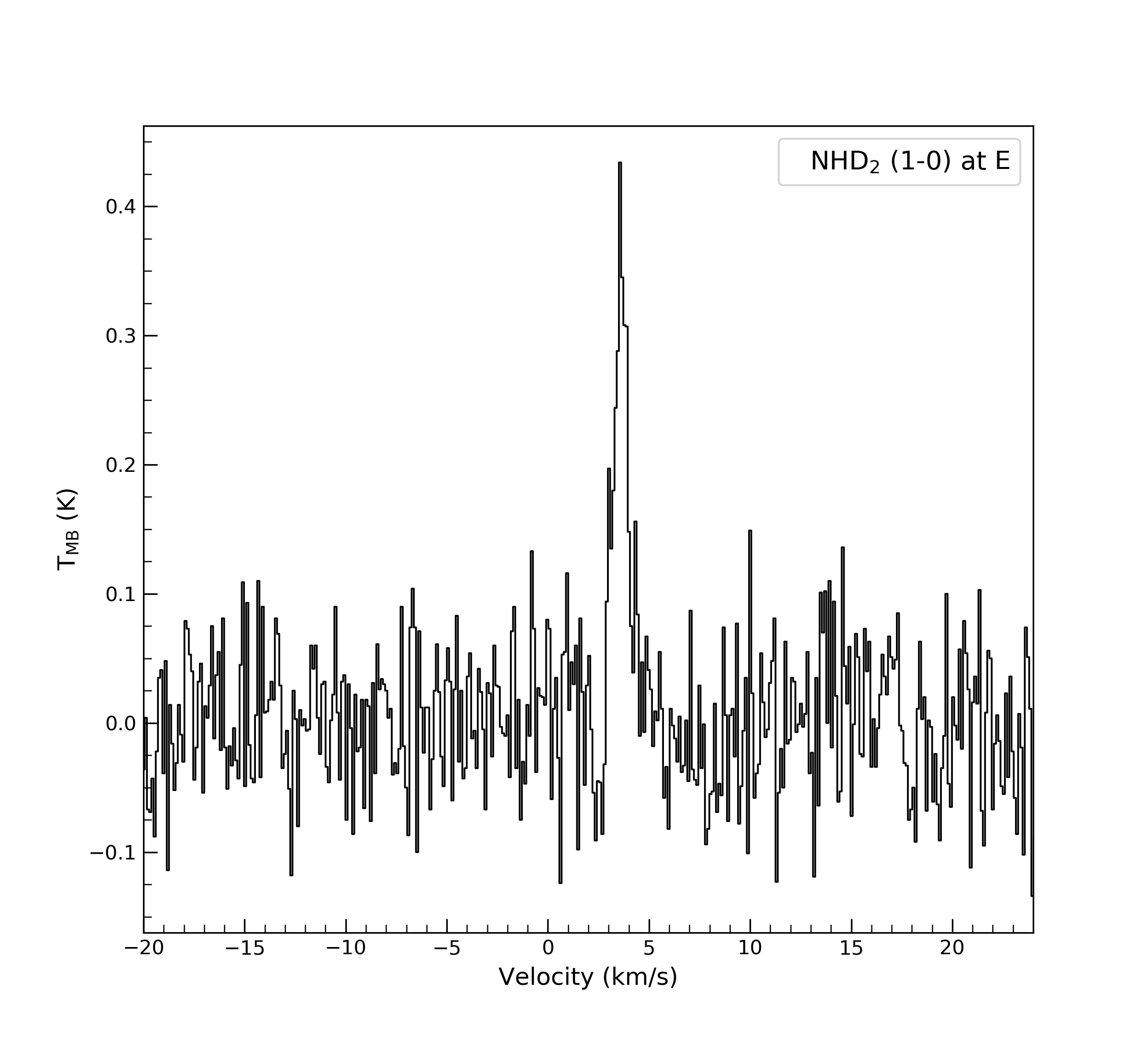}}
    \caption{(a) NHD$_2$ ($1_{0, 1, 0, 2} - 0_{0, 0, 0, 1}$) transition at 335513.793\,MHz. Additional contours are drawn at 1$\sigma$.(b) Averaged spectrum of this transition in a $\SI{10}{\arcsecond}$ radius at the position of 16293E.}
    \label{fig:54c}
\end{figure*}

\begin{figure*}[ht]
	\centering
    \includegraphics[width=0.42\textwidth, trim={0 0.65cm 0 1.18cm},clip]{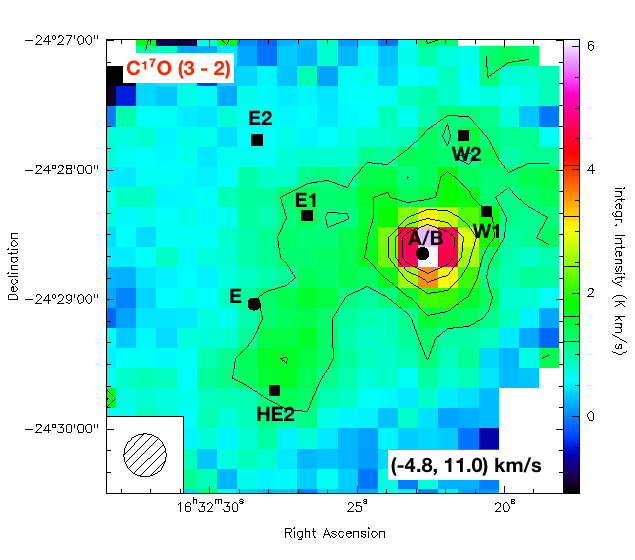}
    \caption{C$^{17}$O ($3 - 2$) transition at 337061.214\,MHz.}
    \label{fig:2}
\end{figure*}

\begin{figure*}[ht]
	\centering
    \includegraphics[width=0.42\textwidth, trim={0 0.65cm 0 1.18cm},clip]{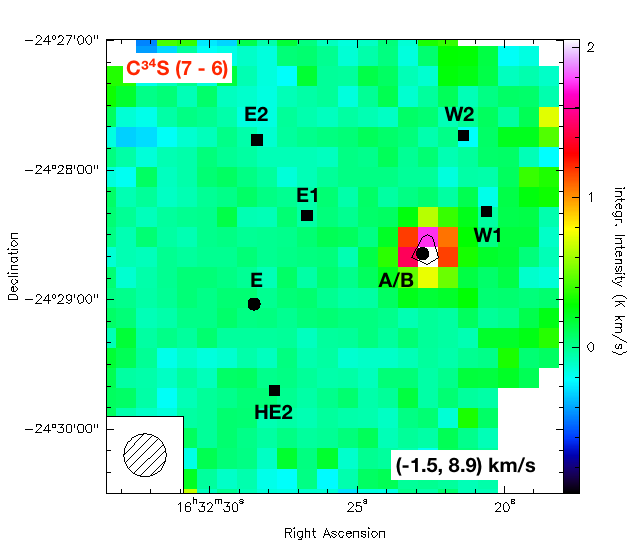}
    \caption{C$^{34}$S ($7 - 6$) transition at 337396.459\,MHz.}
    \label{fig:19}
\end{figure*}

\begin{figure*}[ht]
	\centering
    \subfigure[]{\includegraphics[width=0.42\textwidth, trim={0 0.65cm 5.4cm 1.18cm},clip]{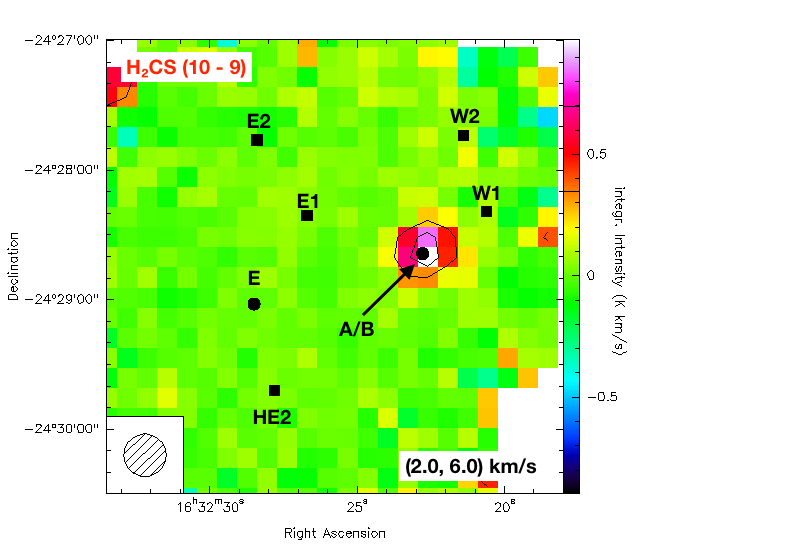}}
    \subfigure[]{\includegraphics[width=0.42\textwidth, trim={0 0.65cm 0 1.18cm},clip]{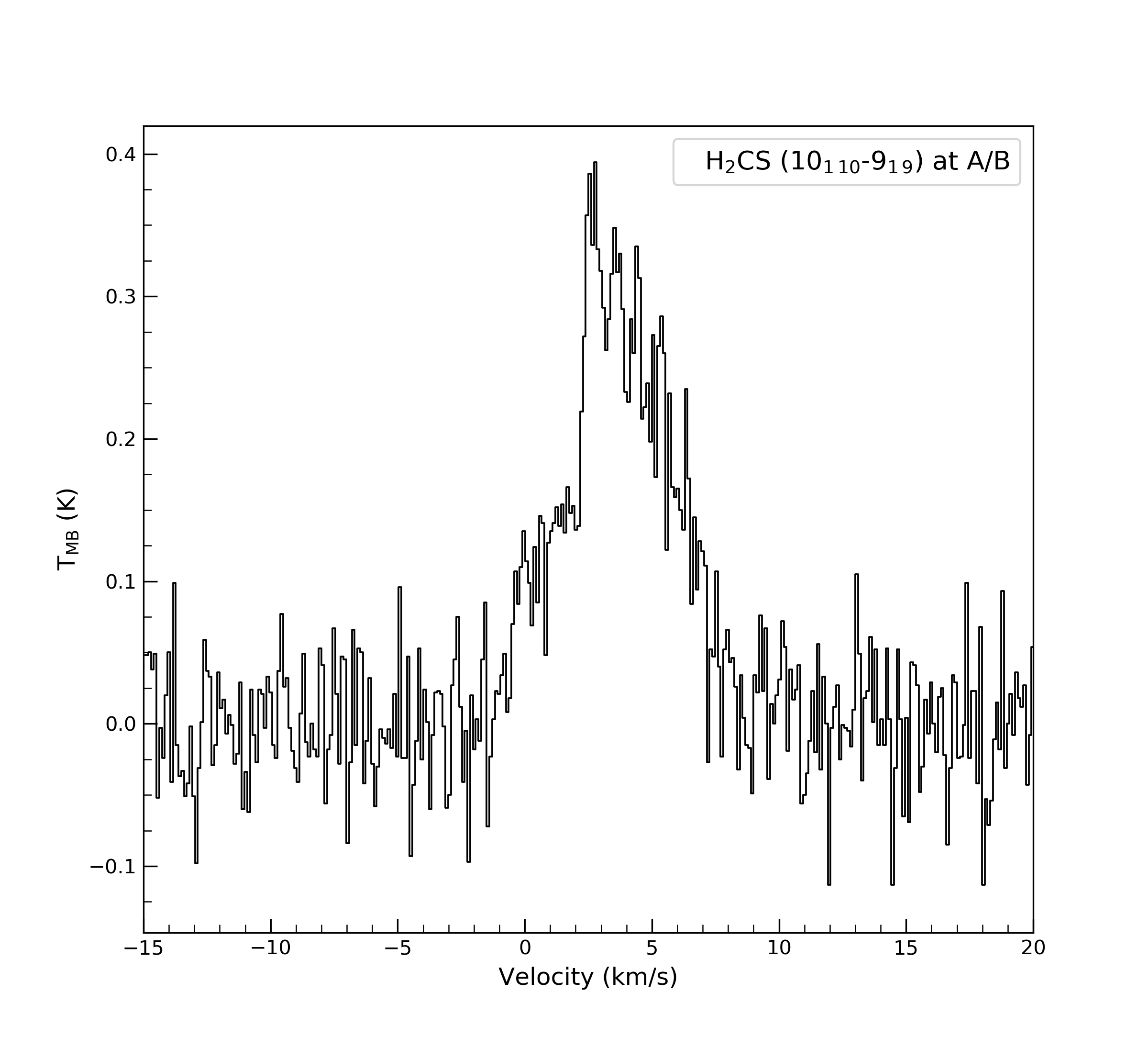}}
    \caption{(a) H$_2$CS $(10_{1, 10} - 9_{1, 9})$) transition at 338083.195\,MHz.  Additional contours are drawn at 1$\sigma$ and 2$\sigma$. (b) Averaged spectrum of this transition in a $\SI{10}{\arcsecond}$ radius at the position of IRAS\,16293 A/B.}
    \label{fig:64}
\end{figure*}

\begin{figure*}[ht]
	\centering
    \includegraphics[width=0.42\textwidth, trim={0 0.65cm 5.4cm 1.18cm},clip]{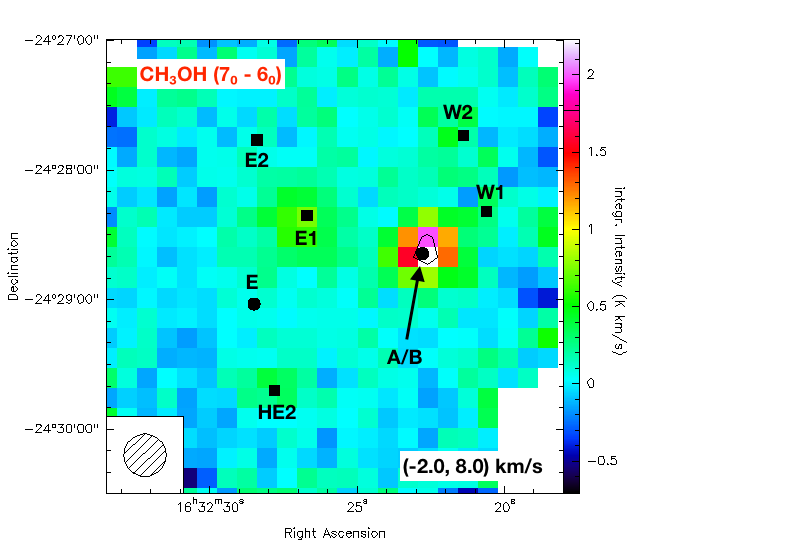}
    \caption{CH$_3$OH-E ($7_{0} - 6_{0}$) transition at 338124.488\,MHz.}
    \label{fig:11}
\end{figure*}

\begin{figure*}[ht]
	\centering
    \subfigure[]{\includegraphics[width=0.42\textwidth, trim={0 0.65cm 5.4cm 1.18cm},clip]{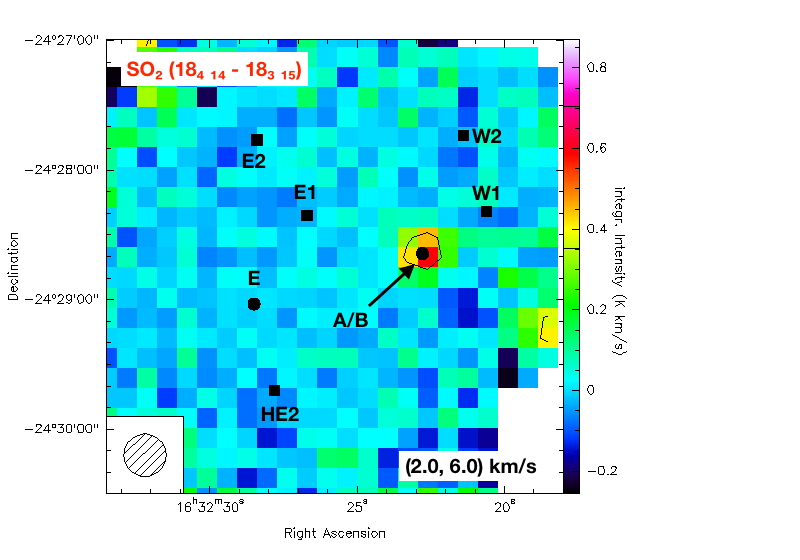}}
    \subfigure[]{\includegraphics[width=0.42\textwidth, trim={0 0.65cm 0 1.18cm},clip]{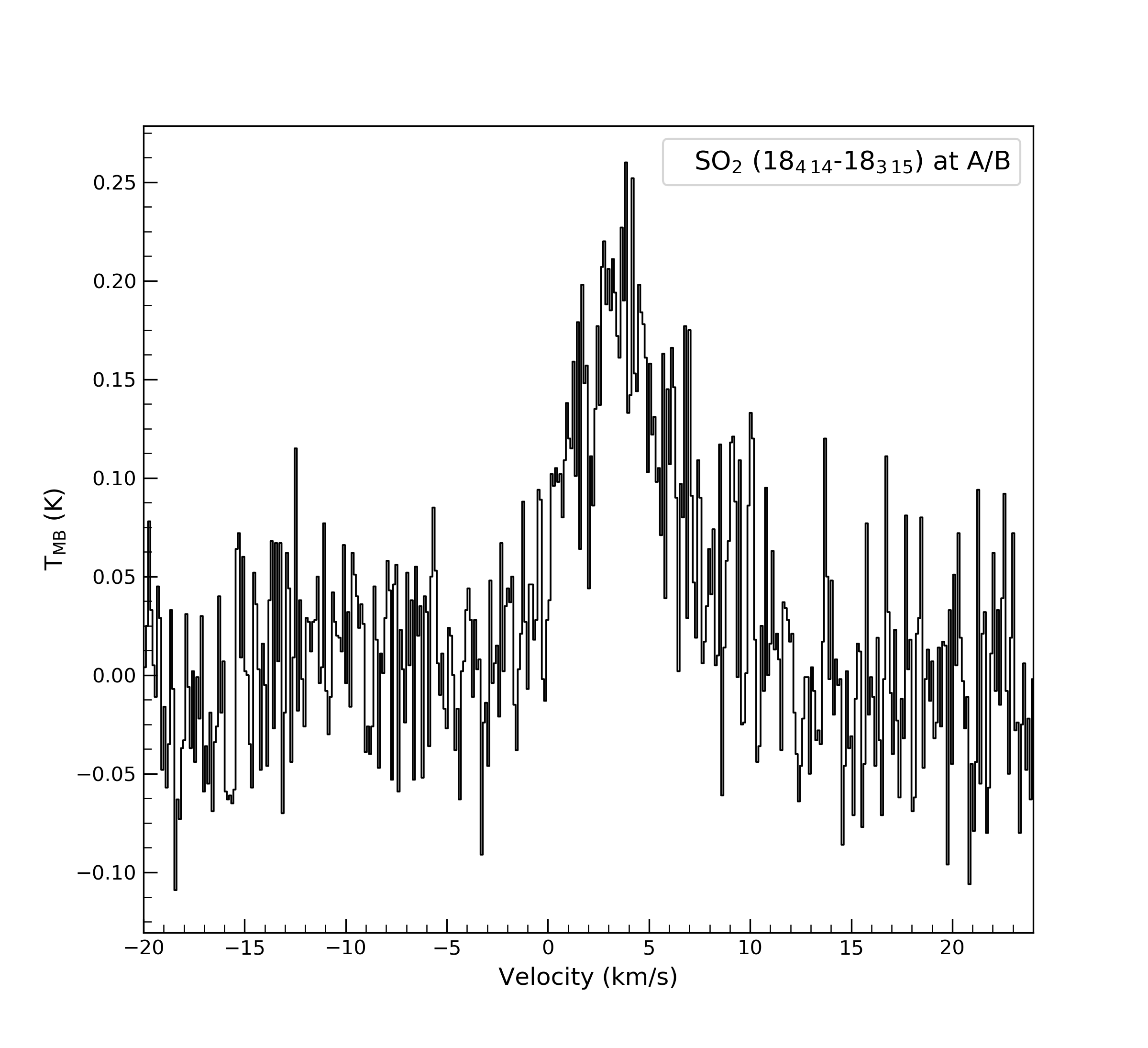}}
    \caption{(a) SO$_2$ $(18_{4, 14} - 18_{3, 15})$ transition at 338305.993\,MHz. Additional contours are drawn at 1$\sigma$. (b) Averaged spectrum of this transition in a $\SI{10}{\arcsecond}$ radius at the position of IRAS\,16293 A/B.}
    \label{fig:70}
\end{figure*}

\begin{figure*}[ht]
	\centering
    \subfigure[]{\includegraphics[width=0.42\textwidth, trim={0 0.65cm 5.4cm 1.18cm},clip]{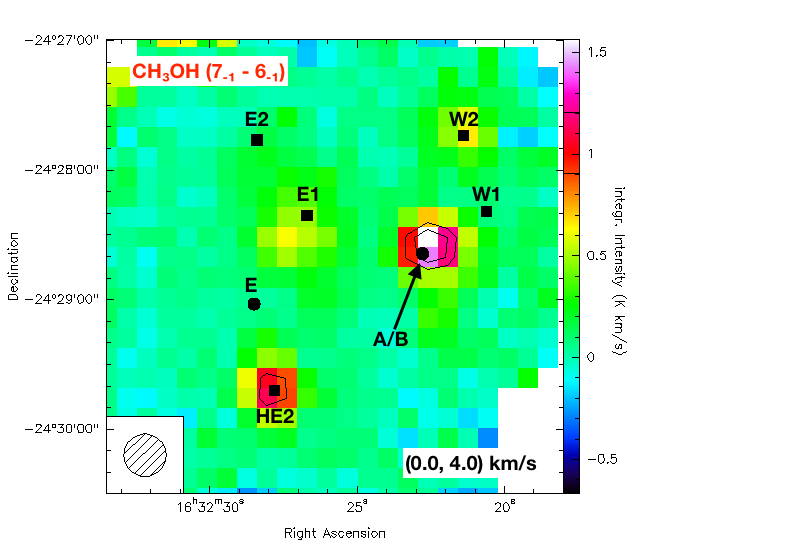}}
    \subfigure[]{\includegraphics[width=0.42\textwidth, trim={0 0.65cm 5.4cm 1.18cm},clip]{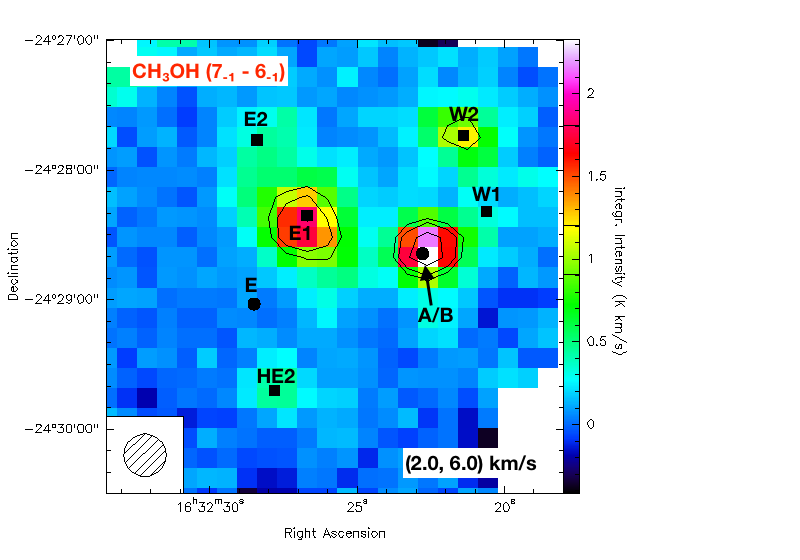}}
    \subfigure[]{\includegraphics[width=0.42\textwidth, trim={0 0.65cm 5.4cm 1.18cm},clip]{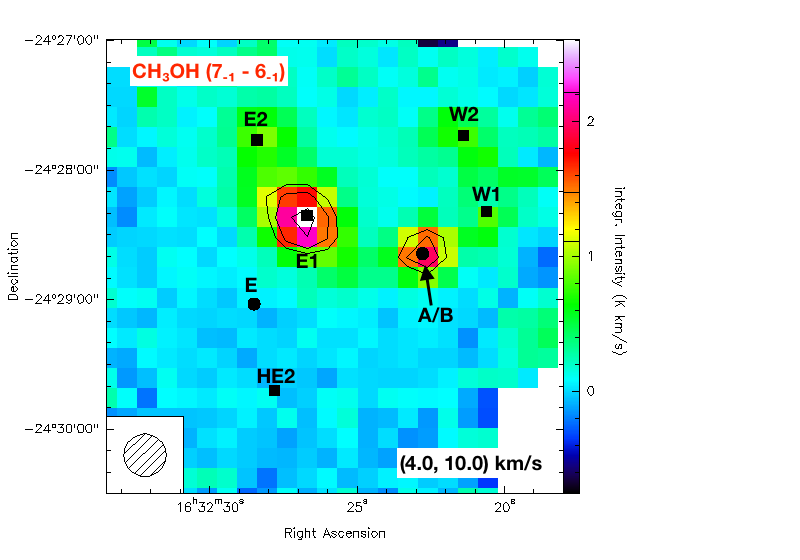}}
    \caption{CH$_3$OH-E ($7_{-1}- 6_{-1}$) transition at 338344.588\,MHz.}
    \label{fig:12}
\end{figure*}

\begin{figure*}[ht]
	\centering
    \subfigure[]{\includegraphics[width=0.42\textwidth, trim={0 0.65cm 5.4cm 1.18cm},clip]{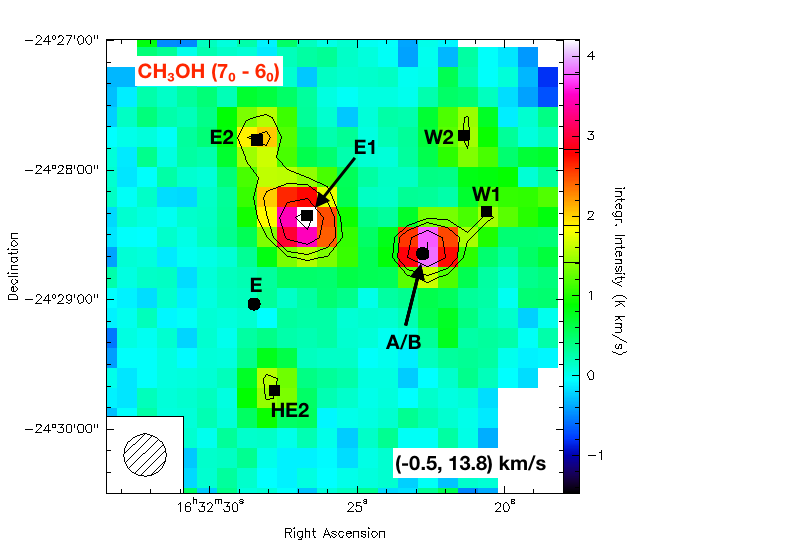}}
    \subfigure[]{\includegraphics[width=0.42\textwidth, trim={0 0.65cm 5.4cm 1.18cm},clip]{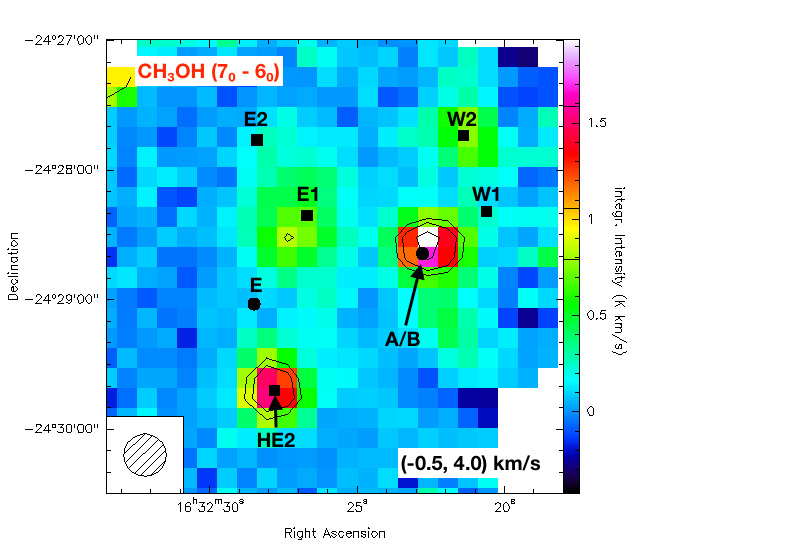}}
    \subfigure[]{\includegraphics[width=0.42\textwidth, trim={0 0.65cm 5.4cm 1.18cm},clip]{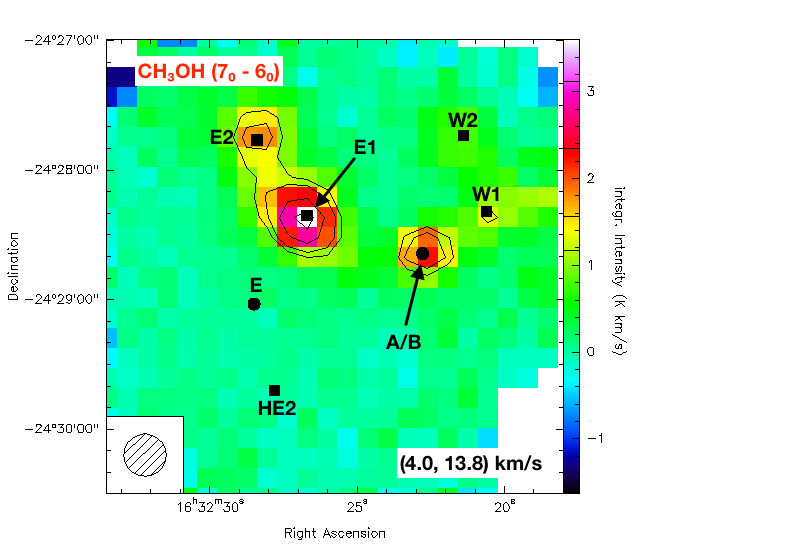}}
    \caption{CH$_3$OH-A$^{+}$ ($7_{0}- 6_{0}$) transition at 338408.698\,MHz.}
    \label{fig:13}
\end{figure*}

\begin{figure*}[ht]
	\centering
    \includegraphics[width=0.42\textwidth, trim={0 0.65cm 5.4cm 1.18cm},clip]{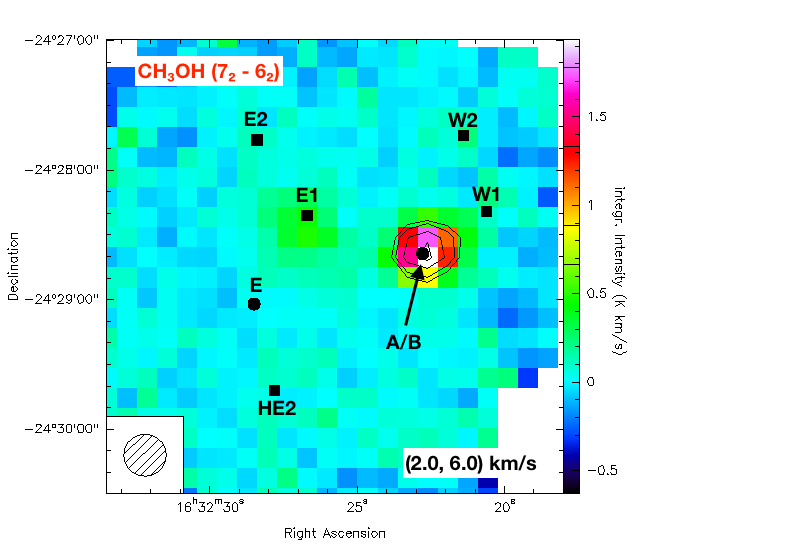}
    \caption{CH$_3$OH-E ($7_{2}- 6_{2}$) and CH$_3$OH-E ($7_{-2}- 6_{-2}$) transition at 338721.693\,MHz and 338722.898\,MHz. The velocity scale is calculated based on a rest frequency of 338721.693\,MHz.}
    \label{fig:14}
\end{figure*}

\begin{figure*}[ht]
	\centering
    \includegraphics[width=0.42\textwidth, trim={0 0.65cm 5.4cm 1.18cm},clip]{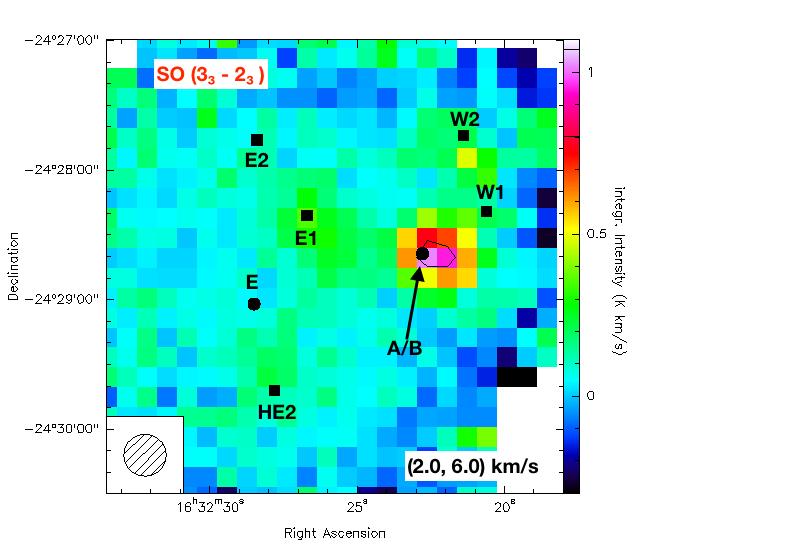}
    \caption{SO ($3_3-2_3$) transition at 339341.459\,MHz.}
    \label{fig:42p5}
\end{figure*}

\begin{figure*}[ht]
	\centering
    \includegraphics[width=0.42\textwidth, trim={0 0.65cm 5.4cm 1.18cm},clip]{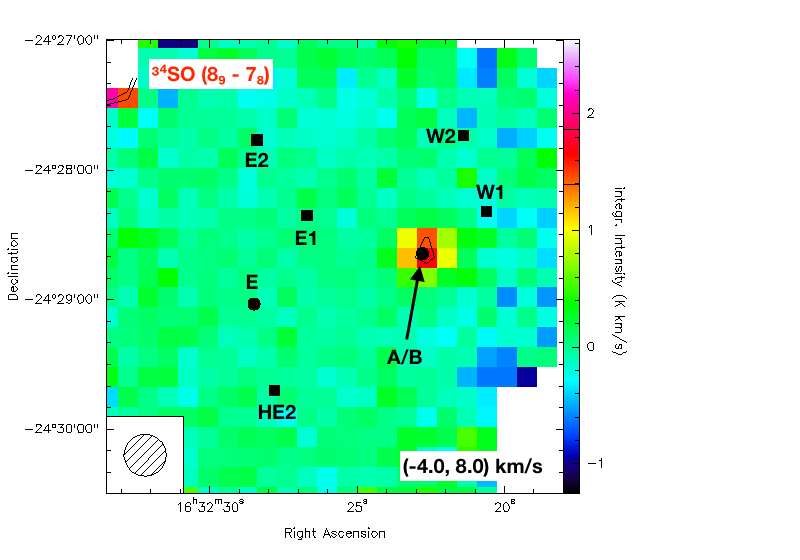}
    \caption{S$^{34}$O ($8_9-7_8$) transition at 339857.269\,MHz.}
    \label{fig:47}
\end{figure*}

\begin{figure*}[ht]
	\centering
    \subfigure[]{\includegraphics[width=0.42\textwidth, trim={0 0.65cm 5.4cm 1.18cm},clip]{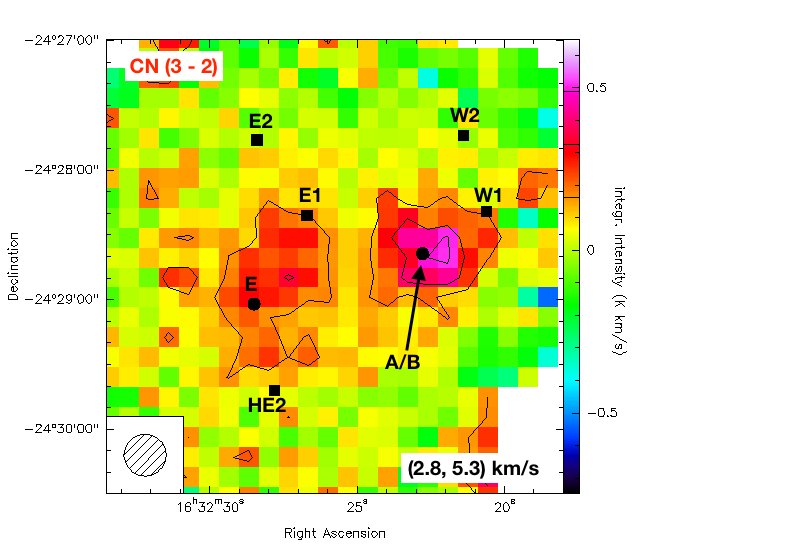}}
    \subfigure[]{\includegraphics[width=0.42\textwidth, trim={0 0.65cm 0 1.18cm},clip]{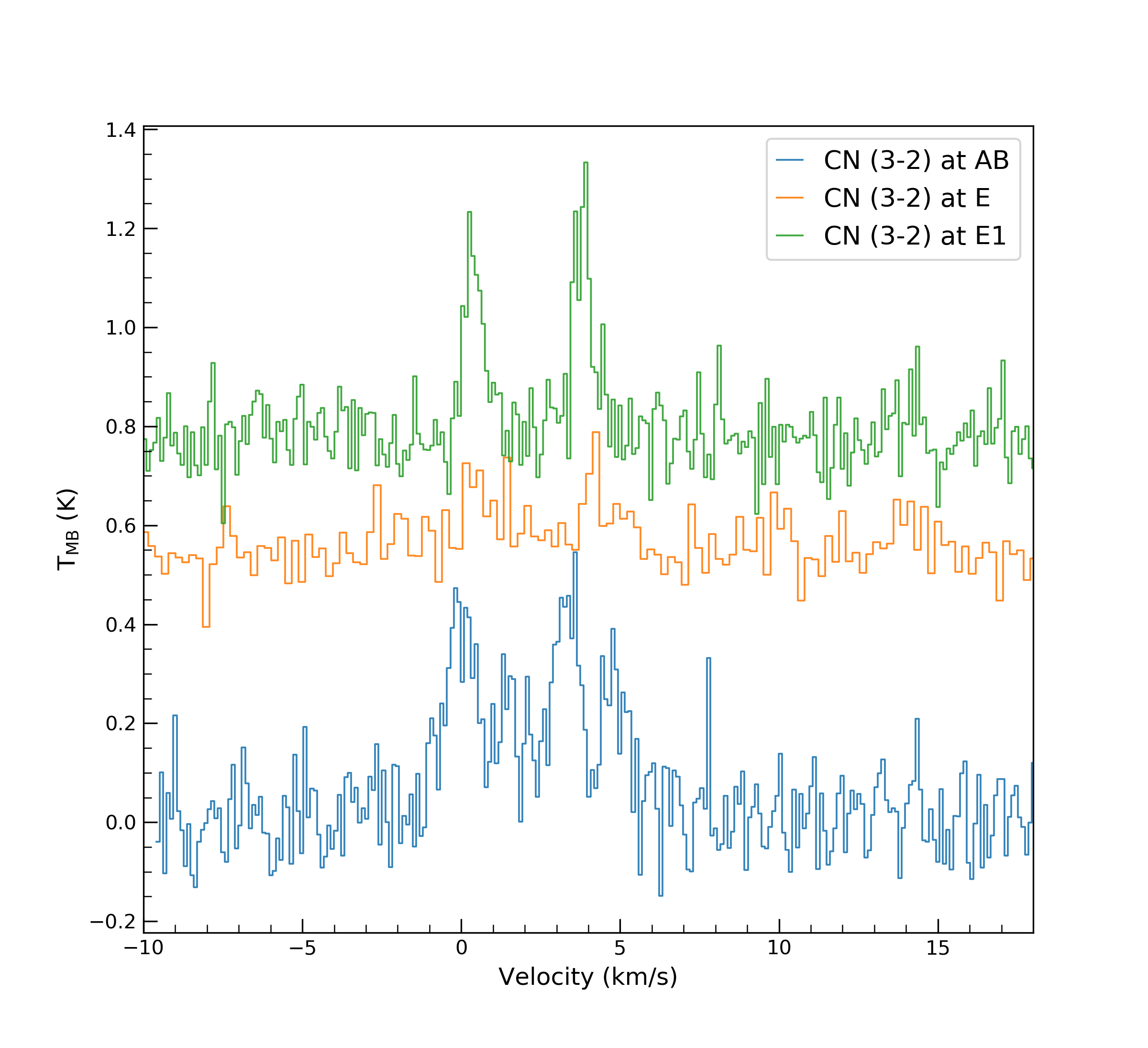}}
    \caption{(a) CN ($3_{0,5/2,7/2}-2_{0,3/2,5/2}$) transition at 340031.549\,MHz. Additional contours are drawn at 1$\sigma$ and 2$\sigma$. (b) Averaged spectrum of this transition in a $\SI{10}{\arcsecond}$ radius at the positions indicated in the upper right corner.}
    \label{fig:55}
\end{figure*}

\begin{figure*}[ht]
	\centering
    \subfigure[]{\includegraphics[width=0.39\textwidth, trim={0 0.65cm 5.4cm 1.18cm},clip]{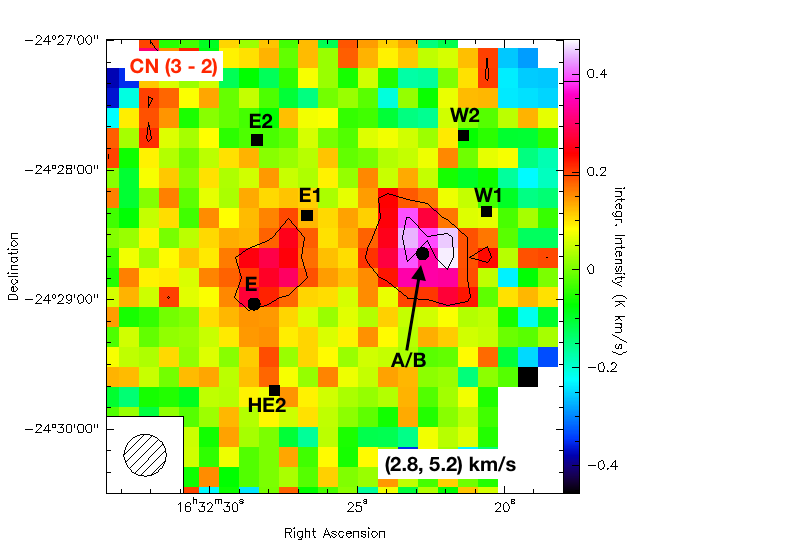}}
    \subfigure[]{\includegraphics[width=0.39\textwidth, trim={0 0.65cm 0 1.18cm},clip]{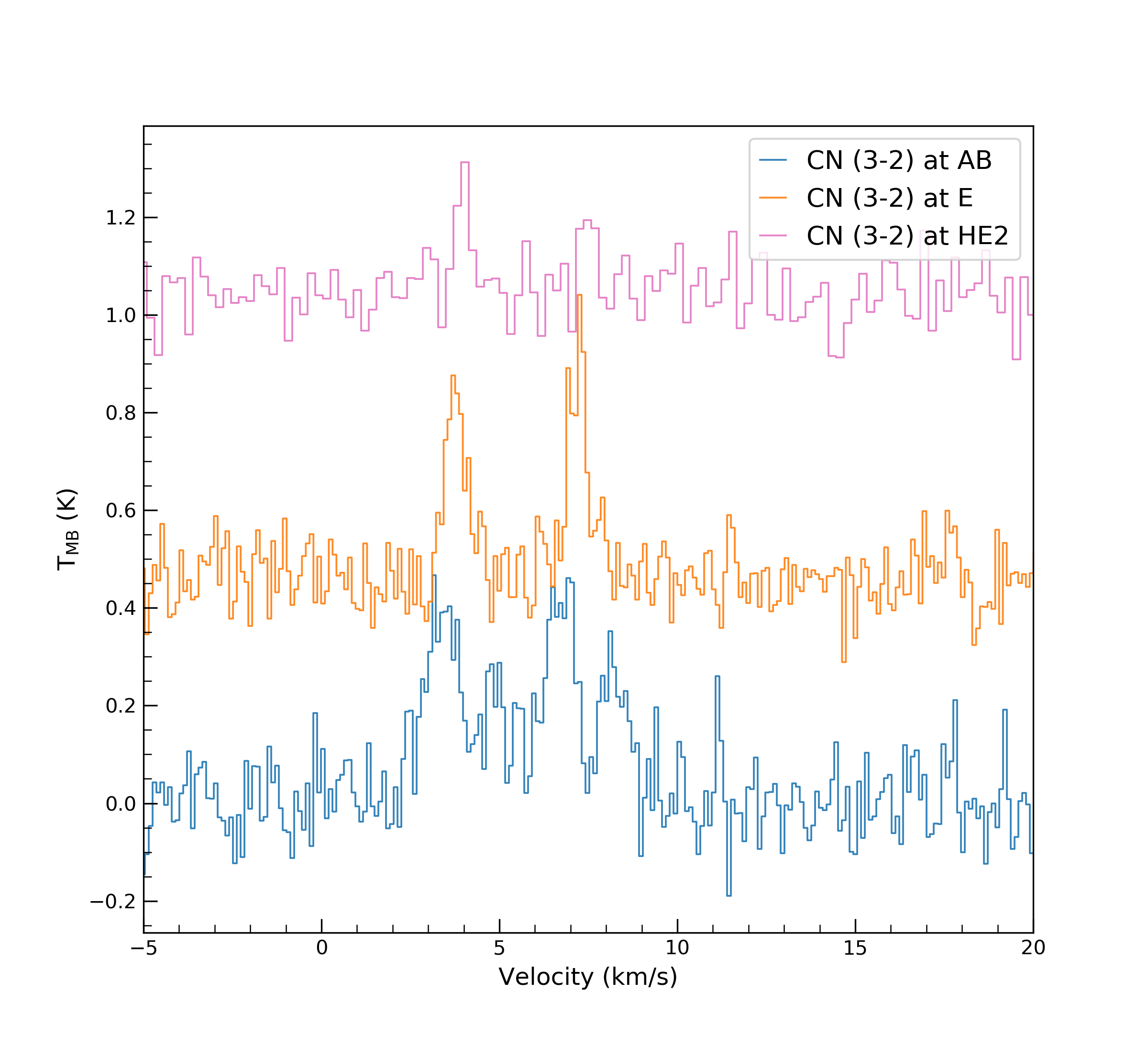}}
    \caption{(a) CN ($3_{0,5/2,5/2}-2_{0,3/2,3/2}$) transition at 340035.408\,MHz. Additional contours are drawn at 1$\sigma$ and 2$\sigma$. (b) Averaged spectrum of this transition in a $\SI{10}{\arcsecond}$ radius at the positions indicated in the upper right corner.}
    \label{fig:56}
\end{figure*}

\begin{figure*}[ht]
	\centering
    \subfigure[]{\includegraphics[width=0.39\textwidth, trim={0 0.65cm 5.4cm 1.18cm},clip]{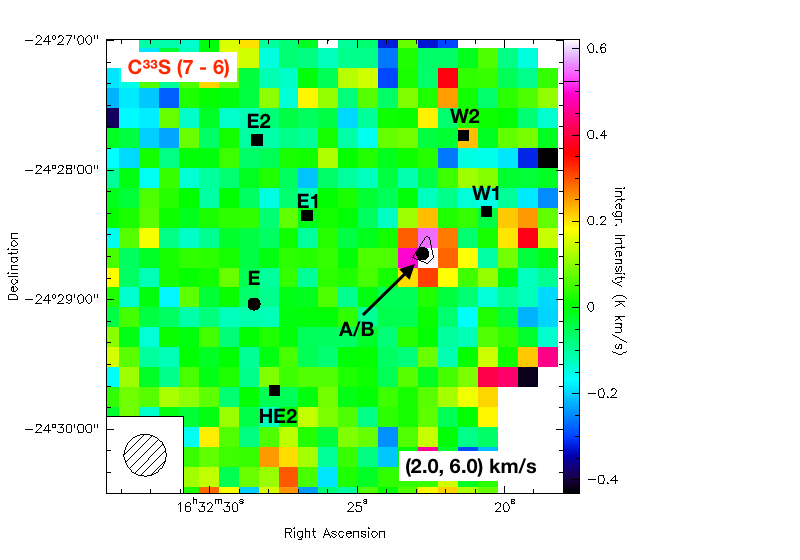}}
    \subfigure[]{\includegraphics[width=0.39\textwidth, trim={0 0.65cm 0 1.18cm},clip]{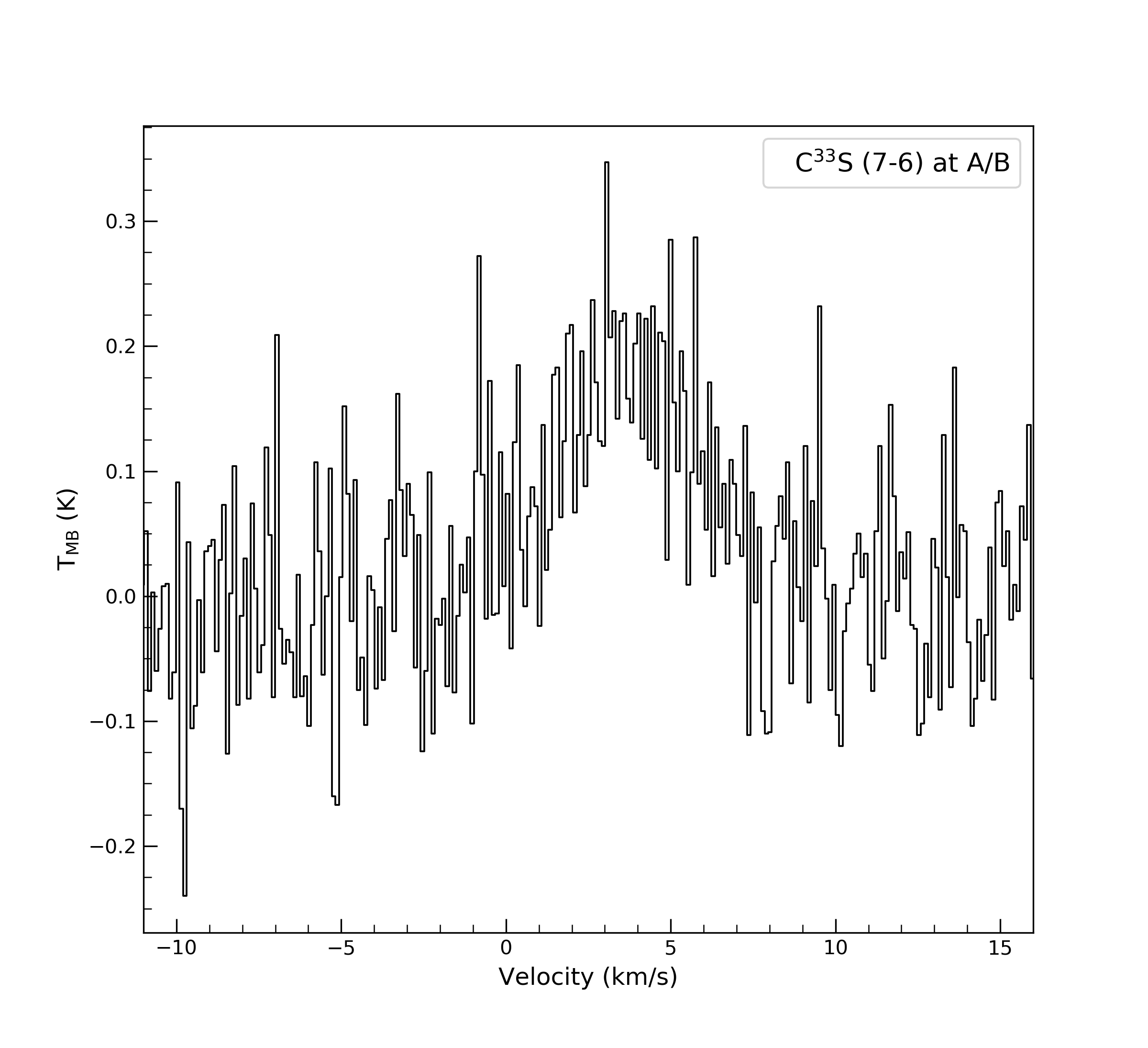}}
    \caption{(a) C$^{33}$S ($7_0 - 6_0$) transition at 340052.575\,MHz. Additional contours are drawn at 1$\sigma$. (b) Averaged spectrum of this transition in a $\SI{10}{\arcsecond}$ radius at the positions indicated in the upper right corner.}
    \label{fig:62}
\end{figure*}

\begin{figure*}[ht]
	\centering
    \subfigure[]{\includegraphics[width=0.39\textwidth, trim={0 0.65cm 5.4cm 1.18cm},clip]{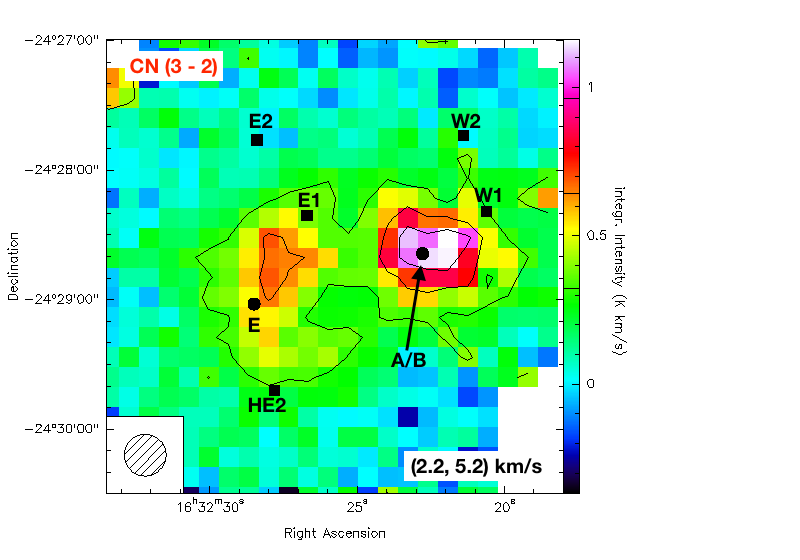}}
    \subfigure[]{\includegraphics[width=0.39\textwidth, trim={0 0.65cm 0 1.18cm},clip]{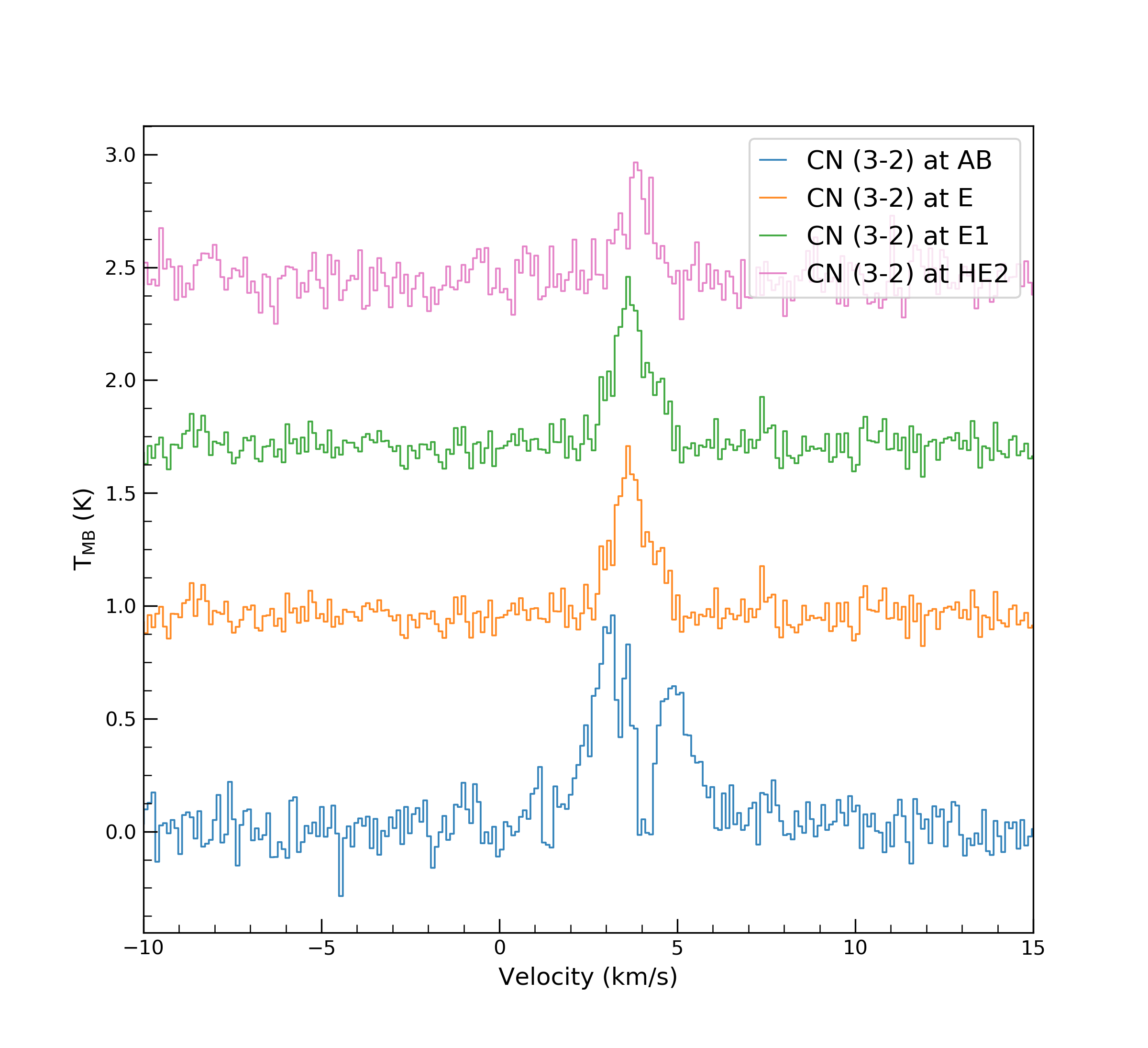}}
    \caption{(a) CN ($3_{0,7/2,9/2} - 2_{0,5/2,7/2}$) and CN ($3_{0,7/2,5/2} - 2_{0,5/2,3/2}$) transitions at 340247.770\,MHz and 340248.544\,MHz. The velocity scale is calculated based on a rest frequency of 340247.770\,MHz. Additional contours are drawn at 1$\sigma$ and 2$\sigma$. (b) Averaged spectrum of this transition in a $\SI{10}{\arcsecond}$ radius at the positions indicated in the upper right corner.}
    \label{fig:57}
\end{figure*}

\begin{figure*}[ht]
	\centering
	\subfigure[]{\includegraphics[width=0.42\textwidth, trim={0 0.65cm 5.4cm 1.18cm},clip]{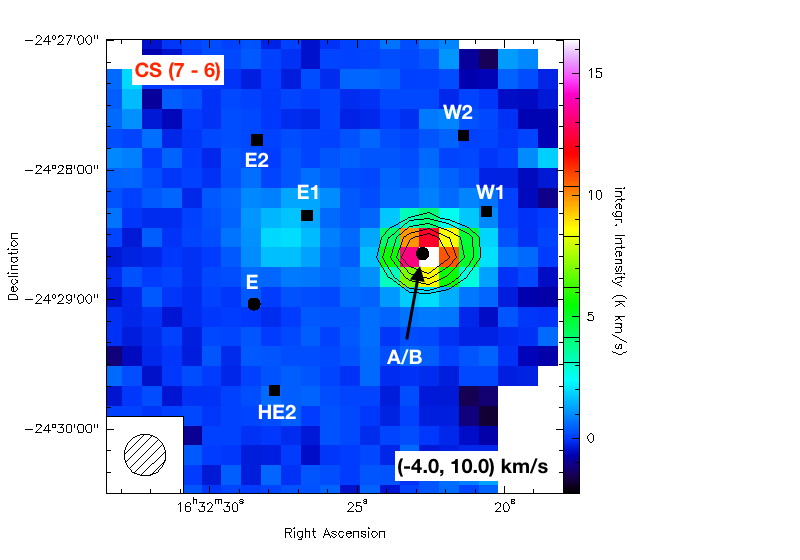} }
    \subfigure[]{\includegraphics[width=0.42\textwidth, trim={0 0.65cm 5.4cm 1.18cm},clip]{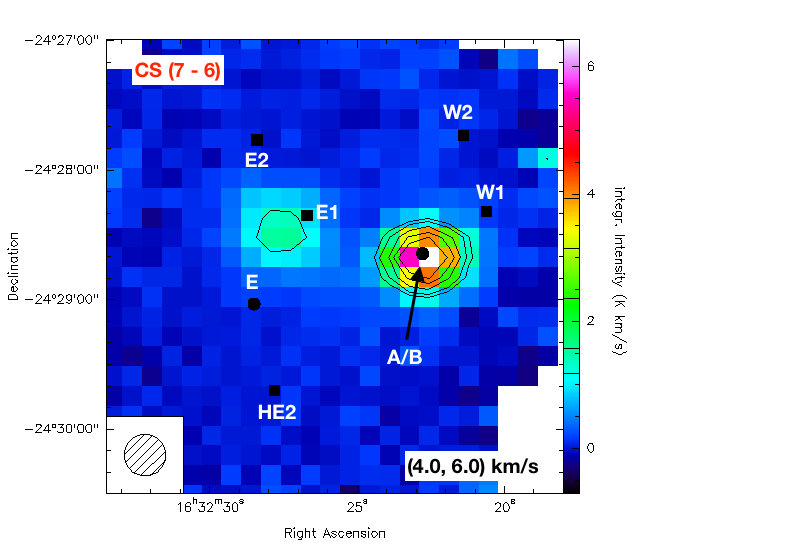}}
    \caption{CS ($7 - 6$) transition at 342882.850\,MHz.}
    \label{fig:17}
\end{figure*}

\begin{figure*}[ht]
	\centering
    \includegraphics[width=0.42\textwidth, trim={0 0.65cm 0 1.18cm},clip]{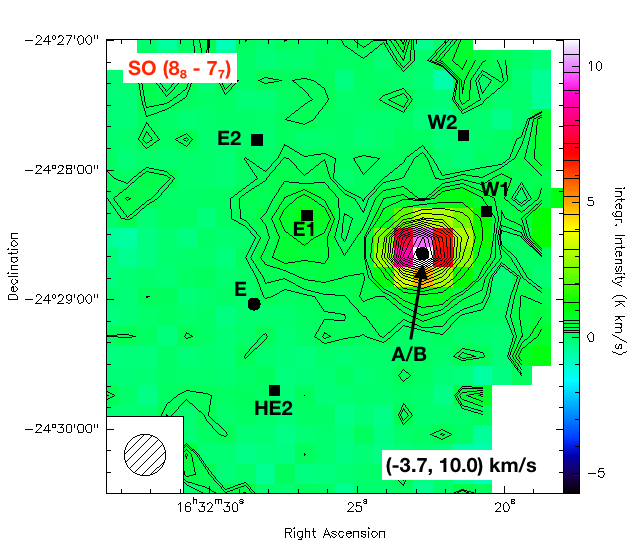}
    \caption{SO ($8_8 - 7_7$) transition at 344310.612\,MHz.}
    \label{fig:44}
\end{figure*}

\begin{figure*}[ht]
	\centering
    \includegraphics[width=0.42\textwidth, trim={0 0.65cm 0 1.18cm},clip]{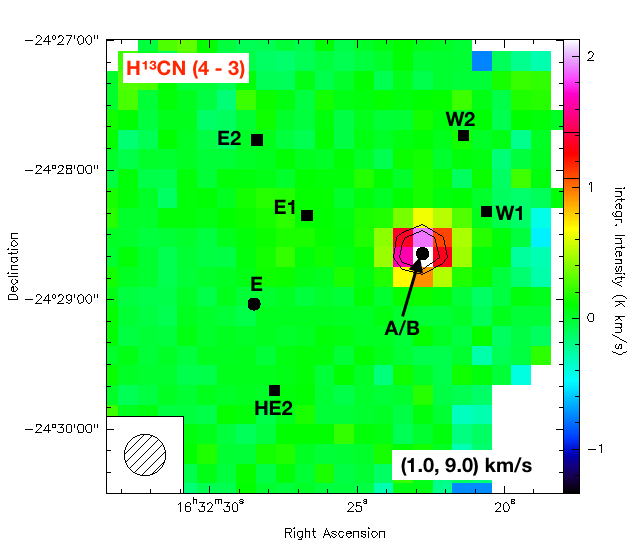}
    \caption{H$^{13}$CN ($4 - 3$) transition at 345339.769\,MHz.}
    \label{fig:23}
\end{figure*}

\begin{figure*}[ht]
	\centering
    \subfigure[]{\includegraphics[width=0.42\textwidth, trim={0 0.01cm 0 0.01cm},clip]{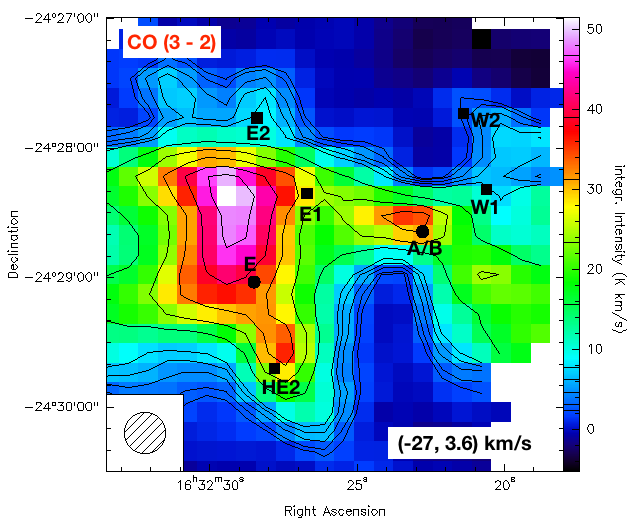}}
    \subfigure[]{\includegraphics[width=0.42\textwidth, trim={0 0.01cm 0 0.01cm},clip]{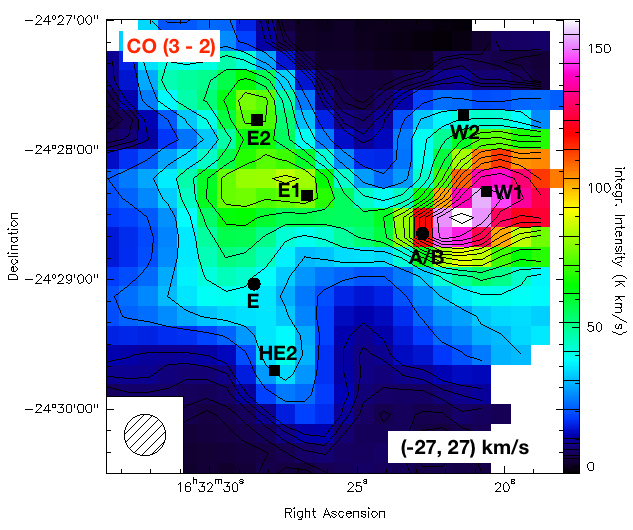}}
    \subfigure[]{\includegraphics[width=0.42\textwidth, trim={0 0.65cm 0 1.18cm},clip]{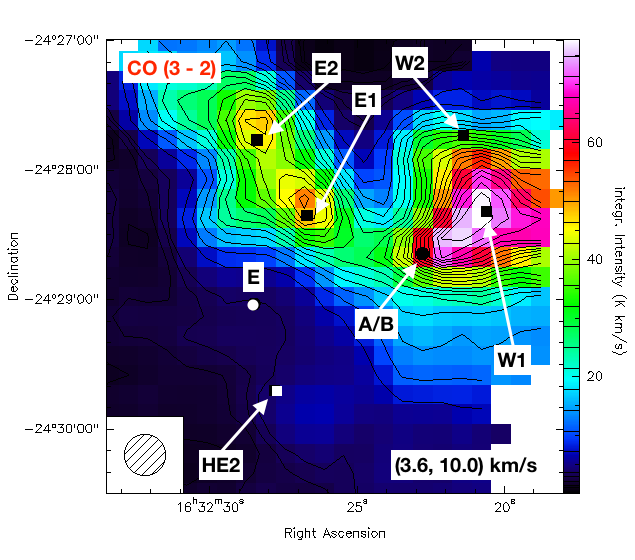}}
    \subfigure[]{\includegraphics[width=0.42\textwidth, trim={0 0.65cm 0 1.18cm},clip]{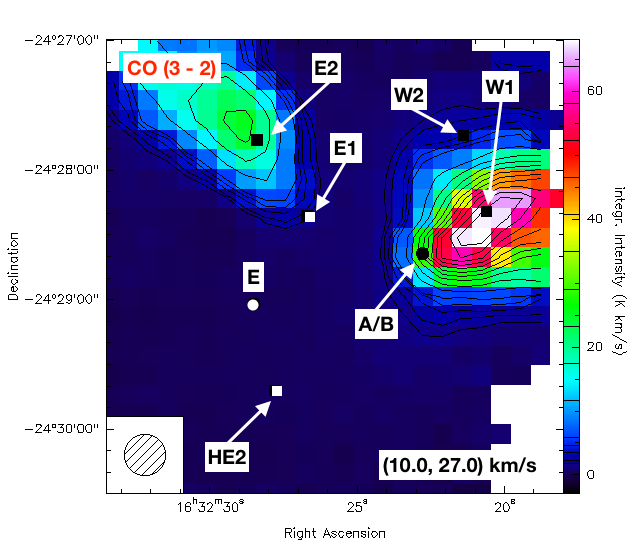}}
    \caption{CO ($3 - 2$) transition at 345795.990\,MHz.}
    \label{fig:1}
\end{figure*}

\begin{figure*}[ht]
	\centering
    \includegraphics[width=0.42\textwidth, trim={0 0.65cm 0 1.18cm},clip]{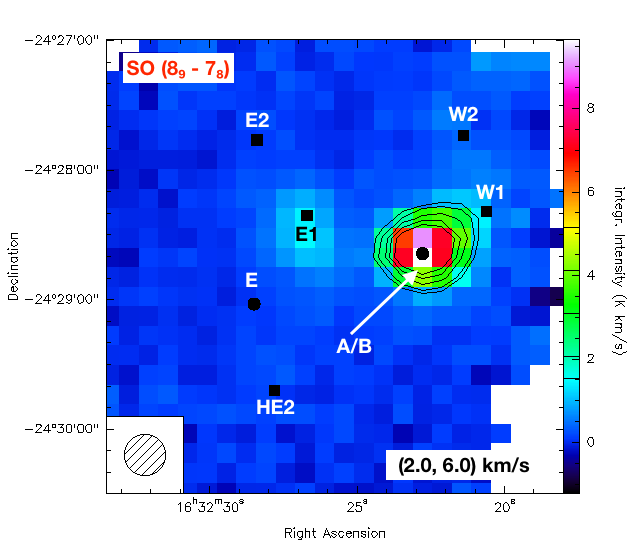}
    \caption{SO ($8_9 - 7_8$) transition at 346528.481\,MHz.}
    \label{fig:45}
\end{figure*}

\clearpage

\begin{figure*}[ht]
	\centering
    \subfigure[]{\includegraphics[width=0.42\textwidth, trim={0 0.65cm 5.4cm 1.18cm},clip]{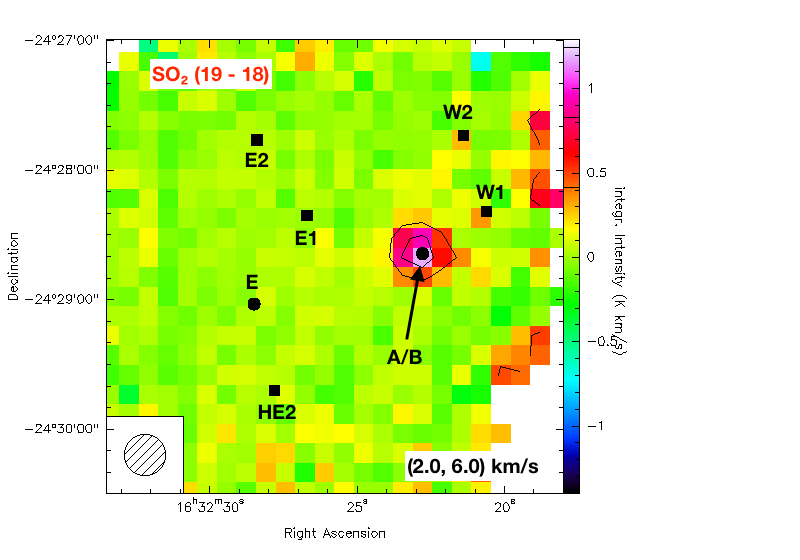}}
    \subfigure[]{\includegraphics[width=0.42\textwidth, trim={0 0.65cm 0 1.18cm},clip]{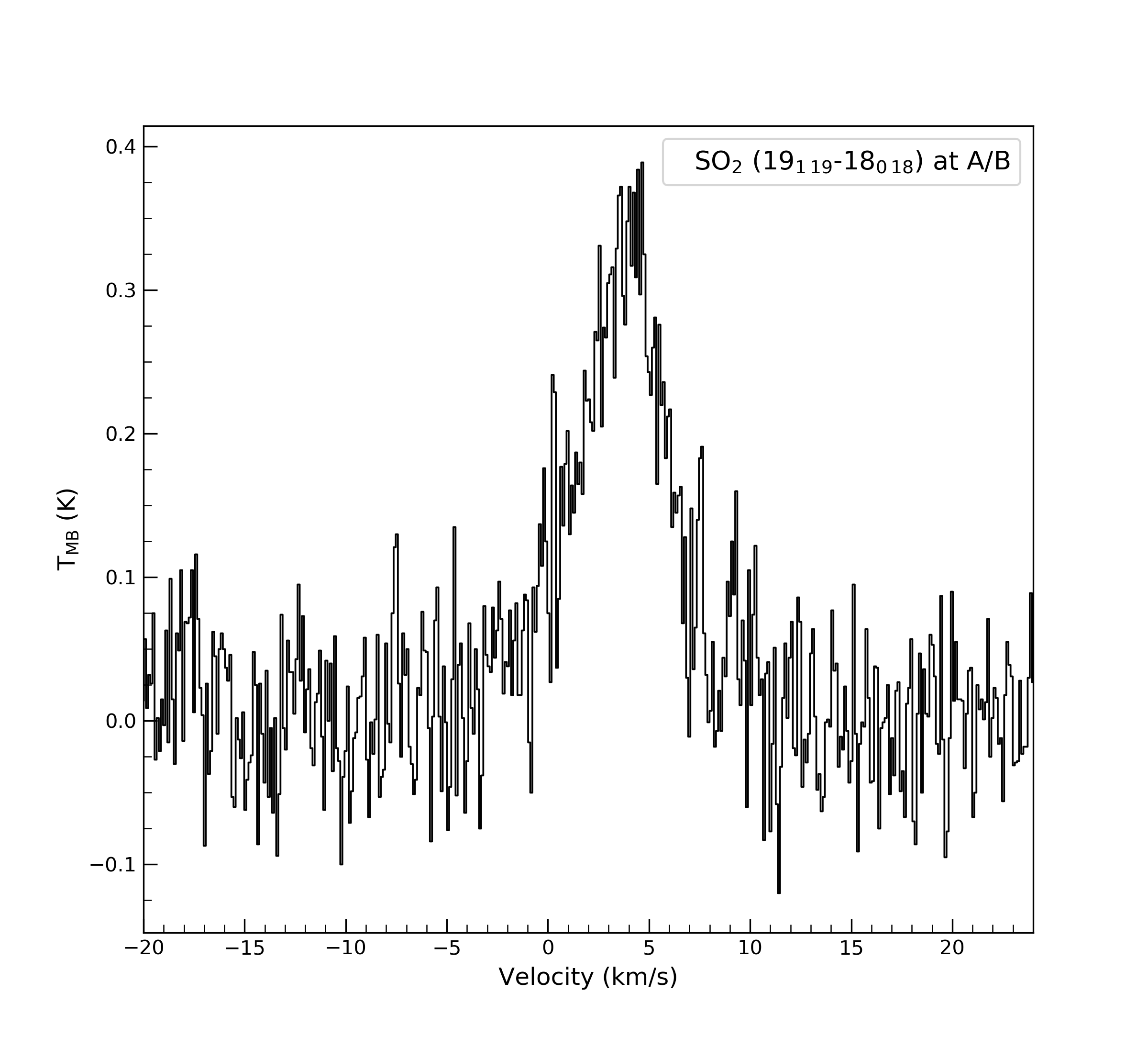}}
    \caption{(a) SO$_2$ ($19_{1, 19}-18_{0, 18}$) transition at 346652.169\,MHz. Additional contours are drawn at 1$\sigma$ and 2$\sigma$. (b) Averaged spectrum of this transition in a $\SI{10}{\arcsecond}$ radius at the position of IRAS\,16293 A/B.}
    \label{fig:52}
\end{figure*}

\begin{figure*}[ht]
	\centering
    \subfigure[]{\includegraphics[width=0.42\textwidth, trim={0 0.65cm 0 1.18cm},clip]{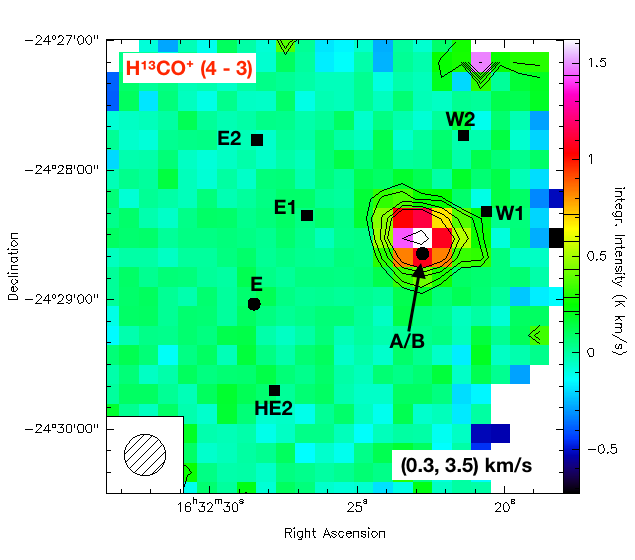}}
    \subfigure[]{\includegraphics[width=0.42\textwidth, trim={0 0.65cm 0 1.18cm},clip]{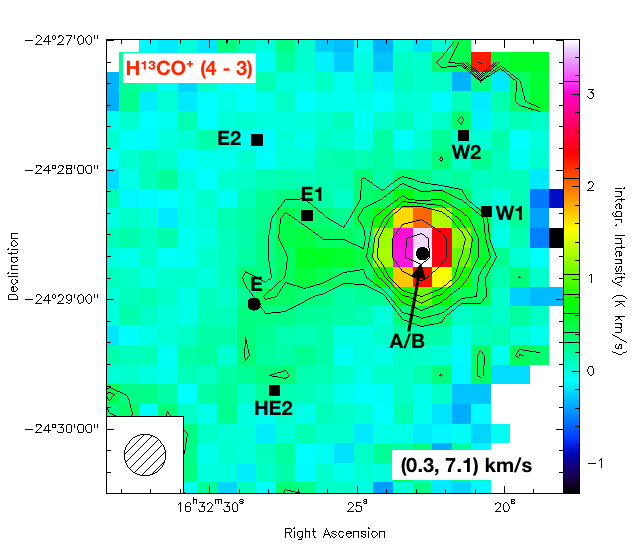}}
    \subfigure[]{\includegraphics[width=0.42\textwidth, trim={0 0.65cm 0 1.18cm},clip]{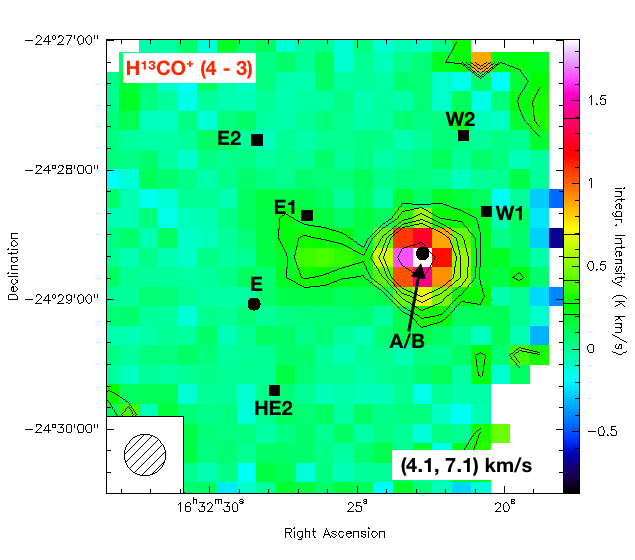}}
    \caption{H$^{13}$CO$^+$ ($4 - 3$) transition at 346998.344\,MHz.}
    \label{fig:26}
\end{figure*}

\begin{figure*}[ht]
	\centering
    \subfigure[]{\includegraphics[width=0.42\textwidth, trim={0 0.65cm 0 1.18cm},clip]{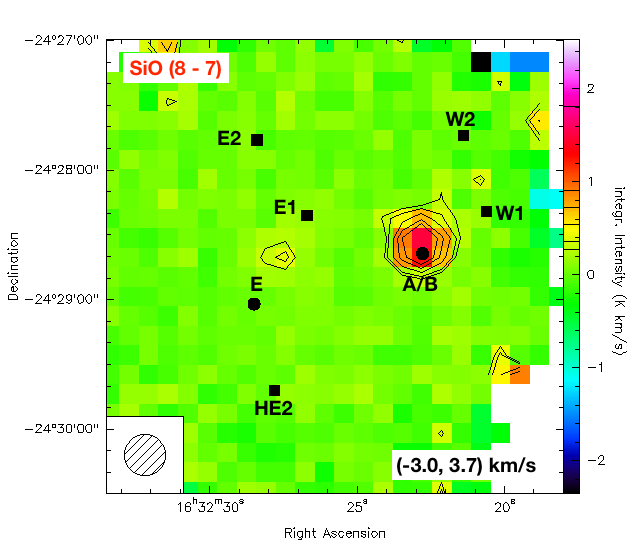}}
    \subfigure[]{\includegraphics[width=0.42\textwidth, trim={0 0.65cm 0 1.18cm},clip]{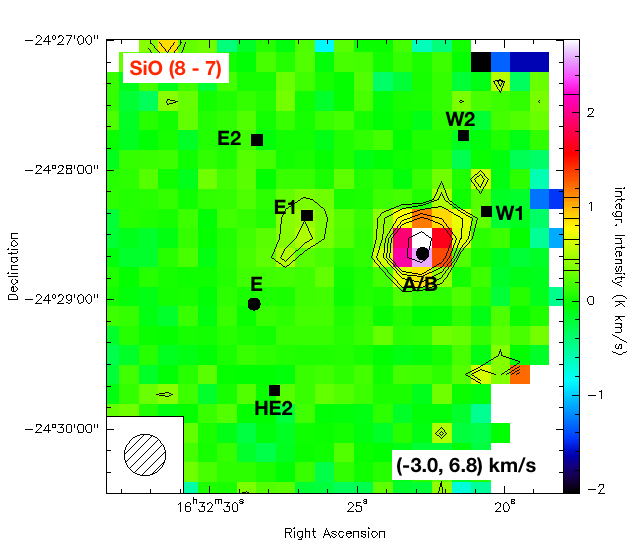}}
    \subfigure[]{\includegraphics[width=0.42\textwidth, trim={0 0.65cm 0 1.18cm},clip]{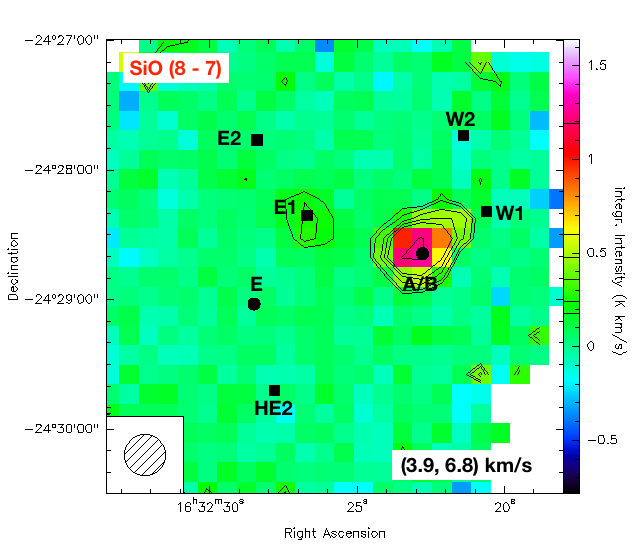}}
    \caption{SiO ($8 - 7$) transition at 347330.581\,MHz.}
    \label{fig:42}
\end{figure*}

\begin{figure*}[ht]
	\centering
    \subfigure[]{\includegraphics[width=0.42\textwidth, trim={0 0.65cm 5.4cm 1.18cm},clip]{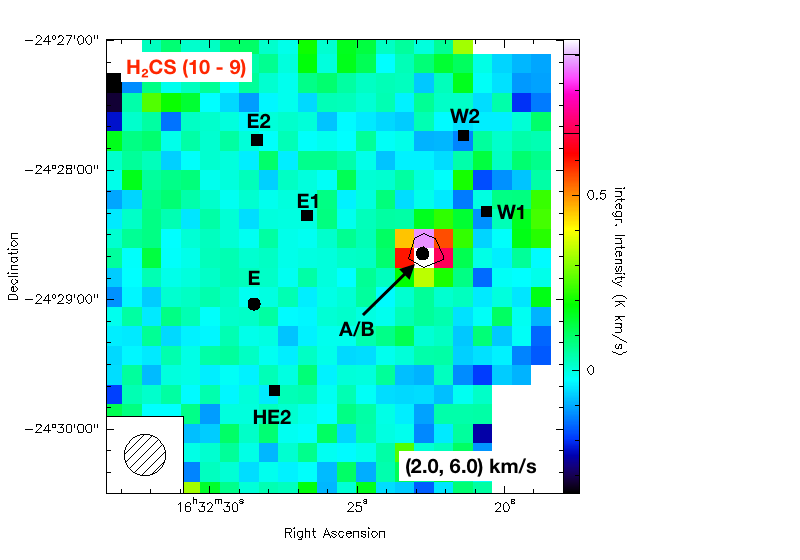}}
    \subfigure[]{\includegraphics[width=0.42\textwidth, trim={0 0.65cm 0 1.18cm},clip]{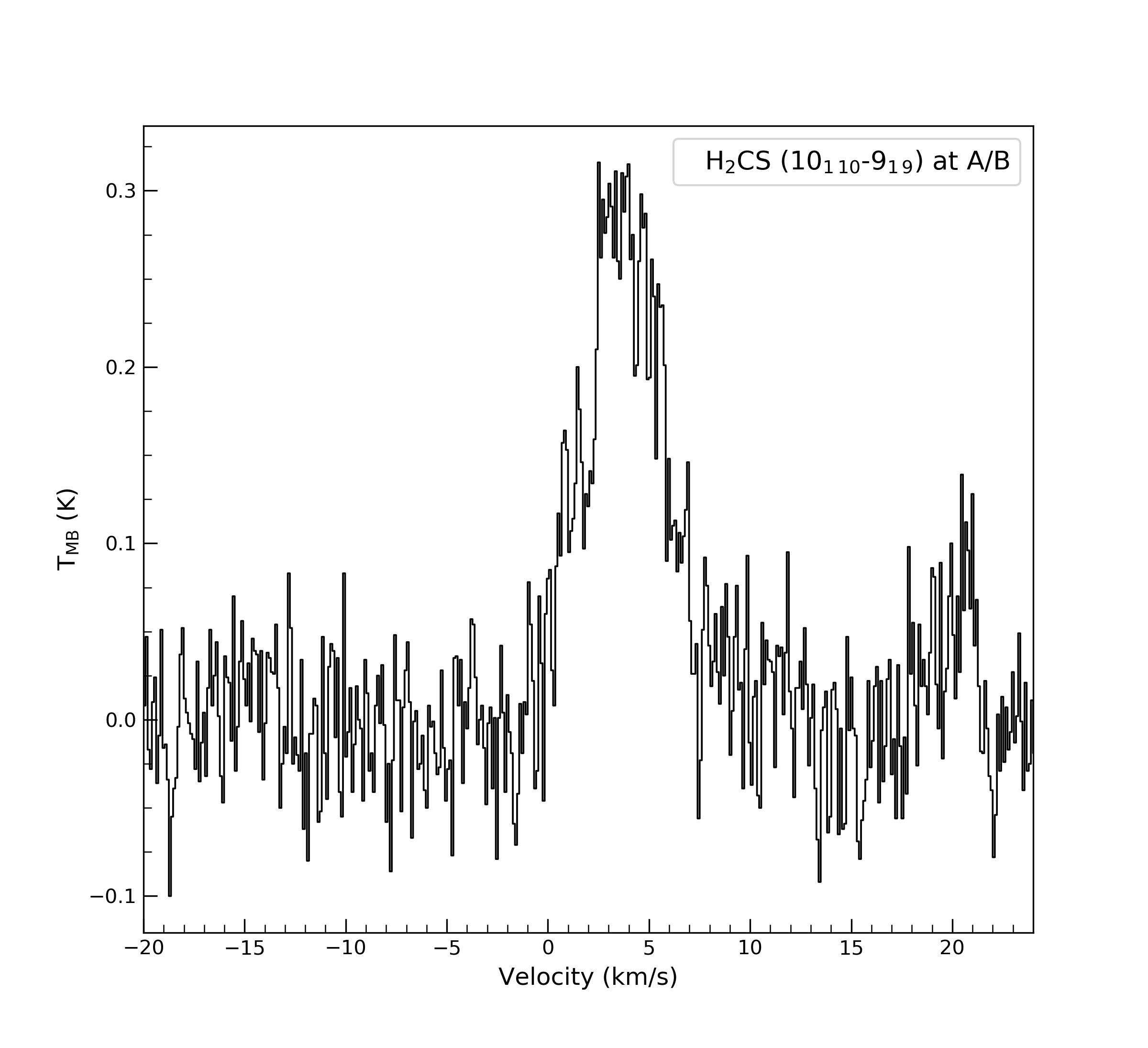}}
    \caption{(a) H$_2$CS ($10_{1, 9} - 9_{1, 8}$) transition at 348534.365\,MHz. (b) Averaged spectrum of this transition in a $\SI{10}{\arcsecond}$ radius at the position of IRAS\,16293 A/B.}
    \label{fig:65}
\end{figure*}

\begin{figure*}[ht]
	\centering
	\subfigure[]{\includegraphics[width=0.42\textwidth, trim={0 0.5cm 4.0cm 0.1cm},clip]{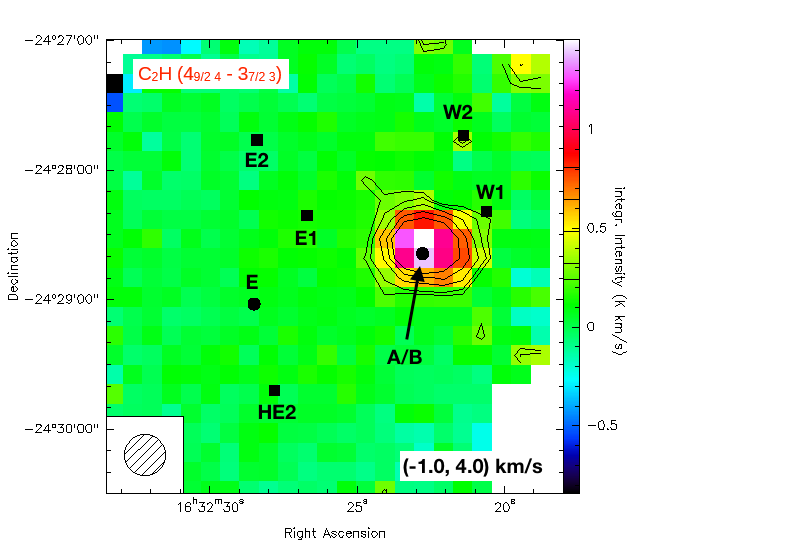}}
    \subfigure[]{\includegraphics[width=0.42\textwidth, trim={0 0.5cm 4.0cm 0.1cm},clip]{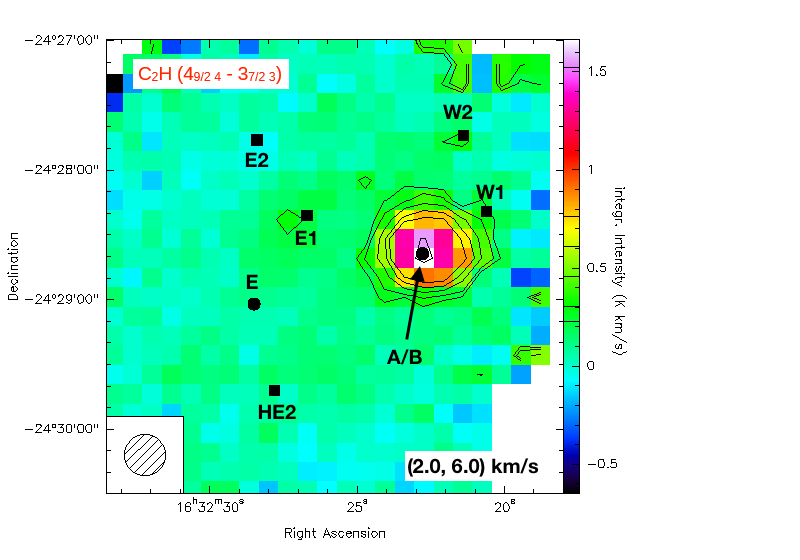}}
    \caption{C$_2$H ($4_{9/2,4} - 3_{7/2,3}$) transition at 349337.706\,MHz.}
    \label{fig:41}
\end{figure*}

\begin{figure*}[ht]
	\centering
	\subfigure[]{\includegraphics[width=0.42\textwidth, trim={0 0.5cm 4.0cm 0.8cm},clip]{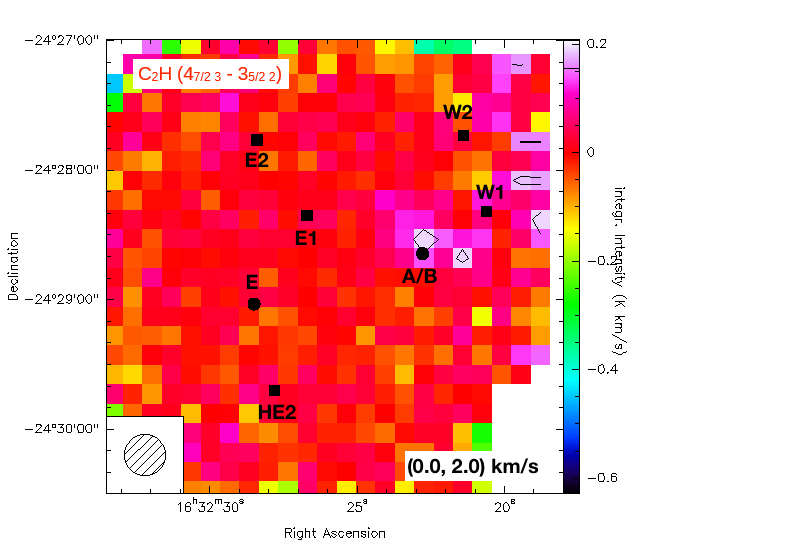}}
    \subfigure[]{\includegraphics[width=0.42\textwidth, trim={0 0.5cm 4.0cm 0.8cm},clip]{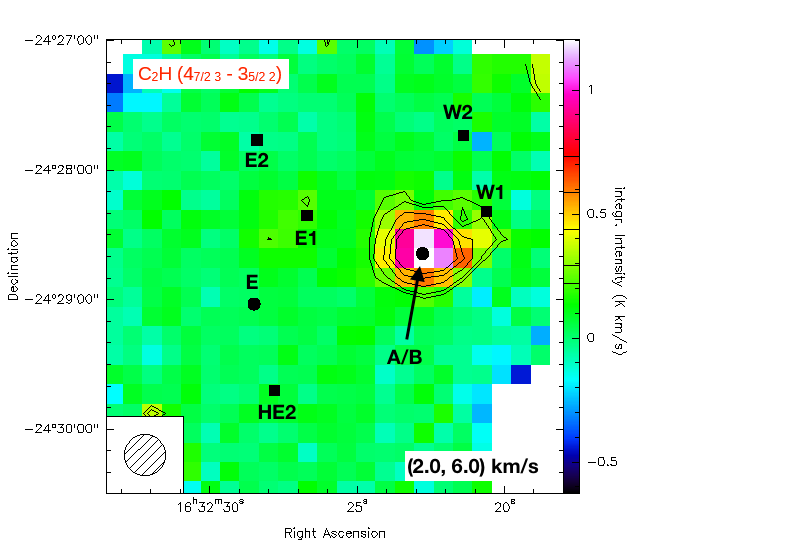}}
    \subfigure[]{\includegraphics[width=0.42\textwidth, trim={0 0.5cm 4.0cm 0.8cm},clip]{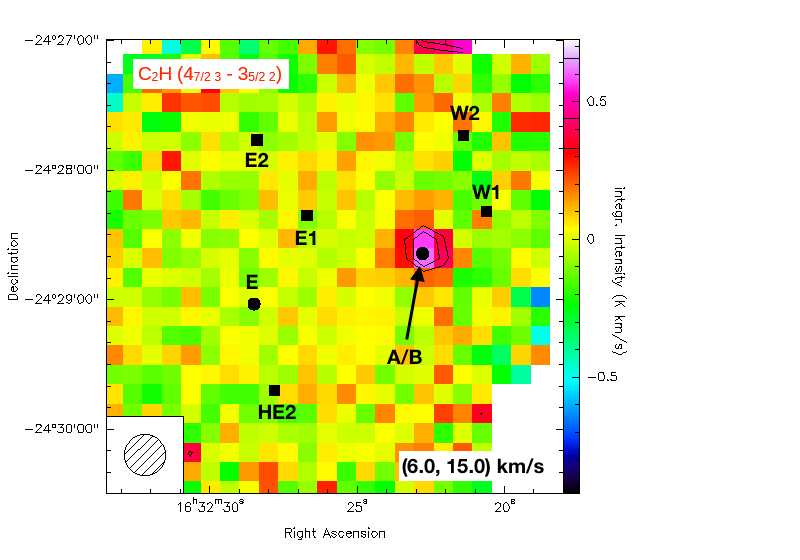}}
    \caption{C$_2$H ($4_{7/2,3}-3_{5/2,2}$) transition at 349399.276\,MHz.}
    \label{fig:40}
\end{figure*}

\begin{figure*}[ht]
	\centering
    \includegraphics[width=0.39\textwidth, trim={0 0.65cm 5.4cm 1.18cm},clip]{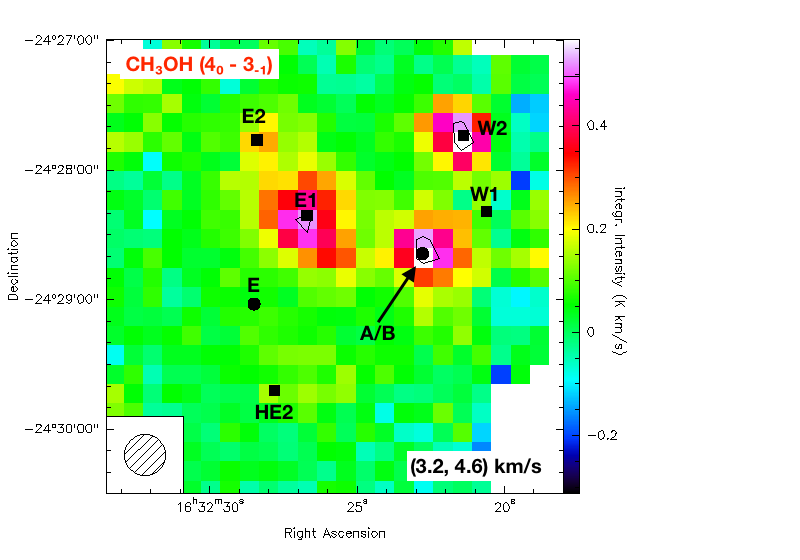}
    \caption{CH$_3$OH-E ($4_{0}- 3_{-1}$) transition at 350687.662\,MHz. The CH$_3$OH-E line is blended with NO transitions at 350689.494\,MHz and 350690.766\,MHz.}
    \label{fig:5a}
\end{figure*}

\begin{figure*}[ht]
	\centering
    \subfigure[]{\includegraphics[width=0.39\textwidth, trim={0 0.65cm 5.4cm 1.18cm},clip]{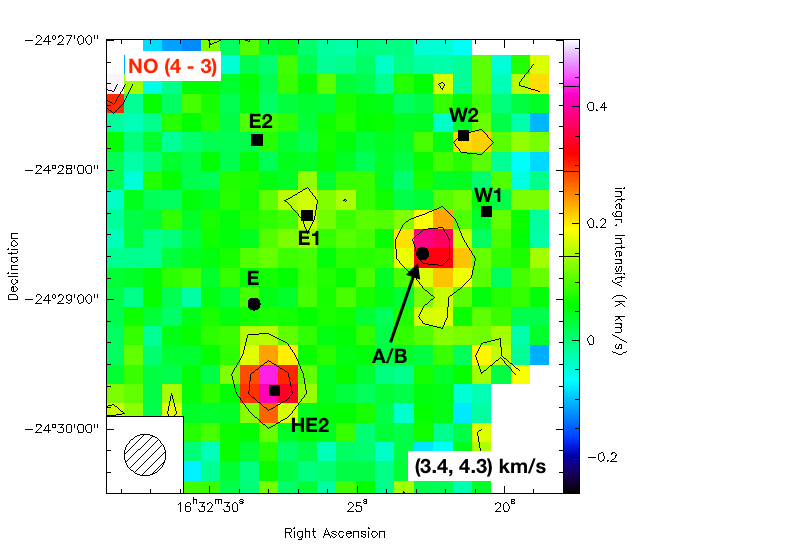}}
    \subfigure[]{\includegraphics[width=0.39\textwidth, trim={0 0.65cm 0 1.18cm},clip]{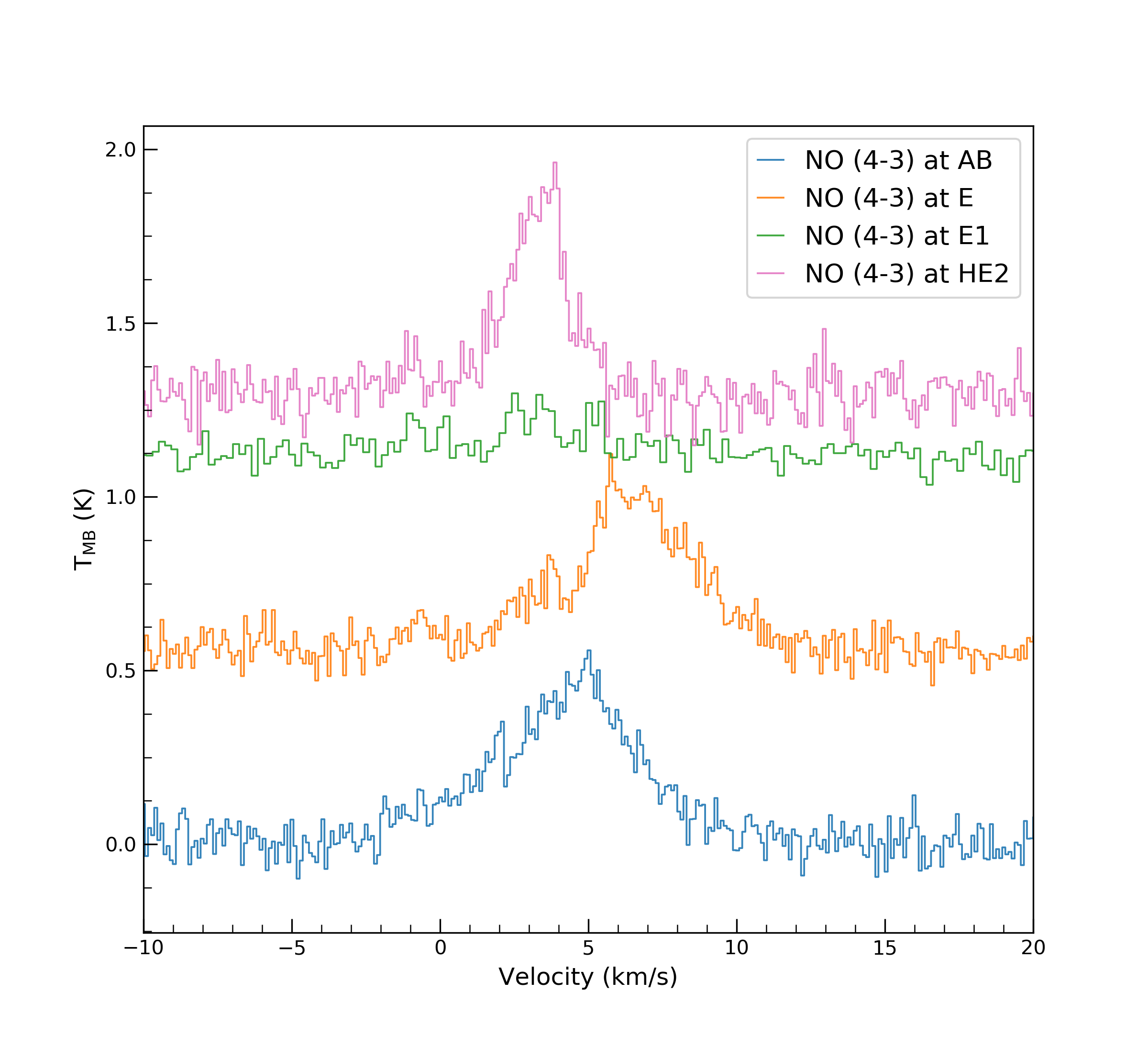}}
    \caption{(a) NO ($4_{-,7/2,9/2}-3_{+,5/2,7/2}$) transition at 350689.494\,MHz. Additional contours are drawn at 1$\sigma$ and 2$\sigma$. (b) Averaged spectrum of this transition in a $\SI{10}{\arcsecond}$ radius at the positions indicated in the upper right corner. The NO line is blended with a CH$_3$OH-E transition at 350787.662\,MHz and a NO transition at 350690.766\,MHz.}
    \label{fig:58}
\end{figure*}

\begin{figure*}[ht]
	\centering
    \subfigure[]{\includegraphics[width=0.39\textwidth, trim={0 0.65cm 5.4cm 1.18cm},clip]{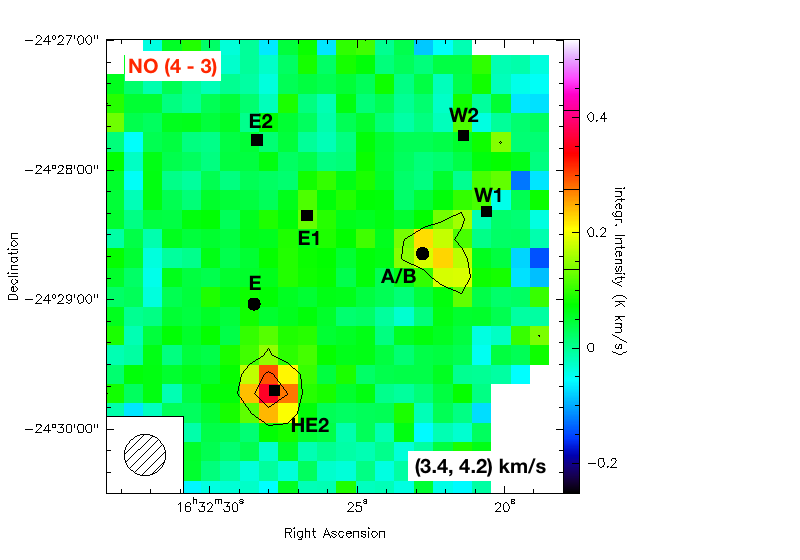}}
    \subfigure[]{\includegraphics[width=0.39\textwidth, trim={0 0.65cm 0 1.18cm},clip]{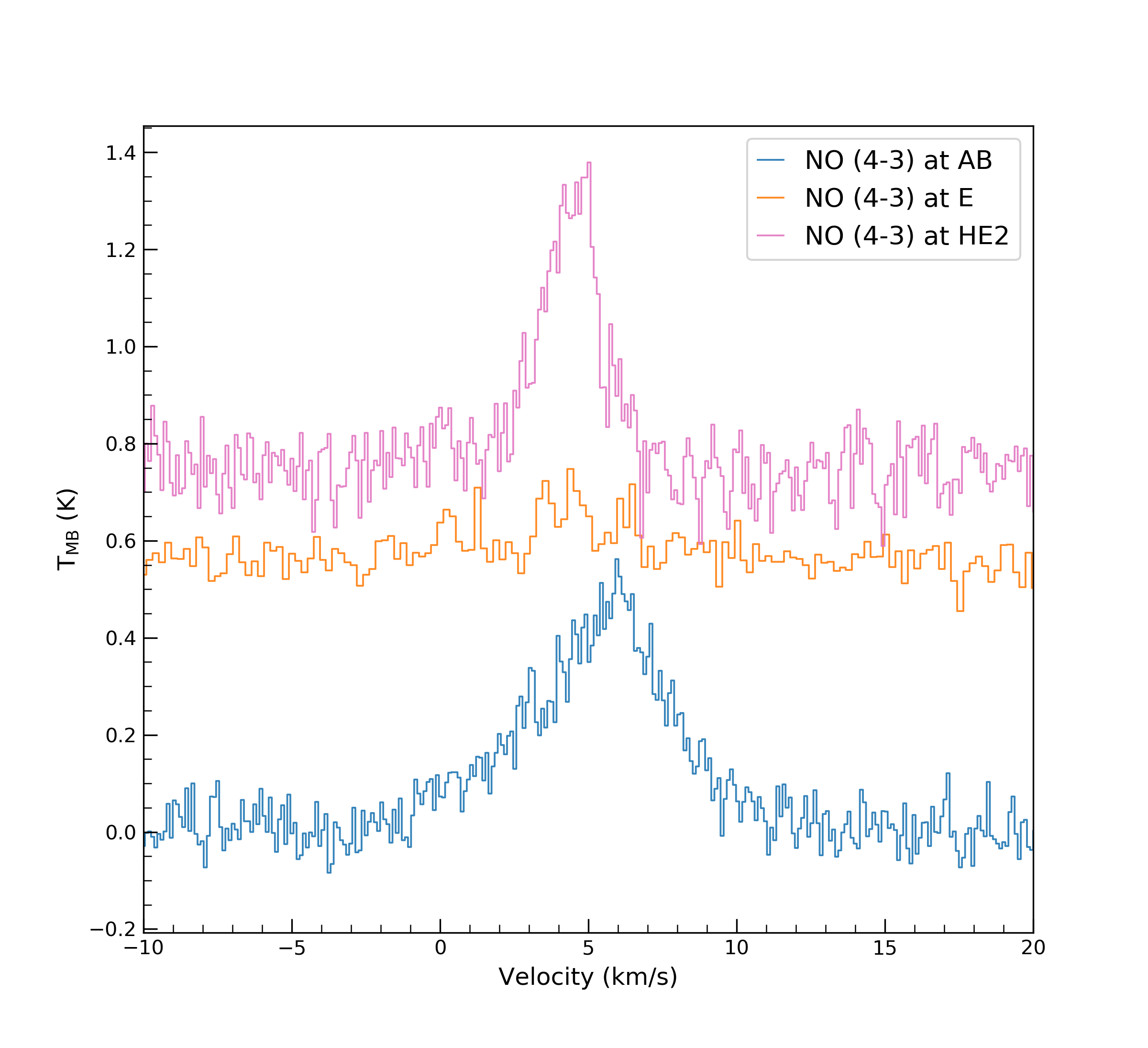}}
    \caption{(a) NO ($4_{-,7/2,7/2}-3_{+,5/2,5/2}$) transition at 350690.766\,MHz. Additional contours are drawn at 1$\sigma$ and 2$\sigma$. (b) Averaged spectrum of this transition in a $\SI{10}{\arcsecond}$ radius at the positions indicated in the upper right corner. The NO line is blended with a CH$_3$OH-E transition at 350787.662\,MHz and a NO transition at 350689.494\,MHz.}
    \label{fig:59}
\end{figure*}

\begin{figure*}[ht]
	\centering
    \subfigure[]{\includegraphics[width=0.42\textwidth, trim={0 0.65cm 5.4cm 1.18cm},clip]{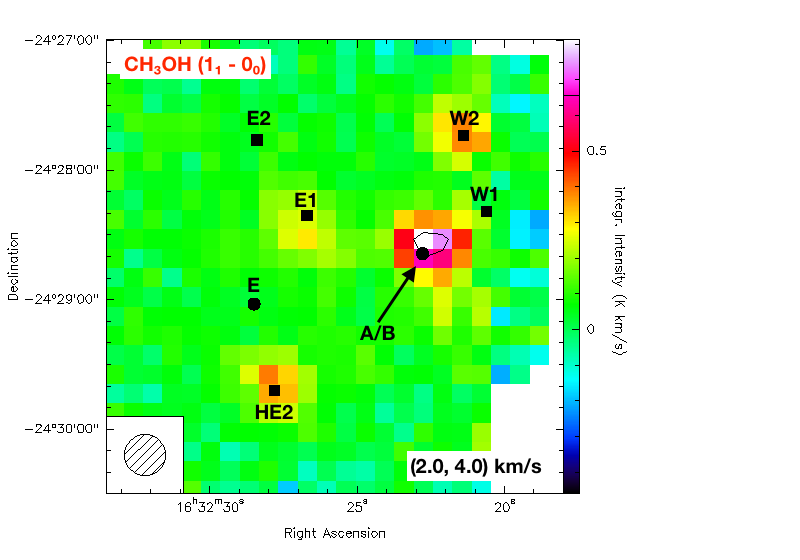}}
    \subfigure[]{\includegraphics[width=0.42\textwidth, trim={0 0.65cm 5.4cm 1.18cm},clip]{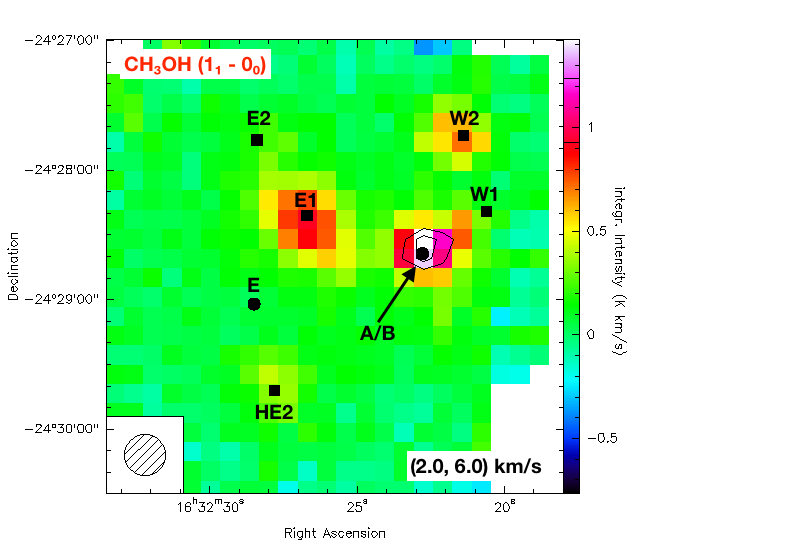}}
    \subfigure[]{\includegraphics[width=0.42\textwidth, trim={0 0.65cm 5.4cm 1.18cm},clip]{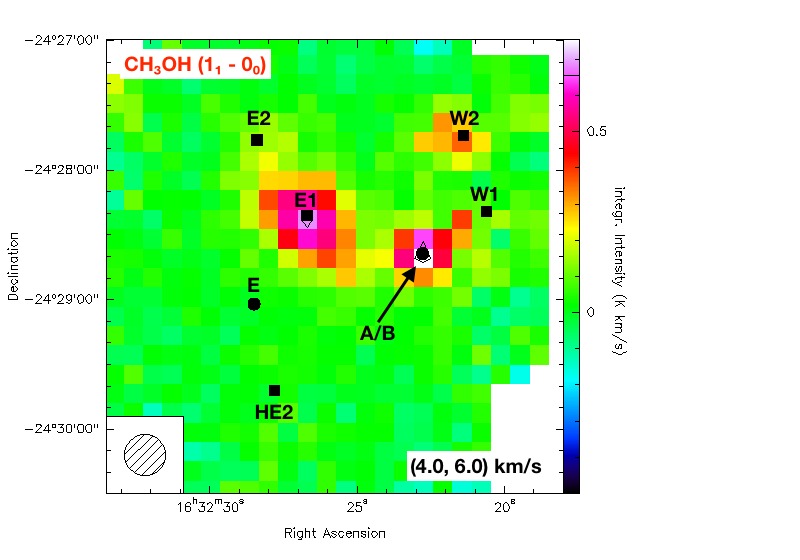}}
    \caption{CH$_3$OH-A$^{+}$ ($1_{1}- 0_{0}$) transition at 350905.100\,MHz.}
    \label{fig:5b}
\end{figure*}

\begin{figure*}[ht]
	\centering
    \subfigure[]{\includegraphics[width=0.42\textwidth, trim={0 0.65cm 5.4cm 1.18cm},clip]{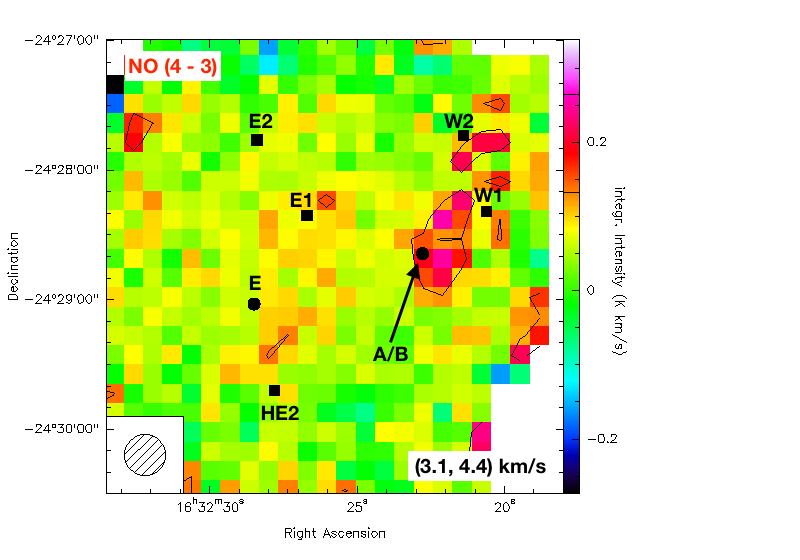}}
    \subfigure[]{\includegraphics[width=0.42\textwidth, trim={0 0.65cm 0 1.18cm},clip]{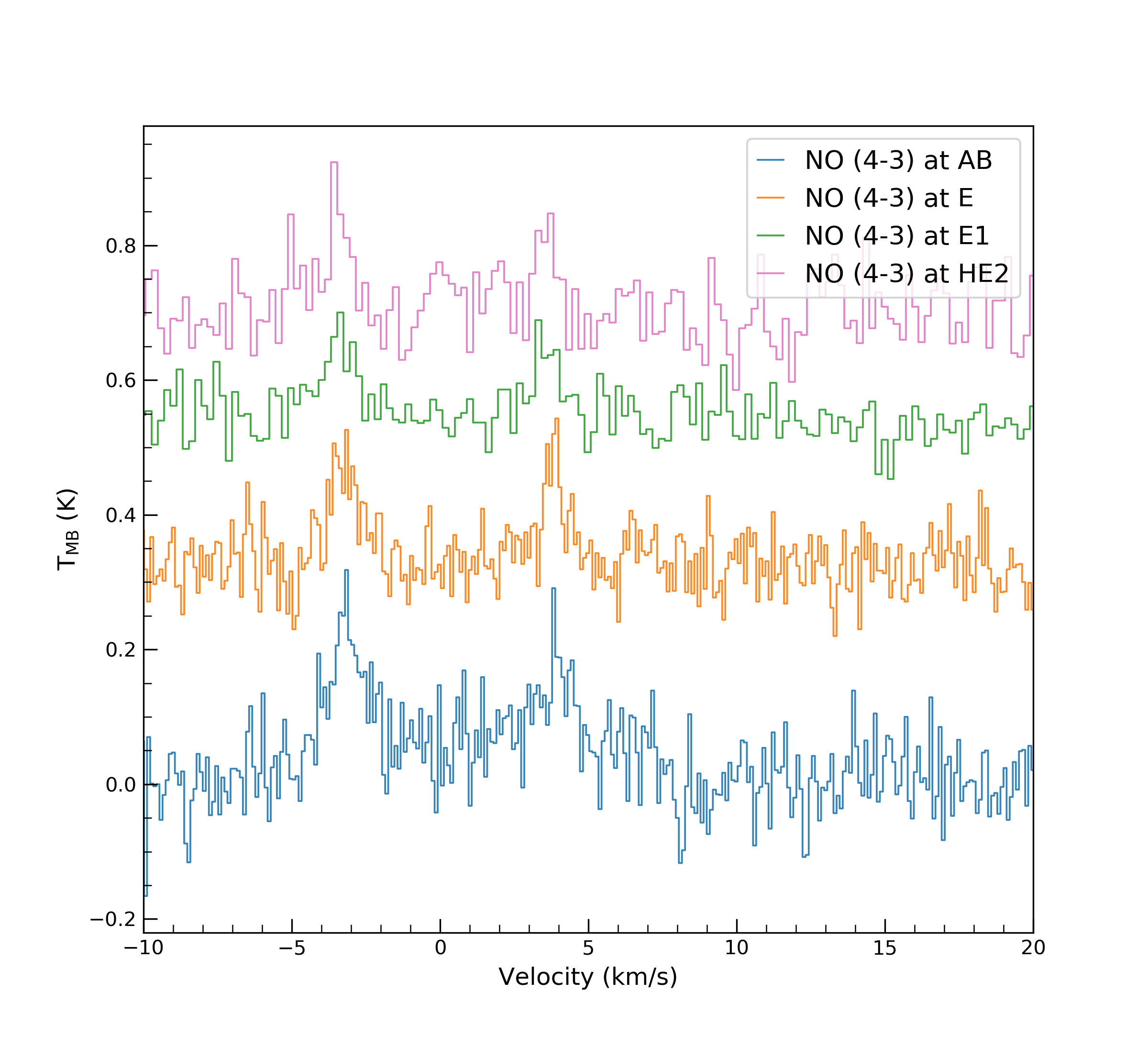}}
    \caption{(a) NO ($4_{+,7/2,9/2}-3_{-,5/2,7/2}$) transition at 351043.524\,MHz. Additional contours are drawn at 1$\sigma$ and 2$\sigma$. (b) Averaged spectrum of this transition in a $\SI{10}{\arcsecond}$ radius at the positions indicated in the upper right corner.}
    \label{fig:60}
\end{figure*}

\begin{figure*}[ht]
	\centering
    \subfigure[]{\includegraphics[width=0.42\textwidth, trim={0 0.65cm 5.4cm 1.18cm},clip]{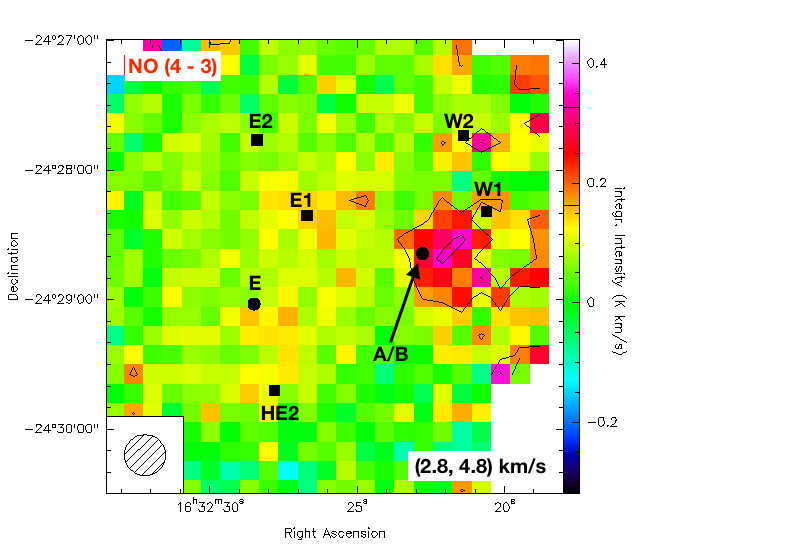}}
    \subfigure[]{\includegraphics[width=0.42\textwidth, trim={0 0.65cm 0 1.18cm},clip]{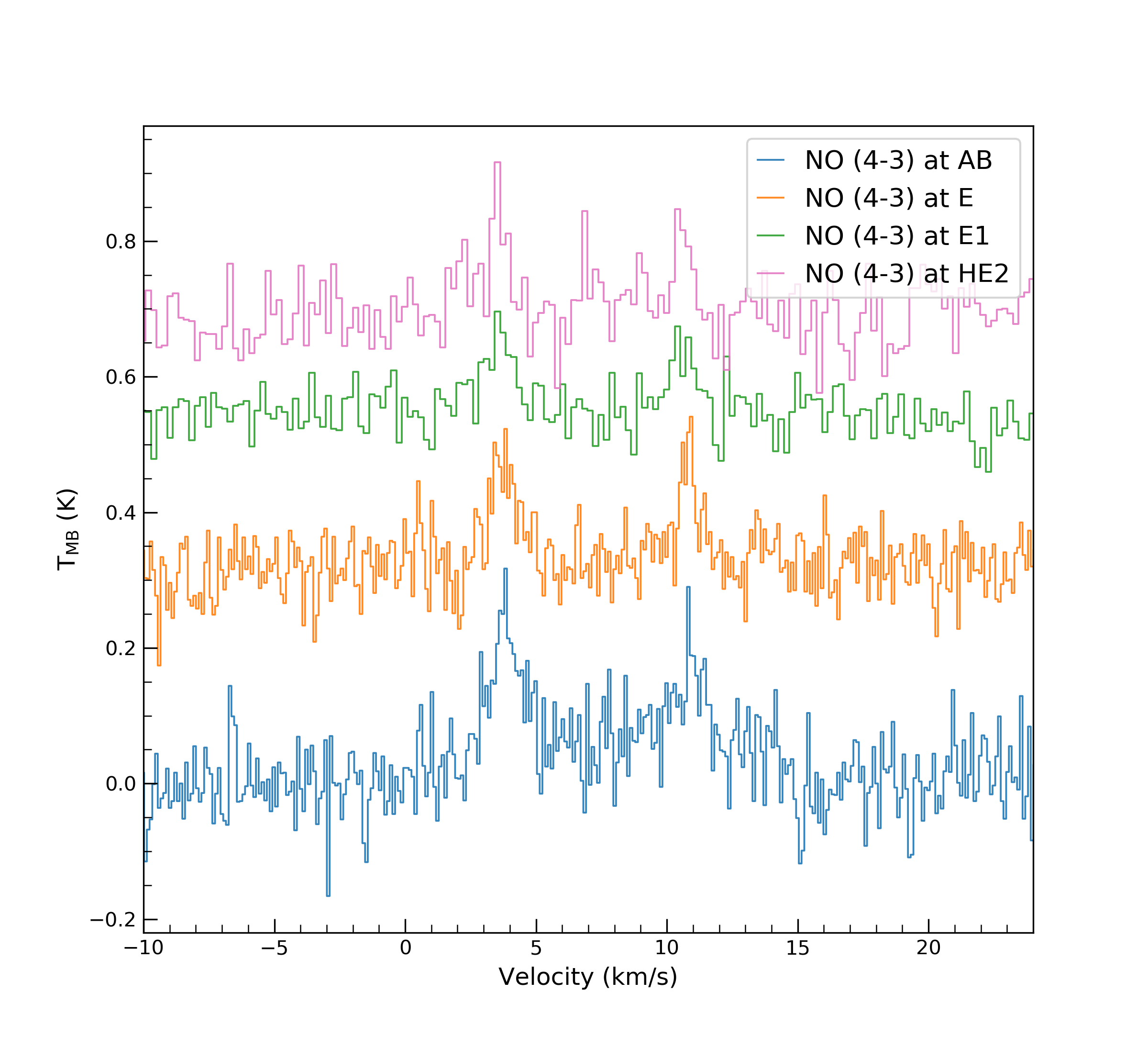}}
    \caption{(a) NO ($4_{+,7/2,7/2}-3_{-,5/2,5/2}$) transition at 351051.705\,MHz. Additional contours are drawn at 1$\sigma$ and 2$\sigma$. (b) Averaged spectrum of this transition in a $\SI{10}{\arcsecond}$ radius at the positions indicated in the upper right corner.}
    \label{fig:61}
\end{figure*}

\begin{figure*}[ht]
	\centering
    \includegraphics[width=0.42\textwidth, trim={0 0.65cm 5.4cm 1.18cm},clip]{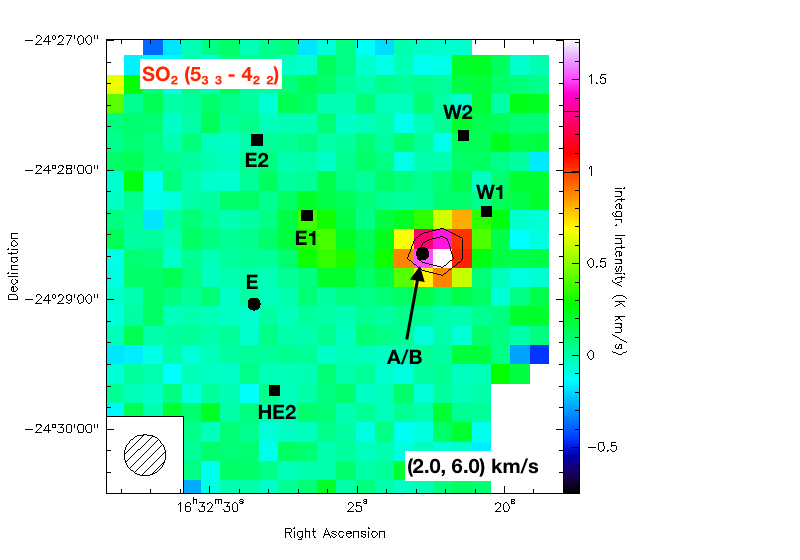}
    \caption{SO$_2$ ($5_{3,3}-4_{2,2}$) transition at 351257.223\,MHz.}
    \label{fig:49}
\end{figure*}

\begin{figure*}[ht]
	\centering
    \subfigure[]{\includegraphics[width=0.42\textwidth, trim={0 0.65cm 5.4cm 1.18cm},clip]{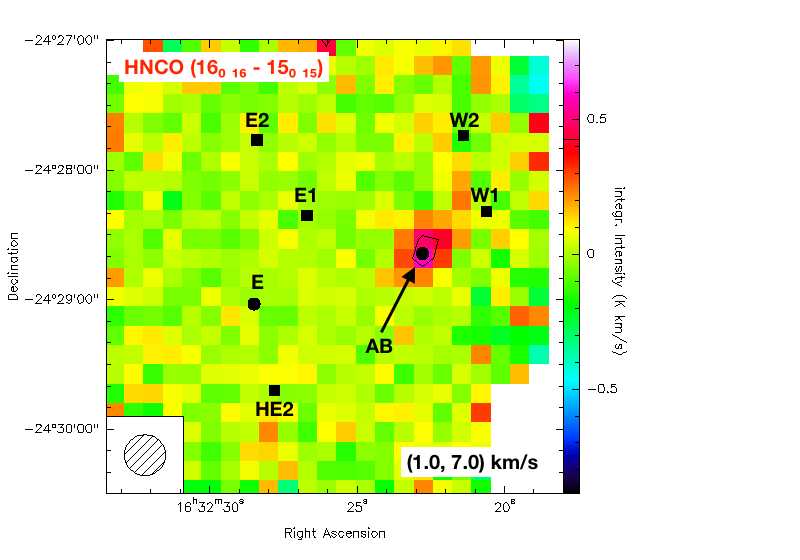}}
    \subfigure[]{\includegraphics[width=0.42\textwidth, trim={0 0.65cm 0 1.18cm},clip]{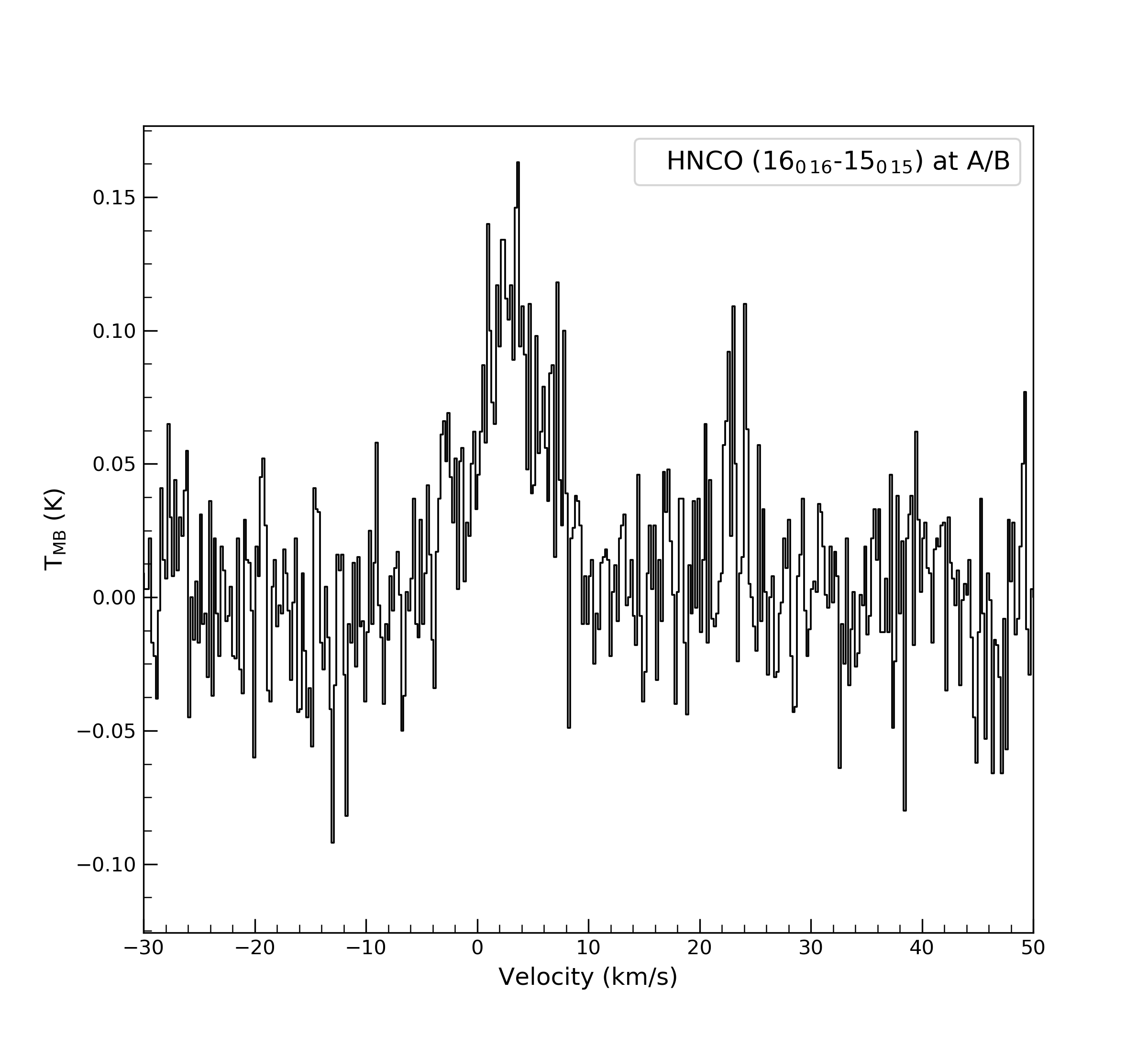}}
    \caption{(a) HNCO $(16_{0, 16} - 15_{0, 15})$ transition at 351633.257\,MHz. Additional contours are drawn at 1$\sigma$. (b) Averaged spectrum of this transition in a $\SI{10}{\arcsecond}$ radius at the position of IRAS\,16293 A/B.}
    \label{fig:71a}
\end{figure*}

\begin{figure*}[ht]
	\centering
    \subfigure[]{\includegraphics[width=0.42\textwidth, trim={0 0.65cm 0 1.18cm},clip]{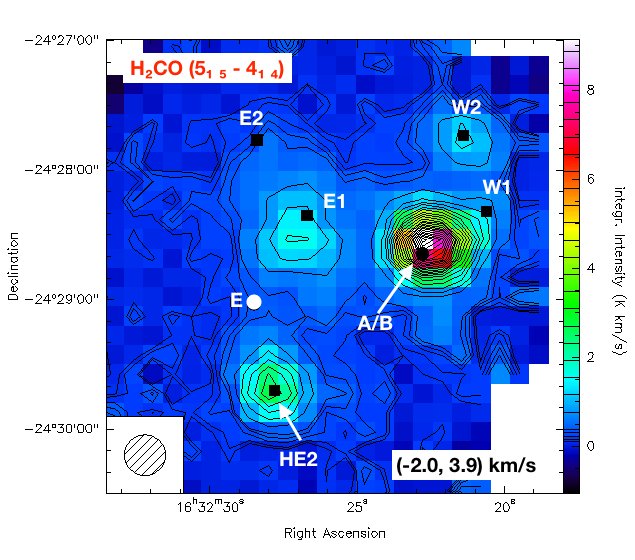}}
    \subfigure[]{\includegraphics[width=0.42\textwidth, trim={0 0.65cm 0 1.18cm},clip]{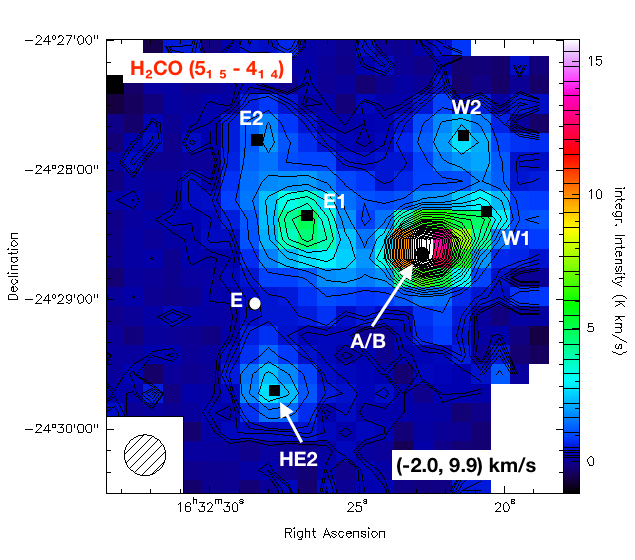}}
    \subfigure[]{\includegraphics[width=0.42\textwidth, trim={0 0.65cm 0 1.18cm},clip]{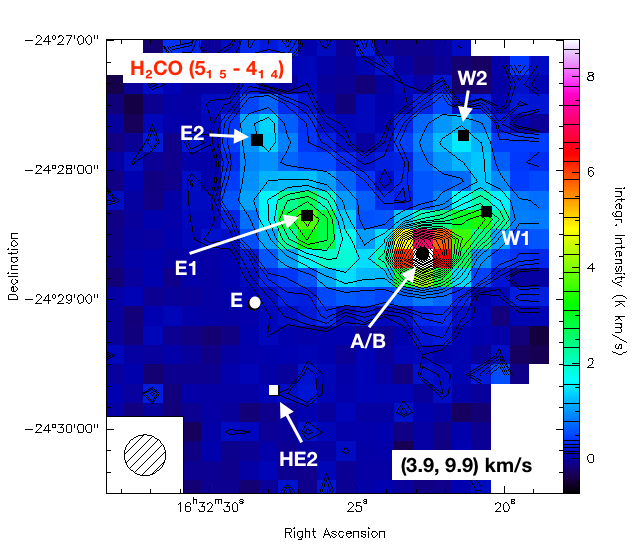}}
    \caption{H$_2$CO ($5_{1,5} - 4_{1,4} $) transition at 351768.645\,MHz}
    \label{fig:34}
\end{figure*}

\begin{figure*}[ht]
	\centering
    \subfigure[]{\includegraphics[width=0.42\textwidth, trim={0 0.65cm 0 1.18cm},clip]{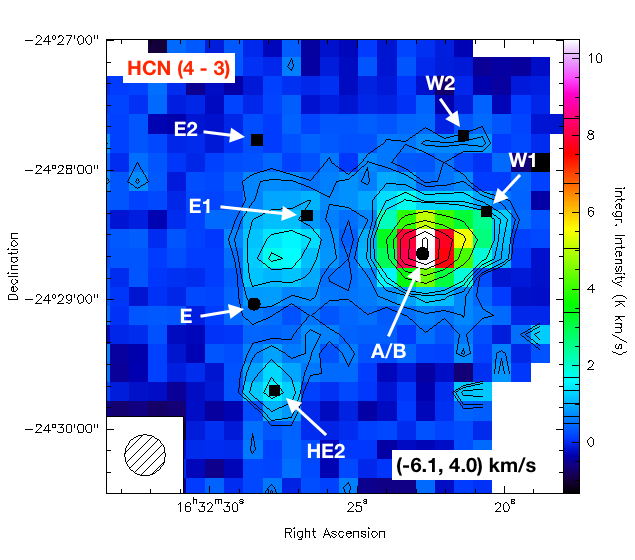}}
    \subfigure[]{\includegraphics[width=0.42\textwidth, trim={0 0.65cm 0 1.18cm},clip]{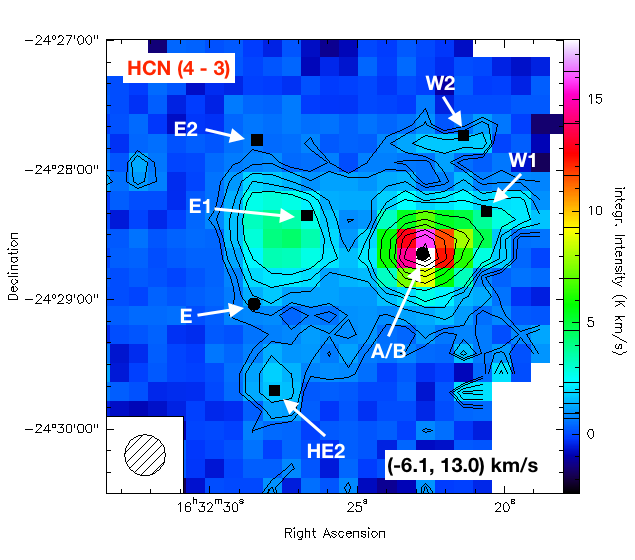}}
    \subfigure[]{\includegraphics[width=0.42\textwidth, trim={0 0.65cm 0 1.18cm},clip]{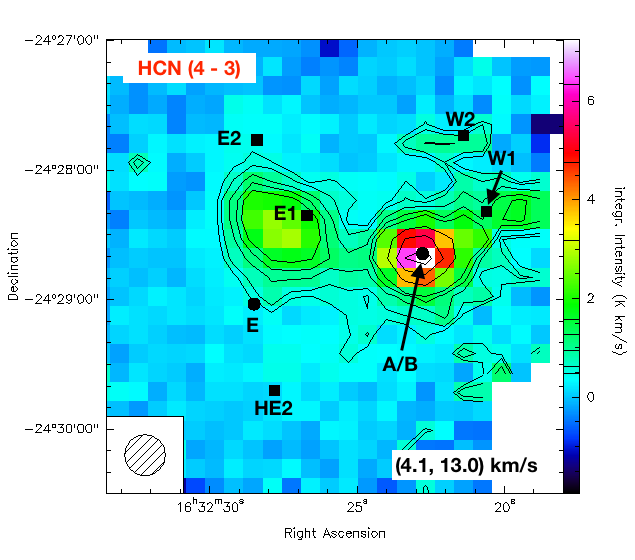}}
    \caption{HCN ($4 - 3$) transition at 354505.477\,MHz.}
    \label{fig:24}
\end{figure*}

\begin{figure*}[ht]
	\centering
    \subfigure[]{\includegraphics[width=0.42\textwidth, trim={0 0.65cm 5.4cm 1.18cm},clip]{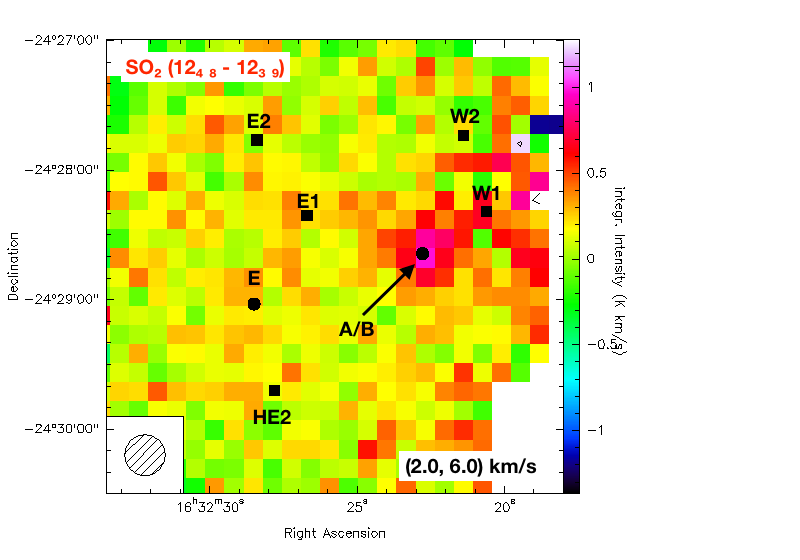}}
    \subfigure[]{\includegraphics[width=0.42\textwidth, trim={0 0.65cm 0 1.18cm},clip]{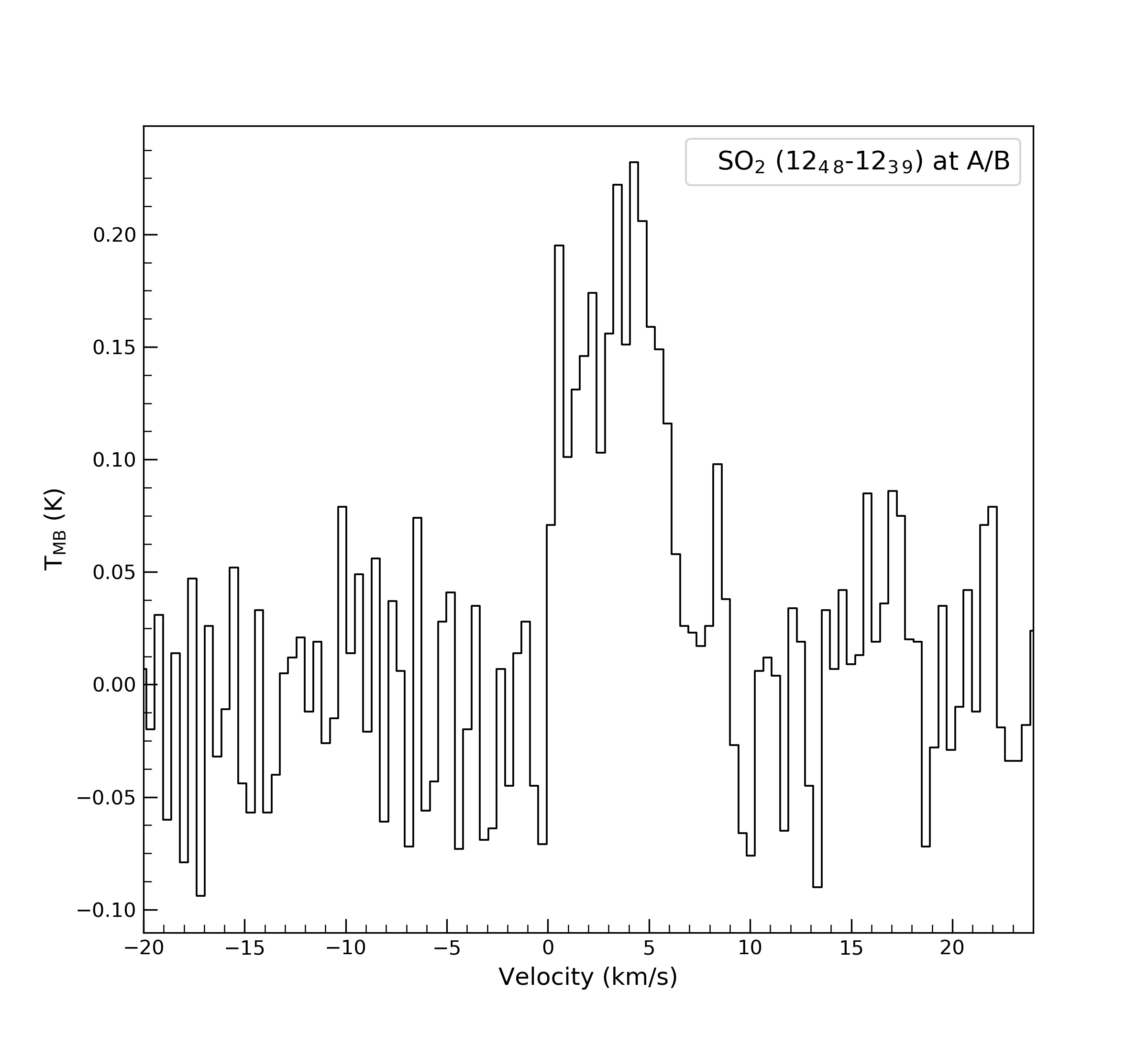}}
    \caption{(a) SO$_2$ $(12_{4, 8} - 12_{3, 9})$ transition at 355045.517\,MHz. (b) Averaged spectrum of this transition in a $\SI{10}{\arcsecond}$ radius at the position of IRAS\,16293 A/B.}
    \label{fig:68}
\end{figure*}

\begin{figure*}[ht]
	\centering
    \subfigure[]{\includegraphics[width=0.42\textwidth, trim={0 0.65cm 0 1.18cm},clip]{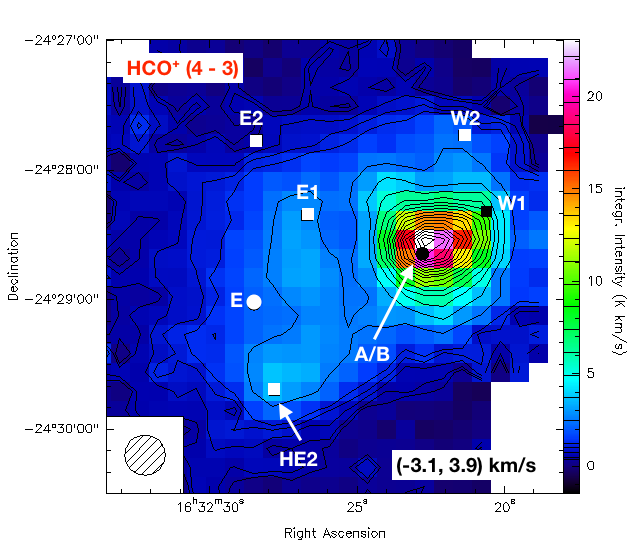}}
    \subfigure[]{\includegraphics[width=0.42\textwidth, trim={0 0.65cm 0 1.18cm},clip]{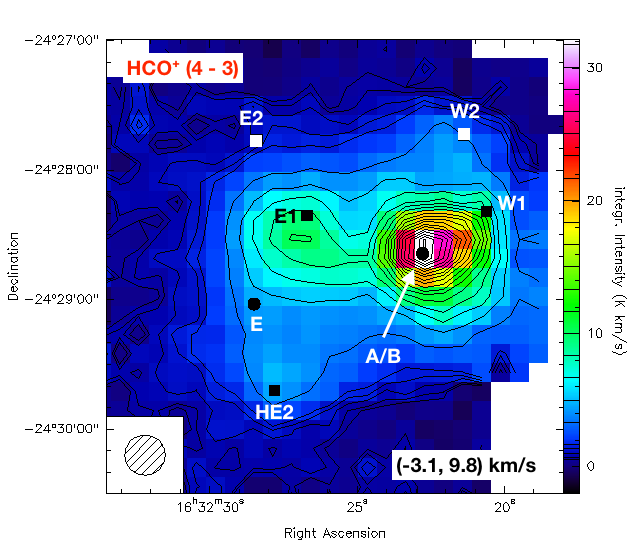}}
    \subfigure[]{\includegraphics[width=0.42\textwidth, trim={0 0.65cm 0 1.18cm},clip]{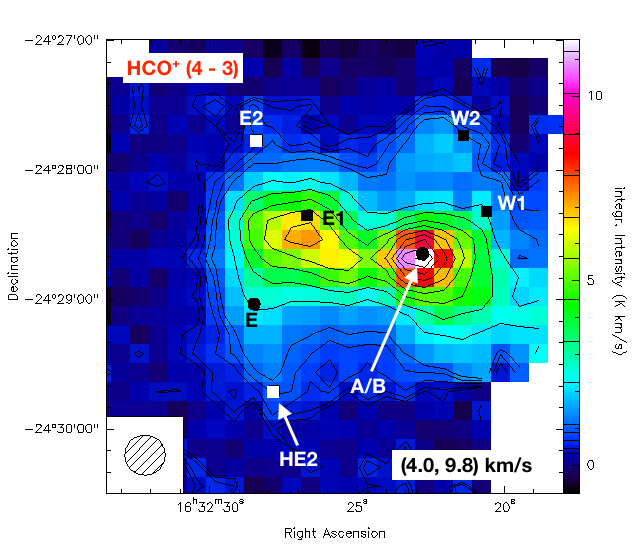}}
    \caption{HCO$^+$ ($4 - 3$) transition at 356734.223\,MHz.}
    \label{fig:27}
\end{figure*}

\begin{figure*}[ht]
	\centering
    \subfigure[]{\includegraphics[width=0.42\textwidth, trim={0 0.65cm 5.4cm 1.18cm},clip]{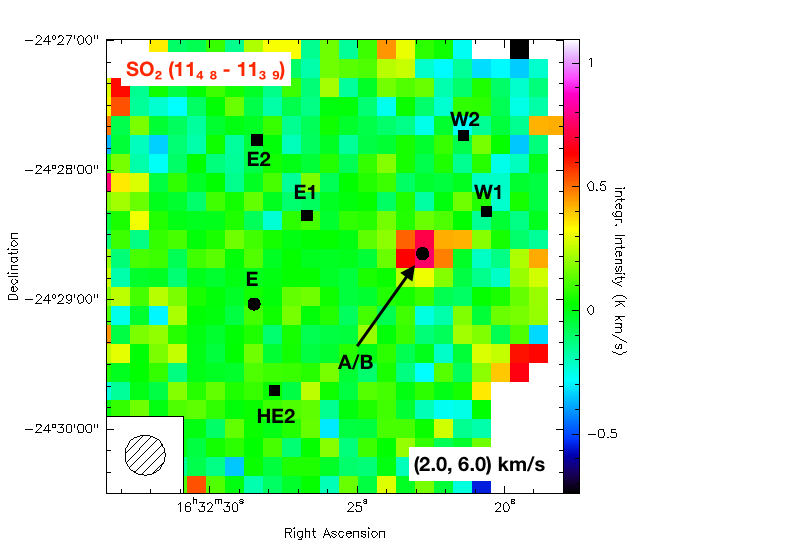}}
    \subfigure[]{\includegraphics[width=0.42\textwidth, trim={0 0.65cm 0 1.18cm},clip]{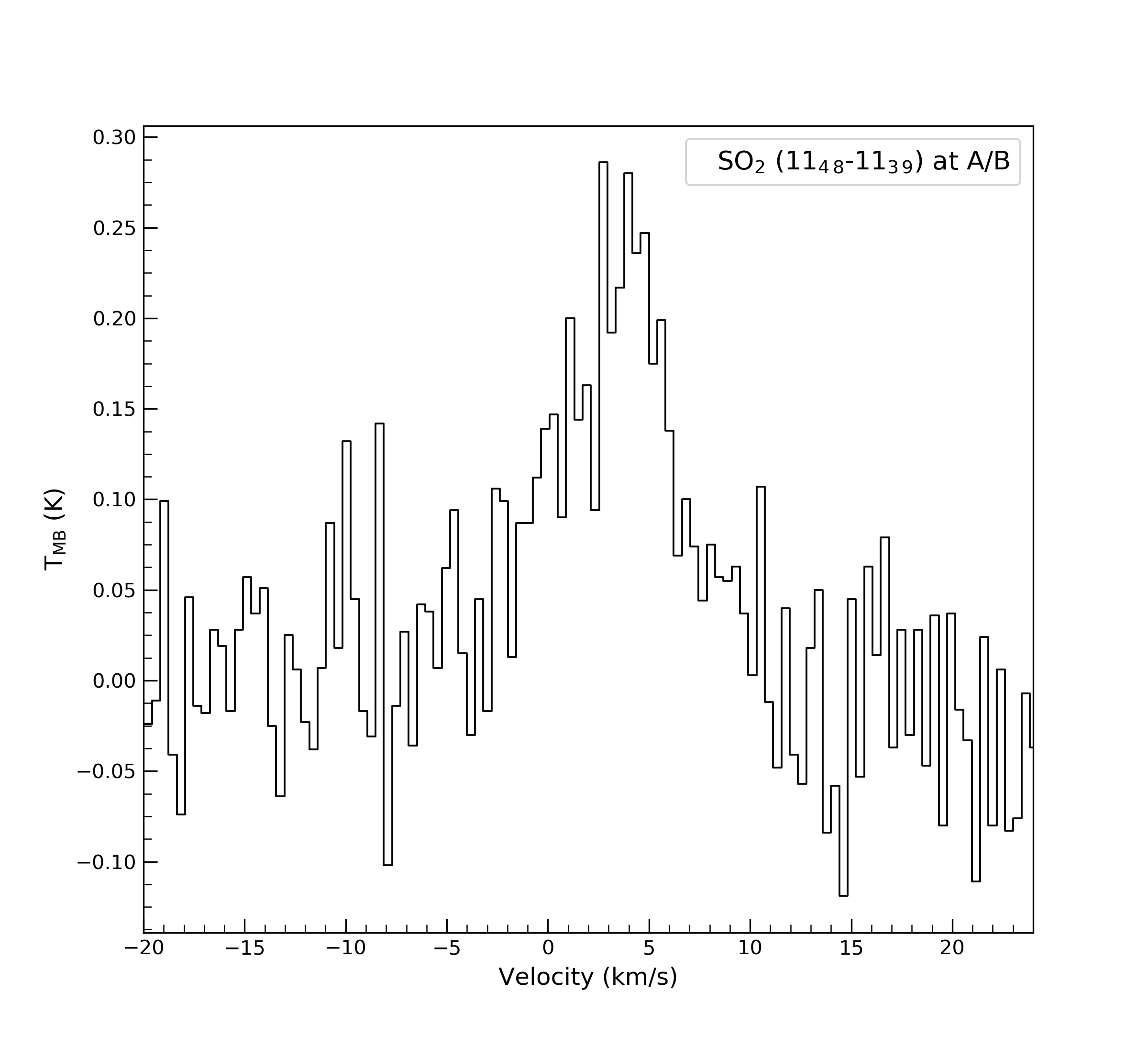}}
    \caption{(a) SO$_2$ $(11_{4, 8} - 11_{3, 9})$ transition at 357387.579\,MHz. (b) Averaged spectrum of this transition in a $\SI{10}{\arcsecond}$ radius at the position of IRAS\,16293 A/B.}
    \label{fig:71c}
\end{figure*}

\begin{figure*}[ht]
	\centering
    \subfigure[]{\includegraphics[width=0.42\textwidth, trim={0 0.65cm 5.4cm 1.18cm},clip]{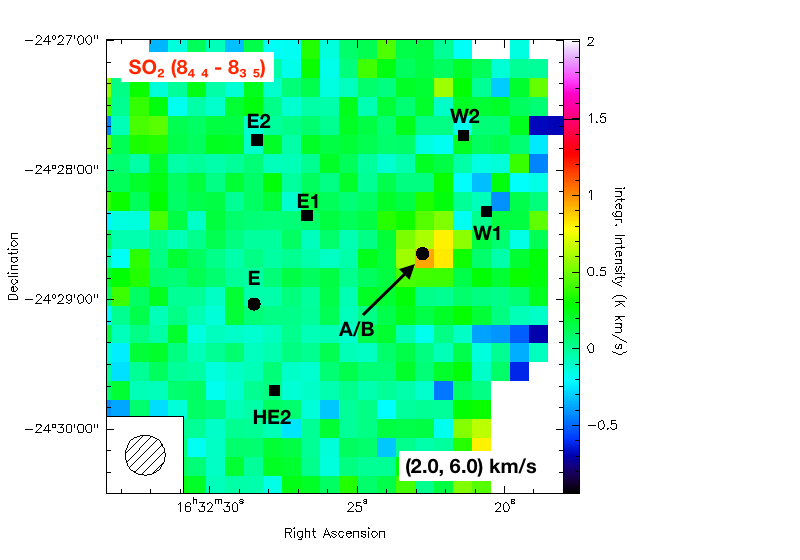}}
    \subfigure[]{\includegraphics[width=0.42\textwidth, trim={0 0.65cm 0 1.18cm},clip]{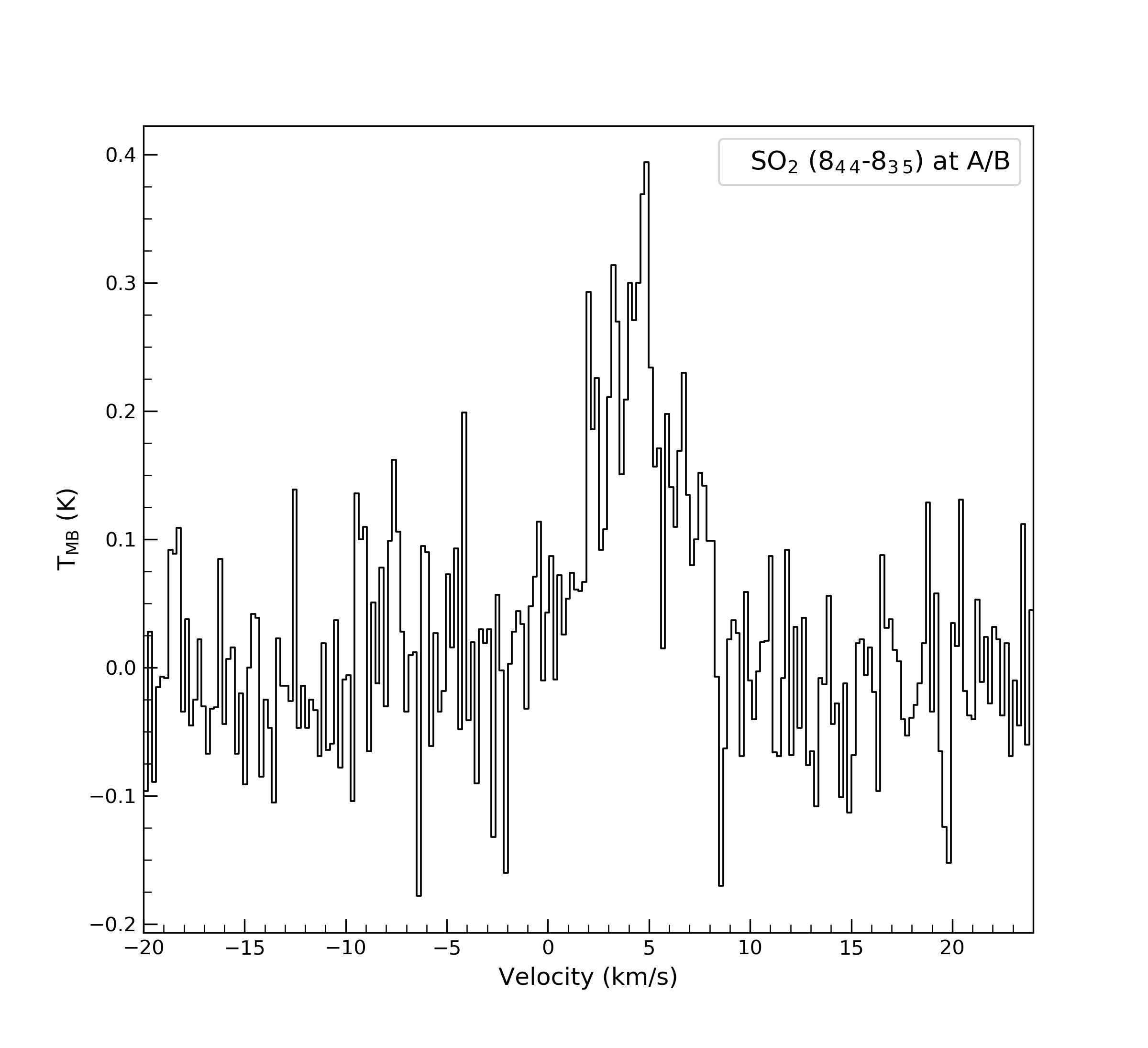}}
    \caption{(a) SO$_2$ $(8_{4, 4} - 8_{3, 5})$ transition at 357581.449\,MHz. (b) Averaged spectrum of this transition in a $\SI{10}{\arcsecond}$ radius at the position of IRAS\,16293 A/B.}
    \label{fig:71b}
\end{figure*}

\begin{figure*}[ht]
	\centering
    \subfigure[]{\includegraphics[width=0.42\textwidth, trim={0 0.65cm 0 1.18cm},clip]{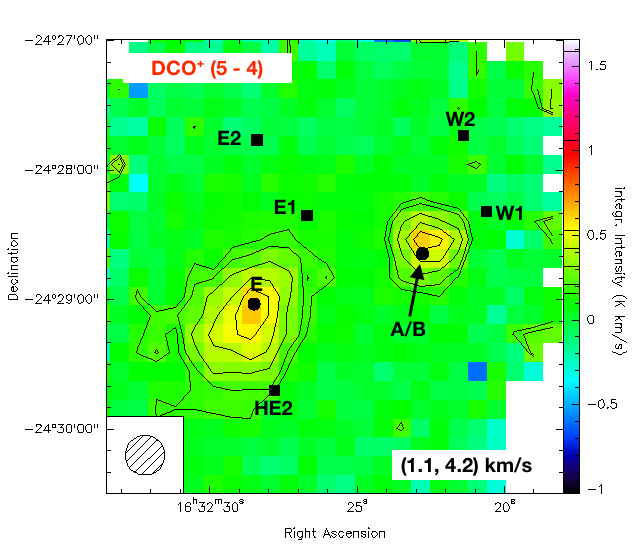}}
    \subfigure[]{\includegraphics[width=0.42\textwidth, trim={0 0.65cm 0 1.18cm},clip]{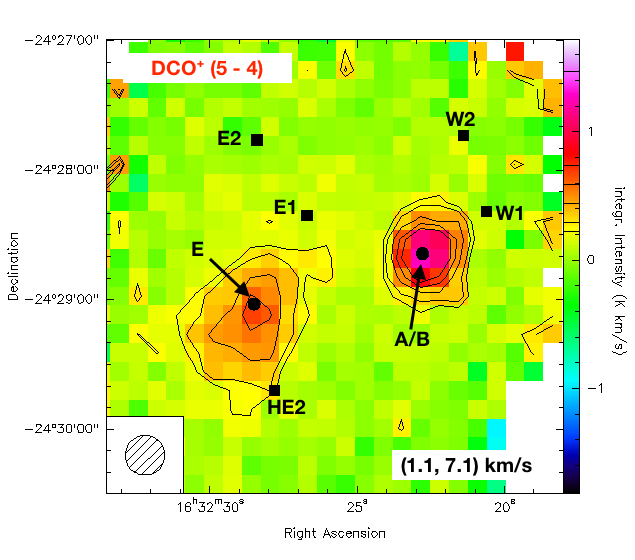}}
    \subfigure[]{\includegraphics[width=0.42\textwidth, trim={0 0.65cm 0 1.18cm},clip]{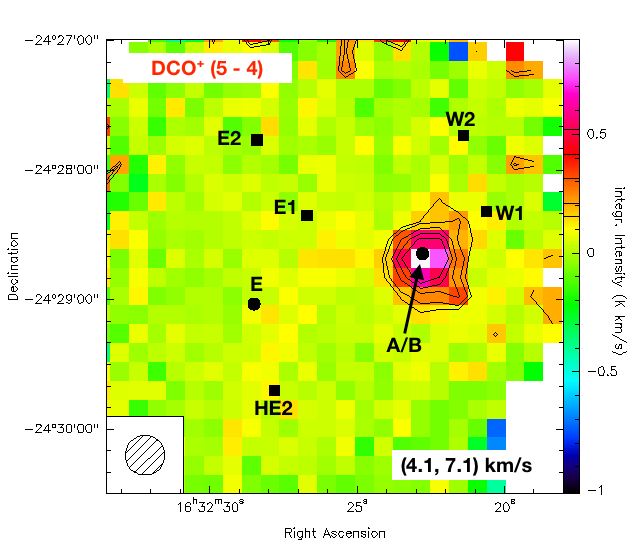}}
    \caption{DCO$^+$ ($5 - 4$) transition at 360169.778\,MHz.}
    \label{fig:21}
\end{figure*}

\begin{figure*}[ht]
	\centering
    \includegraphics[width=0.46\textwidth, trim={0 0.65cm 0 1.18cm},clip]{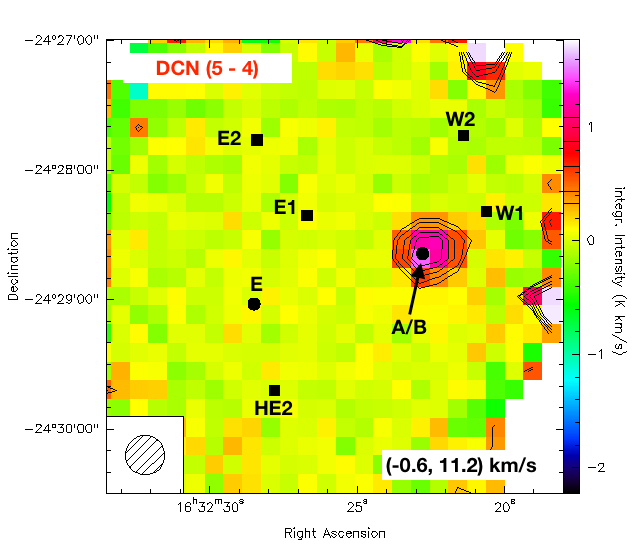}
    \caption{DCN ($5 - 4$) transition at 362045.753\,MHz.}
    \label{fig:20}
\end{figure*}

\begin{figure*}[ht]
	\centering
    \includegraphics[width=0.46\textwidth, trim={0 0.65cm 0 1.18cm},clip]{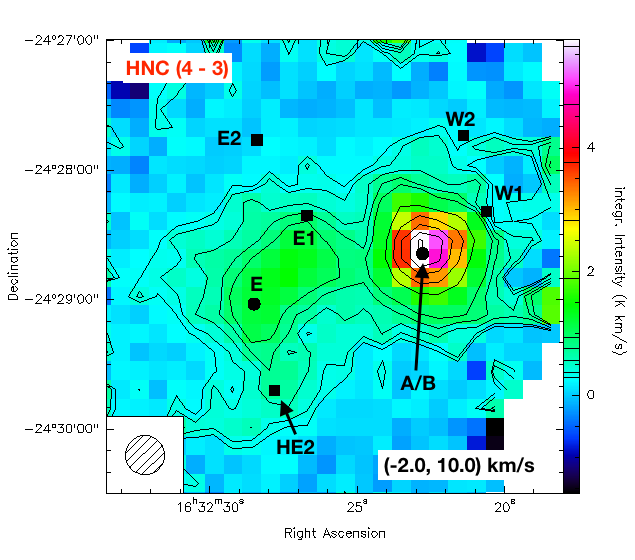}
    \caption{HNC ($4 - 3$) transition at 362630.303\,MHz.}
    \label{fig:25}
\end{figure*}

\begin{figure*}[ht]
	\centering
    \subfigure[]{\includegraphics[width=0.42\textwidth, trim={0 0.65cm 0 1.18cm},clip]{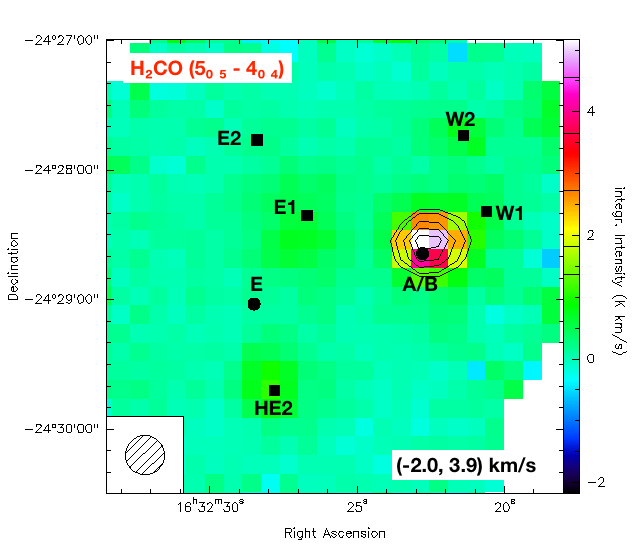}}
    \subfigure[]{\includegraphics[width=0.42\textwidth, trim={0 0.65cm 0 1.18cm},clip]{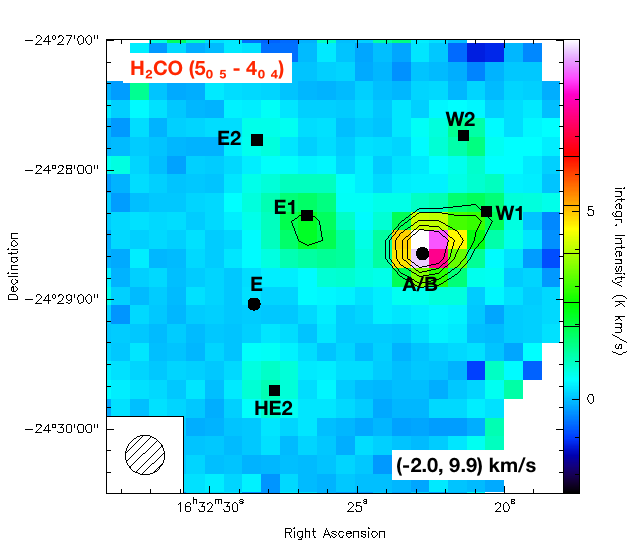}}
    \subfigure[]{\includegraphics[width=0.42\textwidth, trim={0 0.65cm 0 1.18cm},clip]{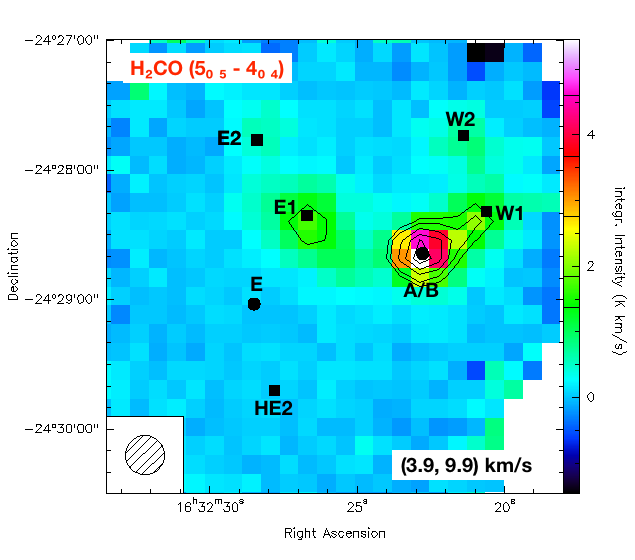}}
    \caption{H$_2$CO ($5_{0,5} - 4_{0,4} $) transition at 362736.048\,MHz.}
    \label{fig:33}
\end{figure*}

\begin{figure*}[ht]
	\centering
    \subfigure[]{\includegraphics[width=0.42\textwidth]{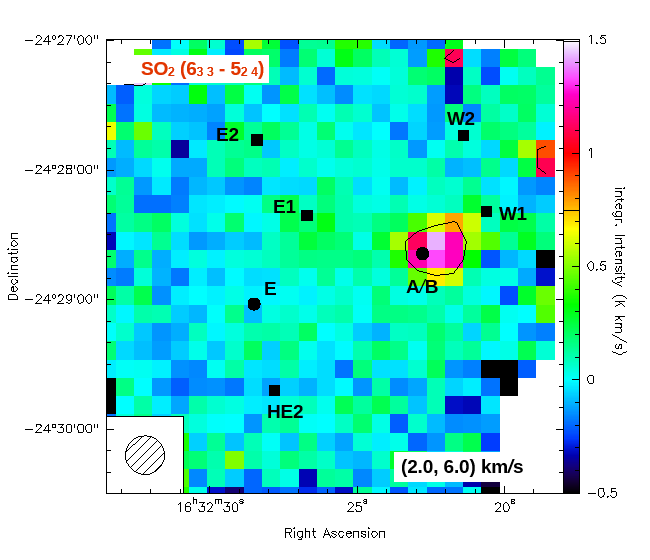}}
    \subfigure[]{\includegraphics[width=0.42\textwidth, trim={0 0.65cm 0 1.18cm},clip]{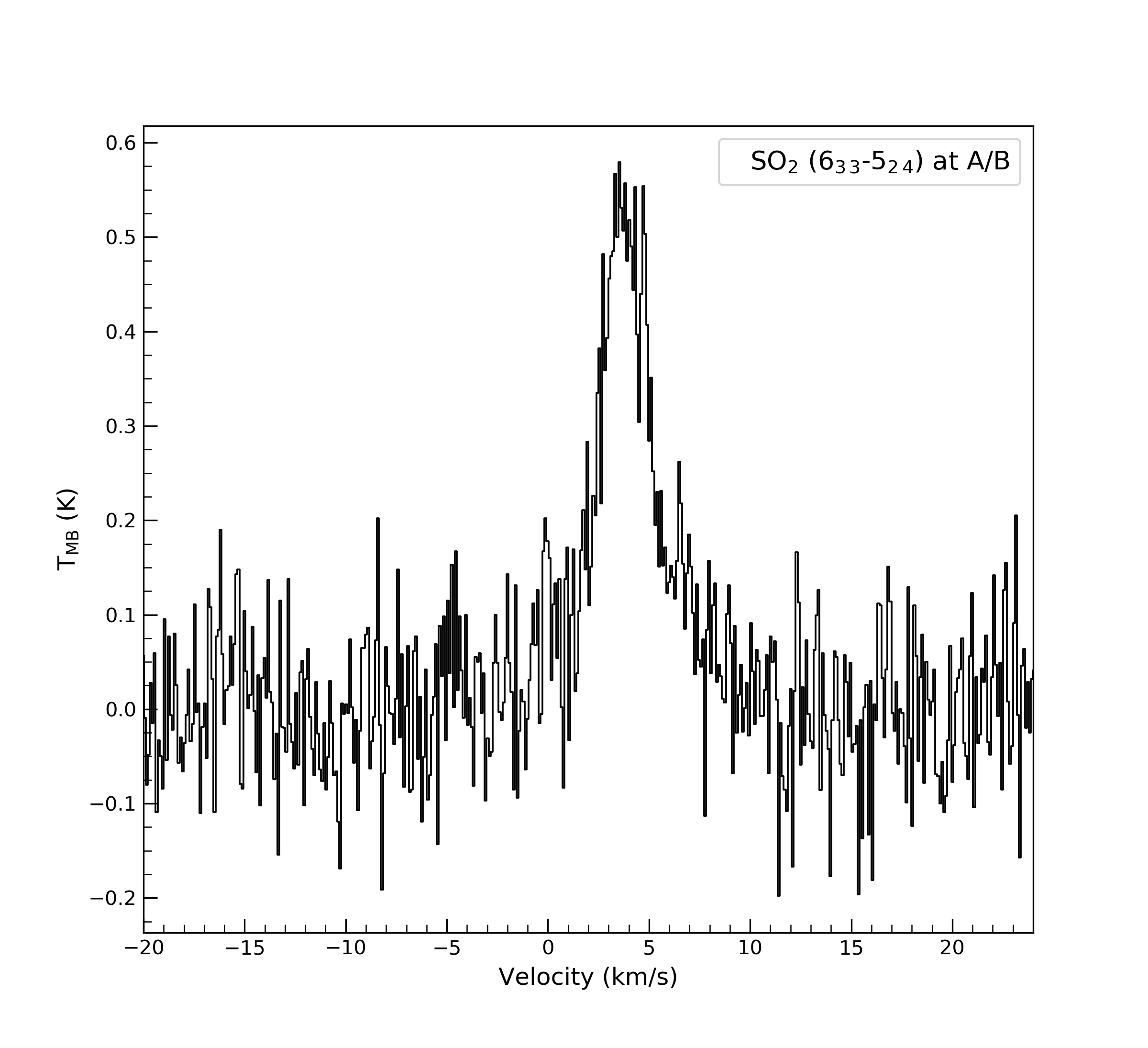}}
    \caption{(a) SO$_2$ ($6_{3,3} - 5_{2,4}$) transition at 371172.451\,MHz. Additional contours are drawn at 1$\sigma$. (b) Averaged spectrum of this transition in a $\SI{10}{\arcsecond}$ radius at the position of IRAS\,16293 A/B.}
    \label{fig:50}
\end{figure*}

\begin{figure*}[ht]
	\centering
    \subfigure[]{\includegraphics[width=0.42\textwidth, trim={0 0.1cm 0 0.1cm},clip]{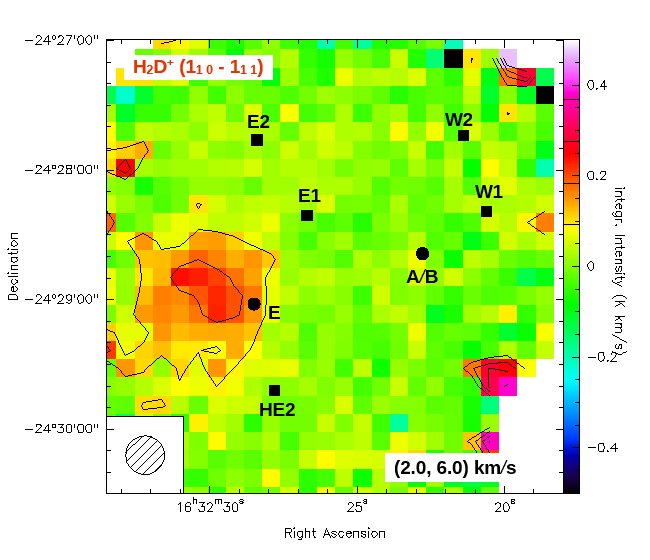}}
    \subfigure[]{\includegraphics[width=0.42\textwidth, trim={0 0.65cm 0 1.18cm},clip]{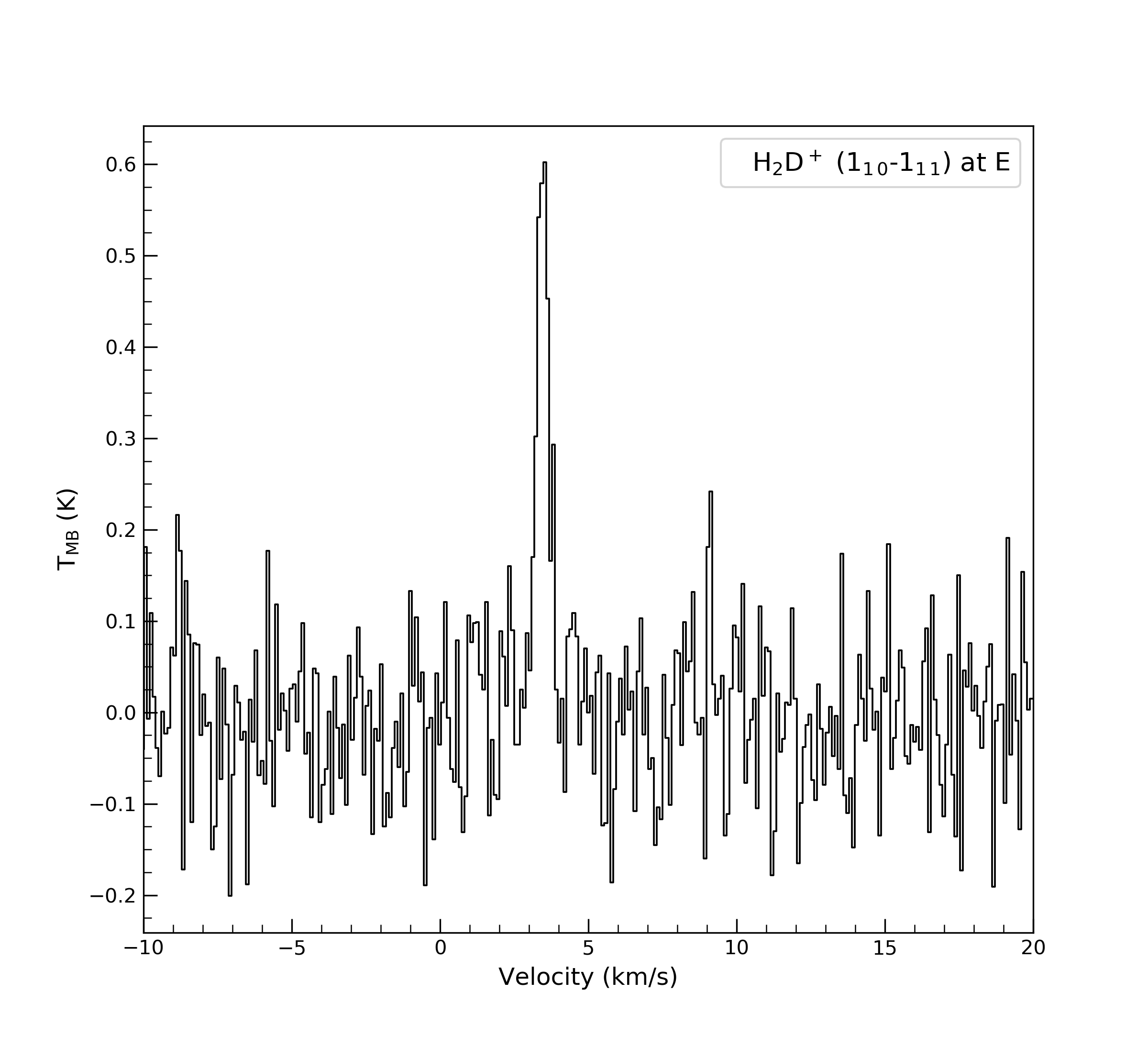}}
    \caption{(a) H$_2$D$^{+}$ ($1_{1,0} - 1_{1,1} $) transition at 372421.356\,MHz. Additional contours are drawn at 1$\sigma$ and 2$\sigma$. (b) Averaged spectrum of this transition in a $\SI{10}{\arcsecond}$ radius at the position of 16293E.}
    \label{fig:36}
\end{figure*}

\begin{figure*}[ht]
	\centering
    \subfigure[]{\includegraphics[width=0.42\textwidth, trim={0 0.65cm 0 1.18cm},clip]{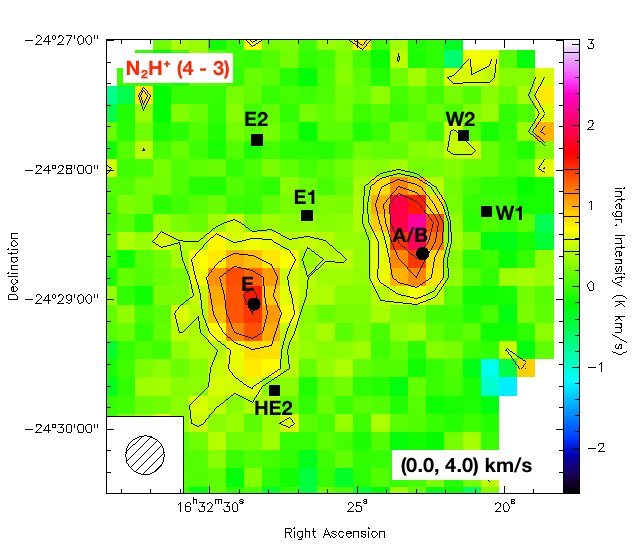}}
    \subfigure[]{\includegraphics[width=0.42\textwidth, trim={0 0.65cm 0 1.18cm},clip]{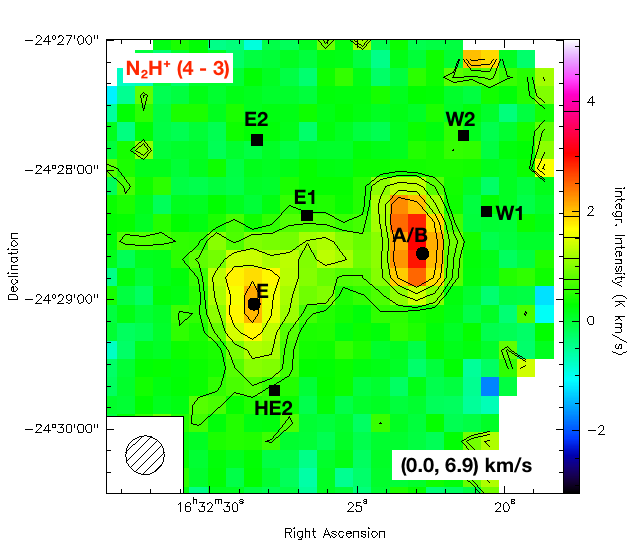}}
    \subfigure[]{\includegraphics[width=0.42\textwidth, trim={0 0.65cm 0 1.18cm},clip]{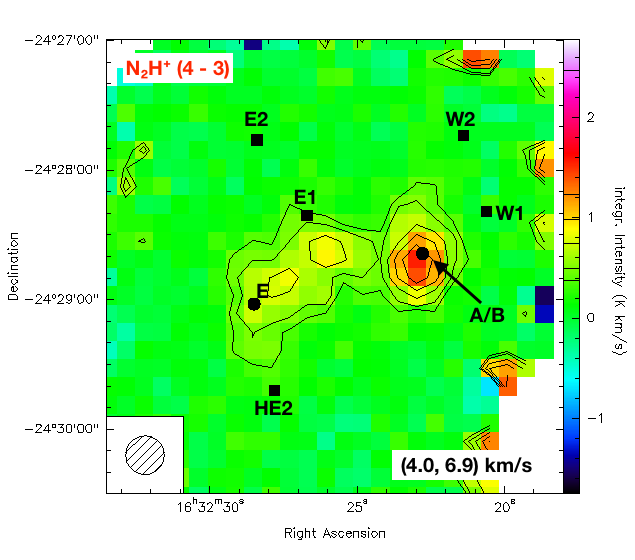}}
    \caption{N$_2$H$^+$($4 - 3 $) transition at 372672.481\,MHz.}
    \label{fig:39}
\end{figure*}

\end{appendix}
\end{document}